\documentclass{article}[10pt]

\usepackage{amsmath,amssymb,graphicx,tikz}

\newdimen\linethick  \linethick=0.4pt
\newdimen\hboxitspace    \hboxitspace=5pt
\newdimen\vboxitspace    \vboxitspace=5pt

\def\beq{\begin{equation}}
\def\eeq{\end{equation}}

\def\fr#1{%
\beq
\vcenter{
\hrule height\linethick
          \hbox{\vrule width\linethick
                \kern\hboxitspace
                \vbox{\kern\vboxitspace
                      \hbox{$\begin{array}{c}\displaystyle#1
         \end{array}$}%
                      \kern\vboxitspace}%
                \kern\hboxitspace
                \vrule width\linethick}%
          \hrule height\linethick}%
\eeq}

\def\rp{[p\,]}
\def\SS{{\!S}}
\newcommand{\qb}{\mathbf{q}}
\newcommand{\tb}{\mathbf{t}}

\newcommand{\ZG}{\mathbb{Z}}

\newcommand{\M}{\mathcal{M}}

\newcommand{\br}[1]{\left(#1\right)}

\newcommand{\summ}[2]{\displaystyle \mathop{\sum}_{#1}^{#2}}
\newcommand{\prodd}[2]{\displaystyle \mathop{\prod}_{#1}^{#2}}

\newcommand{\eq}[1]{\begin{equation} #1 \end{equation}}

\newcommand{\eqn}[1]{\begin{equation*} #1 \end{equation*}}

\newcommand{\ths}{\vartheta}

\newcommand{\ts}[1]{\ifcase#1 \ths_{\br{D_8\oplus D_8}^+}\or \ths_{\ZG\oplus
A_{15}^+}\or\ths_{\ZG^2\oplus \br{E_7\oplus E_7}^+}\or\ths_{\ZG^4\oplus
D_{12}^+}\or \ths_{\ZG^8\oplus E_{8}}\or \ths_{\ZG^{16}} \or \ths_{E_8 \oplus
E_8} \or \ths_{D_{16}^+} \fi}

\newcommand{\ab}{\mathbf{a}}

\def\be{\begin{eqnarray}}
\def\ee{\end{eqnarray}}
\def\nn{\nonumber}

\def\tr{{\rm tr}\,}
\def\Tr{{\rm Tr}\,}

\hoffset= - 1.3in         

\title{{\bf Superpolynomials for torus knots\\
from evolution induced by cut-and-join operators} \vspace{.2cm}}
\author{{\bf P.Dunin-Barkowski}\thanks{{\small
{\it ITEP, Moscow, Russia and MIPT, Dolgoprudny, Russia}};
barkovs@itep.ru}, {\bf A.Mironov}\footnote{ {\small {\it Lebedev
Physics Institute} and {\it ITEP, Moscow, Russia}}; mironov@itep.ru;
mironov@lpi.ru}, {\bf A.Morozov}\thanks{{\small {\it ITEP, Moscow,
Russia}}; morozov@itep.ru}, {\bf A.Sleptsov}\thanks{{\small
{\it ITEP, Moscow, Russia and MIPT, Dolgoprudny, Russia}};
sleptsov@itep.ru}, {\bf A.Smirnov}\thanks{{\small {\it ITEP
Moscow, Russia and MIPT, Dolgoprudny, Russia}}; asmirnov@itep.ru}
\date{ }}

\begin{document}
 \maketitle

\vspace{-5.0cm}

\begin{center}
\hfill FIAN/TD-10/11\\
\hfill ITEP/TH-21/11\\
\end{center}

\vspace{3.5cm}

\centerline{ABSTRACT}

\bigskip

{\footnotesize The colored HOMFLY polynomials, which describe Wilson
loop averages in Chern-Simons theory, possess an especially simple
representation for torus knots, which begins from quantum $R$-matrix
and ends up with a trivially-looking split $W$ representation
familiar from character calculus applications to matrix models and
Hurwitz theory. Substitution of MacDonald polynomials for characters
in these formulas provides a very simple description of
"superpolynomials", much simpler than the recently studied
alternative which deforms relation to the WZNW theory and explicitly
involves the Littlewood-Richardson coefficients. A lot of explicit
expressions are presented for different representations (Young
diagrams), many of them new. In particular, we provide the
superpolynomial ${\cal P}_{[1]}^{[m,km\pm 1]}$ for arbitrary $m$ and
$k$. The procedure is not restricted to the fundamental (all
antisymmetric) representations and the torus knots, still in these
cases some subtleties persist. }

\bigskip

\section{Introduction}

Knot invariants stay among the central subjects of modern
mathematical physics ever since the seminal paper \cite{knots} (see
also latest developments in \cite{knotstoday}). Through variety of
dualities they are related to the main topics of interest in string
theory, in particular, to the stringy avatar of the
Littlewood-Richardson coefficients: the topological vertex $C_{PQR}$
\cite{topver}, defined as a weighted sum over $3d$ partitions with
fixed triple of Young diagrams $P,Q,R$ at the boundaries, which is a
clever group theory toy model of generic string vertices. Of special
interest are concrete expressions and various properties of the (at
least) two-parameter family $C_{PQR}(q|t)$, generalizing the
"refined" McMahon formula
$C_{\centerdot\,\centerdot\,\centerdot}(q|t) = \prod_{i,j}
(1-q^it^j)^{-1}$. Such a double-deformation is related within the
group theory context with the MacDonald polynomials \cite{McD}, i.e.
with the theory of Ruijsenaars integrable system \cite{Ruij}. Its
knot counterparts are the Khovanov-Rozhansky homologies \cite{KhR}
and the "superpolynomials" \cite{GSV,DGR,IGV}

Closer to the Earth,
the knot invariants \cite{ACJK,HOMFLY,knotinv} are non-trivial generalizations
of characters
and, therefore, they begin to attract an increasing attention in all the
(deeply interrelated) fields
of theoretical physics which deal with character calculus:
topological theories,
matrix models,
conformal theory,
integrability theory,
Hurwitz theory,
AGT relations \cite{AGT,TY,3dAGT}.
Knot invariants are usually defined as certain (Wilson loop) averages
in the $3d$ Chern-Simons theory \cite{knots}
in different gauges and
for different descriptions of the knot itself,
see \cite{3dAGT} for a recent review.
The averages, arising in this way, depend on the knot and representation
(Young diagram), and additionally on two parameters:
the Chern-Simons coupling constant and the rank of the group. They are known
as
HOMFLY polynomials \cite{HOMFLY}
(to avoid unnecessary complications we consider only $SU(N)$ groups in this
paper).

These averages can be further deformed by switching from
ordinary to quantum and elliptic groups,
whose characters are MacDonald \cite{McD} and Askey-Wilson (and further Kerov)
polynomials respectively.
The knot invariants, associated with the MacDonald deformation
are known as "superpolynomials" \cite{DGR}.
They depend on three parameters and have an important property that all
the expansion coefficients are positive integers,
thus revealing the hidden structure behind the knot geometry:
that of the Khovanov-Rozhansky homologies.
The additional (third) parameter is that of the quantum group,
and it is also related to the $\beta$-deformation,
which plays an increasingly important role in modern
application of matrix models (for example, it is related to
the central charge in the AGT relations \cite{AGTmamo}).
In principle, any approach to constructing the HOMFLY polynomials
can be straightforwardly deformed and provide the superpolynomials,
however, this is not yet done.
The only notable exception is the very recent paper \cite{ASh},
where a deformation is performed of, in fact, the most complicated
approach to description of the HOMFLY polynomials: that coming
from relation to the WZNW model and relying upon representation of
the fusion rules in terms of the Littlewood-Richardson coefficients.
Still, despite all the technical problems, \cite{ASh} reproduces the
known examples of superpolynomials and provides a working construction,
at least, for all torus knots.

In this paper we do the same in a much simpler way:
by deforming the $R$-matrix representation of the knot and link
invariants, realized as a closed braid.
In the case of torus knots the formulas of this kind
for the HOMFLY invariants are exceptionally simple \cite{chi},
and, as we explain in this paper, they are deformed in an
equally simple way and immediately provide expressions for
the superpolynomials in any desired representations.
Extension to many non-torus knots is also straightforward,
but somewhat more tedious.
The biggest disadvantage of such a construction is that
the $R$-matrix representations are still not derived from
the Chern-Simons theory in the temporal gauge $A_0=0$
(see \cite{MSm} for a recent description of the problems).
From this point of view, more promising can be considerations
in the holomorphic gauge $A_{\bar z} = 0$, which lead
to the Kontsevich integrals and the Vassiliev invariants,
and also can be easily deformed to the $\beta$-ensembles.
This approach will be considered elsewhere.
Another drawback of our construction is that it does not
make the positivity property of the superpolynomials explicit:
one can easily check it in any particular {\it answer},
but the general {\it a priori} reason remains obscure.

\bigskip

The net of different knot invariants is schematically
presented in the following table:

\bigskip

$$
\begin{array}{cccrcclccc}
&&&&&&&&&\\
&&&&&&P_R(A|q|t) = \tilde P({\bf} a|{\bf q}|{\bf t})&&&\\
&&&&&&&&&\\
&&&!!!&\nearrow&\swarrow&{\bf t}=-1\ {\rm or} \ t=q\ \ \ \ \ \ \
&\!\!\!\!\!\!\!\!\! a=1&\searrow&\\
&&&&&&&&&\\
R-matrix&\longrightarrow & &&H_R(a|q) &&&&&F_R({\bf q}|{\bf t})\\
&&&&&&&&&\\
&&&&&\searrow&a=1 \ \ \ \ \ \ \ \ \ \ \ \ \ \ \ {\bf
t}=-1&\swarrow&\nearrow&\cite{EGth}\\
&&&&&&&&&\\
&&&&&&\ \ \ \ \ \ \ \ \ \ A_R(q)&&&\\
&&&&&&&&&\\
&&&&&&&&&\\
&&&&&&&&&\\
&&&&&&&&&\\
\end{array}
$$

\bigskip

\noindent
The superpolynomials $P$ at the top are the most general,
all other, HOMFLY $H$, Heegard-Floer $F$ and Alexander $A$
polynomials are obtained by fixing some of the arguments
at special values: this is shown by descending arrows.
The HOMFLY polynomials $H$ are directly described by Chern-Simons
theory, at least, in principle, and technically
in terms of the quantum $R$-matrix theory.
To obtain a superpolynomial one needs to construct
an ascending arrow, denoted by (!!!): this is the subject
of the present paper for the case of torus knots $T[m,n]$.
Given the $R$-matrix representation of the HOMFLY polynomials,
this arrow involves three deformations:
substitution of the quantum (Schur) dimensions by the MacDonald ones,
deformation of expansion coefficients $c_R^Q$
and deformation of the $W_{[2]}$ operator
(actually, of its eigenvalues on character eigenfunctions).
Another ascending arrow, from the Alexander $A$
to the Heegard-Floer $F$ polynomials is described in \cite{EGth}.
We use it for additional checks.

\bigskip

In our explicit construction of the ascending arrow $(!!!)$
we exploit the fact that the torus knots and links
form the entire series $T[m,mk+p]$ ($p=0,\ldots, m-1$)
where $k$ is an arbitrary positive integer,
and the lowest element of the series with $k=0$ is
\be
T[m,p] = T[p,m] \ \ \ {\rm with}\ \ p<m
\label{boco}
\ee
This allows one to build up the knot invariants recursively:
making an anzatz for the entire series (i.e. for all
$k$ at once)
one can fix the few remaining parameters by imposing the
"initial condition" (\ref{boco}) at $k=0$.
In fact, for the torus knots, the anzatz
is a simple thing to write down:
for example, it is immediately implied by knot invariants
representation through braids
and the universal quantum $R$-matrix.

Torus link $T[m,n]$
(it becomes a knot whenever $m$ and $n$ are mutually prime)
is represented by an $m$-strand braid ${\cal B}_m$
and the knot invariant is schematically given by
\be
{\cal K} = \overline{\Tr}\Big({\cal R}_m\Big)^n
\label{KthrR}
\ee
Here ${\cal R}_m$ is a combination of quantum $R$-matrices
in representation $R_1\otimes \ldots \otimes R_m$
(for a knot all representations are the same $R_i = R$)
and $\overline{\Tr}$ involves the group-theory factor
$\otimes_i t^{\rho_{R_i}}$, so that
$\overline{\Tr}\ I^{\otimes m} = \chi_{R_1}^*\ldots \chi_{R_m}^*$.
Representation $R_1\otimes \ldots \otimes R_m$ can be decomposed
into irreducible representations
\be
R_1\otimes \ldots \otimes R_m \ =\ \oplus_{Q\vdash
|R_1|+\ldots +|R_m|} Q
\ee
and each $Q$ is an eigenspace of ${\cal R}_m$,
with the corresponding eigenvalue $\lambda_Q$.
Then the knot invariant (\ref{KthrR}) is actually equal to ($Q\vdash n$ denotes
all the
Young diagrams of size $n$)
)
\be
P = \sum_{Q\vdash |R_1|+\ldots +|R_m|}c_{R_1\ldots R_m}^Q\lambda_Q^n \chi^*_Q
\label{Ksum}
\ee
thus consisting of three ingredients: 1) the characters calculated in a special
point,
$\chi^*_Q$; 2) the coefficients $c_{R_1\ldots R_m}^Q$ which are specified
by the initial condition (\ref{boco}); 3) and the dependence on $n$ which is
actually
given as evolution from the initial condition at $n=0$ by action of a
cut-and-join
operator. This latter ingredient is nothing but the $W$-representation
of \cite{MMN}, it is ultimately related with matrix model representations
\cite{Wreps} and
Hurwitz theory \cite{MMN}.

In particularly, one obtains in this way a simple general expression for the
colored HOMFLY polynomials for arbitrary torus links \cite{chi}:
what one needs to know for this is that the eigenvalues $\lambda_Q$
are actually expressed through the eigenvalues
\be
\varkappa_Q = \nu_{Q'} - \nu_{Q}, \ \ \ \ \
\hat W_{[2]} s_Q = \varkappa_Q s_Q
\label{evs}
\ee
of the simplest cut-and-join operator $\hat W_{[2]}$ \cite{MMN}
on the Schur eigenfunction $s_Q\{p\}$ corresponding to the Young diagram $Q$
($Q_i$ are the lengths of lines in the diagram, $Q = \{Q_1\geq Q_2
\geq\ldots\geq 0\}$):
\be
\lambda_Q = q^{-2\varkappa_Q/m},
\ \ \ \ \
\nu_Q = \sum_i (i-1)Q_i,
\ \ \ \
\varkappa_Q = {1\over 2}\sum_i Q_i(Q_i-2i+1)=\sum_{(i,j)\in Q}(i-j)
\ee
($Q'$ denotes the transposed Young diagram).

In order to lift HOMFLY to the superpolynomials one needs to deform
from the Schur to MacDonald characters
$\chi_Q = M_Q(q^2|t^2)$ separating the two constituents $\nu_Q$ and $\nu_{Q'}$
of $\varkappa_Q$:
\be
\lambda_Q = t^{2\nu_Q/m} q^{-2\nu_{Q'}/m}
\label{evms}
\ee
As we already noted, (\ref{Ksum}) provides a $W$-representation of the knot
invariant
for the torus link, known to be extremely useful in matrix models,
and it does so for arbitrary values of the time-variables $\{p_k\}$.
To obtain the conventional knot invariants one should restrict
the answer to the special one-parametric family
\be
p_k =p^*_k = \frac{A^k-A^{-k}}{t^k-t^{-k}}
\label{pstar}
\ee
this is how the third argument $A$ of the superpolynomial
appears in the formula.

Formulas (\ref{Ksum}), (\ref{evms}), (\ref{pstar}) and (\ref{boco})
fully describe our construction of the superpolynomial
for the generic torus link $T[m,n]$.
In what follows we provide numerous particular examples.

Note, however, that in order to have the coefficients $c_{R_1\ldots R_m}^Q$ in
(\ref{Ksum}) completely specified by (\ref{boco}), the initial condition should
be
imposed for arbitrary $p_k$, not restricted to (\ref{pstar}). It is not quite
clear
how to achieve this for $|Q|\ge 4$, and at this stage our procedure becomes
more sophisticated.
Actually, we provide a general rule for re-constructing the coefficients $c^Q$
from the Littlewood-Richardson coefficients, i.e. derive explicitly the
solution to
(\ref{boco}). The rule involves a $p$-dependent exchange of $q$ and $t$ at some
places,
see s.\ref{gamma} below.

Another subtlety concerns the choice of unknot superpolynomials for
non-fundamental
representations $R\ne [1^{|R|}]$. Our procedure can be easily applied to {\it
any} choice:
it affects only the form of the initial condition (\ref{boco}). In our examples
we assume the
simplest choice $Unknot_R=M_R^*$, which coincides with the choice in \cite{ASh}
and differs from
that in \cite{IGV,AK}.
This our choice has an advantage of giving rise to a power series with
positive
coefficients which gives the correct Khovanov-Rozhansky homologies.

\bigskip

This construction looks almost obvious,
still it is far from being proved,
thus it is important to ask, what can be actually tested.

Of course, all available examples of superpolynomial are reproduced, including
the original superpolynomials of \cite{DGR}
and the superpolynomial of the trefoil for the symmetric representation of
\cite{ASh}.
We also specially discuss our results for
$T[m,m+1]$ in the fundamental representation which is conjectured to be
expressed through the sums over the Schr\"oder paths
and is related with the Catalan numbers \cite{EG}.
The asymptotic formula for $T[m,n]$ at $m\to\infty$ due to \cite{DGR} is also
immediately reproduced.

By the very construction, the reduction to the colored HOMFLY polynomials
is correct (thus also to the colored Jones and Alexander polynomials),
and, what looks a little less trivial, correct is also
the reduction to the Heegard-Floer polynomials in the fundamental
representation
(found using the construction by \cite{EGth}).

This is, of course, not sufficient, but there is nothing more known
to compare with at the moment.

As for the link superpolynomials, our formulas coincide with the results of
\cite{IGV,AK} for the Hopf link in all fundamental (antisymmetric)
representations and
do not coincide with them in other higher representations because we define
the unknot polynomial differently as we explained above.

In fact, the most impressive are the {\it internal} checks:
the very fact that such a simple construction  works at all,
i.e. always provides a polynomial with positive integer coefficients.
In the MacDonald case the fact, that cancelations occur in
the complicated rational functions of $q,t$-variables
and they turn into relatively simple polynomials,
looks absolutely non-trivial and in practice is quite impressive.

Note that our procedure, in principle, works for non-torus knots too, as an
illustration
we show it working in the example of the series of knots which starts from
$5_2$
(and continues to $10_{139}$). At the same time, for $4_1$ the same procedure
leads to
a superpolynomial with some negative coefficients.

We list in the table below the calculated
superpolynomials indicating the tables where they can be found (and the proper
references to the papers where they were calculated earlier, at least,
in particular cases, if any). Since the first version of this paper has
appeared
there were delivered more papers on the subject by different groups
\cite{GSu,Cor,Ch,Sh,ORSG,MMSS,MMS,FGS,IMMMfe,FGS2}. In some papers the method
developed
here was used in order to
obtained new expressions for knot superpolynomials. We list them for
completeness.

\be
\begin{array}{|c|c|c|c|}
\hline
R&T[1,n]=unknot&T[2,2k+1]&T[2,2k]\\
\hline
[1]&M_{[1]}^*&\ref{2k+1},\cite{DGR}&\ref{2k}\\
\hline
[1,1]&M_{[1,1]}^*&\ref{2k+111}&\\
\hline
[2]&M_{[2]}^*&\ref{2k+12}&\\
\hline
[1]\otimes[1,1]&M_{[1]}^*M_{[1,1]}^*&&\ref{2k1+11}\\
\hline
[1,1]\otimes[1,1]&\left(M_{[1,1]}^*\right)^2&&\ref{2k11+11}\\
\hline
[1]\otimes[2]&M_{[1]}^*M_{[2]}^*&&\ref{2k1+2}\\
\hline
[2]\otimes[1,1]&M_{[2]}^*M_{[1,1]}^*&&\ref{2k2+11}\\
\hline
[2]\otimes[2]&\left(M_{[2]}^*\right)^2&&\ref{2k2+2}\\
\hline
\{R_i\}&&&\cite{AK,IGV}\\
\hline
[p]&&3.3, \cite{FGS}&\\
\hline
[1^p]&&3.3, \cite{FGS}&\\
\hline
\end{array} \ \ \ \ \ \ \ \
\begin{array}{|c||c|c|c|}
\hline
&R=[1]&R=[p]&R=[1^p]\cr
\hline
T[3,3k+1]&\ref{3k+1},\cite{DGR}&&\cr
\hline
T[3,3k+2]&\ref{3k+2},\cite{DGR}&&\cr
\hline
T[3,3k]&\ref{3k}&&\cr
\hline
T[4,4k+1]&\ref{4k+1}&&\cr
\hline
T[4,4k+2]&\ref{4k+2}&&\cr
\hline
T[4,4k+3]&\ref{4k+3}&&\cr
\hline
T[4,4k]&\ref{4k}&&\cr
\hline
T[5,5k+1]&\ref{5k+1}&&\cr
\hline
T[5,5k+2]&\ref{5k+2}&&\cr
\hline
T[5,5k+3]&\ref{5k+3}&&\cr
\hline
T[5,5k+4]&\ref{5k+4}&&\cr
\hline
T[5,5k]&\ref{5k}&&\cr
\hline
T[6,6k+1]&\ref{6k+1}&&\cr
\hline
T[7,7k+1]&\ref{7k+1}&&\cr
\hline
T[7,7k+2]&\ref{7k+2}&&\cr
\hline
T[7,7k+3]&\ref{7k+3}&&\cr
\hline
T[7,7k+4]&\ref{7k+4}&&\cr
\hline
T[7,7k+5]&\ref{7k+5}&&\cr
\hline
T[7,7k+6]&\ref{7k+6}&&\cr
\hline
T[m,km]&\ref{mk}&&\cr
\hline
T[m,mk+1]&\ref{km+1},\cite{EG}&&\cr
\hline
T[m,mk-1]&\ref{km-1}&&\cr
\hline
T[m,mk+2,+3,+4]&\ref{km+2},\cite{MMSS}&&\cr
\hline
5_2,\ 10_{139},\ ...&\ref{52}&&\cr
\hline
4_1,\ ...&\ref{41}&4.3, \cite{IMMMfe}&4.3, \cite{IMMMfe}\cr
\hline
\end{array}
\ee

\paragraph{Notations.} Before proceedings to details, we need to comment on
the
notation, which is pretty messy in the field.
The main {\it a priori} object to study is "the Wilson average"
\be
{\cal K}_R^{\rm knot}=\left<Pe^{\oint_{\rm knot}{\bf A}d{\bf x}}\right>
\ee
We put it in quotes because the $q$-$t$-deformation of the Chern-Simons theory
itself is not fully developed yet, thus, there are even less chances to {\it
derive}
${\cal K}_R^{\rm knot}$ in this way than in the "standard" case of $t=q$
($\beta=1$).
$K_R^{\rm knot}$ is a function of three types of variables: $q$, $t=q^\beta$
and
$A=t^N$, and it is {\it proportional} to a polynomial in all of these
variables.
The proportionality coefficient is a know-dependent power, and we often omit
it.
Still, the polynomial is denoted differently, in general we denote it ${\cal
P}$
(superpolynomial), for the special value of parameters $q=t$ it reduces to the
HOMFLY polynomial ${\cal H}$.

Moreover, the most adequate object in our calculations is somewhat in between
${\cal K}$ and the superpolynomial ${\cal P}$: the linear combination
(\ref{McDd})
of MacDonald dimensions
$M_R^*$ is a polynomial of $A$ and $A^{-1}$ and we denote the corresponding
quasi-polynomial $P$ (or $H$ in the HOMFLY case).
The initial condition (\ref{boco}) is nicely imposed on ${\cal P}$, while in
$P$ it
holds only up to factors of $A$.

One more thing is that ${\cal K}_R$ is proportional to the polynomial $P$
evaluated at the
special point $p_k=p_k^*$ (\ref{pstar}) in the time-variable space (thus, the
proper notation would be $P^*$). The full initial condition should be imposed
at
all values of $p_k$, not restricted to $p_k=p_k^*$, only in this case it would
be
sufficient to define all the coefficients $c^Q$ in (\ref{Ksum}) unambiguously.
Unfortunately, in this text we do not develop such a general formalism, and we
actually denote $P^*$ just through $P$.

We also use the following notation. First of all, throughout
the paper we use the MacDonald polynomials with $q\to q^2$ and $t\to t^2$ as
compared to the standard ones \cite{McD}. We also use the short-cut notation
\be
\{x\}\equiv x-{1\over x},\ \ \ \ [m]_q\equiv {q^m-q^{-m}\over q-q^{-1}},
\ \ \ \ \ \ [1,x]^m_q\equiv\sum^m_{j=0}
{[m]_q!\over [j]_q![m-j]_q!}x^j,
\ee
In particular, $[1,-x]^2_q = 1 -
(q+q^{-1})x + x^2$.

Last but not least, we use $(A,q,t)$ to denote the natural parameters of
MacDonald polynomials
$M_R(q^2,t^2)$, while parameters standard for superpolynomial discussions are
denoted through
$({\bf a,q,t})$. They are related through (\ref{naiveid}).

\section{HOMFLY polynomials for torus knots}

\subsection{General construction \cite{chi}}

The rather formal and abstract construction briefly described in the
Introduction is naturally specified for the
HOMFLY polynomials. Consider torus knot $T[m,n]$ with mutually prime
$m$ and $n$. In this case,
one should use as $\chi_R^*$ the Schur polynomials $s_R^*$ and
define the coefficients $c_R^Q$ from the relation
\be
\chi_R\{p^{(m)}\}\ = \sum_{Q\vdash m|R|} c_R^Q \chi_Q\{p\}
\label{cfromshift}
\ee
where
\be
p^{(m)}_k = p_{mk}
\label{pmk}
\ee
Thus, for the torus knot $T[m,n]$ one has
\be
\boxed{
H_R^{T[m,n]}\{p\} =
\sum_{Q\vdash m|R|} q^{-2\frac{n}{m}\varkappa_Q} c^Q_R s_Q\{p\}
= q^{-2\frac{n}{m}\hat W_{[2]}} \sum_{Q\vdash m|R|} c^Q_R s_Q\{p\}
= q^{-2\frac{n}{m}\hat W_{[2]}} s_R\{p^{(m)}\}
}
\label{Wrep}
\ee
The ordinary knot invariant arises when the special values
are substituted for $p_k$:
\be
p_k^* = \frac{a^k-a^{-k}}{q^k-q^{-k}}
\ee
Then
\be
\boxed{
H_R^{T[m,n]}(a|q) = H_R^{T[m,n]}\{p\,^*\}
}
\ee
Following \cite{chi} we denote the values of quantities at $p_k=p^*_k$
by stars.

However, one can describe in this way only the knot polynomials. In order to
obtain a generic link polynomial, which corresponds to $T[ml,nl]$, $l$
being the maximal common divisor of $m$ and $n$, one has to consider
more general initial condition, with the coefficients $c^Q$
determined instead of (\ref{cfromshift}) from
\be
\prod_{i=1}^l s_{R_i}\{p^{(m)}\}=\sum_{Q\vdash m\sum|R_i|}c^Q_{R_1...R_l}
s_Q\{p\}
\ee
With these coefficients one can still use formula (\ref{Wrep}) for the
polynomial.

\subsection{HOMFLY polynomials from $R$-matrix}

The origin of formula (\ref{Wrep}) is in the $R$-matrix formulation of HOMFLY
polynomials. To see this,
let knot $K$ be presented as a closure of the braid $b\in B_{m}$, such that $b$
is expressed
somehow through the generators of the braid group:
$$
b=g_{i_{1}} g_{i_2}...g_{i_l}
$$
Then the quantum group invariants can be obtained using the well known
universal $R$-matrix
representation of the braid group:
$$
g_{i}=1\otimes 1 \otimes...{R}_{i}\otimes...\otimes 1
$$
such that the invariant is given by the quantum trace
\be
H^{K}_{\lambda}= \tr_{\lambda} \Big( q^{\rho^{\otimes m} } b \Big)
\ee
The quantum dimension is the quantum invariant of the unknot which is the
closure of a single strand
\be
H^{U}_{\lambda} =\tr_{\lambda} q^{\rho } = s_{\lambda}^{\ast}
\ee
where $\rho$ is sum of the positive roots of the algebra.
In this simplest case of the braid group, $B_{2}$ it is generated by the single
element $g$.
The torus knots $T[2,2k+1]$ are the closures of $g^{2k+1}$ and the torus links
$T[2,2k]$ are
the closures of $g^{2k}$.
Consider the simplest case of the fundamental representation of $su(2)$. In
this case one has
\be
g=R=\left( \begin {array}{cccc} q^{-1}&0&0&0\\\noalign{\medskip}0&{\frac
{1-{q}^{2}}{q}}&1&0
\\\noalign{\medskip}0&1&0&0\\\noalign{\medskip}0&0&0&q^{-1}
\end {array} \right),
\ee
The $R$-matrix acts in the product of representations $[1]\otimes[1]=[2] \oplus
[1,1]$.
As the centralizer of the quantum group $U_{q}(su(2))$, it acts as a scalar on
the irreducible
representations $[2]$ and $[1,1]$. Indeed,
the characteristic equation for this $R$-matrix has the form
\be
\det (x- R) =(x-q^{-1})^3(x+q)
\ee
i.e. it has three eigenvalues equal to $q^{-1}$ and one eigenvalue equal to
$-q$. The first three
correspond to the three-dimensional symmetric representation $[2]$ and the last
one to
one-dimensional $[1,1]$. By definition, one has
\be
\tr_{[2]} q^{\rho^{\otimes 2} } = s_{[2]}^{\ast},\ \ \
\tr_{[1,1]} q^{\rho^{\otimes 2} } = s_{[1,1]}^{\ast}
\ee
so that the Jones polynomials (i.e. the HOMFLY polynomial with $a=q^2$) for
knots and links are
\be
H^{T[2,2k]}_{[1]}=\tr_{[1]} \Big( q^{\rho^{\otimes 2}} R^{2 k} \Big) = q^{-2k}
s_{[2]}^{\ast} +
q^{2k} s_{[1,1]}^{\ast}, \ \ \ H^{T[2,2k+1]}_{[1]}=\tr_{[1]} \Big(
q^{\rho^{\otimes 2}}
R^{2 k+1} \Big) = q^{-2k-1} s_{[2]}^{\ast} - q^{2k+1} s_{[1,1]}^{\ast}
\ee
For the generic torus knot $T[m,n]$, the braid representation has the form
$b[m,n]=(g_{1} g_{2}...g_{m-1})^{n}\in B_{m}$, which again has extremely simple
eigenvalues on
the irreducible representations.

\subsection{Fundamental representation}

In particular, for the fundamental representation
\be
H_{[1]}^{T[2,n]}(a|q) = q^{-n}s_2^* - q^{n}s_{11}^*
= q^{-n}\frac{a-a^{-1}}{(q-q^{-1})(q^2-q^{-2})}
\left( aq(1-q^{2n-2}) - \frac{1}{aq}(1-q^{2n+2})\right)
\label{K1T2n}
\ee
\be
H_{[1]}^{T[3,n]}(a|q) = q^{-2n}s_3^* - s_{21}^* + q^{2n}s_{111}^*=
\frac{q^{-2n}[a]_q}{(q-q^{-1})^2 [3]_q!}\left(a^2q^3[1,-q^{2n-3}]^2_q
- [2]_q[1,-q^{2n}]^2_{q^2} + \frac{1}{a^2q^3}[1,-q^{2n+3}]^2_q\right)
\label{K1T3n}
\ee
\be
H_{[1]}^{T[4,n]}(a|q) = q^{-3n}s_4^* -
q^{-n}s_{31}^* + q^{n}s^*_{211} - q^{3n}s^*_{1111}
\ee
Note that $s_{22}$ with
$\varkappa_2 = 0$ does not contribute in this case. Similarly, in the case of
\be
H_{[1]}^{T[5,n]}(a|q) =
q^{-4n} s_5^* - q^{-2n}s_{41}^* + s_{311}^* - q^{2n}s_{2111}^* +
q^{4n}s_{11111}^*
\ee
the two other characters $s^*_{32}$ and
$s^*_{221}$ would enter multiplied by $q^{-4n/5}$ and $q^{4n/5}$
respectively,
 but they do not appear in the sum.
This is the general rule: only terms with {\it integer} values of
$n\varkappa_Q/m$
appear in the sum: for other $Q$ the coefficients $c_{[1]}^Q$
automatically vanish(!).
In fact only the Young diagrams which have no more than one non-unit line and
no
more than one non-unit column do contribute:
\be
H_{[1]}^{T[6,n]}(a|q) = q^{-5n} s_6^* - q^{-3n}s_{51}^* + q^{-n}s_{411}^*
- q^{n}s_{3111}^* + q^{3n}s_{21111}^* - q^{5n}s_{111111}^*
\ee
\be
H_{[1]}^{T[7,n]}(a|q) = q^{-6n} s_7^* - q^{-4n}s_{61}^* + q^{-2n}s_{511}^* -
s^*_{4111}
+ q^{n}s_{31111}^* - q^{4n}s_{211111}^* + q^{6n}s_{111111}^*
\ee
\be
H_{[1]}^{T[8,n]}(a|q) = q^{-7n} s_8^* - q^{-5n}s_{71}^* + q^{-3n}s_{611}^*
- q^{-n}s_{5111}^*
+ q^n s_{41111}^*
- q^{3n}s_{311111}^* + q^{5n}s_{2111111}^* - q^{7n}s_{11111111}^*
\ee
and in general for the fundamental representation
\be
\boxed{
H_{[1]}^{T[m,n]}(a|q) =\sum_{j=0}^{m-1} (-)^j q^{-(m-1-2j)n}s^*_{[m-j,1^j]}}
\label{Wrepfund}
\ee

\subsection{Other representations}

In the next simplest representation
$$
H_{[2]}^{T[2,n]}(a|q) = q^{-6n}s_4^* - q^{-2n}s_{31}^* + s_{22}^*
= q^{-6n} \frac{\{a\}\{aq\}}
{\{q\}\{q^2\}\{q^3\}\{q^4\}}
\cdot\left[ a^2q^5\Big(1-(q^2+1+ q^{-2})q^{4n-4}+(q^2+q^{-2})q^{6n-6}\Big)
\right. -
$$
\be
\left.
- (q+q^{-1})\Big(1 - (q^4+1+q^{-4})q^{4n} + (q^2+q^{-2})q^{6n}\Big)
+ \frac{1}{a^2q^5}\Big(1-(q^2+1+q^{-2})q^{4n+4} +
(q^2+q^{-2})q^{6n+6}\Big)\right]
 \label{K2T2n}
\ee
Note that at $n=1$ this expression (and that of the previous subsection) turns
into
\be
H_{[1]}^{T[2,\,1]}(a|q) = \frac{a-a^{-1}}{q-q^{-1}}\cdot\frac{1}{a}\sim
s^*_{[1]} \nn \\
H_{[2]}^{T[2,\,1]}(a|q) = \frac{(a-a^{-1})\Big((aq) - (aq)^{-1}\Big)}
{(q-q^{-1})(q^{2}-q^{-2})}\cdot{q^2\over a^2}\sim s_{[2]}
\label{atn1}
\ee
i.e. proportional to the unknot polynomials $s_R^*$
in accordance with the initial condition (\ref{boco}).

\subsection{Non-torus knots}

A similar construction persists equally well for non-torus knots. For
instance,
the series of knots which starts from $5_2$ have the HOMFLY polynomials
\be\label{Hnt}
H^{[n]}_{[1]}(a|q)=c^{[3]}q^{-2n}s_{[3]}^*+
c^{[21]}s^*_{[21]}+c^{[111]}q^{2n}s_{[111]}^*
\ee
where
\be
c^{[3]}={1\over q^2},\ \ \ \ \ c^{[21]}=-{q^8-q^6+q^4-q^2+1\over q^2},
\ \ \ \ c^{[111]}=q^6
\ee
and $n$ is multiple of 3, since $5_2$ is described by a 3-strand braid.
Therefore, the series is given by $n=3k$ and, with increasing $k$, one obtains
more and more involved knots: $5_2$ for $k=0$, $10_{139}$ for $k=1$ etc.

\section{Deformation to MacDonald characters.}

\subsection{General construction}

The generalization of HOMFLY to the superpolynomials along the line of the
generic construction described in the
Introduction is quite immediate. That is, the general formula for the
superpolynomial is of the form
\be\label{McDd}
{\cal P}_R^{[m,n]}(A|q,t)\sim\sum c_R^Q\ t^{2\nu_Q/m} q^{-2\nu_{Q'}/m} M_Q^*
\ee
where the coefficients $c^Q_R$ now depend on not only on the number of strands
$m$ and the representation $R$, but also on
$p$. These coefficients are explicitly described in section 6, here we only
illustrate the
calculations with a few examples and discuss their general
structure in the fundamental representation case.

The biggest loophole is the lack of the simple rule (\ref{pmk}): this makes
evaluation of
$c_R^Q$ for $|Q|\ge 4$ a piece of art. There is also uncertainty in the choice
of superpolynomials for
the unknot in non-fundamental representations $R\ne [1^{|R}]$.

\subsection{MacDonald dimensions}

For manifest calculations one needs to know
explicitly the quantities $M_R^*$. These are:
\be
M_1^* = p_1^* = \frac{A-A^{-1}}{t-t^{-1}} \nn \\ \nn \\
M_2^* =   \frac{(A-A^{-1})(Aq-A^{-1}q^{-1})}{(t-t^{-1})(qt-q^{-1}t^{-1})} \nn
\\
M_{11}^*
= \frac{(A-A^{-1})(At^{-1}-A^{-1}t)}{(t-t^{-1})(t^2-t^{-2})}
\label{Mstar12'} \ee

\be M^*_3 = \frac{(A-A^{-1})(Aq-A^{-1}q^{-1})(Aq^2-A^{-1}q^{-2})}
{(t-t^{-1})(qt-q^{-1}t^{-1})(q^2t-q^{-2}t^{-1})} \nn \\
M^*_{21} = \frac{(A-A^{-1})(At^{-1}-A^{-1}t)(Aq- A^{-1}q^{-1})}
{(t-t^{-1})^2(qt^2-q^{-1}t^{-2})} \nn \\
M^*_{111} = \frac{(A-A^{-1})(At^{-1}-A^{-1}t)(At^{-2}-A^{-1}t^2)}
{(t-t^{-1})(t^2-t^{-2})(t^3-t^{-3})} \label{Mstar3} \ee

\be M^*_4 =
\frac{(A-A^{-1})(Aq-A^{-1}q^{-1})(Aq^2-A^{-1}q^{-2})(Aq^3-A^{-1}q^{-3})}
{(t-t^{-1})(qt-q^{-1}t^{-1})(q^2t-q^{-2}t^{-1})(q^3t-q^{-3}t^{-1})} \nn \\
M^*_{31} = \frac{(A-A^{-1})(At^{-1}-A^{-1}t)(Aq-
A^{-1}q^{-1})(Aq^2-A^{-1}q^{-2})}
{(t-t^{-1})^2(qt-q^{-1}t^{-1})(q^2t^2-q^{-2}t^{-2})} \nn \\
M^*_{22} = \frac{(A-A^{-1})(At^{-1}-A^{-1}t)(Aq-
A^{-1}q^{-1})(Aqt^{-1}-A^{-1}q^{-1}t)}
{(t-t^{-1})(t^2-t^{-2})(qt^2-q^{-1}t^{-2})(qt-q^{-1}t^{-1})} \nn \\
M^*_{211} = \frac{(A-A^{-1})(At^{-1}-A^{-1}t)(At^{-2}-A^{-1}t^2)(Aq-
A^{-1}q^{-1}) }
{(t-t^{-1})^2(t^2-t^{-2})(qt^3-q^{-1}t^{-3})} \nn \\
M^*_{1111} =
\frac{(A-A^{-1})(At^{-1}-A^{-1}t)(At^{-2}-A^{-1}t^2)(At^{-3}-A^{-1}t^3)}
{(t-t^{-1})(t^2-t^{-2})(t^3-t^{-3})(t^4-t^{-4})} \label{Mstar4} \ee
and so on.

To understand the structure of these formulas one should keep in
mind the simple picture:
\bigskip

$$
\begin{array}{ccccc}
&&M_R^*(a|q|t)&&\\
&&&&\\
t=q&\swarrow&&\searrow&q=1,\ t=q^\beta\\
&&&&\\
\!\!\!\!\! \!\!\!\!\! \!\!\!\!\! s_R^*(a|q) &&&& J_R^*(a|\beta)\\
&&&&\\
q=1&\searrow && \swarrow &\beta=1 \\
&&&&\\
&& D_R(N) &&
\end{array}
$$

\bigskip

MacDonald dimensions are double deformations of ordinary dimensions
of $SU(N)$ representations $D_R(N)$ in two directions: to $q\neq 1$
and to $\beta\neq 1$. Instead of $\beta$ in MacDonald polynomials one
often uses $t = q^\beta$, but in the limit $q\rightarrow 1$ also
$t\rightarrow 1$, but parameter $\beta$ survives. In all the
deformations one substitutes $N$ by $A = t^N = q^{\beta N}$, and all
dimensions are actually the values of the corresponding characters
at the special point (\ref{pstar}). In the classical case $q=1$ this
corresponds simply to putting $p_k = N$.

The dimension $D_R(N) = s_R(p_k=N)$ is expressed through the Schur
functions (ordinary $SU(\infty)$ characters) and is always a product
of $N$-linear factors. Deformed are actually the
individual factors. Especially simple is the reconstruction of the
ordinary quantum dimensions $s^*_R(a|q)$ from $D_R(N)$: each factor
is substituted by its quantum counterpart, $N \rightarrow [N]_q =
\frac{q^N-q^{-N}}{q-q^{-1}}$, $2 \rightarrow [2]_q$, $3 \rightarrow
[3]_q$. The only tricky point is the deformation of non-prime
integers: $4$ can become either $[4]$ or $[2]^2$, and this is the
only representation-dependent ingredient in the reconstruction of
ordinary quantum dimensions. With $\beta$ deformation things are
somewhat trickier. The rule is that while negative shifts of $N$ are
not deformed, $N-k \rightarrow N-k$ (or, rather, $N-k \rightarrow
\beta N -\beta k$), the positive shifts are: $N+k \rightarrow \beta
N + k$.

To the generic MacDonald dimensions, one can use the nice formula
(see, e.g., \cite{ASh}):
\be
M^*_R = \prod_{k=0}^{\beta-1}\prod_{1\leq
j<i\leq N} \frac{q^{R[j]-R[i]}t^{i-j}q^m -
q^{R[i]-R[j]}t^{j-i}q^{-m}} {t^{i-j}q^m - t^{j-i}q^{-m}}
\ee
valid
for integer values of $\beta$ and $N$. As usual, the result can be
easily continued to arbitrary $\beta$ and $A=t^N$.

Even more useful is the general expression for $M_R^*$ that generalizes the
hook formula for $D_R(N)$ (see Fig.1):
\be\label{McDdh}
\boxed{
M_R^* = \prod_{(i,j)\in R} \frac{Aq^{j-1}/t^{i-1} -
(Aq^{j-1}/t^{i-1})^{-1}} {q^kt^{l+1} - (q^kt^{l+1})^{-1}} }
\ee
where $k = R_i-j-1$ and $l=R_j'-i-1$.

These formulae can be further continued to elliptic
(Askey-Wilson) deformations (and further to the Kerov polynomials), but this is
beyond the scope of the
present paper.

\bigskip

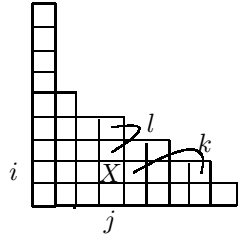
\begin{figure}
\unitlength 1mm 
\linethickness{0.4pt}
\ifx\plotpoint\undefined\newsavebox{\plotpoint}\fi 
\begin{picture}(58.968,38.022)(-45,23)
\put(31.849,31.008){\line(0,1){27.014}}
\put(31.849,58.022){\line(1,0){3.048}}
\put(34.897,58.022){\line(0,-1){27.014}}
\put(34.897,31.008){\line(-1,0){3.048}}
\put(31.849,54.868){\line(1,0){3.0482}}
\put(31.954,51.715){\line(1,0){2.838}}
\put(34.897,31.113){\line(1,0){24.071}}
\put(58.968,31.113){\line(0,1){3.048}}
\put(58.968,34.161){\line(-1,0){27.119}}
\put(37.42,30.903){\line(0,-1){.2102}}
\put(37.63,31.113){\line(0,1){15.031}}
\put(37.63,46.144){\line(-1,0){5.6761}}
\put(31.954,36.999){\line(1,0){23.545}}
\put(55.499,36.999){\line(0,-1){5.781}}
\put(31.954,39.837){\line(1,0){18.079}}
\put(50.033,31.218){\line(0,1){8.514}}
\put(52.661,36.684){\line(0,-1){5.466}}
\put(32.059,42.886){\line(1,0){11.983}}
\put(44.042,42.886){\line(0,-1){11.878}}
\put(46.985,39.417){\line(0,-1){8.304}}
\put(40.678,42.57){\line(0,-1){11.352}}
\put(31.849,48.877){\line(1,0){2.9431}}
\put(42.15,28.906){\makebox(0,0)[cc]{$j$}}
\put(29.326,35.738){\makebox(0,0)[cc]{$i$}}
\put(42.15,35.528){\makebox(0,0)[cc]{$X$}}
\qbezier(45.303,35.528)(55.972,41.677)(54.238,35.423)
\qbezier(42.465,38.261)(49.613,42.728)(42.465,41.519)
\put(54.763,39.417){\makebox(0,0)[cc]{$k$}}
\put(47.616,42.15){\makebox(0,0)[cc]{$l$}}
\end{picture}
\caption{The figure which illustrates the notation in the generalization of the
standard
hook formula to the MacDonald dimensions (\ref{McDdh}).}
\end{figure}

\subsection{From HOMFLY to the superpolynomial: an example}

The simplest superpolynomial is ($n$ is odd here)
\be
{\tilde P}_{[1]}^{T[2,n]}({\bf a}|{\bf q}|{\bf t}) \sim
\left( a\tilde q(1-\tilde q^{2n-2})
- \frac{1}{a\tilde q}(1-\tilde q^{2n+2})\right)
\label{cP1T2n}
\ee
i.e. looks the same as (\ref{K1T2n}), but
with
\be
{\rm for}\ \ P_{[1]}^{T[2,n]}: \ \ \ \ \ \
\tilde q = {\bf q}\sqrt{-{\bf t}}, \ \ \ \ \ \ a = (-{\bf t}){\bf a}
\label{naiveid}
\ee
Such a simple substitution rule does not work
for most of other knots and other representations,
but in this case it does!

At the same time, the first formula in (\ref{K1T2n})
is naturally deformed into
\be
P_{[1]}^{T[2,n]}(A|q|t) \sim \check q^{-n}M_2^*
- \zeta \check q^{n}M_{11}^*
\sim
 \left\{
At\left(1 + \frac{q-t}{t^2(1-qt)} - \frac{\zeta\check q^{2n}}{t^2}\right)
- \frac{1}{At}\left(1 + \frac{t^2(q-t)}{1-qt} - \zeta\check q^{2n}t^2\right)
\right\}
\label{P1T2n}
\ee
with some yet unknown $\check q$ and $\zeta$.

Comparing (\ref{P1T2n}) with (\ref{cP1T2n}) at arbitrary values of $n$
we conclude that they coincide if
\be
\check q = \tilde q,\ \ \ \ \ \ \ \
A = \frac{at}{\tilde q}
\label{Avsatq}
\ee
\be
\tilde q^4  = {\bf q}^4{\bf t}^2 = t^2q^2
\label{tiqvstq}
\ee
and
\be
\zeta^2 = \frac{q^2(1-t^4)^2}{t^2(1-q^2t^2)^2}
\label{gammaP1T2n}
\ee
Relation (\ref{tiqvstq}) is consistent with
the identification suggested in \cite{AK,ASh}
\be
\boxed{
{\bf q} = t, \ \ \ \ \ \ {\bf t} = -q/t,
\ \ \ \ \ \ A \ \stackrel{(\ref{Avsatq})}{=}\ {\bf a}\sqrt{-{\bf t}}
}
\label{GMid}
\ee
Thus, we obtain the following expression for the superpolynomial
\be
\boxed{
P_{[1]}^{T[2,n]}(A|q|t)
= q^{-n}\left(M_2^* - (qt)^n\frac{t^2-t^{-2}}{qt - (qt)^{-1}}M_{11}^*\right)
}
\label{P1T2nM}
\ee
or, more explicitly,
\be
P_{[1]}^{T[2,n]}(A|q|t)
= q^{-n}\frac{\{A\}}{\{t\}\{qt\}}
\left(Aq\Big(1-(qt)^{n-1}\Big) - \frac{1}{Aq}\Big(1-(qt)^{n+1}\Big)\right)
\label{P1T2nMe}
\ee
After substitution of (\ref{GMid}) this turns into the superpolynomial {\it per
se},
with all expansion coefficients positive integers,
which is, of course, not obvious neither from (\ref{P1T2nMe}),
nor, in fact, from (\ref{cP1T2n}).
More precisely,
\be
-A^2q^2\frac{1-(qt)^{n-1}}{1-(qt)^2} + \frac{1-(qt)^{n+1}}{1-(qt)^2}
\ \stackrel{(\ref{GMid})}{=}\
{\bf a}^2{\bf q}^2{\bf t}^3 \frac{1 - ({\bf q}^4{\bf
t}^2)^{\frac{n-1}{2}}}{1-{\bf q}^4{\bf t}^2}
+ \frac{1 - ({\bf q}^4{\bf t}^2)^{\frac{n+1}{2}}}{1-{\bf q}^4{\bf t}^2}
\ee
and for $n=2k+1$, i.e. for links, consisting of a single knot,
all the coefficients in the expansion of the r.h.s. are obviously positive
integers:
\be
{\tilde P}_{[1]}^{T[2,2k+1]}({\bf a}|{\bf q}|{\bf t}) \sim
{\bf a}^2{\bf q}^2{\bf t}^3 \frac{1 - ({\bf q}^4{\bf t}^2)^{k}}{1-{\bf q}^4{\bf
t}^2}
+ \frac{1 - ({\bf q}^4{\bf t}^2)^{k+1}}{1-{\bf q}^4{\bf t}^2}
\ee

\subsection{The rules for $\zeta$ evaluation\label{gamma}}

What is important for other cases,
where there are no known answers for the superpolynomials to compare with,
(\ref{P1T2nMe}) gets drastically simplified
at  $n=1$ due to the initial condition (\ref{boco}): in this case the
knot is equivalent to the unknot (we take into account the different
normalizations
of the polynomials for the unknot and for the $T[2,2k+1]$-series):
\be
P_{[1]}^{T[2,1]}(A|q|t) = \frac{t}{qA}
P_{[1]}^{T[1,2]}(A|q|t)=\frac{A-A^{-1}}{t-t^{-1}}\frac{t}{qA}
\ee
This allows one to define $\zeta$ not from (\ref{gammaP1T2n}),
but directly from (\ref{P1T2nM}) as
\be
\boxed{
\frac{M_2^* - \zeta (qt) M_{11}^*}{M_1^*} \sim A^{-1}
\ \ \ \Longrightarrow \ \ \
\zeta = \frac{t^2-t^{-2}}{qt - (qt)^{-1}}
}
\label{gammafromP1T21}
\ee
In variance with (\ref{gammaP1T2n}), this rule defines $\zeta$
entirely in terms of the MacDonald dimensions $M_R^*$, without any
reference to {\it a priori} knowledge about the superpolynomials.
This makes the MacDonald construction completely self-relied,
and allows one to construct superpolynomials in the situations
where they are not yet known.
As already mentioned, arising expressions do not {\it explicitly}
possess the positivity property of the superpolynomials,
it should be proved separately. However, for every concrete knot one easily
checks this is really the case.

In fact, starting from the initial condition and requiring polynomiality of the
whole
series, one can immediately construct any series of torus knots $T[m,k+p]$. We
list many examples of such series in s.\ref{examples}. Here we would like to
discuss
their general structure.

\subsubsection*{Limit of $n\to\infty$, \cite{DGR}}

First of all, let us note that for the braid with $m$ strands the coefficient
$c_{[m]}=1$
(since in the Abelian case, $A=t$ the superpolynomial reduces to the pure
framing, and
all $M_R^*$ but $M_{[m]}^*$ are equal to zero at $A=t$, see (\ref{McDdh})).
This
is the only term which survives when $n\to\infty$ provided $|q|<1$, $|t|<1$:
\be
q^{(m-1)n}{\cal P}^{[m,n]}_{[1]}\ \stackrel{n\to\infty}{=}\ {M_{[m]}^*\over
M_{[1]}^*}
\ee
One immediately recognizes in this expression formula (118) from \cite{DGR} and
(A2) from
\cite{AK}.

\subsubsection*{General formula for $T[m,km+1]$ superpolynomial in the
fundamental
representation}

Second, in the case of fundamental representation one can construct
the generic formula for knot $T[m,m+1]$:
\be
\boxed{ P^{T[m,k
m+1]}_{[1]}=\sum\limits_{|Q|=m} c^{Q}_{[1]} M_{Q}^{\ast} q^{-2
\nu(Q^{\prime})k} t^{2 \nu(Q)k} }
\ee
the coefficients $c^Q$ in this
case are given by a product of two factors
\be
c^{Q}_{[1]}=\bar{c}^{Q}_{[1]} \gamma^{Q}_{[1]}
\ee
the first of them
being defined from the system of linear equations (cf. the HOMFLY
case, when only this first factor is present)
\be
p_{m}=\sum_{|Q|=m}
\bar{c}^{Q}_{[1]} M_{Q}(p_{k}),
\ee
while the second factor is
manifestly given by the formula
\be
\gamma^{Q}_{[1]}=\dfrac{q^{\alpha_{Q}} }{[m]_{q}}
\sum\limits_{(i,j)\in Q} t^{2 (i-1)} q^{2Q_{1}-2Q_{i}+2 j-2}
\ee
with the proper power $\alpha_{Q}$. The first $\gamma^Q$'s are:
\be
\gamma^{[2]}=\dfrac{1+q^2}{1+q^2}=1, \ \ \ \
\gamma^{[1,1]}=\dfrac{1+t^2}{1+q^2}
\ee
\be
\gamma^{[3]}=\dfrac{1+q^2+q^2 q^2}{1+q^2+q^2 q^2}=1, \ \
\gamma^{[2,1]}=\dfrac{1+q^2+q^2 t^2}{1+q^2+q^2 q^2},\
\ \ \ \ \gamma^{[1,1,1]}
=\dfrac{1+t^2+t^2 t^2}{1+q^2+q^2 q^2}
\ee
\be
\gamma^{[4]}=\dfrac{1+q^2+q^2 q^2+q^2 q^2 q^2}{1+q^2+q^2 q^2 +q^2 q^2 q^2}=1, \
\
\gamma^{[3,1]}=
\dfrac{1+q^2+q^2 q^2+q^2 q^2 t^2}{1+q^2+q^2 q^2+q^2 q^2 q^2},\ \
\gamma^{[2,2]}=\dfrac{(1+q^2)(1+t^2)q}{1+q^2+q^2 q^2+q^2 q^2 q^2}
={\frac { \left( t^2+1 \right) q}{ \left( {q}^{4}+1 \right) }}
\nn
\ee
\be
\gamma^{[2,1,1]}=\dfrac{1+q^2+q^2 t^2+q^2 t^2 t^2}{1+q^2+q^2 q^2 + q^2 q^2
q^2},
\ \
\gamma^{[1,1,1,1]}=\dfrac{1+t^2+t^2 t^2+t^2 t^2 t^2}{1+q^2+q^2 q^2 + q^2 q^2
q^2}
\ee
Similar rules can be easily
written down for other series of the toric knots.

\subsubsection*{$n\to -n$, $A\to A^{-1}$, $q\leftrightarrow t$ symmetry and
$T[m,km-1]$ superpolynomial}

For instance, there is a symmetry between two series $T[m,mk+1]$ and
$T[m,mk+(m-1)]=T[m,mk^{'}-1]$ of the torus knots (below we drop the prime). It
follows from the simple relation:
\be
\boxed{ P^{T[m,k\cdot
m-1]}_{[1]}(A,q|t)=P^{T[m,-k\cdot m+1]}_{[1]}(A^{-1},q^{-1}|t^{-1})
}
\ee
Hence, one suffices to make in the superpolynomial $P^{T[m,k m+1]}_{[1]}$
the substitution $k\rightarrow -k, \ q\rightarrow
q^{-1}, \ t\rightarrow t^{-1}, \ A\rightarrow A^{-1}$ in order to get the
general
formula for series $[m,mk-1]$:
\be
\boxed{ P^{T[m,k m-1]}_{[1]}=\sum\limits_{|Q|=m} \bar{c}^{Q}_{[1]}.
\tilde\gamma^{Q}_{[1]}
M_{Q}^{\ast} q^{2 \nu(Q^{\prime})k} t^{-2 \nu(Q)k} }
\ee
with the
coefficients $\tilde\gamma^{Q}_{[1]}$ manifestly given by the formula
\be
\gamma^{Q}_{[1]}=\dfrac{q^{-\tilde\alpha_{Q}} }{[m]_{q}}
\sum\limits_{(i,j)\in Q} t^{2 (1-i)} q^{2Q_{i}-2Q_{1}+2-2j}
\ee
Note that the MacDonald polynomials do
not change under $q\rightarrow q^{-1}, \ t\rightarrow t^{-1}$.

\section{Comments}

\subsection{Non-fundamental representations. The problem of unknot}

The split $W$-evolution, considered in this paper, describes
knot invariants in terms of vectors in the linear space
of MacDonald polynomials $M_R\{p\}$, which is, at the end of the day,
projected onto the smaller subspace of MacDonald dimensions $M_R^*$.
In this paper we consider the discrete-time "evolution",
describing attachment of any number of "torus braids"
to any original braid. Of course, other types of attachments
can be considered in a similar way, but in this paper
we concentrate on this concrete one.
Such an evolution is defined by the MacDonald counterpart
(splitting or "refinement";)
of the cut-and-join operator $W_{[2]}$ and is universal: it
depends on the choice of $[m,n]$ and on the representation
in a simply controllable way.
However, the result of evolution depends on the starting point:
on "the initial condition". Already in the simplest case
of the series $T[m,mk+1]$ this initial condition
at $k=0$ is given by the unknot, and all the results
for all other knots and links depend on the choice
of this initial condition, i.e. on the formula for
$Unknot_R$.
The naive choice of the MacDonald dimension for the role
of this quantity,
\be
Unknot_R = M_R^*
\label{naiveunknot}
\ee
which naively generalizes the standard choice
of the usual quantum dimensions for the HOMFLY unknot,
is indisputably correct for all the fundamental
(totally antisymmetric) representations $R = [1^{|R|}]$.
However, it is far less obvious for other representations $R$.
The problem is that there is no {\it a priori} definition,
and there are different opinions on what to insist on.
Since this is not a focus of the present paper,
which is devoted to the very idea of representation
in the space of MacDonald polynomial and to consideration
of discrete evolutions in this space,
we do not go into any detail of the unknot problem here.

The simplest example
of what the split $W$-evolution gives is the case of the simplest
symmetric representation $R=[2]$ for torus knot $T[2,2k+1]$, {\it assuming}
the naive choice (\ref{naiveunknot}) for the unknot:
\be
P^{[2,2k+1]}_{[2]}=c^{[4]}_{[2]} M_{[4]}^{\ast} q^{-6(2k+1)} +
\Big(\frac{q}{t}\Big) c^{[3,1]}_{[2]}
M_{[3,1]}^{\ast} q^{-3(2k+1)}t^{2k+1} + \Big(\frac{q}{t}\Big)^2 c^{[2,2]}_{[2]}
M_{[2,2]}^{\ast}
q^{-2(2k+1)}t^{2(2k+1)}
\ee
where the coefficients
\be
c^{[4]}_{[2]}=1,\ \ \ c^{[3,1]}_{[2]}=-\frac{(1-t^2)(1+q^2)(1+t^2q^2)}{1-q^6
t^2}, \ \ \ c^{[2,2]}_{[2]}=\frac{(1-t^4)(1+t^4q^2)}{(1-q^4 t^2)(1-q^2t^2)}
\ee
This result coincides with that of \cite{ASh}, where the choice
(\ref{naiveunknot}) is also implicitly made.
An alternative choice for the unknot is discussed in
\cite{IGV,AK}. The point there is that, for $R \neq [1^{|R|}]$, the
MacDonald dimensions $M_R^*$ are {\it not} polynomials
in $q$ and $t$, even if one substitutes $A=t^N$ with integer $N$.
One can guess (and check) that this has an unpleasant implication
for the torus {\it links}: the split $W$-evolution, starting from
initial condition (\ref{naiveunknot}) gives rise to
non-polynomial expressions for the links.
A simple way out is to change (\ref{naiveunknot})
for some linear combinations,
\be
Unknot_R = \sum_{S \vdash |R|} V_{RS}M_S^*
\ee
which {\it are} polynomials after the substitution $A=t^N$.

Only two examples are explicitly given in \cite{IGV,AK}
\be
Unknot_2 \sim M_2^* + \frac{t^2-q^2}{1-q^2t^2}M_{11}^*q^2t^2 =
\frac{A-A^{-1}}{At(t-t^{-1})(t^2-t^{-2})}\Big(q^2A^2 + t^4-q^2t^2-1\Big)
\ee
and
\be
Unknot_{21} \sim M_{21}^* + \frac{(t^2-q^2)(1+t^2)}{1-q^2t^4}M_{111}^*q^2t^4
= \frac{(A-A^{-1})(At^{-1}-A^{-1}t)}{At(t-t^{-1})^2(t^3-t^{-3})}\Big(q^2A^2 +
t^6-q^2t^4-1\Big)
\ee
The main part of the coefficients in the matrix $V_{RS}$
here comes from the linear transformation between the Schur
and MacDonald functions \cite{AK}:
\be
\left(\begin{array}{c} s_2\\s_{11} \end{array}\right)
= \left(\begin{array}{cc} 1 &\frac{t^2-q^2}{1-q^2t^2} \\0 & 1
\end{array}\right)
\left(\begin{array}{c} M_2\\M_{11} \end{array}\right)
\ee
\be
\left(\begin{array}{c} s_3 \\ s_{21} \\s_{111} \end{array}\right)
= \left(\begin{array}{ccc}
1 & \frac{(t^2-q^2)(1+q^2)}{1-q^4t^2}   &
\frac{(t^2-q^2)(t^4-q^2)}{(1-q^2t^2)(1-q^4t^2)} \\
0 & 1 & \frac{(t^2-q^2)(1+t^2)}{1-q^2t^4} \\0 & 0 & 1 \end{array}\right)
\left(\begin{array}{c} M_3 \\ M_{21} \\M_{111} \end{array}\right)
\ee
The split $W$-evolution preserves this "weak-polynomiality"
condition and cures the problem for the links.
It, however, changes the answers for knots as well.
At the same time, the split $W$-evolution {\it works}
for arbitrary representations.

In fact, one may check that our choice of unknot (\ref{naiveunknot}) leads to
the
"superpolynomials" (which are no longer polynomials at all) which,
being treated as a power series in variables
$({\bf q,t})$ (\ref{GMid}),
have all their coefficients positive. Moreover, they seem to give rise to the
correct
Khovanov-Rozhansky homologies (after the proper reduction of the power
series).
Thus, one may expect our choice of unknot leads to
the {\it correct} superpolynomials. Anyway, our main message here is that
the explicit answer depends not only on the evolution,
but also on the choice of $Unknot_R$ as the initial condition.

\subsection{Unknots from modular matrices \cite{ASh}}

Let us explain what is the unknot superpolynomial in accordance with
\cite{ASh}.
There was developed a formalism of the Hilbert spaces for computations of the
superpolynomials.
In this framework the Hilbert space of
the beta-deformed (refined) Chern-Simons theory coincides with the set of
representations of
$SU_{k}(N)$ at some level $k$. This is the finite dimensional space labeled by
all
Young diagrams $Y$ lying inside the $k\times (N-1)$ box, and corresponding wave
functions
(states) are the MacDonald functions $M_{Y}$ with the parameters
\be
q=\exp\Big(\frac{2 \pi i}{k+\beta N}\Big), \ \ \ t=\exp\Big(\frac{2 \pi i \beta
}{k+\beta N}\Big)
\ee
The superpolynomials can be computed exactly as in the case of the pure
Chern-Simons theory
as the vev of the Wilson loop in $S^{3}$. It is convenient to represent the
sphere $S^3$ as
union of two tori with identification of the cycles $(0,1)$ and $(1,0)$. Then
the answer for
the superpolynomial is given
by
\be
{\cal P}^{K}_{R}=<W_{R}(K)>=\dfrac{<0| K V^{R} K^{-1} S |0>}{<0|S|0>}
\ee
where $V^{R}$ is the operator creating the Wilson loop in the representation
$R$ that wounds
cycle $(1,0)$,
explicitly defined on the set of the wave functions as multiplication on
$M_{R}$:
\be
M_{R} M_{Q}= \sum\limits_{P} V^{R}_{P,Q} M_{P}
\ee
The operator $K=S^{i_1}T S^{i_2}...$ deforms the cycle to the nontrivial torus
knot $T[m,n]$
$(1,0)\rightarrow (m,n)$, the operators $S: (1,0)\rightarrow (0,1)$, identifies
two vacua at
different tori in such a way that the resulting space is $S^3$. All we need is
the explicit
expressions for $S$ and $T$ found in \cite{ASh}:
\be
S_{R,Q}=S_{[],[]}\frac{M_{R}(t^{2 \rho}) M_{Q}(t^{2\rho} q^{2R})}{ q^{2 |R|
|Q|/N}}, \ \ \  T_{R,Q}=\frac{q^{2 ||R||} t^{N|R|}}{t^{ ||R^{\prime}||} q^{2
|R|^2/N}} \delta_{Q,R}
\ee
where $||R||=\sum_{i} R_{i}^2$ and $S_{[],[]}$ is some constant. For example,
for the unknot
$K=S$ and one obtains
\be
<Unknot_{R}>=\dfrac{<0| S V^{R} |0>}{<0|S|0>}=\dfrac{ \sum_{Q} S_{[],Q}
V^{R}_{Q,[]}}{S_{[],[]}}=\frac{\sum_{Q} S_{[],[]} M_{Q}(t^{2\rho})
V^{R}_{Q,[]}}{S_{[],[]}}=M_{R}(t^{2\rho})=M_{R}^{\ast}
\ee
i.e. the vev of the unknot is given by the refined quantum dimension
\be
P^{Unknot}_{R}=M_{R}^{\ast}
\ee
Using this formulas we have computed the superpolynomials for the torus knots
for several first
levels in $k$, $N$, and $R$ and found that the results are in full agreement
with ours.

\subsection{Link polynomials}

A subtle point of our story concerns the links.
Torus links do not seem too much different from
the torus knots, and the only thing
to change in our construction is the initial condition.
Links contain several disconnected
(but intertwined) components, and link invariants
depend on several independent representations.
Thus, the initial condition involves a product of
characters and, thus, non-trivial (but easily
calculable) Littlewood-Richardson coefficients.
The problem, however, is that what one gets
in this way is not a superpolynomial: the
result of the cut-and-join evolution is not a polynomial
in $q,t$-variables.
In fact, this problem is well known: already the
HOMFLY polynomials for links are not really polynomials
if $A$ and $t$ are considered as independent variables,
as in the case of knots.
They become polynomials in ${\bf q}$ only after the
substitution $A=t^N$, i.e. the polynomiality
property is much weaker for links
than it is for knots.
Moreover, even this weaker property survives
$\beta$-deformation only for the fundamental representations
$R = [1^{|R|}]$. If some $R$ are different,
say $R=[2]$, then the result of the evolution,
which starts from the MacDonald dimensions is not a
polynomial in $q,t$, even if one substitutes $A=t^N$.
The problem is, in fact, inherited from
the level of unknots:
the MacDonald dimensions $M_R^*$ for $R \neq [1^{|R|}]$
are not polynomials, even if $A=t^N$.
In \cite{IGV} and further in \cite{AK}
a radical way out was suggested:\footnote{
Note that the recent suggestion in \cite{ASh}
corresponds to the choice $unknot_R = M_R^*$,
not to the choice of \cite{IGV,AK}.
}
to take $unknot_R \neq M_R^*$ for $R=[1^{|R|}]$,
and substitute it by an expression which does
become a polynomial for $A=t^N$.
If our cut-and-join evolution starts from {\it such}
initial condition, it provides answers for links
which are polynomial in the same sense
and coincide with those in \cite{IGV,AK}
up to a linear transformation.
The coefficients of these polynomials,
however, are not always positive, as one would
demand for the true superpolynomials
(see \cite{AK} for discussion and suggestions
about this problem). The solution is the same as in the case of
non-fundamental
representations of knots: to treat the superpolynomials as power series in
$({\bf q,t})$.
The coefficients of this series turn out to be positive, and the result seems
to reproduce the
Khovanov-Rozhansky homologies correctly (after the proper reduction of the
power series).

Let us stress again that this problem has nothing to do with the fundamental
representation
where, for instance, the superpolynomial for link $[m,km]$ takes the form
\be
P^{[m,k m]}_{[1]}=\sum\limits_{|Q|=m} c^{Q}_{[1]} M_{Q}^{\ast} q^{-2
\nu(Q^{\prime}) k}
t^{2 \nu(Q) k}
\ee
with the coefficients $c^{Q}_{[1]}$ being determined as unique solutions to the
system of
linear equations
\be
M_{[1]}^m=p_{1}^m=\sum_{|Q|=m} c^{Q}_{[1]} M_{Q}(p_{k})
\ee
They are given explicitly by
\be
c^{Q}_{[1]}= M_{Q}(\delta_{k,1})\,|Q|!\dfrac{(1-q^2)^{|Q|}}{(1-t^2)^{|Q|}}
\prod\limits_{(i,j)\in Q}
 \frac{1-t^2 t^{2(Q_{j}^{\prime}-i)} q^{2(Q_{i}-j) }}{1-q^2
t^{2(Q_{j}^{\prime}-i)}
 q^{2(Q_{i}-j) }}
\ee
where $M_{Q}(\delta_{k,1})$ is the value of the MacDonald polynomial at the
point
$p_{k}=\delta_{k,1}$.

\subsection{Non-torus knots\label{non-torus}}

Another interesting set of problems concerns
non-torus links.
Our cut-and-join evolution actually describes
what happens when one glues torus braids
${\cal R}_m$ to {\it any} "initial" braid ${\cal B}_m$.
This is true, at least, in the HOMFLY case, when the
quantum $R$-matrix is well defined and well known;
however, as we demonstrate in this paper, it looks
like everything works if one assumes that the same
remains true after the $\beta$-deformation.
This means that if one knows the superpolynomial
\be
\Tr_{R^{\otimes m}} {\cal B}_m =
\sum_{Q} b^Q M_Q^*,
\ee
then one also knows
\be
\Tr_{R^{\otimes m}} {\cal B}_m ({\cal R}_m)^n=
\sum_{Q} b^Q \lambda_Q^n M_Q^*
\ee
i.e., starting from a known HOMFLY or superpolynomial
for some knot, one can reconstruct the same polynomials
for the entire series obtained by the cut-and-join evolution.
It turns out that sometimes this idea works,
but sometimes it fails, at least, partly.
For example, the evolution of the $3$-strand braids
converts the superpolynomial for the figure eight knot $4_1$
into polynomials, but with some coefficients negative. The first term of the
evolution series is HOMFLY equivalent to the composite knot $3_1\# 3_1$.

At the same time, for the next simple $3$-strand knot $5_2$
it produces polynomials with all coefficients
positive, which have chances to be superpolynomials
for some other knots. The first term of this
evolution series is HOMFLY equivalent to $10_{139}$,
and the corresponding superpolynomial coincides with the one in \cite{DGR}.

Speaking in more explicit terms, the HOMFLY series that includes $5_2$ and
$10_{139}$ was constructed in (\ref{Hnt}), while the corresponding series of
the
superpolynomials is
\be
\begin{array}{|c|}
\hline
\\
P^{5_2,n}_{[1]}=c_{[1]}^{[3]} M_{3}^{\ast} q^{-2n} + c_{[1]}^{[2,1]}
M_{2,1}^{\ast} q^{-\frac{2n}{3}} t^{ \frac{2n}{3}}+c_{[1]}^{[1,1,1]}
M_{1,1,1}^{\ast} t^{ 2n }
\\
\\
\hline
\end{array}
\ee
with the coefficients:
\be
c_{[1]}^{[3]}=1, \ \ \ c_{[1]}^{[2,1]}=-{\frac {q \left( -1+t \right)  \left(
t+1 \right)  \left( {q}^{6}{t}^
{6}-{t}^{4}{q}^{2}+{q}^{5}{t}^{3}+{q}^{3}{t}^{3}-{q}^{4}{t}^{2}+{q}^{3
}t+1 \right) }{t \left( -1+{q}^{2}t \right)  \left( {q}^{2}t+1
 \right) }}
\ee
\be
c_{[1]}^{[1,1,1]}={\frac {{q}^{6} \left( 1+{t}^{2} \right)  \left( t+{t}^{2}+1
\right)
 \left( {t}^{2}-t+1 \right)  \left( -1+t \right) ^{2} \left( t+1
 \right) ^{2}{t}^{2}}{ \left( -1+tq \right)  \left( -1+{t}^{2}q
 \right)  \left( {t}^{2}q+1 \right)  \left( tq+1 \right) }}
\ee
First two superpolynomials in the series are:

\noindent
\be
P^{5_2}_{[1]}= \dfrac{\{A\}}{t \{t\} A^2 q^3}\Big( -{q}^{6}{A}^{4}+ \left(
{q}^{6}{t}^{2}-{q}^{5}t+
{q}^{4} \right) {A}^{2}+{q}^{5}{t}^{3}-{q}^{4}{t}^{2}+{q}^{3}t
 \Big)
\ee

\noindent

$P^{5_2,3}_{[1]}=P^{10_{139}}=\dfrac{\{A\}t^2}{ \{t\} A^2q^6 }\Big(
(t^2q^8+q^6-q^7t)A^4+(-q^2-2q^6t^2-q^4-q^6t^4+q^5t-t^4q^8-q^4t^2-t^6q^8+q^7t^3)A^2+q^4t^2-q^5t^3+t^4q^4+q^6t^4+t^8q^8+q^2t^2+q^6t^6+1
\Big)$

\bigskip

\noindent
Note that, in variance with the torus knots, the result depends on odd powers
of $q$ and
the superpolynomial in terms of $(-A,q,t)$ has not only positive coefficients.
The situation
is improved by transition to the variables $({\bf A,q,t})$:
\be
P^{5_2}_{[1]}=
\bf{{\dfrac {1+{a}^{2}t}{ \left( -1+{q}^{2} \right) {a}^{3}{t}^{9/2}{q}^{3}
}}}
 \Big({t}^{8}{q}^{6}{a}^{4}+ ( {q}^{8}{t}^{7}+{t}^{5}{q}^{4}+{q}^{6}{t
}^{6} ) {a}^{2}+{t}^{5}{q}^{8}+{q}^{6}{t}^{4}+{t}^{3}{q}^{4}
 \Big)
\ee

\noindent
$
P^{5_2,3}_{[1]}=P^{10_{139}}_{[1]} =\bf{{\dfrac {1+{a}^{2}t}{ \left( -1+{q}^{2}
\right) {a}^{3}{t}^{15/2}{q}^{3}
}}} \Big(
(t^{10}q^{10}+q^6t^8+t^9q^8)a^4+(t^3q^2+2q^8t^7+t^5q^4+q^{10}t^7+q^6t^6+t^9q^{12}+q^6t^5+t^9q^{14}+q^{10}t^8)a^2+q^6t^4+t^5q^8+t^4q^8+t^6q^{10}+q^{16}t^8+q^4t^2+q^{12}t^6+1
\Big)
$

\bigskip

\noindent
The coefficients here are all positive.

One can repeat the same procedure for the figure eight knot and construct a
series
\be
\boxed{
P^{4_1,n}_{[1]}=c^{[3]}_{[1]} M_{[2]}^{\ast} q^{-2n}+c^{[2,1]}_{[1]}
M_{[2,1,1]}^{\ast} q^{\frac{-2n}{3}}t^{\frac{2n}{3}}+ c^{[1,1,1]}_{[1]}
M_{[1,1,1,1]}^{\ast} t^{2n}
}
\ee
with the coefficients:
\be
c^{[3]}_{[1]}=tq\ \ \
c^{[2,1]}_{[1]}=\frac{(t-1)(1+t)(t^6q^6-t^5q^5-t^4q^2-t^3q^3-q^4t^2-tq+1)}{(-1+q^2t)(1+q^2t)t^2}
\ \\ c^{[1,1,1]}_{[1]} =
\frac{q^3(1+t^2)(t^2+t+1)(t^2-t+1)(t-1)^2(1+t)^2}{t(-1+t^2q)(1+t^2q)(-1+tq)(1+tq)}
\ee
In this case, however, even using the variables $({\bf A,q,t})$ does not make
the polynomial
positive:
\be
\begin{array}{l}
P^{4_1,0}_{[1]}=\dfrac{\{A\}}{\{t\}A^2}\Big(q^2A^4-q^2t^2A^2+tqA^2-A^2+
t^2\Big)
=-\dfrac{\{\bf{a}\}}{\{\bf{q}^2\}a^2t}\Big(\bf{q}^2t^4a^4+ta^2+q^2t^2a^2+q^4t^3a^2+q^2\Big)
   \\ \\
P^{4_1,3}_{[1]}=\dfrac{\{A\}t^2}{\{t\}A^2q^5}\Big(q^5A^4-t^2q^5A^2-t^3q^4A^2-tq^2A^2-q^3A^2+t^2q^3+t^3q^2+t^5q^4+t\Big)\\=-\dfrac{\{\bf{a}\}}{\{\bf{q}^2\}a^2t^6q^6}\Big(\bf{t}^7q^4a^4+q^6t^6a^2-q^6t^5a^2+q^2t^4a^2-q^2t^3a^2-t^4q^8+t^3q^4-q^4t^2-1\Big)
 \\ \\
\end{array}
\ee

\section{Reductions of superpolynomials}

The superpolynomial depends on three parameters, therefore, there are a lot of
various reductions to simpler polynomials. For instance, the HOMFLY polynomial
is obtained by putting $t=q$ or, which is the same, ${\bf t}=-1$. The
HOMFLY polynomial can be further reduced to the Jones polynomial ($t=q$,
$A=q^2$ or
${\bf t}=-1$, ${\bf a}={\bf q}^2$),
Alexander polynomial ($t=q$, $A=1$ or
${\bf t}=-1$, ${\bf a}=1$) or special polynomial ($q=t=1$ or ${\bf t}={\bf
-q}=-1$).
At last, there is yet another
important reduction of the superpolynomial, the Heegard-Floer polynomial, which
is
described below.

\subsection{Alexander polynomial}

The choice of $t=q$ and $A=1$ corresponds to $N=0$ and is
known as the Alexander polynomial
\be
A_{R}(q)= \lim_{a\rightarrow 1} \frac{{H}_R(a|\,q)}{s_R^*}
\ee
This object is interesting, because it also appears
in many other branches of theory and this can be
used for the study of dualities between Chern-Simons theory
and other theories.
In this paper we exploit in the next subsection one of such links:
between the torus knots  and the singularity theory of
Riemann surfaces, relating $T[m,n]$ link and the
complex curve $x^m = y^n$, which allows one to
construct explicitly the Heegard-Floer polynomial in the fundamental
representation,
which is a certain reduction of the superpolynomial at
${\bf a} = 1/{\bf t}$, directly from the Alexander
polynomial.
This is important, because the Alexander polynomial can be
directly read from (\ref{Wrepfund})
in the general form for arbitrary $T[m,n]$.
Indeed, at $A=1$ the ratio
\be
\lim_{A=1}\frac{s^*_{[m-j,1^j]}}{s_1^*}
= [1]_q\lim_{n=0} \frac{[n-j]_q[n-1]_q\ldots[n-1]_q[n+1]_q\ldots [n+m-j-1]_q}
{[j]_q![m-j-1]_q![m]_q} = (-)^j\frac{[1]_q}{[m]_q}
\ee
so that the HOMFLY polynomial (\ref{Wrepfund}) in this limit is equal to
\be
\frac{[1]_q}{[m]_q}q^{-n(m-1)}\sum_{j=0}^{m-1}q^{2nj}
= q^{-n(m-1)}\frac{[1]_q}{[m]_q}\frac{1-q^{2mn}}{1-q^{2n}}
= q^{-(n-1)(m-1)}\frac{(1-q^2)(1-q^{2mn})}{(1-q^{2m})(1-q^{2n})}
\ee
To simplify formulas, from now on we change the normalization of the Alexander
polynomial so that
\be
\boxed{
A_{[1]}^{[m,n]} = \frac{(1-q^2)(1-q^{2mn})}{(1-q^{2m})(1-q^{2n})}}
\ee

\subsection{Heegard-Floer polynomials
\label{Flo}}

In order to construct the Heegard-Floer polynomials, one has first
to omit all pairs of monomials of the superpolynomial, differing by a factor
${\bf a}^2{\bf t}^3$ and obtain the
{\it reduced} superpolynomial (more accurately this is done by calculating
homologies of the operator ${\bf d}_0$ with grading $(-2,0,-3)$ w.r.t. $({\bf
a,q,t})$
\cite{DGR}).
The Heegard-Floer polynomial is then obtained from the {\it reduced}
superpolynomial
by putting ${\bf a}=1/{\bf t}$ practically in the same way as the
Alexander polynomial is obtained from the HOMFLY polynomial:
\be
F_R({\bf q}|{\bf t}) = P^{red}_R({\bf a}={\bf t}^{-1}|{\bf q}|{\bf t})
\label{Fdef}
\ee
What is important, at least, for the fundamental representation and
for the algebraic knots (for the torus knots $T[m,n]$ in particular)
a way is known \cite{EGth}
to construct $F_R$ from the Alexander polynomial,
and, thus, from the HOMFLY polynomial.
The method is based on relation to the singularity theory
of degenerate Riemann surfaces.

Take the Alexander polynomial $A_R(q) = H_R(a=1,q)$.
Then define the Poincare polynomial
\be
\frac{A_R(q)}{1-q^2} = \sum_{i=0}^\infty  q^{2c_i}
\ee
with $c_{i+1}\geq c_i$ being an infinite series.\footnote{
This sequence can be alternatively described
in pure combinatorics terms.
For mutually prime $m,n$ consider the set
of all integers $am+bn$ with non-negative integers $a,b$
and make out of it an ordered sequence $\{c_0,c_1,c_2,\ldots\}$
with $c_i = a_im+b_in$, such that $c_{i+1} > c_i$.
If two combinations coincide, $am+bn = a'm+b'n$, then
it is counted only once.
}
Then, define the reduced Poincare polynomial
\be
(1-Q^2T^2)\sum_{i=0}^\infty  Q^{2c_i}T^{2i}
\ee
This is already a finite-order polynomial.
Next convert all entries with the positive and negative coefficients
in the following way:
\be
\sigma: \ \ \ \ Q^{2a}T^{2b} \longrightarrow {\bf q}^{-2a}{\bf t}^{-2b},
\ \ \ \ \ \ \ \ \
- Q^{2a}T^{2b} \longrightarrow {\bf q}^{-2a}{\bf t}^{1-2b}
\ee
By definition we obtain a polynomial with all coefficients positive
(divided by some power of ${\bf q}$ and ${\bf t}$).
This polynomial coincides with (\ref{Fdef}).

Thus, one has a possibility of checking the superpolynomial by reducing it to
the
Heegard-Floer polynomial, calculating this latter independently via the reduced
Poincare
polynomial and then comparing the results.

\subsection*{Examples:}

\paragraph{Fundamental representation for the trefoil.}
\be
P_1^{[2,3]} = {\bf a^2}{\bf q^2}{\bf t^3} + 1 + {\bf q}^4{\bf t}^2
\ee
The reduced superpolynomial in this case is just the same. Thus,
\be
F_1^{[2,3]} = 1+{\bf q}^2{\bf t} + {\bf q}^4{\bf t}^2
\ee
On the other hand,
\be
H_1^{[2,3]} = -{\bf a}^2{\bf q}^2 + 1 + {\bf q}^4
\ee
and
\be
A_1^{[2,3]} =  1 - {\bf q}^2 + {\bf q}^4
\ee
Next steps convert the Alexander polynomial into
\be
1 + \frac{{\bf q}^4}{1-{\bf q}^2} \ \longrightarrow \ \
(1-Q^2T^2) \left( 1 + \frac{Q^4T^2}{1-Q^2T^2}\right)
= 1- Q^2T^2  + Q^4T^2 \ \longrightarrow \nn \\ \longrightarrow \
1 + \frac{1}{{\bf q}^2{\bf t}} + \frac{1}{{\bf q}^4{\bf t}^2}
\sim {\bf q}^4{\bf t}^2 + {\bf q}^2{\bf t} +1 =F_1^{[2,3]}
\ee

\paragraph{Fundamental representation for $T[3,4]$.}
\be
P_1^{[3,4]} =
\underline{{\bf a}^4{\bf q^6}{\bf t}^8} + {\bf a}^2
\Big({\bf q}^{10}{\bf t}^7 + \underline{\underline{{\bf q}^8{\bf t}^7}}
+ \underline{{\bf q}^6{\bf t}^5}
+ \underline{\underline{\underline{{\bf q}^4{\bf t}^5}}} + {\bf q}^2{\bf
t}^3\Big)
+ \Big({\bf q}^{12}{\bf t}^6 + \underline{\underline{{{\bf q}^8{\bf t}^4}}}
+ {\bf q}^6{\bf t}^4
+ \underline{\underline{\underline{{\bf q}^4{\bf t}^2}}} + 1\Big)
\ee
Underlined are the terms eliminated by reduction. Therefore,
\be
P^{red} = {\bf a}^2
\Big({\bf q}^{10}{\bf t}^7 + {\bf q}^2{\bf t}^3\Big)
+ \Big({\bf q}^{12}{\bf t}^6 + {\bf q}^6{\bf t}^4+1\Big)
\ee
and
\be
F_1^{[3,4]} = {\bf q}^{12}{\bf t}^6 +{\bf q}^{10}{\bf t}^5 + {\bf q}^6{\bf
t}^4+ {\bf q}^2{\bf t}+1
\ee
On the other hand, from the same  $P_1^{[3,4]}$ one obtains
\be
H_1^{[3,4]}
 = a^4q^6 - a^2(q^{10} + q^8 + q^6 + q^4 +q^2) + (q^{12}+q^8 + q^6 +q^4 +1),
\nn\\
A_1^{[3,4]} = q^{12} - q^{10} + q^6 -q^2 + 1 = q^6(q^6 - q^4 + 1 - q^{-4}
+q^{-6})
\ \longrightarrow \
1+ q^6 + q^8 + \frac{q^{12}}{1-q^2} \ \longrightarrow \ \nn \\ \longrightarrow
\
(1-Q^2T^2)(1+Q^6T^2 + Q^8T^4) + Q^{12}T^6 = 1- Q^2T^2+Q^6T^2 - Q^{10}T^6
+Q^{12}T^6
\ \longrightarrow \ \nn \\ \longrightarrow \
1 + \frac{1}{{\bf q}^2{\bf t}} + \underline{\frac{1}{{\bf q}^6{\bf t}^2}} +
\frac{1}{{\bf q}^{10}{\bf t}^5} + \frac{1}{{\bf q}^{12}{\bf t}^6} =
\frac{1}{{\bf q}^{12}{\bf t}^6}F_1^{[3,4]}
\ee
The underlined term in the last transition demonstrates the need to invert
the powers of ${\bf q}$ and ${\bf t}$ (the {\it set} of all other terms is
left
intact by this inversion).

\subsection{Special polynomial and Catalan numbers}

The limit of $q=1$ is singular from the point
of view of the Chern-Simons theory. It is actually the genus zero
limit, with $N\rightarrow \infty$, $\hbar \rightarrow 0$,
t/Hooft coupling $N\hbar = \log a$ fixed.
As usual in this limit much is simplified,
and some non-trivial properties get revealed.
We call this limit of the HOMFLY polynomial
the "special" polynomial
\be
S_R(a) = \lim_{q\rightarrow 1} \frac{{H}_R(A|\,q)}{s_R^*}
\ee
One of its remarkable properties,
at least, for the torus knots,
is a very simple dependence on the representation variable $R$:
\be
S_R(a) = (S_{[1]}(a))^{|R|}
\label{RdepS}
\ee
Among other things this immediately implies
integrability of the knot theory in the genus zero limit:
$d_RS_R(a)$ satisfy the Pl\"ucker relations
\cite{KnotInt}.
Like the Alexander polynomial, $S_1$ can be written down
explicitly for the arbitrary $T[m,n]$:
\be\label{special}
\boxed{
S_{[1]}^{[m,n]}(a) =\sum_{i=0}^{m-1}
(-1)^{i}\dfrac{(m+n-i-1)!}{ m\cdot n\cdot i!\,(m-i-1)!\,(n-i-1)!}a^{2i}}
\ee
For example,
\be
S_{[1]}^{[2,2k+1]}(a) = -a^2k + (k+1),\nn \\
S_{[1]}^{[3,3k+1]}(a)={k(3k-1)\over 2}a^4-k(3k+2)a^2+{(k+1)(3k+2)\over 2}\\
S_{[1]}^{[3,3k+2]}(a)={k(3k+1)\over 2}a^4-(k+1)(3k+1)a^2+{(k+1)(3k+4)\over
2}\nn
\ee
The coefficients in this expression have a remarkable
dual interpretation: they count paths of some special types
on rectangular $2d$ lattice.
In particular, for $n=m+1$ the free coefficient
is the Catalan number:
\be
S_{[1]}^{[m,m+1]}(a=0) = H_1^{[m,m+1]}(a=0|q=1)
= Cat_m={ (2m)!\over m!(m+1)!}
\ee
This duality implies that in general
the superpolynomial $P_1(a|q|t)$ should be a polynomial of $a$
with coefficients, given by $(q,t)$-weighted sums over the
same paths (sometime called $(q,t)$-Catalan numbers).
The explicit statement is known at least for the $n=m+1$ case
\cite{EG}, we used it for testing
our general formulas for the torus knot superpolynomials.

\subsection{From Catalan numbers to superpolynomials\label{paths}}

Now we describe how to extend the dual language of sums over paths to the
superpolynomial case.
First of all, note that the HOMFLY polynomial for the torus knot $T_{m,n}$ (but
not for a link)
with coprime $m$ and $n$ can be written as
\begin{equation}
\label{113}
{\cal H}^{[m,n]}_{[1]}(q)=\sum_{i=0}^{m-1}(-1)^{i}
q^{(m-1)(n-1)}\dfrac{[m+n-i-1]_{q}!}{ [m]_q[n]_q[i]_q!
[m-i-1]_q![n-i-1]_q!}a^{2i}
\end{equation}
i.e. at $q=1$ this formula, indeed, gives (\ref{special}). This formula is
looking as simple as
(\ref{Wrepfund}), however, these two are related via a non-trivial resummation.
The particular
case of the formula at $n=m+1$ is
\begin{equation}
\label{HOMFLY}
{\cal H}^{[m,m+1]}_{[1]}(q)=\sum_{i=0}^{m-1}(-1)^{i}
q^{m(m-1)}\dfrac{[2m-i]_{q}!}{ [m+1]_q[m]_q[i]_q!
[m-i]_q![m-i-1]_q!}a^{2i}
\end{equation}
Now we are going
to rewrite this formula as a sum over paths in the generalized case of
superpolynomial,
following \cite{EG} and private communications with E.Gorsky.

To this end, we define \textit{the $m$-Dyck path} as a lattice path from
$(0,0)$ to $(m,m)$
(i.e. a line from $(0,0)$ to $(m,m)$)
which consists of steps $E=(1,0)$ and $N=(0,1)$ which never go below the
diagonal.
This path can be represented as a word that consists of $E$ and $N$, with $m$
instances of $E$
and $m$ instances of $N$.

Now \textit{the Schr\"oder path} is defined similarly, allowing any number of
$D=(1,1)$ steps in addition to $E$ and $N$ steps. $D$-steps can lie on the
diagonal, but the
path should not go below it. If there are $d$ $D$-steps, then there should be
$m-d$ $N$-steps
and $m-d$ $E$-steps. Denote the set of all diagonal steps of the Schr\"oder
path $\pi$
as $\mathcal{D}(\pi)$.

Let us denote as $\check{\mathcal{S}}_n^i$ the set of all Schr\"oder paths with
$i$ diagonal
steps inside the $m\times m$ square for which no $D$-step lies below the lowest
$E$-step.

Then the number of elements $D_{m}^i := |\check{\mathcal{S}}_m^i|$ of this set
is equal
to
\eq{
D_{m}^i=\dfrac{(2m-i)!}{ (m+1)\cdot m\cdot i!\,(m-i)!\,(m-i-1)!}
}

Let the area $S(\pi)$ of the Schr\"oder path $\pi$ inside the $m \times m$
square be defined
as the usual area of the region inside this $m\times m$ square which lies above
this path.

The following procedure defines the so-called \textit{bounce} of a path inside
the $m\times m$
square. First, we define it for the Dyck path $\pi^{(0)}$.

Let $\pi^{(0)}$ be a Dyck path. Consider a ball starting at point $(m,m)$ and
rolling westward.
Let it change the direction of its movement to the southward one upon touching
a vertical
line of the path, and again to the westward one upon touching the diagonal of
the square.
Then, the path of the ball intersects the diagonal at certain points
\eqn{(j_1,j_1),\;(j_2,j_2),\;\dots}
The bounce is defined as the sum
\eq{b(\pi^{(0)}) := j_1+j_2+\dots}

Now for the Schr\"oder path $\pi^{(i)}$ with $i$ diagonal steps we define the
bounce in the
following way. Let $T(\pi^{(i)})$ be the Dyck path resulting from removal of
all diagonal steps
from the path word. This path will be a path inside the $(m-i)\times (m-i)$
square.

Consider a bounce path for the $T(\pi^{(i)})$ Dyck path as described above.
Denote by
$V(T(\pi^{(i)}))$ the set of all vertical steps of the path $T(\pi^{(i)})$ from
which the ball
bounced (i.e. changed its direction from the westward to the southward one upon
touching
vertical lines corresponding to these steps). Denote by
$T^{-1}(V(T(\pi^{(i)})))$ the set of all
vertical steps of the $\pi^{(i)}$ path which project to steps in
$V(T(\pi^{(i)}))$ upon
applying the $T$ operation. Now for each diagonal step $x$ of the path
$\pi^{(i)}$ define
the function $\nu(x)$ as the number of steps from the set
$T^{-1}(V(T(\pi^{(i)})))$ which lie
to the east of $x$.

Then the bounce for $\pi^{(i)}$ is defined in the following way:
\eq{
b(\pi^{(i)}) := b(T(\pi^{(i)}))+\mathop{\sum}_{x\in
\mathcal{D}(\pi^{(i)})}\nu(x)
}

An example of the Schr\"oder path and the corresponding Dyck path is given in
Fig.\ref{fig:paths}.
\def\xsh{210pt}
\def\ysh{30pt}
\begin{figure}[h!]
\begin{center}
\begin{tikzpicture}
\draw[step=30pt,lightgray,very thin,dashed] (0,0) grid (180pt,180pt);
\draw[lightgray,very thin,dashed] (0,0) --  (180pt,180pt);
\draw[red] (0,0) --  (0pt,30pt) -- (30pt,30pt) -- (30pt,60pt) -- (60pt,90pt)--
(60pt,120pt)-- (90pt,120pt)-- (120pt,150pt)-- (120pt,180pt)-- (180pt,180pt);
\draw[blue,very thick] (30pt,30pt) -- (30pt,60pt);
\draw[blue,very thick] (120pt,150pt)-- (120pt,180pt);

\draw[xshift=\xsh, yshift=\ysh, step=30pt,lightgray,very thin,dashed] (0,0)
grid (120pt,120pt);
\draw[xshift=\xsh, yshift=\ysh, lightgray,very thin,dashed] (0,0) -- 
(120pt,120pt);
\draw[xshift=\xsh, yshift=\ysh, red] (0,0) --  (0pt,30pt) -- (30pt,30pt) --
(30pt,90pt)-- (60pt,90pt)-- (60pt,120pt)-- (120pt,120pt);
\draw[xshift=\xsh, yshift=\ysh, blue,very thick] (30pt,30pt) -- (30pt,60pt);
\draw[xshift=\xsh, yshift=\ysh, blue,very thick] (60pt,90pt)-- (60pt,120pt);
\draw[xshift=\xsh, yshift=\ysh, black,dashdotted,very thick] (3pt,3pt) -- 
(3pt,27pt) -- (33pt,27pt) -- (33pt,57pt)-- (63pt,57pt)-- (63pt,117pt)--
(117pt,117pt);
\end{tikzpicture}
\end{center}
\caption{The Schr\"oder path $\pi$ (on the left) and the corresponding Dyck
path $T(\pi)$
(on the right) drawn in red, with steps from the sets $T^{-1}(V(T(\pi)))$ and
$V(T(\pi))$
drawn in thick blue, and the bounce path corresponding to $T(\pi)$ drawn in
dashdotted black.
$S(\pi)=12$, $b(\pi)=5$}
\label{fig:paths}
\end{figure}
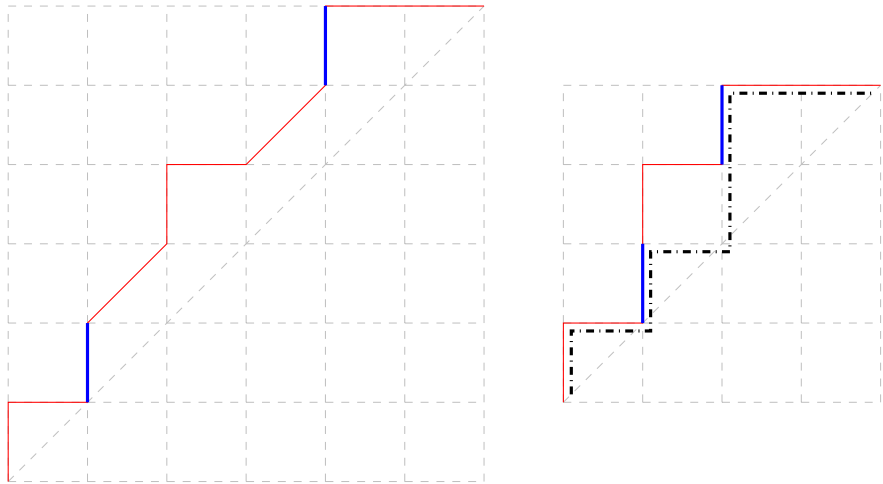

Now define $\sigma_{m}^i(\qb,\tb)$ as
\eq{
\label{dexpr}
\sigma_{m}^i(\qb,\tb) := \mathop{\sum}_{\pi\in \check{\mathcal{S}}_m^i} \qb^{2
S(\pi) + 2
b(\pi) -i}\; \tb^{2 S(\pi)}
}
Now one can formulate a conjecture
for the superpolynomial for torus knot $T[m,m+1]$:
\eq{
{\cal P}_{[1]}^{[m,m+1]}(\ab,\qb,\tb)=\summ{i=0}{m-1} \ab^{2i}
\sigma_{m}^i(\qb,\tb)
}

We computed this expression for a number of values of $m$ and it turned out to
coincide with
the formulas in the Tables.

We would also like to describe briefly another construction of this kind,
proposed by E.Gorsky in the Appendix of \cite{ORSG}. It is based on a
generalization of
statistics on the paths known as the ``dinv'' statistics, which is in a sense
dual to the bounce
statistics. This construction conjecturally works for any torus link $T[m,n]$
with arbitrary
$m$ and $n$.

First note that the Dyck path can be identified with the Young diagram by
including all the boxes in
the square which lie above the path. The (generalization of) dinv statistics is
defined in
terms of the Young diagrams, and we use this terminology.

To be more exact, let $D$ be a Young diagram. In what follows we assume $|D|\in
\mathbb{Z}_{\geq 0}$ to be its norm
(the number of boxes) and $|D|=d_1+\dots+d_{l(D)},\; d_1\geq\dots\geq
d_{l(D)},$ to be the
corresponding integer partition. Here $l(D)$ is the number of lines of the
diagram.

Fix $m$ and $n$. Then, let $\mathcal{D}^{m,n}$ be the set of all Young diagrams
$D$,
$|D|=d_1+\dots+d_{l(D)}$, (including the empty one) such that
\be
\dfrac{d_i}{m}+\dfrac{i}{n} \leq 1,\quad i=1\ldots l(D)
\ee

Choose the box $(x,y)$ inside the Young Diagram $D$,
($1\leq y \leq {l(D)}$ and $1\leq x \leq d_y$) and define the \textit{arm} and
the
\textit{leg} respectively as
\be
\mathfrak{a}((x,y)) = d_y - x,
\ee
\be
\mathfrak{l}((x,y)) = \tilde{d}_x - y
\ee
where the tilde refers to the transposed Young diagram.
Then define
\be
h^+_{\frac{m}{n}} (D) := \left|\left\{ (x,y) \in D \bigg|
\dfrac{\mathfrak{a}((x,y))}{\mathfrak{l}((x,y))+1} \leq \dfrac{m}{n} <
\dfrac{\mathfrak{a}((x,y))+1}{\mathfrak{l}((x,y))} \right\}\right|,
\ee
where $||$ stands for the number of elements in the set. This $h^+$ is the
mentioned
generalization of the dinv statistics.

For all $i$ s.t. $1\leq i\leq l(D)$ define $\alpha_i(D)$ as the number of such
$i',\; 1\leq i'\leq l(D),$ that
\be
\dfrac{d_i-d_{i'}}{m}>\dfrac{i'+1-i}{n}
\ee
and define $\gamma_i(D)$ as the number of such $i',\; 1\leq i'\leq l(D)+1,$
that
\be
\dfrac{d_i-d_{i'}}{m}>\dfrac{{i'}-i}{n},
\ee
where we assume that $d_{l(D)+1}=0$. Then, defining
\be
\beta_i(D) := \gamma_i(D) - \alpha_i(D)
\ee
one can formulate the conjecture for the $T[m,n]$ superpolynomial in the
fundamental
representation:
\eq{
{\cal P}_{[1]}^{[m,n]}(\ab,\qb,\tb)=\mathop{\sum}_{D\in\mathcal{D}^{m,n}}
\qb^{2(|D|+h^+_{\frac{m}{n}}(D))}\tb^{2|D|}\prodd{i=1}{l(D)}\br{1+\ab^2\qb^{-2\beta_i(D)}\tb}
}
This was also tested for a number of values of $m$ and $n$ and it turned out to
coincide
with formulas from the Tables.

\section{Conclusion}

To conclude, we formulated a very simple and practical
prescription for recursive evaluation of the superpolynomial
$\tilde P_R({\bf a}|{\bf q}|{\bf t})$ for all torus links $T[m,n]$
(they are single knots when $m$ and $n$ are mutually prime) and sometimes even
non-torus.
That is, we introduce the split $W$-evolution which describes
knot invariants in terms of vectors in the linear space
of MacDonald polynomials $M_R\{p\}$, which is, at the end of the day,
projected onto the smaller subspace of MacDonald dimensions $M_R^*$.
The prescription is a straightforward modification of the known
generic expression (\ref{Wrep}) for the HOMFLY polynomials,
which is an example of the $W$-representation,
and is absolutely explicit: see, for example, (\ref{Wrepfund}),
where the r.h.s. contains only the quantum dimensions of $SU(N)$
representations.
The only difference in the superpolynomial case is that one
should substitute the ordinary quantum dimensions
by the explicitly known MacDonald dimension and deform the
expansion coefficients in (\ref{Wrep}).
The rule is that the HOMFLY polynomial
\be
K_R(a|q) =
\sum_{Q\vdash |R|} q^{-2n\varkappa_Q/m}C_R^Q s_Q^*
\ee
is substituted by the superpolynomial
\be
P_R(A|q|t) =
\sum_{Q\vdash |R|} q^{-2n\nu(Q')/m}t^{2\nu(Q)/m}c_R^Q
M_Q^*
\label{PRpre}
\ee
The new coefficients $c_R^Q$, which are rational functions of $q$ and $t$,
are recursively defined from the initial condition
(\ref{boco}). These numbers depend
on the residue $p = n\ {\rm mod}(m)$, i.e. on the choice
of a particular series $T[m,mk+n_m]$ of the torus knots.
Finally, the conventional superpolynomials ${\cal P}_R({\bf a}|{\bf q}|{\bf
t})$
of \cite{DGR}
are obtained by the change of variables (\ref{GMid}),
suggested earlier in \cite{AK,ASh}
which describe somewhat parallel, but far more sophisticated
algorithms for constructing the superpolynomials.
Unfortunately, we have not yet managed to raise the initial condition from the
subspace
(\ref{pstar}) to the entire $p_k$-space. Therefore, it is insufficient to
determine all
$c^Q$ with $|Q|\ge 4$. The remaining coefficients are evaluated in a less
algorithmic way:
from the condition that the r.h.s. of (\ref{PRpre}) is indeed (proportional to)
a polynomial.

We listed some examples of the superpolynomials, some known, some new,
see the table below.
We also tested consistency with the construction of the Heegard-Floer
polynomials
which was recently suggested in \cite{EGth} and with
expressions for the $T[m,m+1]$ knots (and the fundamental
representation $R=[1]$) found in \cite{EG}.
For non-fundamental representations $R\ne [1^{|R|}]$ the final expressions
depend on the choice
of the superpolynomials for unknots which is still a subject of dispute in the
literature.
We address the issue in more detail elsewhere.

A conceptual meaning of the entire construction,
its relation to deformations of quantum $R$-matrices
(to dynamical $R$-matrices of the Ruijsenaars model,
underlying the theory of MacDonald functions),
to integrability, to $2d$ and $3d$ AGT relations
(including relation to the WZNW models already exploited in \cite{ASh}),
to matrix models and to their $W$-representations {\it a la}
\cite{Wreps}
will be discussed elsewhere.

\section*{Acknowledgements}

We are indebted to Ivan Cherednik, Eugene Gorsky, Sergei Gukov,
Alexei Oblomkov and Shamil Shakirov for very useful and inspiring
communications.

Our work is partly supported by Ministry of Education and Science of
the Russian Federation under contract 14.740.11.0081 (P.DB., A.Mir., A.Mor.,
A.Sl.)
and 14.740.11.0347 (A.S.), by RFBR
grants 10-02-00509 (A.Mir., A.Sl.), 10-02-00499 (A.Mor., P.DB.) and 09-02-00393
(A.S.),
by joint grants 11-02-90453-Ukr, 09-02-93105-CNRSL, 09-02-91005-ANF,
10-02-92109-Yaf-a, 11-01-92612-Royal Society.

\newpage



\setcounter{section}{0}

\newpage

\section*{Tables of superpolynomials/power series for torus
knots/links\label{examples}}

\vspace{1cm}

Through the tables, $P$ denotes unnormalized superpolynomial/power
series, while ${\cal P}$ is normalized by the unknot. Hence, $P$ is never
a polynomial, and ${\cal P}$ is a polynomial (in the case of the torus knot)
or a power series (in the case of the torus link) with positive coefficients.

\vspace{1cm}

\contentsline {section}{\numberline {1}Unknots}{25}
\contentsline {section}{\numberline {2}Torus knots: fundamental
representations}{25}
\contentsline {subsection}{\numberline {2.1}Case $(2,n)$, series $n=2 k$
fundamental representation}{25}
\contentsline {subsection}{\numberline {2.2}Case $(2,n)$, series $n=2 k+1$
fundamental representation}{26}
\contentsline {subsection}{\numberline {2.3}Case $(3,n)$, series $n=3 k$
fundamental representation}{28}
\contentsline {subsection}{\numberline {2.4}Case $(3,n)$, series $n=3 k+1$
fundamental representation}{30}
\contentsline {subsection}{\numberline {2.5}Case $(3,n)$, series $n=3 k+2$
fundamental representation}{31}
\contentsline {subsection}{\numberline {2.6}Case $(4,n)$, $n=4 k$ fundamental
representation}{33}
\contentsline {subsection}{\numberline {2.7}Case $(4,n)$, $n=4 k+1$ fundamental
representation}{35}
\contentsline {subsection}{\numberline {2.8}Case (4,n), n=4k+2 fundamental
representation}{36}
\contentsline {subsection}{\numberline {2.9}Case $(4,n)$, $n=4 k+3$ fundamental
representation}{37}
\contentsline {subsection}{\numberline {2.10}Case $(5,n)$, $n=5 k$ fundamental
representation}{39}
\contentsline {subsection}{\numberline {2.11}Case $(5,n)$, $n=5 k+1$
fundamental representation}{40}
\contentsline {subsection}{\numberline {2.12}Case $(5,n)$, $n=5 k+2$
fundamental representation}{42}
\contentsline {subsection}{\numberline {2.13}Case $(5,n)$, $n=5 k+3$
fundamental representation}{43}
\contentsline {subsection}{\numberline {2.14}Case $(5,n)$, $n=5 k+4$
fundamental representation}{45}
\contentsline {subsection}{\numberline {2.15}Case $(6,n)$, $n=6 k+1$
fundamental representation}{47}
\contentsline {subsection}{\numberline {2.16}Case $(7,n)$, $n=7 k+1$
fundamental representation}{49}
\contentsline {subsection}{\numberline {2.17}Case $(7,n)$, $n=7 k+2$
fundamental representation}{52}
\contentsline {subsection}{\numberline {2.18}Case $(7,n)$, $n=7 k+3$
fundamental representation}{52}
\contentsline {subsection}{\numberline {2.19}Case $(7,n)$, $n=7 k+4$
fundamental representation}{53}
\contentsline {subsection}{\numberline {2.20}Case $(7,n)$, $n=7 k+5$
fundamental representation}{54}
\contentsline {subsection}{\numberline {2.21}Case $(7,n)$, $n=7 k+6$
fundamental representation}{54}
\contentsline {subsection}{\numberline {2.22}Case $(m,km)$, fundamental
representation}{55}
\contentsline {subsection}{\numberline {2.23}Case $(m,km+1)$, fundamental
representation}{57}
\contentsline {subsection}{\numberline {2.24}Case $(m,km-1)$, fundamental
representation}{57}
\contentsline {subsection}{\numberline {2.25}Cases $(m,mk+2), \ (m,mk+3), \
(m,mk+4)$, fundamental representation \cite {MMSS}}{58}
\contentsline {section}{\numberline {3}Torus knots: higher
representations}{58}
\contentsline {subsection}{\numberline {3.1}Case $(2,n)$, series $n=2 k+1$,
symmetric representation [2] }{58}
\contentsline {subsection}{\numberline {3.2}Case $(2,n)$, series $n=2 k+1$,
antisymmetric representation [1,1] }{60}
\contentsline {subsection}{\numberline {3.3}Case $(2,2k+1)$, all symmetric and
antisymmetric representation \cite {FGS}}{61}
\contentsline {subsection}{\numberline {3.4}Case $(2,n)$ series $n=2 k$,
representations [1]$\otimes $[1,1]}{62}
\contentsline {subsection}{\numberline {3.5}Case $(2,n)$ series $n=2 k$,
representations [1]$\otimes $[2]}{64}
\contentsline {subsection}{\numberline {3.6}Case $(2,n)$ series $n=2 k$,
representations [1,1]$\otimes $[1,1]}{65}
\contentsline {subsection}{\numberline {3.7}Case $(2,n)$ series $n=2 k$,
representations [2]$\otimes $[1,1]}{67}
\contentsline {subsection}{\numberline {3.8}Case $(2,n)$ series $n=2 k$,
representations [2]$\otimes $[2]}{69}
\contentsline {section}{\numberline {4}Non-torus knots}{70}
\contentsline {subsection}{\numberline {4.1}$5_2$ and descendants: fundamental
representation }{70}
\contentsline {subsection}{\numberline {4.2}Figure eight knot and descendants:
fundamental representation}{72}
\contentsline {subsubsection}{\numberline {4.2.1}Symmetric and antisymmetric
representations for figure eight knot \cite {IMMMfe}}{73}

\begin{small}
\section{Unknots\label{unkn}} \be
\begin{array}{l}
P^{Unknot}_{[1]}=M_{1}^{\ast}=\dfrac{A-A^{-1}}{t-t^{-1}}=
\bf{\dfrac {q(a^2t+1)}{at^{1/2} (1-{q}^{2}) }}
\\
 \\
P^{Unknot}_{[2]}=M_{2}^{\ast}=\dfrac{(A-A^{-1})(Aq-A^{-1}q^{-1})}{(t-t^{-1})(qt-q^{-1}t^{-1})}=
\bf{\dfrac {(a^2t+1)(a^2q^2t^3+1)}{a^2t^2 \{q^2t\}(1-{q}^{2})}}
\\
\\
P^{Unknot}_{[1,1]}=M_{11}^{\ast}=\dfrac{(A-A^{-1})(At^{-1}-A^{-1}t)}{(t-t^{-1})(t^2-t^{-2})}=
\bf{\dfrac {(a^2t+1)(a^2t+q^2)}{a^2t \{q^2\}(1-{q}^{2})}}
\\
\\
P^{Unknot}_{[3]}=M_{3}^{\ast}=\dfrac{(A-A^{-1})(Aq-A^{-1}q^{-1})(Aq^2-A^{-1}q^{-2})}
{(t-t^{-1})(qt-q^{-1}t^{-1})(q^2t-q^{-2}t^{-1})}=
\bf{\dfrac {(a^2t+1)(a^2t^3q^2+1)(a^2t^5q^4+1)}{a^3t^{9/2}q^2
\{q^3t^2\}\{q^2t\}(1-{q}^{2})}}
\\
\\
P^{Unknot}_{[2,1]}=M_{21}^{\ast}=\dfrac{(A-A^{-1})(At^{-1}-A^{-1}t)(Aq-
A^{-1}q^{-1})}
{(t-t^{-1})^2(qt^2-q^{-1}t^{-2})}=\bf{\dfrac
{(a^2t+1)(a^2t^3q^2+1)(a^2t+q^2)}{a^3t^{5/2}q \{q^3t^2\}\{q\}(1-q^2)}}
\\
\\
P^{Unknot}_{[1,1,1]}=M_{111}^{\ast}=
\dfrac{(A-A^{-1})(At^{-1}-A^{-1}t)(At^{-2}-A^{-1}t^2)}
{(t-t^{-1})(t^2-t^{-2})(t^3-t^{-3})}=
\bf{\dfrac {(a^2t+1)(a^2t+q^4)(a^2t+q^2)}{a^3t^{3/2}q^2
\{q^3\}\{q^2\}(1-q^2)}}
\\ \\
....
\end{array}
\ee
\end{small}

The r.h.s. of all these expressions being expanded into the power series in
${\bf q,t}$
are series with positive coefficients. This is always correct.

\section{Torus knots: fundamental representations}

\begin{small}
\subsection{Case $(2,n)$, series $n=2 k$ fundamental representation\label{2k}}
\be
\begin{array}{|c|}
\hline\\
P^{T[2,n]}_{[1]}=c^{[2]}_{[1]} M_{[2]}^{\ast} q^{-n}+ c^{[1,1]}_{[1]}
M_{[1,1]}^{\ast} t^{n}
\\
\\
\hline
\end{array}
\ee
with the coefficients:
\be
\boxed{
c^{[2]}_{[1]}=1,\ \ \ c^{[1,1]}_{[1]}=\frac{(1-q^2)(1+t^2)}{1-q^2 t^2}}
\ee
Similarly to the case of unknot, the obtained expressions for
$P^{T[2,n]}_{[1]}$ are not polynomials
with positive coefficients. Moreover, even ${\cal P}^{T[2,n]}_{[1]}$
(i.e. the expression normalized by unknot) is not a polynomial with positive
coefficients. However, it is a power series in ${\bf q,t}$ with positive
coefficients
as is seen from the examples below.

\be
\begin{array}{l}
P^{T[2,0]}_{[1]}=-\dfrac{\{A\} }{\{t\}^2 A} \Big(1-A^2\Big) =\bf{ -{\dfrac {
\left( {a}^{2}t+1 \right) {q}^{2}}{ \left( -1+{q}^{2}
 \right) ^{2}{a}^{2}t}}\Big( {a}^{2}t+1 \Big)= \dfrac{{a}^{2}t+1}{
-1+{q}^{2}}\cdot\dfrac{({a}^{2}t+1)q^2}{a^2t}\Big(\summ{j=0}{\infty}q^{2j}\Big)}
 \\
 \\
P^{T[2,2]}_{[1]}=-\dfrac{\{A\} }{\{t\}^2 A q^2}
\Big(-{t}^{2}+{q}^{2}{t}^{2}-{q}^{2}{A}^{2}+1\Big)=\bf{  -{\dfrac {{a}^{2}t+1}{
\left( -1+{q}^{2} \right) ^{2}{a}^{2}{t}^{3}}} \Big(
-{q}^{2}+{q}^{4}{t}^{2}+{q}^{2}{t}^{3}{a}^{2}+1 \Big)
}=\\=\bf{\dfrac{{a}^{2}t+1}{
-1+{q}^{2}}\Big(\dfrac{1}{a^2t^3}+q^2+\dfrac{({a}^{2}t+1)q^4}{a^2t}\summ{j=0}{\infty}q^{2j}\Big)}\\
\\
P^{T[2,4]}_{[1]}=-\dfrac{\{A\} }{\{t\}^2 A q^4} \Big( \left(
-{q}^{4}{t}^{2}-{q}^{2}+{q}^{2}{t}^{2} \right)
{A}^{2}-{t}^{4}{q}^{2}+{t}^{4}{q}^{4}+1+{q}^{2}{t}^{2}-{t}^{2}
 \Big)=\\
\bf{ -{\dfrac {{a}^{2}t+1}{ \left( -1+{q}^{2} \right) ^{2}{a}^{2}{t}^{5}{q}^
{2}}} \Big(\left( {q}^{6}{t}^{5}+{t}^{3}{q}^{2}-{t}^{3}{q}^{4} \right)
{a}^{2}-{q}^{6}{t}^{2}+{t}^{4}{q}^{8}+1+{q}^{4}{t}^{2}-{q}^{2}
\Big)}=\\=\bf{\dfrac{{a}^{2}t+1}{
-1+{q}^{2}}\Big(\dfrac{1}{a^2t^5q^2}+\dfrac{1}{t^2}+\dfrac{1}{a^2t^3}q^2+q^4+\dfrac{({a}^{2}t+1)q^6}{a^2t}\summ{j=0}{\infty}q^{2j}\Big)}
\\
\\
P^{T[2,6]}_{[1]}=-\dfrac{\{A\} }{\{t\}^2 A q^6} \Big(\left(
{t}^{4}{q}^{4}-{q}^{4}{t}^{2}-{q}^{6}{t}^{4}-{q}^{2}+{q}^{2}{t}^{2} \right)
{A}^{2}-{q}^{4}{t}^{6}+{q}^{6}{t}^{6}+1-{t}^{4}{q}^{2}+{
t}^{4}{q}^{4}+{q}^{2}{t}^{2}-{t}^{2}
\Big)=
\\
\bf{ -{\dfrac {{a}^{2}t+1}{ \left( -1+{q}^{2} \right) ^{2}{a}^{2}{t}^{7}{q}^
{4}}}
\Big(\left(
-{t}^{5}{q}^{8}+{q}^{6}{t}^{5}+{q}^{10}{t}^{7}+{t}^{3}{q}^{2}-{t}^{3}{q}^{4}
\right) {a}^{2}-{t}^{4}{q}^{10}+{q}^{12}{t}^{6}+1-{q}^{
6}{t}^{2}+{t}^{4}{q}^{8}+{q}^{4}{t}^{2}-{q}^{2} \Big)
}=\\=\bf{\dfrac{{a}^{2}t+1}{
-1+{q}^{2}}\Big(\dfrac{1}{q^4a^2t^7}+\dfrac{1}{q^2t^4}+\dfrac{1}{a^2t^5}+\dfrac{1}{t^2}q^2+\dfrac{1}{a^2t^3}q^4+q^6+\dfrac{({a}^{2}t+1)q^8}{a^2t}\summ{j=0}{\infty}q^{2j}\Big)}
\\
....\\
\boxed{P^{T[2,2k]}_{[1]}=\dfrac{\{A\} }{\{t\}^2 A
q^{2k}}{\cal{P}}^{T[2,2k]}_{[1]}=\bf{\dfrac{{a}^{2}t+1}{
-1+{q}^{2}}\dfrac{q^{2k+2}}{a^2t}\Big((1+a^2t^3q^2)\summ{j=1}{k}\dfrac{1}{q^{4j}t^{2j}}+({a}^{2}t+1)\summ{j=0}{\infty}q^{2j}\Big)}}
\end{array}
\ee
Our answer for the Hopf link $P^{[2,2]}$ is in full agreement with the known
results, for example  \cite{IGV} and generalize it to $m>2$.
\begin{itemize}
\item{HOMFLY case}
\end{itemize}
\be
\begin{array}{l}
H^{T[2,0]}_{[1]}=-\dfrac{\{A\} }{\{q\}^2 A} \Big( 1-A^2 \Big)  \\
H^{T[2,2]}_{[1]}=-\dfrac{\{A\} }{\{q\}^2 A q^2}
\Big(-{q}^{2}+{q}^{4}+1-{q}^{2}{A}^{2}\Big)\\
H^{T[2,4]}_{[1]}=-\dfrac{\{A\}}{\{q\}^2 A q^4}
\Big(-{q}^{6}+{q}^{8}-{q}^{6}{A}^{2}+{q}^{4}{A}^{2}+{q}^{4}-{q}^{2}+1-{q}^{
2}{A}^{2}\Big)\\
H^{T[2,6]}_{[1]}=-\dfrac{\{A\}}{\{q\}^2 A q^6}
\Big({q}^{12}-{q}^{10}-{q}^{10}{A}^{2}+{q}^{8}{A}^{2}+{q}^{8}-{q}^{6}-{q}^{
6}{A}^{2}+{q}^{4}{A}^{2}+{q}^{4}-{q}^{2}-{q}^{2}{A}^{2}+1
\Big)\\
....\\
\boxed{H^{T[2,2k]}_{[1]}=\dfrac{\{A\}}{\{q\}^2 A q^{2k}}
{\cal{H}}^{T[2,2k]}_{[1]}=q^{-2k}s_{[2]}^{*}+q^{2k}s_{[1,1]}^{*}}
\end{array}
\ee
\begin{itemize}
\item{Alexander case}
\end{itemize}
\be
\begin{array}{l}
A^{T[2,0]}_{[1]}=2\\
\\
A^{T[2,2]}_{[1]}={\textbf{q}}^{4}+1\\
\\
A^{T[2,4]}_{[1]}={\textbf{q}}^{8}+1\\
\\
A^{T[2,6]}_{[1]}={\textbf{q}}^{12}+1\\
....\\
\end{array}
\ee

\subsection{Case $(2,n)$, series $n=2 k+1$ fundamental
representation\label{2k+1}}
\be
\begin{array}{|c|}
\hline\\
P^{T[2,n]}_{[1]}=c^{[2]}_{[1]} M_{[2]}^{\ast} q^{-n}+ c^{[1,1]}_{[1]}
M_{[1,1]}^{\ast} t^{n}
\\
\\
\hline
\end{array}
\ee
with the coefficients:
\be
\boxed{
c^{[2]}_{[1]}=1,\ \ \ c^{[1,1]}_{[1]}=-\frac{1-t^4}{1-q^2 t^2}
\Big(\frac{q}{t}\Big)
=-{\{t^2\}\over \{qt\}}}
\ee
such that one obtains:
\be
\begin{array}{l}
P^{T[2,1]}_{[1]}=\dfrac{\{A\} t}{\{t\} A q} =\bf{\dfrac { \left( {a}^{2}t+1
\right) q}{ \left(-1+{q}^{2} \right) {a}^{
2}{t}^{2}}}\\
\\
P^{T[2,3]}_{[1]}=\dfrac{\{A\} t}{\{t\} A q^3}
\Big({q}^{2}(-A^{2})+{q}^{2}{t}^{2}+1\Big)=\bf{{\dfrac {{a}^{2}t+1}{ \left(
-1+{q}^{2} \right) {a}^{2}{t}^{4}q}} \Big(
{q}^{4}{t}^{2}+{q}^{2}{t}^{3}{a}^{2}+1 \Big)} \\
P^{T[2,5]}_{[1]}=\dfrac{\{A\} t}{\{t\} A q^5} \Big( \left(
{q}^{4}{t}^{2}+{q}^{2} \right) (-A^{2})+{t}^{4}{q}^{4}+{q}^{2}
{t}^{2}+1
\Big)=\bf{ {\dfrac {{a}^{2}t+1}{ \left( -1+{q}^{2} \right)
{a}^{2}{t}^{6}{q}^{3}}}\Big( \left( {q}^{6}{t}^{5}+{t}^{3}{q}^{2} \right)
{a}^{2}+{t}^{4}{q}^{8}+{q}^{4}{t}^{2}+1
\Big) }\\
\\
P^{T[2,7]}_{[1]}=\dfrac{\{A\} t}{\{t\} A q^7} \Big(\left(
q^6t^4+{q}^{4}{t}^{2}+{q}^{2} \right) (-A^{2})+q^6t^6+{t}^{4}{q}^{4}+{q}^{2}{t}^{2}+1
\Big)=\\
\bf{  {\dfrac {{a}^{2}t+1}{ \left( -1+{q}^{2} \right) {a}^{2}{t}^{8}{q}^{5}}}
\Big( \left( {q}^{10}{t}^{7}+{q}^{6}{t}^{5}+{t}^{3}{q}^{2} \right)
{a}^{2}+{q}^{12}{t}^{6}+{t}^{4}{q}^{8}+{q}^{4}{t}^{2}+1
\Big) \bf}\\
\\
P^{T[2,9]}_{[1]}=\dfrac{\{A\} t}{\{t\} A q^9}
\Big(({t}^{6}{q}^{8}+{q}^{6}{t}^{4}+{q}^{4}{t}^{2}+{q}^{2})(-A^2)+{t}^{8}{q}^{8}+{q}^{6}{t}^{6}+{t}^{4}{q}^{4}+{q}^{2}{t}^{2}+1
\Big)=\\
\bf{{\dfrac {{a}^{2}t+1}{ \left( -1+{q}^{2} \right) {a}^{2}{t}^{10}{q}^{7}}
}\Big( \left( {q}^{14}{t}^{9}+{q}^{10}{t}^{7}+{q}^{6}{t}^{5}+{t}^{3}{q}^{2}
\right) {a}^{2}+{q}^{16}{t}^{8}+{q}^{12}{t}^{6}+{t}^{4}{q}^{8}+{q}^{4
}{t}^{2}+1
 \Big)}
\\
....\\
\boxed{P^{T[2,2k+1]}_{[1]}=\dfrac{\{A\} t}{\{t\} A
q^{2k+1}}{\cal{P}}^{T[2,2k+1]}_{[1]}}
\end{array}
\ee
one can check that the normalized polynomial ${\cal{P}}_{2,2k+1}$ here coincide
exactly with one obtained in \cite{DGR}.
\begin{itemize}
\item{HOMFLY case}
\end{itemize}
At the point $t=q$ one has:
\be
H^{T[2,n]}_{[1]}=s_{[2]}^{\ast} q^{-n}- s_{[1,1]}^{\ast} q^{n}
\ee
Several first answers are:
\be
\begin{array}{l}
H^{T[2,1]}_{[1]}=\dfrac{\{A\}}{\{q\} A} \\
H^{T[2,3]}_{[1]}=\dfrac{\{A\} }{\{q\} A q^2} \Big( {q}^{4}-{q}^{2}{A}^{2}+1
\Big)\\
H^{T[2,5]}_{[1]}=\dfrac{\{A\} }{\{q\} A q^4} \Big( 
{q}^{8}-{q}^{6}{A}^{2}+{q}^{4}-{q}^{2}{A}^{2}+1 \Big)\\
H^{T[2,7]}_{[1]}=\dfrac{\{A\} }{\{q\} A q^6} \Big(
{q}^{12}-{q}^{10}{A}^{2}+{q}^{8}-{q}^{6}{A}^{2}+{q}^{4}-{q}^{2}{A}^{2}
+1
 \Big)\\
H^{2,9}_{[1]}=\dfrac{\{A\} }{\{q\} A q^8} \Big(
{q}^{16}-{q}^{14}{A}^{2}+{q}^{12}-{q}^{10}{A}^{2}+{q}^{8}-{q}^{6}{A}^{
2}+{q}^{4}-{q}^{2}{A}^{2}+1
  \Big)\\
....\\
\boxed{H^{2,2k+1}_{[1]}=\dfrac{\{A\} }{\{q\} A q^{2k}}
{\cal{H}}^{T[2,2k+1]}_{[1]}=q^{-2k-1}s_{[2]}^{*}-q^{2k+1}s_{[1,1]}^{*}}=\\
\\=\boxed{\dfrac
{{q}^{12\,k+6}-{q}^{6+8\,k}-{q}^{2+8\,k}+{q}^{2+4\,k}-{q}^{8\,k+4}+{q}^{4+4\,k}+{q}^{4\,k}-1-\Big({q}^{12\,k+4}+2\,{q}^{8\,k+4}
-{q}^{4+4\,k}+{q}^{2+8\,k}-2\,{q}^{2+4\,k}+{q}^{2}\Big){a}^{2}}{ ( {q}^{4}-1 ) 
( {q}^{2+4\,k}-1 )
 ( {q}^{4\,k}-1 ) }}
\end{array}
\ee
and the results coincides with the well known HOMFLY polynomials, see
(\ref{Wrepfund}) and
(\ref{113}).
\begin{itemize}
\item{Heegard-Floer case}
\end{itemize}
 \be
\begin{array}{l}
F^{T[2,1]}_{[1]}=1\\
\\
F^{T[2,3]}_{[1]}={\textbf{q}}^{4}{\textbf{t}}^{2}+{\textbf{t}}^{3}{\textbf{q}}^{2}+1\\
\\
F^{T[2,5]}_{[1]}={\textbf{t}}^{4}{\textbf{q}}^{8}+{\textbf{q}}^{6}{\textbf{t}}^{5}+
{\textbf{q}}^{4}{\textbf{t}}^{2}+{\textbf{t}}^{3}{\textbf{q}}^{2}+1\\
\\
F^{T[2,7]}_{[1]}={\textbf{q}}^{12}{\textbf{t}}^{6}+{\textbf{q}}^{10}{\textbf{t}}^{7}+{\textbf{t}}^{4}{\textbf{q}}^{8}+
{\textbf{q}}^{6}{\textbf{t}}^{5}+{\textbf{q}}^{4}{
\textbf{t}}^{2}+{\textbf{t}}^{3}{\textbf{q}}^{2}+1
\\
\\
F^{T[2,9]}_{[1]}={\textbf{q}}^{16}{\textbf{t}}^{8}+{\textbf{q}}^{14}{\textbf{t}}^{9}+{\textbf{q}}^{12}{\textbf{t}}^{6}+{\textbf{q}}^{10}{\textbf{t}}^{7}+{\textbf{t}}^{4
}{\textbf{q}}^{8}+{\textbf{q}}^{6}{\textbf{t}}^{5}+{\textbf{q}}^{4}{\textbf{t}}^{2}+{\textbf{t}}^{3}{\textbf{q}}^{2}+1\\
....\\
\end{array}
\ee
\begin{itemize}
\item{Alexander case}
\end{itemize}
  \be
\begin{array}{l}
A^{T[2,1]}_{[1]}=1\\
\\
A^{T[2,3]}_{[1]}={\textbf{q}}^{4}-{\textbf{q}}^{2}+1\\
\\
A^{T[2,5]}_{[1]}={\textbf{q}}^{8}-{\textbf{q}}^{6}+{\textbf{q}}^{4}-{\textbf{q}}^{2}+1\\
\\
A^{T[2,7]}_{[1]}={\textbf{q}}^{12}-{\textbf{q}}^{10}+{\textbf{q}}^{8}-{\textbf{q}}^{6}+{\textbf{q}}^{4}-{\textbf{q}}^{2}+1\\
\\
A^{T[2,9]}_{[1]}={\textbf{q}}^{16}-{\textbf{q}}^{14}+{\textbf{q}}^{12}-{\textbf{q}}^{10}+{\textbf{q}}^{8}
-{\textbf{q}}^{6}+{\textbf{q}}^{4}-{\textbf{q}}^{2}+1\\
....\\
\end{array}
\ee
\subsection{Case $(3,n)$, series $n=3 k$ fundamental representation\label{3k}}
\be
\begin{array}{|c|}
\hline\\
P^{T[3,n]}_{[1]}=c^{[3]}_{[1]} M_{[3]}^{\ast} q^{-2n} +c^{[2,1]}_{[1]}
M_{[2,1]}^{\ast} q^{-\frac{2n}{3}} t^{\frac{2n}{3}}+c^{[1,1,1]}_{[1]}
M_{[1,1,1]}^{\ast} t^{2n}
\\
\\
\hline
\end{array}
\ee

the coefficients in this case are:
\be
\boxed{
c^{[3]}_{[1]}=1, \ \ \ c^{[2,1]}_{[1]}=\frac{(1-q^2)(q^2+2t^2q^2+2+t^2)}{(1-q^4
t^2)},
\ \ \ c^{[1,1,1]}_{[1]}=\dfrac{(1+t^2)(1+t^2+t^4)(1-q^2)^2}{(1-t^2
q^2)(1-t^4q^2)}}
\ee

Similarly to the case of unknot, the obtained expressions for
$P^{T[3,n]}_{[1]}$ are not polynomials
with positive coefficients. Moreover, even ${\cal P}^{T[3,n]}_{[1]}$
(i.e. the expression normalized by unknot) is not a polynomial with positive
coefficients. However, it is a power series in ${\bf q,t}$ with positive
coefficients
as is seen from the examples below.

\be
\begin{array}{l}
P^{T[3,0]}_{[1]}=\dfrac{\{A\} }{\{t\}^3  A^2} \Big( 1+{A}^{4}-2\,{A}^{2}
\Big)=\bf{ {\dfrac { \left( {a}^{2}t+1 \right) {q}^{3}}{ \left( -1+{q}^{2}
 \right) ^{3}{a}^{3}{t}^{3/2}}}
\Big(  1+{t}^{2}{a}^{4}+2\,{a}^{2}t  \Big)= \dfrac{{a}^{2}t+1}{
-1+{q}^{2}}\cdot\dfrac{({a}^{2}t+1)^2q^3}{a^3t^{3/2}}\Big(\summ{j=0}{\infty}(1+j)q^{2j}\Big)}
 \\
\\
P^{T[3,3]}_{[1]}=\dfrac{\{A\} }{\{t\}^3 A^2 q^6} \Big({q}^{6}{A}^{4}+ (
-{q}^{6}{t}^{4}-{t}^{4}{q}^{2}+2\,{q}^{2}{t}^{2
}-{q}^{6}{t}^{2}-{q}^{4}{t}^{2}-{q}^{4}-{q}^{2}+2\,{t}^{4}{q}^{4}
 ) {A}^{2}+{q}^{6}{t}^{6}-2\,{q}^{4}{t}^{6}+\\
 {t}^{6}{q}^{2}+{q}^{
4}{t}^{2}+{t}^{4}{q}^{4}+1-2\,{t}^{4}{q}^{2}+{t}^{4}+{q}^{2}{t}^{2}-2
\,{t}^{2}
 \Big)=\\
 \\
 =\bf{ {\dfrac {{a}^{2}t+1}{ \left( -1+{q}^{2} \right) ^{3}{a}^{3}{t}^{15/2}{q
}^{3}}}
 \left({q}^{6}{t}^{8}{a}^{4}+ ( {q}^{10}{t}^{7}+{q}^{8}{t}^{7}+{q}^{6}{t
}^{3}-2\,{t}^{5}{q}^{8}-2\,{t}^{3}{q}^{4}+{t}^{5}{q}^{4}+{t}^{3}{q}^{2
}+{q}^{6}{t}^{5}) {a}^{2}+{q}^{12}{t}^{6}\right.}\\ \bf{\left.
-2\,{t}^{4}{q}^{10} +{t
}^{2}{q}^{8}+{q}^{6}{t}^{4}+{q}^{4}+{t}^{4}{q}^{8}-2\,{q}^{6}{t}^{2}+{
q}^{4}{t}^{2}-2\,{q}^{2}+1 \right) }=\bf{\dfrac{a^2t+1}{-1+q^2}}\Big({\dfrac
{1}{{a}^{3}{t}^{15/2}{q}^{3}}}+{\dfrac {1}{a{t}^{9/2}q}}+{\dfrac { (
{a}^{2}{t}^{3}+1 ) q}{{a}^{3}{t}^{11/2}}}+\\+\bf{\dfrac { (
3\,{a}^{2}t+{a}^{4}{t}^{4}+1 ) {q}^{3}}{{a}^{3}{t}^{7/2}}}
+{\dfrac { ( 2\,{a}^{4}{t}^{4}+3+3\,{a}^{2}t+{a}^{2}{t}^{3} )
{q}^{5}}{{a}^{3}{t}^{7/2}}}+\dfrac{(1+a^2t)q^{7}}{a^3t^{7/2}}\summ{j=0}{\infty}q^{2j}\Big[3(1+a^2t^3)+jt^2(1+a^2t)\Big]\Big)\\
\\
P^{T[3,6]}_{[1]}=\dfrac{\{A\} }{\{t\}^3 A^2 q^{12}} (
1+{t}^{4}-2\,{t}^{2}+{q}^{2}{t}^{2}-{q}^{4}{t}^{6}-2\,{t}^{4}{q}^{2}-{
t}^{4}{q}^{4}+{q}^{4}{t}^{2}-{q}^{6}{t}^{6}+{t}^{6}{q}^{2}+{q}^{6}{t}^
{4}-{q}^{6}{t}^{8}+{q}^{4}{t}^{8}+{q}^{6}{t}^{10}+\\{t}^{4}{q}^{8}+
 ( 2\,{q}^{2}{t}^{2}-{q}^{6}{t}^{8}-{q}^{4}-2\,{q}^{6}{t}^{2}-2\,
{t}^{4}{q}^{8}-2\,{t}^{6}{q}^{10}+{t}^{8}{q}^{8}-{t}^{4}{q}^{10}+{q}^{
4}{t}^{2}+{t}^{4}{q}^{4}-{t}^{2}{q}^{8}-{q}^{12}{t}^{8}+{q}^{10}{t}^{8
}-{q}^{4}{t}^{6}+\\2\,{t}^{10}{q}^{10}-{q}^{2}-{q}^{12}{t}^{10}-{t}^{4}{
q}^{2}-{t}^{10}{q}^{8}+3\,{t}^{6}{q}^{8}+3\,{q}^{6}{t}^{4} ) {A}
^{2}+{t}^{6}{q}^{8}-{t}^{8}{q}^{8}+{q}^{10}{t}^{8}+ ( {q}^{6}{t}^
{4}-2\,{t}^{6}{q}^{10}-2\,{q}^{6}{t}^{2}+{t}^{2}{q}^{8}+\\{t}^{2}{q}^{10
}+{q}^{6}-2\,{t}^{4}{q}^{8}+{t}^{6}{q}^{8}+{q}^{12}{t}^{6}+{t}^{4}{q}^
{10} ) {A}^{4}-2\,{t}^{10}{q}^{8}+{q}^{12}{t}^{12}+{t}^{12}{q}^{
8}-2\,{t}^{12}{q}^{10}+{t}^{10}{q}^{10} )\\
\\ =\bf{ {\dfrac {{a}^{2}t+1}{ ( -1+{q}^{2} ) ^{3}{a}^{3}{t}^{27/2}{q
}^{9}}}} \Big( (
{q}^{18}{t}^{14}-2\,{q}^{16}{t}^{12}+{q}^{14}{t}^{12}+{q}^{12}{t}^{12}+{q}^{14}{t}^{10}-2\,{q}^{12}{t}^{10}+
{q}^{10}{t}^{10}+{q}^{10}{t}^{8}-2\,{q}^{8}{t}^{8}+\\+\bf{q}^{6}{t}^{8} )
{a}^{4}+ (
{q}^{22}{t}^{13}+{q}^{20}{t}^{13}-2\,{q}^{20}{t}^{11}-{q}^{18}{t}^{11}+2\,{q}^{16}{t}^{11}+{q}^{14}{t}^{11}+{q}^{18}{t}^{9}
-{q}^{16}{t}^{9}+{q}^{10}{t}^{9}-3\,{q}^{14}{t}^{9}+2\,{q}^{12}{t}^{9}-\\-3\,\bf{q}^{10}{t}^{7}+{q}^{14}{t}^{7}+2\,{q}^{8}{t}^{7}+{q}^{10}{t}^{5}-{q}^{8}{t}^{5}-{q}^{6}{t}^{5}+{q}^{4}{t}^{5}+{q}^{6}{t}^{3}-2\,{q}^{4}{t}^{3}+{q}^{2}{t}^{3}
) {a}^{2}
+{q}^{24}{t}^{12}+{q}^{20}{t}^{10}+{q}^{20}{t}^{8}-\\-2\,\bf{q}^{22}{t}^{10}-{q}^{16}{t}^{8}+{q}^{18}{t}^{10}+{q}^{12}{t}^{8}-2\,{q}^{18}{t}^{8}-{q}^{14}{t}^{6}+{q}^{14}{t}^{8}+{q}^{10}{t}^{6}+{q}^{12}{t}^{4}-{q}^{10}{t}^{4}-{q}^{8}{t}^{4}+{q}^{6}{t}^{4}+{q}^{8}{t}^{2}-\\-2\,\bf{q}^{6}{t}^{2}+{q}^{4}{t}^{2}
+{q}^{4}-2\,{q}^{2}+{q}^{16}{t}^{6}+1-{q}^{12}{t}^{6}\Big)=\\=   
\bf{\dfrac{a^2t+1}{-1+q^2}}\Big(\dfrac{1}{{a}^{3}{t}^{27/2}{q}^{9}}+{\dfrac
{1}{a{t}^{21/2}{q}^{7}}}+{\dfrac
{{a}^{2}{t}^{3}+1}{{a}^{3}{t}^{23/2}{q}^{5}}}+{\dfrac
{{a}^{4}{t}^{4}+{a}^{2}t+1}{{a}^{3}{t}^{19/2}{q}^{3}}}+{\dfrac
{2\,{a}^{2}{t}^{3}+1}{{a}^{3}{t}^{19/2}q}}
+{\dfrac { ( {a}^{4}{t}^{4}+1+{a}^{2}t+{a}^{2}{t}^{3} )
q}{{a}^{3}{t}^{15/2}}}+\\+\bf{\dfrac { (
{a}^{4}{t}^{6}+4\,{a}^{2}{t}^{3}+1+{t}^{2} ) {q}^{3}}{{a}^{3}{t}^{15/2}}}
+{\dfrac { ( 3+4\,{a}^{2}t+{a}^{2}{t}^{3}+3\,{a}^{4}{t}^{4} )
{q}^{5}}{{a}^{3}{t}^{11/2}}}+{\dfrac { (
4+3\,{a}^{2}t+3\,{a}^{4}{t}^{4}+4\,{a}^{2}{t}^{3} )
{q}^{7}}{{a}^{3}{t}^{11/2}}}
+\\+\bf{\dfrac { (
3+{t}^{2}+3\,{a}^{2}t+3\,{a}^{4}{t}^{4}+{a}^{4}{t}^{6}+6\,{a}^{2}{t}^{3} )
{q}^{9}}{{a}^{3}{t}^{11/2}
}}+{\dfrac { (
3+2\,{a}^{4}{t}^{6}+3\,{t}^{2}+3\,{a}^{2}t+3\,{a}^{4}{t}^{4}+6\,{a}^{2}{t}^{3}+{t}^{5}{a}^{2}
)
{q}^{11}}{{a}^{3}{t}^{11/2}}}+\\+\bf{\dfrac{(1+a^2t)q^{13}}{a^3t^{11/2}}}\summ{j=0}{\infty}q^{2j}\Big[3(1+a^2t^3)(1+t^2)+jt^4(1+a^2t)\Big]\Big)\\
....\\
\boxed{P^{T[3,3k]}_{[1]}=\dfrac{\{A\} }{\{t\}^3 A^2
q^{6k}}{\cal{P}}^{T[3,3k]}_{[1]}}

\end{array}
\ee
\begin{itemize}
\item{HOMFLY case}
\end{itemize}
\be
\begin{array}{l}
H^{T[3,0]}_{[1]}=\dfrac{\{A\} }{\{q\}^3  A^2} \Big(\left( A-1 \right) ^{2}
\left( A+1 \right) ^{2}\Big)  \\
H^{T[3,3]}_{[1]}=\dfrac{\{A\} }{\{q\}^3 A^2 q^6}
\Big({q}^{12}-2\,{q}^{10}+2\,{q}^{8}-{q}^{10}{A}^{2}+{q}^{8}{A}^{2}-{q}^{6}
-2\,{q}^{6}{A}^{2}+2\,{q}^{4}+{q}^{4}{A}^{2}-2\,{q}^{2}+{q}^{6}{A}^{4}
-{q}^{2}{A}^{2}+1
\Big)\\
H^{T[3,6]}_{[1]}=\dfrac{\{A\} }{\{q\}^3 A^2 q^{12}} \Big(
1-{q}^{6}-2\,{q}^{2}+2\,{q}^{4}+{q}^{12}-{q}^{18}+2\,{q}^{20}-2\,{q}^{
22}+{q}^{24}+{q}^{14}{A}^{2}+{q}^{4}{A}^{2}-{q}^{8}{A}^{2}-\\{q}^{2}{A}^
{2}+{q}^{6}{A}^{4}-2\,{q}^{12}{A}^{2}+{q}^{18}{A}^{4}-2\,{q}^{16}{A}^{
4}-{q}^{22}{A}^{2}+2\,{q}^{14}{A}^{4}-{q}^{16}{A}^{2}+{q}^{10}{A}^{2}-
{q}^{12}{A}^{4}-2\,{q}^{8}{A}^{4}+2\,{q}^{10}{A}^{4}+{q}^{20}{A}^{2}
\Big)\\
....\\
\boxed{H^{T[3,3k]}_{[1]}=\dfrac{\{A\} q}{\{q\}^3 A^2 q^{6k}}
{\cal{H}}^{T[3,3k]}_{[1]}=q^{-6k}s_{[3]}^{*}+2s_{[2,1]}^{*}+q^{6k}s_{[1,1,1]}^{*}}
\end{array}
\ee

\begin{itemize}
\item{Alexander case}
\end{itemize}
\be
\begin{array}{l}
A^{T[3,0]}_{[1]}=4\\
\\
A^{T[3,3]}_{[1]}={\textbf{q}}^{12}-{\textbf{q}}^{10}+{\textbf{q}}^{8}+2\,{\textbf{q}}^{6}+{\textbf{q}}^{4}-{\textbf{q}}^{2}+1\\
\\
A^{T[3,6]}_{[1]}=1-{\textbf{q}}^{2}+{\textbf{q}}^{4}-{\textbf{q}}^{8}+{\textbf{q}}^{10}+2\,{\textbf{q}}^{12}
+{\textbf{q}}^{14}-{\textbf{q}}^{16}+{\textbf{q}}^{
20}-{\textbf{q}}^{22}+{\textbf{q}}^{24}
\\
....\\
\end{array}
\ee

\subsection{Case $(3,n)$, series $n=3 k+1$ fundamental
representation\label{3k+1}}
\be
\begin{array}{|c|}
\hline\\
P^{T[3,n]}_{[1]}=c^{[3]}_{[1]} M_{[3]}^{\ast} q^{-2n} +
c^{[2,1]}_{[1]} M_{[2,1]}^{\ast} q^{-\frac{2(n-1)}{3}} t^{\frac{2(n-1)}{3}}
+c^{[1,1,1]}_{[1]} M_{[1,1,1]}^{\ast} t^{2n}
\\
\\
\hline
\end{array}
\ee
the coefficients in this case are:
\be
\boxed{
c^{[3]}_{[1]}=1, \ \ \ c^{[2,1]}_{[1]}=-\frac{(1-t^2)(1+q^2+t^2q^2)}{(1-q^4
t^2)},
\ \ \ c^{[1,1,1]}_{[1]}=\dfrac{(1-t^4)(1-t^6)}{(1-t^2 q^2)(1-t^4q^2)}
\Big(\frac{q}{t}\Big)^{2}=
{\{t^2\}\{t^3\}\over\{tq\}\{t^2q\}}}
\ee
such that the first several superpolynomials are:
\be
\begin{array}{l}
P^{T[3,1]}_{[1]}=\dfrac{\{A\} t^2}{\{t\} A^2 q^2}=\bf{ {\dfrac { \left(
{a}^{2}t+1 \right) q}{{a}^{3} \left( -1+{q}^{2}
 \right) {t}^{7/2}}}}   \\
\\
P^{T[3,4]}_{[1]}=\dfrac{\{A\} t^2}{\{t\} A^2 q^8} \Big( {q}^{6}{A}^{4}+ \left(
{q}^{6}{t}^{2}+{q}^{4}{t}^{2}+{q}^{6}{t}^{4}+{
q}^{4}+{q}^{2} \right) (-A^{2})+{q}^{6}{t}^{6}+{q}^{4}{t}^{2}+{t}^{4}{q
}^{4}+{q}^{2}{t}^{2}+1\Big)\\
=\bf{ {\dfrac {{a}^{2}t+1}{{a}^{3}{q}^{5} \left( -1+{q}^{2} \right) {t}^{19/2
}}} \Big( {q}^{6}{t}^{8}{a}^{4}+ \left(
{q}^{8}{t}^{7}+{q}^{6}{t}^{5}+{q}^{10}{t
}^{7}+{t}^{5}{q}^{4}+{t}^{3}{q}^{2} \right) {a}^{2}+{q}^{12}{t}^{6}+{q
}^{6}{t}^{4}+{t}^{4}{q}^{8}+{q}^{4}{t}^{2}+1
 \Big)}
\\
P^{T[3,7]}_{[1]}=\dfrac{\{A\} t^2}{\{t\} A^2 q^{14}} \Big( (
{t}^{2}{q}^{8}+{t}^{2}{q}^{10}+{q}^{6}+{t}^{4}{q}^{10}+{q}^{12
}{t}^{6} ) {A}^{4}+ ( {q}^{2}+{q}^{4}+{q}^{10}{t}^{8}+{q}^
{4}{t}^{2}+{q}^{6}{t}^{4}+2\,{q}^{6}{t}^{2}+2\,{t}^{4}{q}^{8}+\\2\,{t}^{
6}{q}^{10}+{t}^{4}{q}^{10}+{t}^{2}{q}^{8}+{t}^{6}{q}^{8}+{q}^{12}{t}^{
10}+{q}^{12}{t}^{8} ) (-A^{2})+1+{q}^{4}{t}^{2}+{q}^{12}{t}^{12}+
{q}^{2}{t}^{2}+{t}^{4}{q}^{4}+{q}^{6}{t}^{6}+{q}^{6}{t}^{4}+{t}^{4}{q}
^{8}+{t}^{6}{q}^{8}+\\+{t}^{8}{q}^{8}+{q}^{10}{t}^{8}+{t}^{10}{q}^{10}
\Big)=\\
=\bf{ {\dfrac {{a}^{2}t+1}{{a}^{3}{q}^{11} \left( -1+{q}^{2} \right) {t}^{21/2
}}} \Big(( {q}^{18}{t}^{14}+{q}^{12}{t}^{12}+{q}^{6}{t}^{8}+{t}^{10}{q}^{
10}+{q}^{14}{t}^{12} ) {a}^{4}+ ( {q}^{6}{t}^{5}+{q}^{10}{t
}^{7}+{t}^{5}{q}^{4}+2\,{q}^{8}{t}^{7}+ }\\ \bf{{
{q}^{20}{t}^{13}+{q}^{14}{t}^{
11}+2\,{q}^{16}{t}^{11}+2\,{t}^{9}{q}^{12}+{q}^{14}{t}^{9}+ {q}^{18}{t}
^{11}+{q}^{22}{t}^{13}+{t}^{3}{q}^{2}+{t}^{9}{q}^{10} ) {a}^{2}+
1+{q}^{12}{t}^{6}+{q}^{4}{t}^{2}+}}\\
\bf{{{q}^{6}{t}^{4}+{t}^{4}{q}^{8}+{q}^{14
}{t}^{8}+{t}^{6}{q}^{10}+{q}^{12}{t}^{8}+{q}^{16}{t}^{8}+{q}^{18}{t}^{
10}+{q}^{24}{t}^{12}+{q}^{20}{t}^{10} \Big)}}
\\
P^{T[3,10]}_{[1]}=\dfrac{\{A\} t^2}{\{t\} A^2 q^{20}} \Big( (
{t}^{6}{q}^{14}+{t}^{4}{q}^{14}+{q}^{6}+{q}^{12}{t}^{6}+{q}^{
14}{t}^{8}+{t}^{12}{q}^{18}+{t}^{10}{q}^{16}+{t}^{4}{q}^{12}+{t}^{2}{q
}^{8}+{t}^{2}{q}^{10}+{t}^{4}{q}^{10}+{q}^{16}{t}^{8} ) {A}^{4}\\+
 ( {q}^{4}+2\,{q}^{14}{t}^{8}+{t}^{4}{q}^{12}+{t}^{6}{q}^{8}+2\,
{t}^{10}{q}^{14}+2\,{t}^{4}{q}^{8}+{q}^{4}{t}^{2}+{t}^{10}{q}^{16}+2\,
{t}^{4}{q}^{10}+2\,{q}^{12}{t}^{8}+{t}^{10}{q}^{12}+2\,{t}^{12}{q}^{16
}+{q}^{10}{t}^{8}+\\2\,{q}^{12}{t}^{6}+{q}^{6}{t}^{4}+{t}^{2}{q}^{8}+2\,
{q}^{6}{t}^{2}+{q}^{2}+{t}^{6}{q}^{14}+{t}^{16}{q}^{18}+{t}^{14}{q}^{
16}+{t}^{12}{q}^{14}+{t}^{14}{q}^{18}+2\,{t}^{6}{q}^{10} ) (-A^
2)+1+{t}^{4}{q}^{4}+{q}^{4}{t}^{2}+{q}^{6}{t}^{6}+\\{q}^{6}{t}^{4}+{t}^{
4}{q}^{8}+{t}^{8}{q}^{8}+{t}^{6}{q}^{8}+{q}^{12}{t}^{6}+{t}^{6}{q}^{10
}+{t}^{14}{q}^{14}+{t}^{12}{q}^{14}+{t}^{14}{q}^{16}+{q}^{2}{t}^{2}+{t
}^{10}{q}^{10}+{t}^{16}{q}^{16}+{t}^{10}{q}^{12}+{t}^{12}{q}^{12}+\\{q}^
{18}{t}^{18}+{t}^{10}{q}^{14}+{q}^{12}{t}^{8}+{q}^{10}{t}^{8}
\Big)=\\=
\bf{ {\dfrac {{a}^{2}t+1}{{a}^{3}{q}^{17} \left( -1+{q}^{2} \right) {t}^{43/2
}}} \Big( (
{t}^{16}{q}^{20}+{t}^{18}{q}^{24}+{q}^{6}{t}^{8}+{q}^{26}{t}^{18}+{q}^{22}{t}^{16}+
{t}^{16}{q}^{18}+{t}^{14}{q}^{18}+{t}^{14}{q}^{16
}+{t}^{12}{q}^{14}+}\\\bf{{{q}^{12}{t}^{12}+{t}^{10}{q}^{10}+ 
{q}^{30}{t}^{20}
 ) {a}^{4}+ ( {t}^{19}{q}^{34}+{t}^{15}{q}^{26}+{t}^{17}{q}
^{26}+{t}^{13}{q}^{22}+2\,{t}^{13}{q}^{18}+{t}^{13}{q}^{16}+2\,{t}^{11
}{q}^{14}+2\,{t}^{13}{q}^{20}+}}\\\bf{{ {t}^{11}{q}^{18}+2\,{t}^{15}{q}^{24}+
{t}
^{9}{q}^{10}+2\,{t}^{17}{q}^{28}+{t}^{15}{q}^{20}+ {q}^{6}{t}^{5}+{t}^{
5}{q}^{4}+{t}^{3}{q}^{2}+2\,{q}^{8}{t}^{7}+{t}^{17}{q}^{30}+2\,{t}^{11
}{q}^{16}+{q}^{10}{t}^{7}+}}\\\bf{{ 2\,{t}^{15}{q}^{22}+{q}^{14}{t}^{9}+
2\,{t}^{
9}{q}^{12}+{t}^{19}{q}^{32} ) {a}^{2}+1+{t}^{10}{q}^{18}+{q}^{6}
{t}^{4}+{t}^{4}{q}^{8}+{t}^{10}{q}^{16}+{q}^{12}{t}^{6}+{t}^{18}{q}^{
36}+{t}^{14}{q}^{28}+{t}^{14}{q}^{26}+}}\\\bf{{
{t}^{16}{q}^{32}+{t}^{12}{q}^{18
}+{t}^{14}{q}^{24}+{t}^{12}{q}^{20}+{t}^{12}{q}^{22}+{q}^{4}{t}^{2}+{t
}^{6}{q}^{10}+{t}^{16}{q}^{30}+{q}^{12}{t}^{8}+{q}^{16}{t}^{8}+{q}^{14
}{t}^{8}+{q}^{24}{t}^{12}+{t}^{10}{q}^{20} \Big)}}\\
....\\
\boxed{P^{T[3,3k+1]}_{[1]}=\dfrac{\{A\} t^2}{\{t\} A^2
q^{6k+2}}{\cal{P}}^{T[3,3k+1]}_{[1]}}
\end{array}
\ee
\begin{itemize}
\item{HOMFLY case}
\end{itemize}
\be
\begin{array}{l}
H^{T[3,1]}_{[1]}=\dfrac{\{A\} }{\{q\} A^2}   \\
\\
H^{T[3,4]}_{[1]}=\dfrac{\{A\} }{\{q\} A^2 q^6}
\Big({q}^{12}-{q}^{10}{A}^{2}+{q}^{8}-{q}^{8}{A}^{2}-{q}^{6}{A}^{2}+{q}^{6}
+{q}^{4}+{q}^{6}{A}^{4}-{q}^{4}{A}^{2}-{q}^{2}{A}^{2}+1\Big)\\
\\
H^{T[3,7]}_{[1]}=\dfrac{\{A\} }{\{q\} A^2 q^{12}}
\Big(1+{q}^{6}+{q}^{4}+{q}^{8}+{q}^{10}+2\,{q}^{12}+{q}^{14}+{q}^{16}+{q}^{
18}+{q}^{20}+{q}^{24}-{q}^{22}{A}^{2}-{q}^{18}{A}^{2}+{q}^{18}{A}^{4}-\\
{q}^{2}{A}^{2}-{q}^{4}{A}^{2}+{q}^{10}{A}^{4}+{q}^{14}{A}^{4}-2\,{q}^{
12}{A}^{2}+{q}^{6}{A}^{4}+{q}^{12}{A}^{4}-2\,{q}^{10}{A}^{2}-{q}^{6}{A
}^{2}-{q}^{20}{A}^{2}-2\,{q}^{14}{A}^{2}-2\,{q}^{16}{A}^{2}-\\2\,{q}^{8}
{A}^{2}
\Big)\\
\\
H^{T[3,10]}_{[1]}=\dfrac{\{A\}}{\{q\} A^2 q^{18}}
\Big(1+{q}^{6}+{q}^{4}+{q}^{8}+{q}^{10}+2\,{q}^{12}+{q}^{14}+2\,{q}^{16}+2
\,{q}^{18}+2\,{q}^{20}+{q}^{22}+2\,{q}^{24}+{q}^{26}+{q}^{28}-\\3\,{q}^{
22}{A}^{2}-3\,{q}^{18}{A}^{2}+2\,{q}^{18}{A}^{4}-{q}^{2}{A}^{2}-{q}^{4
}{A}^{2}-{q}^{32}{A}^{2}+{q}^{30}{A}^{4}+{q}^{24}{A}^{4}+{q}^{32}-{q}^
{30}{A}^{2}-2\,{q}^{28}{A}^{2}-{q}^{34}{A}^{2}+\\{q}^{10}{A}^{4}+{q}^{14
}{A}^{4}-2\,{q}^{12}{A}^{2}+{q}^{6}{A}^{4}+{q}^{12}{A}^{4}+{q}^{20}{A}
^{4}-2\,{q}^{26}{A}^{2}+{q}^{16}{A}^{4}+{q}^{22}{A}^{4}+{q}^{26}{A}^{4
}-2\,{q}^{10}{A}^{2}-{q}^{6}{A}^{2}-\\2\,{A}^{2}{q}^{24}-3\,{q}^{20}{A}^
{2}-3\,{q}^{14}{A}^{2}-3\,{q}^{16}{A}^{2}+{q}^{30}+{q}^{36}-2\,{q}^{8}
{A}^{2}
\Big)\\
....\\
\boxed{H^{T[3,3k+1]}_{[1]}=\dfrac{\{A\} }{\{q\} A^2 q^{6k}}
{\cal{H}}^{T[3,3k+1]}_{[1]}=q^{-6k-2}s_{[3]}^{*}-s_{[2,1]}^{*}+q^{6k+2}s_{[1,1,1]}^{*}}
\end{array}
\ee
and the results coincides with the well known HOMFLY polynomials, see
(\ref{Wrepfund}) and
(\ref{113}).

\begin{itemize}
\item{Floer case}
\end{itemize}
\be
\begin{array}{l}
F^{T[3,1]}_{[1]}=1   \\
\\
F^{T[3,4]}_{[1]}={q}^{6}{t}^{8}+{q}^{6}{t}^{5}+{q}^{10}{t}^{7}+{t}^{3}{q}^{2}+{q}^{12}{
t}^{6}+{q}^{6}{t}^{4}+1
\\
\\
F^{T[3,7]}_{[1]}=1+{q}^{16}{t}^{11}+{q}^{22}{t}^{13}+{q}^{6}{t}^{4}+{q}^{12}{t}^{8}+{q}
^{18}{t}^{10}+{q}^{24}{t}^{12}+{q}^{8}{t}^{7}+{t}^{3}{q}^{2}
\\
\\
F^{T[3,10]}_{[1]}=1+{q}^{22}{t}^{15}+{q}^{18}{t}^{12}+{q}^{24}{t}^{14}+{q}^{30}{t}^{16}+
{q}^{34}{t}^{19}+{q}^{36}{t}^{18}+{q}^{14}{t}^{11}+{q}^{28}{t}^{17}+{q
}^{6}{t}^{4}+{q}^{12}{t}^{8}+{q}^{8}{t}^{7}+{t}^{3}{q}^{2}

\\
....\\
\end{array}
\ee

\begin{itemize}
\item{Alexander case}
\end{itemize}
\be
\begin{array}{l}
A^{T[3,1]}_{[1]}=1   \\
\\
A^{T[3,4]}_{[1]}={\textbf{q}}^{12}-{\textbf{q}}^{10}+{\textbf{q}}^{6}-{\textbf{q}}^{2}+1\\
\\
A^{T[3,7]}_{[1]}=1+{\textbf{q}}^{6}-{\textbf{q}}^{2}-{\textbf{q}}^{8}+{\textbf{q}}^{12}-{\textbf{q}}^{16}+{\textbf{q}}^{18}
-{\textbf{q}}^{22}+{\textbf{q}}^{24}
\\
\\
A^{T[3,10]}_{[1]}=1+{\textbf{q}}^{6}-{\textbf{q}}^{2}-{\textbf{q}}^{8}+{\textbf{q}}^{12}-{\textbf{q}}^{14}+{\textbf{q}}^{18}-{\textbf{q}}^{22}+
{\textbf{q}}^{24}
-{\textbf{q}}^{28}+{\textbf{q}}^{36}-{\textbf{q}}^{34}+{\textbf{q}}^{30}

\\
....\\
\end{array}
\ee
\subsection{Case $(3,n)$, series $n=3 k+2$ fundamental
representation\label{3k+2}}
\be
\begin{array}{|c|}
\hline\\
P^{T[3,n]}_{[1]}=c^{[3]}_{[1]} M_{[3]}^{\ast} q^{-2n} +
c^{[2,1]}_{[1]} M_{[2,1]}^{\ast} q^{-\frac{2(n-2)}{3}} t^{\frac{2(n-2)}{3}}
+c^{[1,1,1]}_{[1]} M_{[1,1,1]}^{\ast} t^{2n}
\\
\\
\hline
\end{array}
\ee
the coefficients in this case are:
\be
\boxed{
c^{[3]}_{[1]}=1, \ \ \ c^{[2,1]}_{[1]}=-\frac{(1-t^2)(1+t^2+t^2q^2)}{(1-q^4
t^2)},
 \ \ \ c^{[1,1,1]}_{[1]}=\dfrac{(1-t^4)(1-t^6)}{(1-t^2 q^2)(1-t^4q^2)}
\Big(\frac{q}{t}\Big)^{2}}
\ee
such that the first several superpolynomials are:
\be
\begin{array}{l}
P^{T[3,2]}_{[1]}=\dfrac{\{A\} t^2}{\{t\} A^2 q^4}\Big(
{q}^{2}{t}^{2}+1+{q}^{2}(-A^{2}) \Big)=\bf{{\dfrac {{a}^{2}t+1}{ \left(
-1+{q}^{2} \right) q{a}^{3}{t}^{11/2}}} \Big(
{q}^{4}{t}^{2}+{q}^{2}{t}^{3}{a}^{2}+1 \Big)}   \\
\\
P^{T[3,5]}_{[1]}=\dfrac{\{A\} t^2}{\{t\} A^2 q^{10}}\Big((
{q}^{6}+{t}^{2}{q}^{8}) {A}^{4}+( {q}^{6}{t}^{4}
+{t}^{6}{q}^{8}+2\,{q}^{6}{t}^{2}+{q}^{4}{t}^{2}+{t}^{4}{q}^{8}+{q}^{4
}+{q}^{2}) (-A^{2})+{t}^{8}{q}^{8}+{q}^{6}{t}^{4}+\\{q}^{6}{t}^{6}
+{q}^{2}{t}^{2}+{t}^{4}{q}^{4}+1+{q}^{4}{t}^{2}
\Big)=\\
\bf{  {\dfrac {{a}^{2}t+1}{ \left( -1+{q}^{7} \right) q{a}^{3}{t}^{23/2}}}
\Big( \left( {q}^{6}{t}^{8}+{t}^{10}{q}^{10} \right) {a}^{4}+ \left(
{q}^{10}{t}^{7}+{q}^{14}{t}^{9}+2\,{q}^{8}{t}^{7}+{q}^{6}{t}^{5}+{t}^{9}{q}
^{12}+{t}^{5}{q}^{4}+{t}^{3}{q}^{2} \right) {a}^{2}+}\\ \bf{{
{q}^{16}{t}^{8}+{t
}^{6}{q}^{10}+{q}^{12}{t}^{6}+{q}^{4}{t}^{2}+{t}^{4}{q}^{8}+1+{q}^{6}{
t}^{4} \Big) }}
\\
\\
P^{T[3,8]}_{[1]}=\dfrac{\{A\} t^2}{\{t\} A^2 q^{16}} \Big( \left(
{t}^{4}{q}^{12}+{q}^{6}+{t}^{4}{q}^{10}+{t}^{2}{q}^{10}+{t}^{2}{q}^{8}+{q}^{14}{t}^{8}+{q}^{12}{t}^{6}
\right) {A}^{4}+ ( {q}^
{4}+{q}^{2}+{t}^{6}{q}^{8}+{q}^{4}{t}^{2}+\\2\,{t}^{4}{q}^{10}+2\,{t}^{4
}{q}^{8}+2\,{q}^{12}{t}^{8}+2\,{t}^{6}{q}^{10}+{q}^{6}{t}^{4}+2\,{q}^{
6}{t}^{2}+{q}^{12}{t}^{6}+{q}^{10}{t}^{8}+{t}^{2}{q}^{8}+{q}^{14}{t}^{
12}+{t}^{10}{q}^{14}+{t}^{10}{q}^{12}) (-A^{2})+\\1+{q}^{12}{t}^{
12}+{q}^{2}{t}^{2}+{t}^{4}{q}^{4}+{q}^{4}{t}^{2}+{q}^{6}{t}^{6}+{q}^{6
}{t}^{4}+{t}^{6}{q}^{10}+{t}^{4}{q}^{8}+{q}^{10}{t}^{8}+{t}^{6}{q}^{8}
+{t}^{14}{q}^{14}+{t}^{8}{q}^{8}+{t}^{10}{q}^{12}+{t}^{10}{q}^{10}
 \Big)=\\
 =\bf{ {\dfrac {{a}^{2}t+1}{ \left( -1+{q}^{13} \right) q{a}^{3}{t}^{35/2}}} 
\Big(
(
{q}^{14}{t}^{12}+{t}^{10}{q}^{10}+{q}^{6}{t}^{8}+{q}^{22}{t}^{16}+{q}^{12}{t}^{12}
+{q}^{18}{t}^{14}+{q}^{16}{t}^{14} ) {a}^{4}
+}\\ \bf{{ ( {q}^{18}{t}^{13}+2\,{q}^{16}{t}^{11}+{t}^{9}{q}^{10}+2\,{q}^{
20}{t}^{13}+{q}^{18}{t}^{11}+2\,{q}^{8}{t}^{7}+{q}^{6}{t}^{5}+{q}^{14}
{t}^{9}+2\,{q}^{14}{t}^{11}+2\,{t}^{9}{q}^{12}+{q}^{24}{t}^{15}+}}\\\bf{{{q}^{
10}{t}^{7}+{q}^{22}{t}^{13}+{q}^{26}{t}^{15}+{t}^{5}{q}^{4}+{t}^{3}{q}
^{2} ) {a}^{2}+1+{t}^{6}{q}^{10}+{q}^{12}{t}^{8}+{q}^{16}{t}^{10
}+{q}^{22}{t}^{12}+{q}^{24}{t}^{12}+{q}^{28}{t}^{14}+{q}^{6}{t}^{4}+}}\\\bf{{
{t
}^{4}{q}^{8}+{q}^{12}{t}^{6}+{q}^{4}{t}^{2}+{q}^{20}{t}^{10}+{q}^{14}{
t}^{8}+{q}^{18}{t}^{10}+{q}^{16}{t}^{8} \Big)}}\\
....\\
\boxed{P^{T[3,3k+2]}_{[1]}=\dfrac{\{A\} t^2}{\{t\} A^2
q^{6k+4}}{\cal{P}}^{T[3,3k+2]}_{[1]}}
\end{array}
\ee
\begin{itemize}
\item{HOMFLY case}
\end{itemize}
\be
\begin{array}{l}
H^{T[3,2]}_{[1]}=\dfrac{\{A\}}{\{q\} A^2 q^2}\Big({q}^{4}-{q}^{2}{A}^{2}+1
\Big)   \\
\\
H^{T[3,5]}_{[1]}=\dfrac{\{A\}}{\{q\} A^2 q^8} \Big(
{q}^{16}-{q}^{14}{A}^{2}+{q}^{12}-{q}^{12}{A}^{2}-{q}^{10}{A}^{2}+{q}^
{10}+{q}^{8}+{q}^{10}{A}^{4}-2\,{q}^{8}{A}^{2}-{q}^{6}{A}^{2}+{q}^{6}+
{q}^{4}+{q}^{6}{A}^{4}\\-{q}^{4}{A}^{2}-{q}^{2}{A}^{2}+1
 \Big)\\
\\
H^{T[3,8]}_{[1]}=\dfrac{\{A\}}{\{q\} A^2 q^{14}} \Big(
1+{q}^{6}+{q}^{4}+{q}^{8}+{q}^{10}+2\,{q}^{12}+{q}^{14}+2\,{q}^{16}+{q
}^{18}+{q}^{20}+{q}^{22}+{q}^{24}+{q}^{28}+{q}^{6}{A}^{4}-{q}^{2}{A}^{
2}+{q}^{18}{A}^{4}+\\{q}^{16}{A}^{4}+{q}^{22}{A}^{4}-{q}^{26}{A}^{2}-{q}
^{24}{A}^{2}-{q}^{22}{A}^{2}-2\,{q}^{20}{A}^{2}-2\,{q}^{18}{A}^{2}+{q}
^{12}{A}^{4}+{q}^{14}{A}^{4}-2\,{q}^{16}{A}^{2}-{q}^{4}{A}^{2}-2\,{q}^
{12}{A}^{2}-2\,{q}^{10}{A}^{2}-\\3\,{q}^{14}{A}^{2}-2\,{q}^{8}{A}^{2}-{q
}^{6}{A}^{2}+{q}^{10}{A}^{4}
\Big)\\
....\\
\boxed{H^{T[3,3k+2]}_{[1]}=\dfrac{\{A\} }{\{q\} A^2 q^{6k+2}}
{\cal{H}}_{3,3k+2}=q^{-6k-4}s_{[3]}^{*}-s_{[2,1]}^{*}+q^{6k+4}s_{[1,1,1]}^{*}}
\\
\end{array}
\ee
and the results coincides with the well known HOMFLY polynomials, see
(\ref{Wrepfund}) and
(\ref{113}).

\begin{itemize}
\item{Floer case}
\end{itemize}
\be
\begin{array}{l}
F^{T[3,2]}_{[1]}={q}^{4}{t}^{2}+{t}^{3}{q}^{2}+1\\
\\
F^{T[3,5]}_{[1]}={q}^{14}{t}^{9}+{q}^{8}{t}^{7}+{t}^{3}{q}^{2}+{q}^{16}{t}^{8}+{t}^{6}{
q}^{10}+1+{q}^{6}{t}^{4}
\\
\\
F^{T[3,8]}_{[1]}=1+{q}^{14}{t}^{11}+{q}^{20}{t}^{13}+{q}^{6}{t}^{4}+{q}^{12}{t}^{8}+{q}
^{16}{t}^{10}+{q}^{22}{t}^{12}+{q}^{8}{t}^{7}+{q}^{28}{t}^{14}+{q}^{26
}{t}^{15}+{t}^{3}{q}^{2}
\\
....
\end{array}
\ee
\begin{itemize}
\item{Alexander case}
\end{itemize}
\be
\begin{array}{l}
A^{T[3,2]}_{[1]}={\textbf{q}}^{4}-{\textbf{q}}^{2}+1\\
\\
A^{T[3,5]}_{[1]}={\textbf{q}}^{16}-{\textbf{q}}^{14}+{\textbf{q}}^{10}-{\textbf{q}}^{8}+{\textbf{q}}^{6}-{\textbf{q}}^{2}+1
\\
\\
A^{T[3,8]}_{[1]}=1+{\textbf{q}}^{6}-{\textbf{q}}^{2}-{\textbf{q}}^{8}+{\textbf{q}}^{12}-{\textbf{q}}^{14}+
{\textbf{q}}^{16}-{\textbf{q}}^{20}+{\textbf{q}}^{22}
-{\textbf{q}}^{26}+{\textbf{q}}^{28}
\\
....
\end{array}
\ee

\subsection{Case $(4,n)$, $n=4 k$ fundamental representation\label{4k}}
\be
\begin{array}{|c|}
\hline\\
P^{T[4,n]}_{[1]}=c^{[4]}_{[1]} M_{[4]}^{\ast} q^{-3n}+c^{[3,1]}_{[1]}
M_{[3,1]}^{\ast} q^{-\frac{3n}{2}} t^{\frac{n}{2}}
+c^{[2,2]}_{[1]} M_{[2,2]}^{\ast} q^{-n} t^{n} + c^{[2,1,1]}_{[1]}
M_{[2,1,1]}^{\ast} q^{-\frac{n}{2}} t^{\frac{3n}{2}} + c^{[1,1,1,1]}_{[1]}
M_{[1,1,1,1]}^{\ast} t^{3n}
\\
\\
\hline
\end{array}
\ee
with the coefficients
\fr
{\nonumber
c^{[4]}_{[1]}=1, \ \ \ c^{[3,1]}_{[1]}={\frac { \left( -1+{q}^{2} \right) 
\left( 3\,{q}^{4}{t}^{2}+{q}^{4}+2
\,{q}^{2}{t}^{2}+2\,{q}^{2}+3+{t}^{2} \right) }{-1+{q}^{6}{t}^{2}}}, \ \ \
c^{[2,2]}_{[1]}={\frac { \left( 1+{t}^{2} \right)  \left( -1+{q}^{2} \right)
^{2}
 \left( {q}^{2}+2\,{q}^{2}{t}^{2}+2+{t}^{2} \right) }{ \left( -1+{q}^{
2}{t}^{2} \right)  \left( -1+{q}^{4}{t}^{2} \right) }}\\ \nonumber
c^{[2,1,1]}_{[1]}={\frac { \left( 1+{t}^{2} \right)  \left( -1+{q}^{2} \right)
^{2}
 \left( 3\,{t}^{4}{q}^{2}+2\,{q}^{2}{t}^{2}+{q}^{2}+{t}^{4}+2\,{t}^{2}
+3 \right) }{ \left( -1+{q}^{2}{t}^{2} \right) ^{2} \left( {q}^{2}{t}^
{2}+1 \right) }}, \ \ \ c^{[1,1,1,1]}_{[1]}={\frac { \left( {t}^{4}+1 \right) 
\left( {t}^{2}+{t}^{4}+1 \right)
 \left( 1+{t}^{2} \right) ^{2} \left( -1+{q}^{2} \right) ^{3}}{
 \left( -1+{t}^{6}{q}^{2} \right)  \left( -1+{q}^{2}{t}^{2} \right)
 \left( -1+{t}^{4}{q}^{2} \right) }}}

Similarly to the case of unknot, the obtained expressions for
$P^{T[4,n]}_{[1]}$ are not polynomials
with positive coefficients. Moreover, even ${\cal P}^{T[4,n]}_{[1]}$
(i.e. the expression normalized by unknot) is not a polynomial with positive
coefficients. However, it is a power series in ${\bf q,t}$ with positive
coefficients
as is seen from the examples:
$$ P^{T[4,0]}_{[1]}=\frac{\{A\}}{\{t\}^4 A^3}(1-A^2)^3 = \bf{{\dfrac { \left(
{a}^{2}
t+1 \right) {q}^{4}}{ \left( -1+{q}^{2} \right)
^{4}{a}^{4}{t}^{2}}}}=\bf{\dfrac{a^2t+1}{1-q^2}}\cdot\dfrac{(a^2t+1)^3}{a^4t^2}\summ{j=0}{\infty}
\dfrac{1}{2}(j+1)(j+2)q^{2j}$$

\bigskip
\noindent
$P^{T[4,4]}_{[1]}=\frac{\{A\}}{\{t\}^4  q^{12} A^3}\Big(
1+{q}^{6}{t}^{4}+{q}^{8}{t}^{6}-3\,{t}^{2}+3\,{t}^{4}-{t}^{6}+{q}^{10}
{t}^{10}+3\,{q}^{8}{t}^{12}-3\,{q}^{10}{t}^{12}+{q}^{12}{t}^{12}+{t}^{
6}{q}^{10}-2\,{t}^{8}{q}^{8}+{q}^{10}{t}^{8}-{q}^{12}{A}^{6}+{t}^{4}{q
}^{8}+ ( -{q}^{6}{t}^{6}+3\,{q}^{6}{t}^{4}+{t}^{2}{q}^{10}+{q}^{6
}+{t}^{2}{q}^{12}+{q}^{12}{t}^{6}+{q}^{8}{t}^{2}-5\,{t}^{4}{q}^{8}+{q}
^{8}+{q}^{10}-3\,{q}^{6}{t}^{2}+3\,{q}^{8}{t}^{6}-3\,{t}^{6}{q}^{10}+{
t}^{4}{q}^{10}+{t}^{4}{q}^{12}) {A}^{4}+( 7\,{q}^{6}{t}^{
4}+7\,{q}^{8}{t}^{6}+3\,{q}^{10}{t}^{10}-{q}^{12}{t}^{6}-2\,{t}^{6}{q}
^{10}+2\,{q}^{10}{t}^{8}-{q}^{12}{t}^{8}-{q}^{12}{t}^{10}-{q}^{4}-2\,{
t}^{4}{q}^{10}-{t}^{2}{q}^{10}-2\,{t}^{4}{q}^{8}-2\,{q}^{8}{t}^{2}-{q}
^{2}-2\,{q}^{4}{t}^{6}-3\,{t}^{4}{q}^{2}-3\,{t}^{10}{q}^{8}+2\,{q}^{4}
{t}^{2}-2\,{q}^{6}{t}^{2}+3\,{q}^{2}{t}^{2}-3\,{q}^{6}{t}^{6}+{t}^{6}{
q}^{2}-2\,{q}^{6}{t}^{8}-{q}^{6}+{q}^{4}{t}^{8}+{q}^{6}{t}^{10})
{A}^{2}-{q}^{6}{t}^{12}-3\,{t}^{4}{q}^{2}-3\,{t}^{10}{q}^{8}+
{q}^{4}{t}^{2}+{q}^{6}{t}^{2}+{q}^{2}{t}^{2}-2\,{t}^{4}{q}^{4}-4\,{q}^
{6}{t}^{6}-{t}^{8}{q}^{2}+3\,{t}^{6}{q}^{2}+2\,{q}^{4}{t}^{8}+3\,{q}^{
6}{t}^{10}-{t}^{10}{q}^{4} \Big)=\\ = \bf{\dfrac {{a}^{2}t+1}{ ( -1+{q}^{2} )
^{4}{a}^{4}{t}^{14}{q}^{8}}} \Big( 1-3\,{q}^{2}+3\,{q}^{4}-{q}^{6}+ (
{q}^{18}{t}^{14}+{t}^{10}{q}^{
8}+{q}^{14}{t}^{12}-3\,{q}^{16}{t}^{12}+3\,{q}^{14}{t}^{10}-5\,{q}^{12
}{t}^{10}-3\,{t}^{8}{q}^{8}+{q}^{14}{t}^{14}+{q}^{10}{t}^{12}+{q}^{16}
{t}^{14}+3\,{q}^{10}{t}^{8}-{q}^{12}{t}^{8}+{q}^{6}{t}^{8}+{t}^{10}{q}
^{10}+{q}^{12}{t}^{12} ) {a}^{4}+ ( -3\,{q}^{20}{t}^{11}+{q
}^{20}{t}^{13}+{t}^{3}{q}^{2}-3\,{t}^{3}{q}^{4}+{t}^{5}{q}^{4}+3\,{q}^
{6}{t}^{3}-2\,{q}^{6}{t}^{5}+{q}^{6}{t}^{7}-{t}^{3}{q}^{8}+2\,{t}^{5}{
q}^{10}-{t}^{5}{q}^{12}+2\,{q}^{8}{t}^{7}+3\,{t}^{7}{q}^{12}-7\,{q}^{
10}{t}^{7}-{t}^{7}{q}^{16}+2\,{t}^{7}{q}^{14}-7\,{q}^{14}{t}^{9}+2\,{t
}^{9}{q}^{10}+2\,{t}^{9}{q}^{12}+{q}^{12}{t}^{11}+2\,{q}^{14}{t}^{11}-
2\,{q}^{18}{t}^{11}+{q}^{22}{t}^{13}+3\,{q}^{18}{t}^{9}+{q}^{18}{t}^{
13}+2\,{q}^{16}{t}^{11} ) {a}^{2}+{q}^{4}{t}^{2}-3\,{q}^{6}{t}^{
2}+{q}^{6}{t}^{4}-{t}^{2}{q}^{10}+3\,{t}^{2}{q}^{8}-2\,{t}^{4}{q}^{8}+
2\,{t}^{4}{q}^{12}-{t}^{4}{q}^{14}+{t}^{6}{q}^{8}-4\,{q}^{12}{t}^{6}-{
q}^{18}{t}^{6}+3\,{t}^{6}{q}^{16}+{t}^{6}{q}^{10}+{q}^{16}{t}^{10}+{q}
^{12}{t}^{8}-2\,{q}^{16}{t}^{8}+{q}^{14}{t}^{8}+{q}^{12}{t}^{15}{a}^{6
}+{q}^{20}{t}^{10}+3\,{q}^{20}{t}^{8}+{q}^{18}{t}^{10}-3\,{q}^{18}{t}^
{8}-3\,{q}^{22}{t}^{10}+{q}^{24}{t}^{12} \Big) =
\bf{\dfrac{a^2t+1}{1-q^2}}\cdot\Big( {\dfrac
{1}{{t}^{14}{a}^{4}{q}^{8}}}+{\dfrac {1}{{t}^{11}{a}^{2}{q}^{6}}}+{\dfrac
{1+{t}^{3}{a}^{2}}{{t}^{12}{a}^{4}{q}^{4}}}+{\dfrac
{(a^2t+1)(1+t^3a^2)}{{t}^{10}{a}^{4}{q}^{2}}}
+{\dfrac {1+{t}^{2}+{t}^{6}{a}^{4}+5\,{t}^{3}{a}^{2}}{{t}^{10}{a}^{4}}}+{\dfrac
{ ( 4+4\,{t}^{4}{a}^{4}+2\,{t}^{3}{a}^{2}+{t}^{6}{a}^{4}+5\,{a}^{2}t )
{q}^{2}}{{t}^{8}{a}^{4}}}
+{\dfrac { (1+t^3a^2)(t^6a^4+3t^3a^2+4a^2t+5+t^2)
{q}^{4}}{{t}^{8}{a}^{4}}}+\dots \Big)$

 \bigskip
 \noindent
 $ P^{T[4,8]}_{[1]}=\frac{\{A\}}{\{t\}^4  q^{24}
A^3}\Big(1+3\,{q}^{12}{t}^{12}-2\,{q}^{10}{t}^{12}-{q}^{12}{t}^{18}-2\,{q}^{20}
{t}^{20}-2\,{q}^{6}{t}^{4}-5\,{q}^{8}{t}^{6}-3\,{q}^{20}{t}^{22}+{q}^{
22}{t}^{22}-{q}^{14}{t}^{20}+2\,{q}^{14}{t}^{18}+2\,{t}^{16}{q}^{12}-{
t}^{16}{q}^{10}+2\,{q}^{16}{t}^{20}-3\,{t}^{2}+3\,{t}^{4}-{t}^{6}-2\,{
t}^{14}{q}^{16}+2\,{q}^{12}{t}^{6}+{t}^{6}{q}^{14}-{t}^{6}{q}^{10}+2\,
{t}^{12}{q}^{18}+4\,{t}^{8}{q}^{8}-2\,{q}^{10}{t}^{8}-6\,{q}^{12}{t}^{
8}+2\,{t}^{10}{q}^{16}+{q}^{22}{t}^{20}-2\,{t}^{14}{q}^{12}+{t}^{12}{q
}^{14}-5\,{t}^{16}{q}^{18}-6\,{t}^{12}{q}^{16}+{t}^{4}{q}^{10}+2\,{t}^
{4}{q}^{8}-{q}^{18}{t}^{14}+{q}^{20}{t}^{14}+2\,{q}^{20}{t}^{16}+2\,{q
}^{16}{t}^{8}+2\,{q}^{14}{t}^{8}-{t}^{12}{q}^{6}-6\,{t}^{10}{q}^{14}+3
\,{t}^{14}{q}^{14}+{t}^{12}{q}^{20}-3\,{t}^{4}{q}^{2}+4\,{t}^{16}{q}^{
16}+2\,{t}^{12}{q}^{8}+{t}^{10}{q}^{18}+{q}^{24}{t}^{24}-3\,{q}^{22}{t
}^{24}+{q}^{22}{t}^{18}-2\,{t}^{18}{q}^{20}-{t}^{24}{q}^{18}+3\,{t}^{
22}{q}^{18}-{t}^{22}{q}^{16}-2\,{q}^{14}{t}^{16}+3\,{t}^{24}{q}^{20}+{
t}^{18}{q}^{18}-{q}^{16}{t}^{18}-2\,{t}^{10}{q}^{8}+{t}^{4}{q}^{12}+
 ( 3\,{t}^{10}{q}^{20}-{q}^{24}{t}^{12}+3\,{q}^{22}{t}^{12}+2\,{q
}^{20}{t}^{8}-{t}^{2}{q}^{16}-{q}^{18}{t}^{2}-{q}^{18}{t}^{4}+{q}^{12}
{t}^{6}+4\,{q}^{18}{t}^{6}-{t}^{6}{q}^{20}-3\,{t}^{6}{q}^{14}+2\,{t}^{
4}{q}^{16}-{t}^{4}{q}^{20}+{t}^{12}{q}^{18}+{t}^{10}{q}^{16}-2\,{q}^{
16}{t}^{8}+{q}^{14}{t}^{8}-{q}^{12}-3\,{t}^{12}{q}^{20}-{q}^{22}{t}^{8
}-3\,{t}^{10}{q}^{18}-{t}^{2}{q}^{14}+3\,{t}^{2}{q}^{12}-3\,{t}^{4}{q}
^{12}+3\,{t}^{4}{q}^{14}-{q}^{22}{t}^{10}-{t}^{6}{q}^{22} ) {A}^
{6}+ ( -{q}^{12}{t}^{12}+3\,{q}^{6}{t}^{4}+2\,{q}^{8}{t}^{6}+{t}^
{10}{q}^{20}+3\,{q}^{22}{t}^{12}+3\,{q}^{20}{t}^{8}+{q}^{8}+{t}^{14}{q
}^{16}+{t}^{2}{q}^{16}+2\,{q}^{18}{t}^{4}+4\,{q}^{12}{t}^{6}+4\,{q}^{
18}{t}^{6}+2\,{t}^{6}{q}^{20}-4\,{t}^{6}{q}^{16}-9\,{t}^{6}{q}^{14}+3
\,{t}^{4}{q}^{16}+2\,{t}^{6}{q}^{10}+4\,{t}^{12}{q}^{18}-{t}^{8}{q}^{8
}+{q}^{10}{t}^{8}+6\,{t}^{10}{q}^{16}-2\,{q}^{22}{t}^{16}+{t}^{12}{q}^
{14}+2\,{t}^{16}{q}^{18}-2\,{t}^{4}{q}^{10}-{t}^{2}{q}^{10}+2\,{q}^{18
}{t}^{14}-2\,{q}^{8}{t}^{2}-2\,{q}^{20}{t}^{14}-6\,{q}^{16}{t}^{8}+6\,
{q}^{14}{t}^{8}-{t}^{14}{q}^{14}-7\,{t}^{12}{q}^{20}-{t}^{16}{q}^{16}+
{q}^{10}+{q}^{22}{t}^{8}-4\,{t}^{8}{q}^{18}-9\,{t}^{10}{q}^{18}-3\,{q}
^{22}{t}^{18}+3\,{t}^{18}{q}^{20}-{t}^{18}{q}^{18}+2\,{t}^{2}{q}^{14}+
3\,{t}^{2}{q}^{12}-7\,{t}^{4}{q}^{12}+{t}^{4}{q}^{14}+{q}^{12}{t}^{10}
-{q}^{10}{t}^{10}+2\,{q}^{22}{t}^{10}-3\,{q}^{6}{t}^{2}+{q}^{24}{t}^{
14}-{q}^{22}{t}^{14}-{q}^{6}{t}^{6}+{q}^{24}{t}^{18}+{q}^{24}{t}^{16}+
{q}^{6} ) {A}^{4}+ ( {q}^{12}{t}^{12}-{q}^{10}{t}^{12}+2\,{
q}^{6}{t}^{4}-4\,{q}^{8}{t}^{6}-3\,{t}^{10}{q}^{20}-3\,{q}^{20}{t}^{22
}+3\,{q}^{22}{t}^{22}+{q}^{14}{t}^{18}-2\,{q}^{18}{t}^{20}+{t}^{16}{q}
^{12}+{q}^{16}{t}^{20}-{q}^{22}{t}^{12}-{q}^{20}{t}^{8}-5\,{t}^{14}{q}
^{16}+10\,{q}^{12}{t}^{6}-{q}^{18}{t}^{6}-4\,{t}^{6}{q}^{16}-2\,{t}^{6
}{q}^{14}-{t}^{4}{q}^{16}+4\,{t}^{6}{q}^{10}+10\,{t}^{12}{q}^{18}-5\,{
q}^{10}{t}^{8}-2\,{q}^{12}{t}^{8}+12\,{t}^{10}{q}^{16}+2\,{q}^{22}{t}^
{20}-3\,{q}^{22}{t}^{16}-{t}^{14}{q}^{12}-4\,{t}^{12}{q}^{14}-4\,{t}^{
16}{q}^{18}-{q}^{4}-2\,{t}^{12}{q}^{16}+2\,{t}^{4}{q}^{10}-2\,{t}^{2}{
q}^{10}+7\,{t}^{4}{q}^{8}+4\,{q}^{18}{t}^{14}-3\,{q}^{8}{t}^{2}+2\,{q}
^{20}{t}^{14}+7\,{q}^{20}{t}^{16}+12\,{q}^{14}{t}^{8}-4\,{t}^{10}{q}^{
14}-{q}^{2}+{t}^{14}{q}^{14}-4\,{t}^{12}{q}^{20}-2\,{q}^{4}{t}^{6}-3\,
{t}^{4}{q}^{2}+{t}^{12}{q}^{8}-4\,{t}^{8}{q}^{18}-2\,{t}^{10}{q}^{18}+
{q}^{22}{t}^{18}+2\,{t}^{18}{q}^{20}+{t}^{22}{q}^{18}-{q}^{14}{t}^{16}
-2\,{t}^{18}{q}^{18}-{q}^{16}{t}^{18}-{t}^{10}{q}^{8}-{t}^{2}{q}^{12}-
4\,{t}^{4}{q}^{12}-3\,{t}^{4}{q}^{14}+2\,{q}^{4}{t}^{2}-4\,{q}^{12}{t}
^{10}+{q}^{10}{t}^{10}+{q}^{6}{t}^{2}-{q}^{24}{t}^{22}+{t}^{14}{q}^{10
}+3\,{q}^{2}{t}^{2}-2\,{q}^{22}{t}^{14}-{q}^{24}{t}^{20}-2\,{q}^{6}{t}
^{6}+{t}^{6}{q}^{2}-{q}^{6}{t}^{8}-{q}^{24}{t}^{18}-{q}^{6}+{q}^{4}{t}
^{8}+{q}^{6}{t}^{10} ) {A}^{2}+{q}^{4}{t}^{2}+{q}^{12}{t}^{10}+3
\,{q}^{10}{t}^{10}+{q}^{6}{t}^{2}+2\,{t}^{14}{q}^{10}-{t}^{14}{q}^{8}+
{q}^{2}{t}^{2}-2\,{t}^{4}{q}^{4}+{q}^{6}{t}^{6}-{t}^{8}{q}^{2}+3\,{t}^
{6}{q}^{2}-{q}^{6}{t}^{8}+2\,{q}^{4}{t}^{8}+2\,{q}^{6}{t}^{10}-{t}^{10
}{q}^{4}
\Big) =\\ = \bf{{\dfrac {{a}^{2
}t+1}{ ( -1+{q}^{2} ) ^{4}{a}^{4}{t}^{26}{q}^{20}}}} \Big(
1-3\,{q}^{2}+3\,{q}^{4}-{q}^{6}+{q}^{22}{t}^{12}+{t}^{14}{q}^{20}-6\,{
q}^{24}{t}^{14}-{q}^{42}{t}^{18}+{q}^{44}{t}^{22}-{q}^{30}{t}^{12}-{q}
^{22}{t}^{8}-2\,{q}^{30}{t}^{14}+{q}^{26}{t}^{14}+2\,{q}^{24}{t}^{10}-
{q}^{26}{t}^{10}-2\,{q}^{26}{t}^{12}+2\,{q}^{28}{t}^{12}+3\,{q}^{28}{t
}^{14}+2\,{q}^{26}{t}^{16}-{q}^{34}{t}^{14}+2\,{q}^{32}{t}^{14}+2\,{q}
^{30}{t}^{18}-6\,{q}^{28}{t}^{16}+{q}^{28}{t}^{18}-{q}^{32}{t}^{18}-2
\,{q}^{30}{t}^{16}+{q}^{32}{t}^{20}-2\,{q}^{38}{t}^{20}+{q}^{36}{t}^{
18}+{q}^{40}{t}^{22}-5\,{q}^{34}{t}^{18}+4\,{q}^{32}{t}^{16}+{q}^{34}{
t}^{20}-{q}^{34}{t}^{16}+2\,{q}^{36}{t}^{20}+( -{q}^{20}{t}^{11}
+2\,{q}^{20}{t}^{13}+4\,{q}^{32}{t}^{21}-12\,{q}^{22}{t}^{15}+2\,{q}^{
28}{t}^{17}-10\,{q}^{30}{t}^{19}+2\,{q}^{36}{t}^{19}+4\,{q}^{24}{t}^{
15}-2\,{q}^{42}{t}^{23}+4\,{q}^{26}{t}^{15}+2\,{q}^{28}{t}^{19}-12\,{q
}^{26}{t}^{17}+{q}^{44}{t}^{25}-7\,{q}^{36}{t}^{21}-{q}^{40}{t}^{23}+3
\,{q}^{38}{t}^{23}+{q}^{46}{t}^{25}+4\,{q}^{34}{t}^{19}+5\,{q}^{30}{t}
^{17}+2\,{q}^{36}{t}^{23}+{q}^{42}{t}^{25}-4\,{q}^{32}{t}^{19}-2\,{q}^
{34}{t}^{21}+{q}^{34}{t}^{23}+{q}^{28}{t}^{21}+3\,{q}^{42}{t}^{21}-{q}
^{40}{t}^{19}-2\,{q}^{38}{t}^{21}+3\,{q}^{30}{t}^{21}+{t}^{3}{q}^{2}-3
\,{t}^{3}{q}^{4}+{t}^{5}{q}^{4}+3\,{q}^{6}{t}^{3}-2\,{q}^{6}{t}^{5}+{q
}^{6}{t}^{7}-{t}^{3}{q}^{8}+2\,{t}^{5}{q}^{10}-{t}^{5}{q}^{12}-{q}^{8}
{t}^{7}+2\,{t}^{7}{q}^{12}-2\,{q}^{10}{t}^{7}-{t}^{7}{q}^{16}+{t}^{7}{
q}^{14}+4\,{q}^{14}{t}^{9}+3\,{t}^{9}{q}^{10}-7\,{t}^{9}{q}^{12}+{t}^{
13}{q}^{14}+2\,{q}^{12}{t}^{11}-2\,{q}^{14}{t}^{11}+4\,{q}^{22}{t}^{17
}+2\,{q}^{20}{t}^{15}+2\,{q}^{38}{t}^{19}+4\,{q}^{26}{t}^{19}-{q}^{36}
{t}^{17}+{q}^{24}{t}^{19}+{q}^{34}{t}^{17}+{q}^{20}{t}^{17}-3\,{q}^{44
}{t}^{23}-{q}^{28}{t}^{15}-{q}^{32}{t}^{15}+{q}^{30}{t}^{15}+{q}^{22}{
t}^{11}+{q}^{26}{t}^{13}-{q}^{28}{t}^{13}-{q}^{24}{t}^{13}-{q}^{20}{t}
^{9}-{q}^{24}{t}^{11}+5\,{q}^{18}{t}^{11}+4\,{q}^{22}{t}^{13}+{q}^{18}
{t}^{9}+4\,{t}^{13}{q}^{16}+3\,{t}^{15}{q}^{18}-10\,{q}^{18}{t}^{13}-4
\,{q}^{16}{t}^{11} ) {a}^{2}+ ( -{q}^{24}{t}^{14}+{q}^{42}{
t}^{26}-3\,{q}^{40}{t}^{24}-2\,{q}^{34}{t}^{22}+{q}^{30}{t}^{24}+3\,{q
}^{14}{t}^{14}+3\,{q}^{34}{t}^{24}+2\,{q}^{32}{t}^{24}-6\,{q}^{24}{t}^
{18}+{q}^{26}{t}^{16}+{q}^{30}{t}^{18}-{q}^{28}{t}^{16}-{q}^{32}{t}^{
18}+2\,{q}^{32}{t}^{20}+2\,{q}^{34}{t}^{20}-{q}^{36}{t}^{20}+{q}^{6}{t
}^{8}+3\,{t}^{18}{q}^{20}-4\,{q}^{22}{t}^{18}+6\,{q}^{22}{t}^{16}-{q}^
{16}{t}^{10}-3\,{t}^{8}{q}^{8}+3\,{q}^{10}{t}^{8}-{q}^{12}{t}^{8}+{q}^
{10}{t}^{12}+2\,{t}^{16}{q}^{16}-{q}^{12}{t}^{12}+{t}^{12}{q}^{18}+{t}
^{16}{q}^{18}+{t}^{10}{q}^{8}+{q}^{40}{t}^{26}-2\,{q}^{38}{t}^{24}+2\,
{q}^{14}{t}^{10}+3\,{q}^{28}{t}^{22}-2\,{q}^{14}{t}^{12}-{q}^{36}{t}^{
24}+{q}^{22}{t}^{14}+{q}^{38}{t}^{26}+4\,{q}^{30}{t}^{20}-4\,{q}^{26}{
t}^{20}-7\,{q}^{32}{t}^{22}+6\,{q}^{26}{t}^{18}-9\,{q}^{28}{t}^{20}+{q
}^{30}{t}^{22}+2\,{q}^{26}{t}^{22}+4\,{q}^{24}{t}^{20}+3\,{q}^{38}{t}^
{22}-{q}^{20}{t}^{12}-2\,{t}^{10}{q}^{10}+4\,{q}^{18}{t}^{14}-9\,{q}^{
20}{t}^{16}+2\,{q}^{22}{t}^{20}+{q}^{18}{t}^{18}+2\,{q}^{16}{t}^{12}-7
\,{q}^{16}{t}^{14} ) {a}^{4}+ ( {q}^{26}{t}^{23}+{q}^{24}{t
}^{23}-2\,{q}^{28}{t}^{23}+{q}^{22}{t}^{21}-4\,{q}^{24}{t}^{21}+{q}^{
30}{t}^{25}-3\,{q}^{30}{t}^{23}+{q}^{32}{t}^{25}-3\,{q}^{34}{t}^{25}+{
q}^{36}{t}^{27}+3\,{q}^{28}{t}^{21}-2\,{q}^{20}{t}^{19}+{q}^{12}{t}^{
15}-{q}^{30}{t}^{21}-3\,{q}^{18}{t}^{17}+{q}^{18}{t}^{19}+{q}^{20}{t}^
{21}+{q}^{16}{t}^{17}-{q}^{22}{t}^{17}+3\,{t}^{15}{q}^{16}-{q}^{26}{t}
^{19}+3\,{q}^{32}{t}^{23}+2\,{q}^{24}{t}^{19}+3\,{q}^{20}{t}^{17}+{q}^
{28}{t}^{25}-3\,{t}^{15}{q}^{14}-{t}^{15}{q}^{18} ) {a}^{6}+{q}^
{4}{t}^{2}-3\,{q}^{6}{t}^{2}+{q}^{6}{t}^{4}+{q}^{48}{t}^{24}+3\,{q}^{
44}{t}^{20}+2\,{q}^{36}{t}^{16}-{t}^{2}{q}^{10}+3\,{t}^{2}{q}^{8}-2\,{
t}^{4}{q}^{8}+2\,{t}^{4}{q}^{12}-{t}^{4}{q}^{14}+{t}^{6}{q}^{8}+{q}^{
12}{t}^{6}-{q}^{18}{t}^{6}+2\,{t}^{6}{q}^{16}-{t}^{6}{q}^{14}-2\,{t}^{
6}{q}^{10}-{q}^{16}{t}^{10}+2\,{q}^{12}{t}^{8}+4\,{q}^{16}{t}^{8}-5\,{
q}^{14}{t}^{8}+2\,{t}^{12}{q}^{18}+{q}^{14}{t}^{10}-3\,{q}^{46}{t}^{22
}+3\,{q}^{40}{t}^{18}+{q}^{42}{t}^{22}-{q}^{38}{t}^{16}+2\,{q}^{22}{t}
^{14}-3\,{q}^{42}{t}^{20}-2\,{q}^{40}{t}^{20}+2\,{q}^{24}{t}^{16}-6\,{
q}^{20}{t}^{12}+3\,{q}^{20}{t}^{10}+2\,{q}^{20}{t}^{8}-2\,{q}^{18}{t}^
{10}-2\,{q}^{18}{t}^{8}-2\,{q}^{22}{t}^{10}+3\,{q}^{24}{t}^{12}+{q}^{
16}{t}^{12}
\Big)=\bf{\dfrac{a^2t+1}{1-q^2}}\cdot\Big( {\dfrac
{1}{{t}^{26}{a}^{4}{q}^{20}}}+{\dfrac {1}{{t}^{23}{a}^{2}{q}^{18}}}+{\dfrac
{1+{t}^{3}{a}^{2}}{{t}^{24}{a}^{4}{q}^{16}}}+{\dfrac
{{a}^{2}t+{t}^{4}{a}^{4}+{t}^{3}{a}^{2}+1}{{t}^{22}{a}^{4}{q}^{14}}}
+{\dfrac
{{t}^{6}{a}^{4}+1+{t}^{2}+2\,{t}^{3}{a}^{2}}{{t}^{22}{a}^{4}{q}^{12}}}+{\dfrac
{3\,{t}^{3}{a}^{2}+{t}^{6}{a}^{4}+{t}^{4}{a}^{4}+1+{a}^{2}t}{{t}^{20}{a}^{4}{q}^{10}}}+{\dfrac
{2\,{t}^{3}{a}^{2}+2\,{t}^{6}{a}^{4}+2\,{t}^{5}{a}^{2}+{t}^{9}{a}^{6}+1+2\,{t}^{2}}{{t}^{20}{a}^{4}{q}^{8}}}
+{\dfrac {(3a^2t^3+a^2t+t^2+1)(a^2t^3+1)}{{t}^{18}{a}^{4}{q}^{6}}}+\\+{\dfrac
{7\,{t}^{5}{a}^{2}+2\,{a}^{4}{t}^{8}+{t}^{9}{a}^{6}+2\,{t}^{6}{a}^{4}+2\,{t}^{3}{a}^{2}+2\,{t}^{2}+1+{t}^{4}}{{t}^{18}{a}^{4}{q}^{4}}}
+{\dfrac
{8\,{t}^{3}{a}^{2}+7\,{t}^{6}{a}^{4}+3\,{t}^{5}{a}^{2}+{t}^{4}{a}^{4}+{a}^{2}t+{t}^{9}{a}^{6}+{a}^{4}{t}^{8}+1+5\,{t}^{2}}{{t}^{16}{a}^{4}{q}^{2}}}+{\dfrac
{(1+t^3a^2)(a^4t^8+t^6a^4+5t^5a^2+5t^3a^2+6t^2+t^4+1)}{{t}^{16}{a}^{4}}}
+{\dfrac { (1+t^3a^2)(4t^6a^4+2t^5a^2+7t^3a^2+5a^2t+5t^2+5)
{q}^{2}}{{t}^{14}{a}^{4}}}+\dots \Big)$

\bigskip
\noindent
$\boxed{P^{T[4,4k]}_{[1]}=\dfrac{\{A\}}{\{t\}^4 A^3
q^{12k}}{\cal{P}}^{T[4,4k]}_{[1]}}$

\begin{itemize}
\item{HOMFLY case}
\end{itemize}
$$ H^{T[4,0]}_{[1]}=\frac{\{A\}}{\{q\}^4 A^3}(1-A^2)^3 $$

\bigskip
\noindent
$H^{T[4,4]}_{[1]}=\frac{\{A\}}{\{q\}^4  q^{12} A^3}\Big( -{q}^{12}{A}^{6}+ (
-5\,{q}^{12}+5\,{q}^{14}+{q}^{6}-2\,{q}^{8}+5
\,{q}^{10}+{q}^{18}-2\,{q}^{16} ) {A}^{4}+( -5\,{q}^{12}+2
\,{q}^{20}-{q}^{2}+2\,{q}^{4}+3\,{q}^{10}-2\,{q}^{18}-{q}^{22}-{q}^{16
}+3\,{q}^{14}-2\,{q}^{6}-{q}^{8} ) {A}^{2}+1-3\,{q}^{18}+2\,{q}^
{8}-3\,{q}^{6}+2\,{q}^{16}-3\,{q}^{2}+4\,{q}^{4}-3\,{q}^{22}-{q}^{12}+
{q}^{24}+4\,{q}^{20}
 \Big)$

\bigskip
\noindent
$H^{T[4,8]}_{[1]}=\frac{\{A\}}{\{q\}^4  q^{24} A^3}\Big((
-{q}^{36}+3\,{q}^{14}-4\,{q}^{16}-2\,{q}^{20}-4\,{q}^{32}+3\,{q}^{18}+3\,{q}^{30}+{q}^{24}-{q}^{12}-2\,{q}^{28}+3\,{q}^{34})
{A}^{6}+( 3\,{q}^{34}+{q}^{6}+3\,{q}^{14}+2\,{q}^{10}-4\,{q}^{32
}+5\,{q}^{22}-7\,{q}^{20}+7\,{q}^{18}-7\,{q}^{28}-4\,{q}^{16}-2\,{q}^{
12}-2\,{q}^{40}+{q}^{42}-2\,{q}^{8}+2\,{q}^{38}-2\,{q}^{36}+7\,{q}^{30
}-3\,{q}^{24}+5\,{q}^{26}) {A}^{4}+ ( -3\,{q}^{10}+2\,{q}^
{4}-3\,{q}^{28}-{q}^{46}-4\,{q}^{14}+4\,{q}^{36}-3\,{q}^{38}-{q}^{2}-3
\,{q}^{20}+2\,{q}^{44}+{q}^{16}-3\,{q}^{24}+2\,{q}^{8}-4\,{q}^{34}-2\,
{q}^{42}-2\,{q}^{6}+4\,{q}^{12}+3\,{q}^{26}+{q}^{32}+3\,{q}^{22}+{q}^{
18}+{q}^{30}+2\,{q}^{40} ) {A}^{2}+1+2\,{q}^{8}+5\,{q}^{36}-3\,{
q}^{42}+4\,{q}^{4}-3\,{q}^{30}-3\,{q}^{38}-3\,{q}^{10}+6\,{q}^{16}-3\,
{q}^{18}-3\,{q}^{2}+4\,{q}^{44}+5\,{q}^{12}+{q}^{24}+6\,{q}^{32}+{q}^{
48}-3\,{q}^{46}-6\,{q}^{34}+2\,{q}^{40}-6\,{q}^{14}-3\,{q}^{6}
\Big)$

\bigskip
\noindent
$\boxed{H^{T[4,4k]}_{[1]}=\dfrac{\{A\}}{\{q\}^{4} A^3 q^{12k}}
{\cal{H}}^{T[4,4k]}_{[1]}}$

\begin{itemize}
\item{Alexander case}
\end{itemize}
\be
\begin{array}{l}
A^{T[4,0]}_{[1]}=8,\\
\\\
A^{T[4,4]}_{[1]}=1+{\textbf{q}}^{24}-2\,{\textbf{q}}^{22}+2\,{\textbf{q}}^{20}+{\textbf{q}}^{16}+2\,{\textbf{q}}^{14}
+2\,{\textbf{q}}^{10}+{\textbf{q}
}^{8}+2\,{\textbf{q}}^{4}-2\,{\textbf{q}}^{2}\\
\\\
A^{T[4,8]}_{[1]}=1-2\,{\textbf{q}}^{8}+2\,{\textbf{q}}^{4}+2\,{\textbf{q}}^{38}-2\,{\textbf{q}}^{14}
-2\,{\textbf{q}}^{2}-2\,{\textbf{q}}^{46}
+2\,{\textbf{q}}^{10}+5\,{\textbf{q}}^{16}-2\,{\textbf{q}}^{20}+2\,{\textbf{q}}^{22}+2\,{\textbf{q}}^{26}-\\2\,{\textbf{q}}^{
28}+2\,{\textbf{q}}^{44}+{\textbf{q}}^{48}-2\,{\textbf{q}}^{34}+5\,{\textbf{q}}^{32}-2\,{\textbf{q}}^{40}
\end{array}
\ee
\subsection{Case $(4,n)$, $n=4 k+1$ fundamental representation\label{4k+1}}
\be
\begin{array}{|c|}
\hline\\
P^{T[4,n]}_{[1]}=c^{[4]}_{[1]} M_{[4]}^{\ast} q^{-3n}+c^{[3,1]}_{[1]}
M_{[3,1]}^{\ast} q^{-\frac{3n}{2}} t^{\frac{n}{2}}
+c^{[2,2]}_{[1]} M_{[2,2]}^{\ast} q^{-n} t^{n} + c^{[2,1,1]}_{[1]}
M_{[2,1,1]}^{\ast} q^{-\frac{n}{2}} t^{\frac{3n}{2}} + c^{[1,1,1,1]}_{[1]}
M_{[1,1,1,1]}^{\ast} t^{3n}
\\
\\
\hline
\end{array}
\ee
\fr{
c^{[4]}_{[1]}=1,\ \ \ c^{[3,1]}_{[1]}=-{\frac {\sqrt {q} \left(
{q}^{4}{t}^{2}+{q}^{2}+1+{q}^{4} \right)
 \left( -1+{t}^{2} \right) }{\sqrt {t} \left( -1+{q}^{6}{t}^{2}
 \right) }}\ \ \ c^{[2,2]}_{[1]}={\frac {{q}^{2} \left( -1+{t}^{2} \right) 
\left( 1+{t}^{2} \right)
 \left( {t}^{2}-{q}^{2} \right) }{t \left( -1+{q}^{4}{t}^{2} \right)
 \left( -1+{q}^{2}{t}^{2} \right) }} \ \ \ \ \ \ \ \ \ \ \ \  \ \ \ \
 \nonumber \\ \nonumber
 c^{[2,1,1]}_{[1]}={\frac {{q}^{3/2} \left( 1+{t}^{2} \right)  \left(
-1+{t}^{2} \right)
^{2} \left( {t}^{4}{q}^{2}+{q}^{2}{t}^{2}+{q}^{2}+1 \right) }{{t}^{3/2
} \left( {q}^{2}{t}^{2}+1 \right)  \left( -1+{q}^{2}{t}^{2} \right) ^{
2}}}
,\ \ \ c^{[1,1,1,1]}_{[1]}=-{\frac {{q}^{3} \left( {t}^{4}+1 \right)  \left(
{t}^{2}+{t}^{4}+1
 \right)  \left( 1+{t}^{2} \right) ^{2} \left( -1+{t}^{2} \right) ^{3}
}{{t}^{3} \left( -1+{q}^{2}{t}^{2} \right)  \left( -1+{t}^{6}{q}^{2}
 \right)  \left( -1+{t}^{4}{q}^{2} \right) }}}
Several first superpolynomials are:
$$P^{T[4,1]}_{[1]}=\dfrac{\{A\} t^3}{\{t\} A^3 q^3}=\bf{-{\frac { \left(
{a}^{2}t+1 \right) q}{ \left( -1+{q}^{2} \right) {a}^
{4}{t}^{5}}}
}$$
\bigskip \\
$P^{T[4,5]}_{[1]}=\dfrac{\{A\} t^3}{\{t\} A^3 q^{15}}\Big( {q}^{12}(-A^{2})^3+
( {q}^{12}{t}^{6}+{t}^{2}{q}^{12}+{t}^{2}{q}^{8
}+{q}^{6}+{q}^{8}+{t}^{2}{q}^{10}+{q}^{10}+{t}^{4}{q}^{10}+{t}^{4}{q}^
{12} ) {A}^{4}+ ({t}^{10}{q}^{12}+{q}^{4}+2\,{q}^{6}{t}^{
2}+{t}^{2}{q}^{10}+{q}^{6}{t}^{4}+2\,{t}^{4}{q}^{10}+{q}^{12}{t}^{6}+2
\,{t}^{4}{q}^{8}+2\,{t}^{6}{q}^{10}+{q}^{6}+{t}^{6}{q}^{8}+{q}^{12}{t}
^{8}+2\,{t}^{2}{q}^{8}+{q}^{10}{t}^{8}+{q}^{2}+{q}^{4}{t}^{2} )
(-A^{2})+1+{q}^{6}{t}^{2}+{q}^{6}{t}^{6}+{q}^{6}{t}^{4}+{t}^{6}{q}^{8}+
{t}^{4}{q}^{8}+{t}^{8}{q}^{8}+{q}^{2}{t}^{2}+{t}^{4}{q}^{4}+{t}^{6}{q}
^{10}+{t}^{12}{q}^{12}+{q}^{10}{t}^{8}+{q}^{4}{t}^{2}+{t}^{10}{q}^{10}
 \Big)=\\ = -{\dfrac {{\textbf{a}}^{2}\textbf{t}+1}{ ( -1+{\textbf{q}}^{2} )
{\textbf{a}}^{4}{\textbf{t}}^{17}{\textbf{q}}^{11
}}}  ( {\textbf{q}}^{12}{\textbf{t}}^{15}{\textbf{a}}^{6}+ (
{\textbf{q}}^{16}{\textbf{t}}^{14}+{\textbf{q}}^{18}{\textbf{t}}^{14}+{\textbf{q}}^
{10}{\textbf{t}}^{12}+{\textbf{t}}^{14}{\textbf{q}}^{14}+{\textbf{t}}^{10}{\textbf{q}}^{10}+{\textbf{t}}^{10}{\textbf{q}}^{8}
+{\textbf{t}}^{12
}{\textbf{q}}^{14}+{\textbf{q}}^{6}{\textbf{t}}^{8}+{\textbf{t}}^{12}{\textbf{q}}^{12}
) {\textbf{a}}^{4}+ ( {\textbf{q}}^
{18}{\textbf{t}}^{13}+2\,{\textbf{q}}^{8}{\textbf{t}}^{7}+2\,{\textbf{t}}^{9}{\textbf{q}}^{10}+
{\textbf{t}}^{11}{\textbf{q}}^{12}+{\textbf{t}}
^{9}{\textbf{q}}^{14}+{\textbf{q}}^{10}{\textbf{t}}^{7}+2\,{\textbf{q}}^{16}{\textbf{t}}^{11}+{\textbf{t}}^{5}{\textbf{q}}^{4}
+{\textbf{t}}^{3
}{\textbf{q}}^{2}+2\,{\textbf{t}}^{11}{\textbf{q}}^{14}+2\,{\textbf{t}}^{9}{\textbf{q}}^{12}+{\textbf{q}}^{20}{\textbf{t}}^{13}+
{\textbf{q}}^{
18}{\textbf{t}}^{11}+{\textbf{q}}^{6}{\textbf{t}}^{5}+{\textbf{q}}^{22}{\textbf{t}}^{13}+{\textbf{q}}^{6}{\textbf{t}}^{7}
) {\textbf{a}}
^{2}+1+{\textbf{q}}^{4}{\textbf{t}}^{2}+{\textbf{q}}^{6}{\textbf{t}}^{4}+{\textbf{t}}^{4}{\textbf{q}}^{8}+{\textbf{t}}^{6}{\textbf{q}}^{8}
+{\textbf{q}}
^{12}{\textbf{t}}^{6}+{\textbf{t}}^{6}{\textbf{q}}^{10}+{\textbf{q}}^{12}{\textbf{t}}^{8}+{\textbf{q}}^{16}{\textbf{t}}^{8}
+{\textbf{q}}^{14}{
\textbf{t}}^{8}+{\textbf{q}}^{20}{\textbf{t}}^{10}+{\textbf{q}}^{18}{\textbf{t}}^{10}+{\textbf{q}}^{24}{\textbf{t}}^{12}
+{\textbf{t}}^{10}{\textbf{q}}^
{16}
 )$\\

\bigskip
\noindent
$P^{T[4,9]}_{[1]}=\dfrac{\{A\} t^3}{\{t\} A^3 q^{27}}
\Big(1+{t}^{4}{q}^{10}+{q}^{20}{t}^{20}+{q}^{24}{t}^{24}+{t}^{18}{q}^{18}+{
q}^{22}{t}^{18}+{t}^{16}{q}^{18}+{q}^{16}{t}^{14}+{t}^{18}{q}^{20}+{t}
^{22}{q}^{22}+{q}^{16}{t}^{16}+2\,{q}^{20}{t}^{16}+{t}^{20}{q}^{22}+{t
}^{14}{q}^{20}+{t}^{12}{q}^{20}+2\,{t}^{12}{q}^{16}+2\,{q}^{18}{t}^{12
}+2\,{t}^{10}{q}^{14}+{q}^{2}{t}^{2}+{t}^{4}{q}^{4}+{q}^{4}{t}^{2}+{q}
^{6}{t}^{2}+{q}^{6}{t}^{6}+{q}^{6}{t}^{4}+2\,{t}^{4}{q}^{8}+{t}^{4}{q}
^{12}+{t}^{6}{q}^{8}+2\,{q}^{12}{t}^{6}+{t}^{6}{q}^{14}+2\,{t}^{6}{q}^
{10}+{t}^{8}{q}^{8}+{q}^{10}{t}^{8}+2\,{q}^{12}{t}^{8}+ ( {q}^{6}
+{t}^{4}{q}^{10}+{q}^{8}+{q}^{10}+{t}^{14}{q}^{20}+2\,{t}^{12}{q}^{20}
+{q}^{18}{t}^{12}+{t}^{8}{q}^{22}+2\,{t}^{2}{q}^{10}+{t}^{2}{q}^{8}+{t
}^{16}{q}^{24}+{t}^{14}{q}^{24}+2\,{t}^{2}{q}^{14}+3\,{t}^{2}{q}^{12}+
2\,{t}^{4}{q}^{12}+4\,{t}^{4}{q}^{14}+{t}^{2}{q}^{16}+2\,{q}^{18}{t}^{
4}+{q}^{12}{t}^{6}+4\,{q}^{18}{t}^{6}+2\,{t}^{6}{q}^{20}+4\,{t}^{6}{q}
^{16}+2\,{t}^{6}{q}^{14}+3\,{t}^{4}{q}^{16}+{t}^{18}{q}^{24}+3\,{t}^{
12}{q}^{22}+{t}^{16}{q}^{22}+2\,{t}^{14}{q}^{22}+2\,{t}^{10}{q}^{22}+2
\,{q}^{16}{t}^{8}+{q}^{14}{t}^{8}+3\,{q}^{20}{t}^{8}+4\,{t}^{8}{q}^{18
}+4\,{q}^{20}{t}^{10}+2\,{q}^{18}{t}^{10}+{t}^{10}{q}^{16}) {A}
^{4}+ ({t}^{6}{q}^{20}+{t}^{4}{q}^{16}+{q}^{18}{t}^{4}+{t}^{6}{
q}^{22}+{t}^{8}{q}^{22}+{t}^{2}{q}^{16}+{t}^{4}{q}^{20}+{q}^{18}{t}^{6
}+{t}^{2}{q}^{14}+{t}^{10}{q}^{22}+{q}^{18}{t}^{2}+{q}^{20}{t}^{8}+{q}
^{24}{t}^{12}+{q}^{12} ) (-A^{2})^3+ ({q}^{2}+{q}^{4}+{q}^{6
}+4\,{t}^{4}{q}^{10}+2\,{q}^{22}{t}^{18}+{t}^{16}{q}^{18}+{q}^{16}{t}^
{14}+{t}^{18}{q}^{20}+2\,{q}^{20}{t}^{16}+{t}^{20}{q}^{22}+4\,{t}^{14}
{q}^{20}+4\,{t}^{12}{q}^{20}+2\,{t}^{12}{q}^{16}+4\,{q}^{18}{t}^{12}+2
\,{t}^{10}{q}^{14}+{q}^{4}{t}^{2}+2\,{q}^{6}{t}^{2}+{q}^{6}{t}^{4}+2\,
{t}^{2}{q}^{10}+3\,{t}^{2}{q}^{8}+2\,{t}^{4}{q}^{8}+{t}^{2}{q}^{12}+4
\,{t}^{4}{q}^{12}+3\,{t}^{4}{q}^{14}+{t}^{6}{q}^{8}+4\,{q}^{12}{t}^{6}
+{q}^{18}{t}^{6}+4\,{t}^{6}{q}^{16}+5\,{t}^{6}{q}^{14}+{t}^{4}{q}^{16}
+2\,{t}^{6}{q}^{10}+{t}^{22}{q}^{24}+{t}^{18}{q}^{24}+{t}^{20}{q}^{24}
+{t}^{12}{q}^{22}+3\,{t}^{16}{q}^{22}+{q}^{10}{t}^{8}+2\,{q}^{12}{t}^{
8}+2\,{t}^{14}{q}^{22}+5\,{q}^{16}{t}^{8}+4\,{q}^{14}{t}^{8}+{q}^{20}{
t}^{8}+4\,{t}^{8}{q}^{18}+3\,{q}^{20}{t}^{10}+5\,{q}^{18}{t}^{10}+{t}^
{10}{q}^{12}+4\,{t}^{10}{q}^{16}+2\,{q}^{18}{t}^{14}+{t}^{12}{q}^{14}
 ) (-A^{2})+2\,{q}^{16}{t}^{8}+2\,{q}^{14}{t}^{8}+{q}^{18}{t}^{10
}+{t}^{10}{q}^{12}+2\,{t}^{10}{q}^{16}+{t}^{10}{q}^{10}+2\,{q}^{18}{t}
^{14}+{t}^{14}{q}^{14}+{t}^{12}{q}^{14}+{t}^{12}{q}^{12} \Big) =\\ = -{\dfrac
{{\textbf{a}}^{2}\textbf{t}+1}{ ( -1+{\textbf{q}}^{2} )
{\textbf{a}}^{4}{\textbf{t}}^{29}{\textbf{q}}^{23
}}} \Big(
1+{\textbf{q}}^{40}{\textbf{t}}^{20}+{\textbf{t}}^{14}{\textbf{q}}^{20}+2\,{\textbf{t}}^{12}{\textbf{q}}^{20}
+{\textbf{t}}^{12}{\textbf{q}}^{
16}+2\,{\textbf{q}}^{18}{\textbf{t}}^{12}+{\textbf{t}}^{10}{\textbf{q}}^{14}+{\textbf{q}}^{4}{\textbf{t}}^{2}+
{\textbf{q}}^{6}{\textbf{t}}^{4}
+{\textbf{t}}^{4}{\textbf{q}}^{8}+2\,{\textbf{t}}^{16}{\textbf{q}}^{24}+2\,{\textbf{t}}^{14}{\textbf{q}}^{24}
+{\textbf{t}}^{6}{\textbf{q}}^{8}
+{\textbf{q}}^{12}{\textbf{t}}^{6}+{\textbf{t}}^{6}{\textbf{q}}^{10}+{\textbf{t}}^{12}{\textbf{q}}^{22}
+{\textbf{q}}^{26}{\textbf{t}}^{14}+{\textbf{q}}
^{28}{\textbf{t}}^{14}+2\,{\textbf{q}}^{12}{\textbf{t}}^{8}+2\,{\textbf{t}}^{14}{\textbf{q}}^{22}+
{\textbf{q}}^{48}{\textbf{t}}^{24}+
{\textbf{q}}^{32}{\textbf{t}}^{16}+2\,{\textbf{q}}^{26}{\textbf{t}}^{16}+2\,{\textbf{q}}^{28}{\textbf{t}}^{16}
+2\,{\textbf{q}}^{30}{\textbf{t}
}^{18}+{\textbf{q}}^{32}{\textbf{t}}^{20}+{\textbf{q}}^{30}{\textbf{t}}^{16}+2\,{\textbf{q}}^{32}{\textbf{t}}^{18}+{\textbf{q}}^{34}{
\textbf{t}}^{20}+{\textbf{q}}^{28}{\textbf{t}}^{18}+{\textbf{q}}^{16}{\textbf{t}}^{8}+{\textbf{q}}^{14}{\textbf{t}}^{8}
+{\textbf{q}}^{34}{\textbf{t}}^{
18}+{\textbf{q}}^{40}{\textbf{t}}^{22}+{\textbf{q}}^{36}{\textbf{t}}^{18}+{\textbf{q}}^{38}{\textbf{t}}^{20}
+2\,{\textbf{q}}^{36}{\textbf{t}}^
{20}+{\textbf{q}}^{44}{\textbf{t}}^{22}+{\textbf{q}}^{20}{\textbf{t}}^{10}+{\textbf{q}}^{18}{\textbf{t}}^{10}
+{\textbf{q}}^{24}{\textbf{t}}^{
12}+ (
{\textbf{t}}^{3}{\textbf{q}}^{2}+{\textbf{t}}^{5}{\textbf{q}}^{4}+{\textbf{q}}^{6}{\textbf{t}}^{5}+{\textbf{q}}^{6}{\textbf{t}}^{7
}+{\textbf{q}}^{14}{\textbf{t}}^{13}+{\textbf{q}}^{34}{\textbf{t}}^{23}+{\textbf{q}}^{28}{\textbf{t}}^{21}
+{\textbf{q}}^{46}{\textbf{t}}^{25}+
{\textbf{q}}^{38}{\textbf{t}}^{21}+3\,{\textbf{q}}^{30}{\textbf{t}}^{21}+2\,{\textbf{q}}^{36}{\textbf{t}}^{23}
+5\,{\textbf{q}}^{20}{\textbf{t}
}^{15}+4\,{\textbf{q}}^{26}{\textbf{t}}^{19}+2\,{\textbf{q}}^{40}{\textbf{t}}^{23}+{\textbf{q}}^{24}{\textbf{t}}^{19}
+{\textbf{q}}^{
42}{\textbf{t}}^{25}+{\textbf{q}}^{20}{\textbf{t}}^{17}+2\,{\textbf{q}}^{8}{\textbf{t}}^{7}+\bf{{q}^{10}{t}^{7}+{q}^{44
}{t}^{25}+4\,{q}^{22}{t}^{17}+3\,{t}^{9}{q}^{10}+2\,{t}^{9}{q}^{12}}+4
\,{q}^{18}{t}^{13}+4\,{q}^{22}{t}^{15}+3\,{q}^{18}{t}^{15}+4\,{q}^{16}
{t}^{13}+4\,{q}^{32}{t}^{21}+2\,{q}^{28}{t}^{17}+4\,{q}^{30}{t}^{19}+3
\,{q}^{38}{t}^{23}+2\,{q}^{24}{t}^{15}+5\,{q}^{28}{t}^{19}+4\,{q}^{26}
{t}^{17}+{q}^{26}{t}^{15}+{q}^{42}{t}^{23}+4\,{q}^{34}{t}^{21}+2\,{q}^
{32}{t}^{19}+5\,{q}^{24}{t}^{17}+{q}^{30}{t}^{17}+{q}^{34}{t}^{19}+2\,
{q}^{36}{t}^{21}+2\,{t}^{11}{q}^{12}+{t}^{9}{q}^{14}+2\,{q}^{16}{t}^{
11}+2\,{q}^{20}{t}^{13}+{q}^{18}{t}^{11}+{q}^{22}{t}^{13}+4\,{t}^{11}{
q}^{14} ) {a}^{2}+ ( {t}^{18}{q}^{18}+4\,{q}^{22}{t}^{18}+4
\,{t}^{16}{q}^{18}+{q}^{10}{t}^{12}+2\,{q}^{16}{t}^{14}+3\,{t}^{18}{q}
^{20}+2\,{q}^{16}{t}^{16}+2\,{q}^{20}{t}^{16}+2\,{t}^{20}{q}^{22}+{q}^
{6}{t}^{8}+2\,{q}^{32}{t}^{24}+{q}^{38}{t}^{24}+3\,{q}^{34}{t}^{24}+2
\,{q}^{36}{t}^{24}+{q}^{38}{t}^{26}+{q}^{42}{t}^{26}+2\,{t}^{18}{q}^{
24}+{q}^{30}{t}^{24}+4\,{t}^{20}{q}^{24}+{q}^{40}{t}^{26}+{t}^{16}{q}^
{22}+2\,{q}^{26}{t}^{22}+2\,{q}^{28}{t}^{20}+4\,{q}^{30}{t}^{22}+{t}^{
10}{q}^{8}+{q}^{26}{t}^{18}+{q}^{30}{t}^{20}+4\,{q}^{26}{t}^{20}+2\,{q
}^{32}{t}^{22}+3\,{q}^{28}{t}^{22}+{q}^{34}{t}^{22}+{t}^{10}{q}^{10}+{
q}^{18}{t}^{14}+3\,{t}^{14}{q}^{14}+{t}^{12}{q}^{14}+2\,{t}^{12}{q}^{
12} ) {a}^{4}+ ( {q}^{28}{t}^{25}+{q}^{36}{t}^{27}+{q}^{16}
{t}^{17}+{q}^{12}{t}^{15}+{q}^{24}{t}^{21}+{q}^{26}{t}^{23}+{q}^{22}{t
}^{21}+{q}^{18}{t}^{19}+{q}^{30}{t}^{25}+{q}^{32}{t}^{25}+{q}^{24}{t}^
{23}+{q}^{20}{t}^{21}+{q}^{28}{t}^{23}+{q}^{20}{t}^{19} ) {a}^{6
}+2\,{t}^{10}{q}^{16}+{q}^{42}{t}^{22} \Big) $

\bigskip
\noindent
$\boxed{P^{T[4,4k+1]}_{[1]}=\dfrac{\{A\} t^3}{\{t\} A^3
q^{12k+3}}{\cal{P}}^{T[4,4k+1]}_{[1]}}$

\begin{itemize}
\item{HOMFLY case}
\end{itemize}
$$H^{T[4,1]}_{[1]}=\dfrac{\{A\} }{\{q\} A^3 }$$
\bigskip\\
$ H^{T[4,5]}_{[1]}=\dfrac{\{A\} }{\{q\} A^3 q^{12} }\Big(-{q}^{12}{A}^{6}+(
{q}^{6}+{q}^{18}+{q}^{16}+2\,{q}^{14}+{q}^{8}
+{q}^{12}+2\,{q}^{10} ) {A}^{4}+( -{q}^{20}-2\,{q}^{16}-2
\,{q}^{18}-2\,{q}^{8}-3\,{q}^{12}-2\,{q}^{6}-{q}^{4}-{q}^{2}-3\,{q}^{
14}-3\,{q}^{10}-{q}^{22} ) {A}^{2}+1+2\,{q}^{16}+2\,{q}^{8}+{q}^
{20}+{q}^{24}+{q}^{10}+{q}^{6}+{q}^{4}+{q}^{14}+2\,{q}^{12}+{q}^{18}
\Big) $

\bigskip
\noindent
$ H^{T[4,9]}_{[1]}=\dfrac{\{A\} }{\{q\} A^3 q^{24}
}\Big(1+3\,{q}^{36}+2\,{q}^{34}+{q}^{38}+4\,{q}^{32}+2\,{q}^{8}+{q}^{48}+{q}
^{42}+{q}^{4}+2\,{q}^{40}+3\,{q}^{12}+2\,{q}^{14}+{q}^{44}+3\,{q}^{30}
+{q}^{10}+4\,{q}^{16}+3\,{q}^{18}+4\,{q}^{20}+3\,{q}^{22}+5\,{q}^{24}+
3\,{q}^{26}+4\,{q}^{28}+{q}^{6}+ ( -2\,{q}^{24}-{q}^{32}-{q}^{12}
-{q}^{36}-{q}^{16}-2\,{q}^{28}-{q}^{22}-{q}^{18}-{q}^{30}-{q}^{26}-2\,
{q}^{20} ) {A}^{6}+ ( 2\,{q}^{10}+5\,{q}^{28}+7\,{q}^{22}+4
\,{q}^{34}+7\,{q}^{26}+4\,{q}^{32}+2\,{q}^{36}+5\,{q}^{20}+6\,{q}^{18}
+6\,{q}^{30}+2\,{q}^{38}+{q}^{42}+2\,{q}^{12}+6\,{q}^{24}+{q}^{40}+{q}
^{6}+{q}^{8}+4\,{q}^{14}+4\,{q}^{16} ) {A}^{4}+ ( -6\,{q}^{
32}-6\,{q}^{14}-{q}^{44}-9\,{q}^{22}-{q}^{4}-2\,{q}^{6}-4\,{q}^{38}-6
\,{q}^{34}-4\,{q}^{10}-6\,{q}^{16}-8\,{q}^{24}-2\,{q}^{40}-8\,{q}^{28}
-{q}^{46}-4\,{q}^{36}-8\,{q}^{18}-2\,{q}^{42}-9\,{q}^{26}-{q}^{2}-4\,{
q}^{12}-2\,{q}^{8}-8\,{q}^{20}-8\,{q}^{30} ) {A}^{2} \Big) $

\bigskip
\noindent
$\boxed{H^{T[4,4k+1]}_{[1]}=\dfrac{\{A\}}{\{q\} A^3 q^{12k}}
{\cal{H}}^{T[4,4k+1]}_{[1]}=
q^{-12k-3}s_{[4]}^{*}-q^{-4k-1}s_{[3,1]}^{*}+q^{4k+1}s_{[2,1,1]}^{*}-q^{12k+3}s_{[1,1,1,1]}^{*}}$

\bigskip

\noindent
and the results coincides with the well known HOMFLY polynomials, see
(\ref{Wrepfund}) and
(\ref{113}).

\begin{itemize}
\item{Floer case}
\end{itemize}
$
F^{T[4,5]}_{[1]}=1+{\textbf{q}}^{12}{\textbf{t}}^{15}+{\textbf{q}}^{24}{\textbf{t}}^{12}+{\textbf{t}}^{6}{\textbf{q}}^{8}
+{\textbf{q}}^{16}{\textbf{t}}^{10}+{\textbf{q}
}^{22}{\textbf{t}}^{13}+{\textbf{t}}^{3}{\textbf{q}}^{2} $

\bigskip
\noindent
$F^{T[4,9]}_{[1]}=1+{\textbf{q}}^{16}{\textbf{t}}^{12}+{\textbf{q}}^{24}{\textbf{t}}^{16}
+{\textbf{q}}^{32}{\textbf{t}}^{20}+{\textbf{q}}^{48}{\textbf{t}}^{24}+
{\textbf{q}}^{40}{\textbf{t}}^{22}+{\textbf{q}}^{20}{\textbf{t}}^{15}+{\textbf{q}}^{28}{\textbf{t}}^{19}
+{\textbf{q}}^{38}{\textbf{t}}^{23}+{\textbf{q}
}^{46}{\textbf{t}}^{25}+{\textbf{t}}^{6}{q}^{8}+{\textbf{t}}^{9}{q}^{10}+{\textbf{t}}^{3}{q}^{2}$
\begin{itemize}
\item{Alexander case}
\end{itemize}

\be
\begin{array}{l}
 A^{T[4,1]}_{[1]}=1\\
 \\
A^{T[4,5]}_{[1]}=
1+{\textbf{q}}^{8}-{\textbf{q}}^{12}-{\textbf{q}}^{2}+{\textbf{q}}^{16}-{\textbf{q}}^{22}+{\textbf{q}}^{24}\\
\\
A^{T[4,9]}_{[1]}=1+{\textbf{q}}^{8}-{\textbf{q}}^{38}+{\textbf{q}}^{32}-{\textbf{q}}^{46}
+{\textbf{q}}^{40}-{\textbf{q}}^{2}-{\textbf{q}}^{10}+{\textbf{q}}^{16
}-{\textbf{q}}^{20}+{\textbf{q}}^{24}-{\textbf{q}}^{28}+{\textbf{q}}^{48}
\end{array}
\ee

\subsection{Case (4,n), n=4k+2 fundamental representation\label{4k+2}}
\be
\begin{array}{|c|}
\hline\\
P^{T[4,n]}_{[1]}=c^{[4]}_{[1]} M_{[4]}^{\ast} q^{-3n}+c^{[3,1]}_{[1]}
M_{[3,1]}^{\ast} q^{-\frac{3n}{2}} t^{\frac{n}{2}}
+c^{[2,2]}_{[1]} M_{[2,2]}^{\ast} q^{-n} t^{n} + c^{[2,1,1]}_{[1]}
M_{[2,1,1]}^{\ast} q^{-\frac{n}{2}} t^{\frac{3n}{2}} + c^{[1,1,1,1]}_{[1]}
M_{[1,1,1,1]}^{\ast} t^{3n}
\\
\\
\hline
\end{array}
\ee
\fr{
c^{[4]}_{[1]}=1,\ \ \ c^{[3,1]}_{[1]}=-{\frac {q \left(
{q}^{4}{t}^{2}+{q}^{2}t^2+q^2+1 \right)  \left( -1+{t}^{2} \right) }{t \left(
-1+{q}^{6}{t}^{2}  \right) }}\ \ \ c^{[2,2]}_{[1]}={\frac { \left( 1+{t}^{2}
\right)  \left( t^6q^4-2q^4t^4+q^6t^4+t^2-2t^2q^2+q^2 \right) }{t^2 \left(
-1+{q}^{4}{t}^{2} \right)  \left( -1+{q}^{2}{t}^{2} \right) }} \ \\ \nonumber
\\ \nonumber
c^{[2,1,1]}_{[1]}=-{\frac {q \left( 1+{t}^{2} \right)^2  \left( -1+{t}^{2}
\right)\left( -1+{q}^{2} \right)  }{t  \left( -1+{q}^{2}{t}^{2} \right) ^{2}}},
\ \ \ c^{[1,1,1,1]}_{[1]}={\frac{{q}^{2} \left( {t}^{4}+1 \right)  \left(
{t}^{2}+{t}^{4}+1  \right)  \left( 1+{t}^{2} \right) ^{2} \left( -1+{t}^{2}
\right) ^{2}\left( -1+{q}^{2} \right) }{{t}^{2} \left( -1+{q}^{2}{t}^{2}
\right)  \left( -1+{t}^{6}{q}^{2}  \right)  \left( -1+{t}^{4}{q}^{2} \right)
}}}

Similarly to the case of unknot, the obtained expressions for
$P^{T[4,n]}_{[1]}$ are not polynomials
with positive coefficients. Moreover, even ${\cal P}^{T[4,n]}_{[1]}$
(i.e. the expression normalized by unknot) is not a polynomial with positive
coefficients. However, it is a power series in ${\bf q,t}$ with positive
coefficients
as is seen from the examples:
$P^{T[4,2]}_{[1]}=\dfrac{\{A\} t^2}{\{t\}^2 A^3 q^6}\Big(
t^2q^4A^2-t^2q^2A^2+q^2A^2-t^4q^4+t^4q^2-t^2q^2+t^2-1
\Big)\bf{=\dfrac{a^2t+1}{1-q^2}}\cdot\Big(\dfrac
{1}{{t}^{8}{a}^{4}{q}^{2}}+{\dfrac {1}{{a}^{2}{t}^{5}}}+{\dfrac
{{q}^{2}}{{a}^{4}{t}^{6}}}+{\dfrac
{{q}^{4}}{{a}^{2}{t}^{3}}}+\dfrac{a^2t+1}{t^4a^4}q^6\summ{j=0}{\infty}q^{2j}\Big)$
\bigskip \\
$P^{T[4,6]}_{[1]}=\dfrac{\{A\} t^2}{\{t\}^2 A^3 q^{18}}\Big(
q^{12}A^6+t^4q^{16}A^6+t^2q^{14}A^6-t^4q^{14}A^6-t^2q^{12}A^6+t^2q^6A^4+t^6q^{12}A^4-q^8A^4-t^8q^{16}A^4-t^6q^{16}A^4-t^{10}q^{16}A^4+t^6q^{10}A^4+t^4q^8A^4-2t^4q^{14}A^4-q^6A^4+t^{10}q^{14}A^4-q^{10}A^4+t^8q^{12}A^4-t^6q^{14}A^4-2t^2q^{12}A^4+t^4q^{10}A^4-t^2q^{10}A^4-t^{14}q^{14}A^2-t^8q^{12}A^2-t^8q^{10}A^2-t^4q^4A^2+t^{10}q^{14}A^2+t^{12}q^{16}A^2+t^6q^{14}A^2+t^4q^{12}A^2-t^4q^8A^2-t^{10}q^{12}A^2-t^6q^8A^2+3t^4q^{10}A^2+t^{14}q^{16}A^2+3t^8q^{14}A^2-t^4q^6A^2+q^4A^2-2t^6q^{10}A^2+q^2A^2-t^{10}q^{10}A^2+t^2q^{10}A^2-t^8q^8A^2+t^{10}q^{16}A^2+3t^6q^{12}A^2-t^{12}q^{12}A^2+t^2q^6A^2+q^6A^2+3t^2q^8A^2-t^2q^2A^2-t^6q^6A^2-t^{10}q^{14}-2t^6q^{10}+t^6q^8+t^{12}q^{10}-t^{14}q^{14}-t^{16}q^{16}+t^8q^{10}-2t^8q^{12}+t^{16}q^{14}-t^2q^6+t^8q^6-t^2q^4-2t^4q^8+t^6q^4+t^{14}q^{12}+t^{10}q^8+t^4q^2-t^2q^2-t^{12}q^{14}+t^2-1
\Big) =\bf{\dfrac{a^2t+1}{1-q^2}}\cdot \Big( {\dfrac
{1}{{t}^{20}{a}^{4}{q}^{14}}}+{\dfrac {1}{{t}^{17}{a}^{2}{q}^{12}}}+{\dfrac
{{a}^{
2}{t}^{3}+1}{{t}^{18}{a}^{4}{q}^{10}}}+{\dfrac
{{t}^{4}{a}^{4}+{a}^{2}{t}^{3}+1+{a}^{2}t}{{t}^{16}{a}^{4}{q}^{8}}}
+{\dfrac
{{t}^{6}{a}^{4}+2\,{a}^{2}{t}^{3}+{t}^{2}+1}{{t}^{16}{a}^{4}{q}^{6}}}+{\dfrac
{{t}^{6}{a}^{4}+3\,{a}^{2}{t}^{3}+{a}^{2}t+{t}^{4}{a}^{4}+1}{{t}^{14}{a}^{4}{q}^{4}}}+{\dfrac
{{t}^{9}{a}^{6}+2\,{t}^{6}{a}^{4}+{t}^{5}{a}^{2}+2\,{a}^{2}{t}^{3}+2\,{t}^{2}+1}{{t}^{14}{a}^{4}{q}^{2}}}
+{\dfrac
{2\,{t}^{6}{a}^{4}+{t}^{4}{a}^{4}+4\,{a}^{2}{t}^{3}+{a}^{2}t+1}{{t}^{12}{a}^{4}}}+{\dfrac
{ (
{t}^{9}{a}^{6}+2\,{t}^{6}{a}^{4}+{t}^{5}{a}^{2}+2\,{a}^{2}{t}^{3}+2\,{t}^{2}+1
) {q}^{2}}{{t}^{12}
{a}^{4}}}+{\dfrac { (
2\,{t}^{6}{a}^{4}+{t}^{4}{a}^{4}+4\,{a}^{2}{t}^{3}+{a}^{2}t+1 )
{q}^{4}}{{t}^{10}{a}^{4}}}
+{\dfrac { (
{t}^{9}{a}^{6}+3\,{t}^{6}{a}^{4}+{t}^{5}{a}^{2}+3\,{a}^{2}{t}^{3}+2\,{t}^{2}+1
) {q}^{6}}{{t}^{10}
{a}^{4}}}+{\dfrac { (
{t}^{6}{a}^{4}+4\,{a}^{2}{t}^{3}+2\,{a}^{2}t+{t}^{7}{a}^{6}+3\,{t}^{4}{a}^{4}+2
) {q}^{8}
}{{t}^{8}{a}^{4}}}+\\+{\dfrac { (
2\,{t}^{6}{a}^{4}+2\,{t}^{4}{a}^{4}+5\,{a}^{2}{t}^{3}+{t}^{2}+{a}^{2}t+{t}^{7}{a}^{6}+2
) {q}^{10}}{{t}^{8}{a}^{4}}}+{\dfrac { (
3\,{t}^{6}{a}^{4}+{t}^{5}{a}^{2}+2\,{t}^{2}+1+2\,{t}^{4}{a}^{4}+5\,{a}^{2}{t}^{3}+{a}^{2}t+{t}^{7}{a}^{6}
) {q}^{12}}{{t}^{8}{a}^{4}}}+{\dfrac { (
2\,{t}^{5}{a}^{2}+4\,{a}^{2}{t}^{3}+3\,{t}^{2}+3\,{t}^{6}{a}^{4}+1+2\,{t}^{4}{a}^{4}+{a}^{2}t+{t}^{7}{a}^{6}
) {q}^{14}}{{t}^{8}{a}^{4}}}
+\\+{\dfrac { (
3\,{t}^{5}{a}^{2}+2\,{t}^{2}+4\,{a}^{2}{t}^{3}+3\,{t}^{6}{a}^{4}+1+2\,{t}^{4}{a}^{4}+{a}^{2}t+{t}^{7}{a}^{6}
) {q}^{16}}{{t}^{8}{a}^{4}}} + \dfrac { ( {a}^{2}t+1 )  (
1+{t}^{2}+{a}^{2}{t}^{3} )^{2}q^{18}}{{t}^{8}{a}^{4}}\summ{j=0}{\infty}q^{2j}
\Big) $

\bigskip
\noindent
$\boxed{P^{T[4,4k+2]}_{[1]}=\dfrac{\{A\} t^2}{\{t\}^2 A^3
q^{12k+6}}{\cal{P}}^{T[4,4k+2]}_{[1]}}$

\begin{itemize}
\item{HOMFLY case}
\end{itemize}
\bigskip
\noindent
$H^{T[4,2]}_{[1]}=\frac{\{A\} }{ \{q\}^2 A^3 q^{4}}\Big(
-q^8+q^6-q^4+q^2-1+(q^6-q^4+q^2)A^2 \Big)$

 \bigskip
\noindent
$H^{T[4,6]}_{[1]}=\frac{\{A\} }{ \{q\}^2 A^3 q^{16}}\Big(
-q^{32}+q^{30}-q^{28}-q^{24}+q^{22}-2q^{20}+2q^{18}-2q^{16}+2q^{14}-2q^{12}+q^{10}-q^{8}-q^{4}+q^{2}+(-q^{26}-q^{22}-q^{18}+q^{16}-q^{14}-q^{10}-q^{6})A^{4}+(q^{30}+q^{26}+2q^{22}-q^{20}+2q^{18}-2q^{16}+2q^{14}-q^{12}+2q^{10}+q^{6}+q^{2})A^{2}-1+(q^{20}-q^{18}+q^{16}-q^{14}+q^{12})A^{6}
\Big)$

\bigskip
\noindent
$H^{T[4,10]}_{[1]}=\frac{\{A\} }{ \{q\}^2 A^3 q^{28}}\Big(
-q^{56}+q^{54}-q^{52}-q^{48}+q^{46}-2q^{44}+q^{42}-2q^{40}+q^{38}-2q^{36}+2q^{34}-3q^{32}+3q^{30}-3q^{28}+3q^{26}-3q^{24}+2q^{22}-2q^{20}+q^{18}-2q^{16}+q^{14}-2q^{12}+q^{10}-q^{8}-q^{4}+q^{2}-1+(q^{44}-q^{42}+q^{40}+q^{36}-q^{34}+2q^{32}-2q^{30}+2q^{28}-2q^{26}+2q^{24}-q^{22}+q^{20}+q^{16}-q^{14}+q^{12})A^{6}+(-q^{50}-q^{46}-2q^{42}-2q^{38}-2q^{34}+q^{32}-2q^{30}+2q^{28}-2q^{26}+q^{24}-2q^{22}-2q^{18}-2q^{14}-q^{10}-q^{6})A^{4}+(q^{54}+q^{50}+2q^{46}+2q^{42}+3q^{38}-q^{36}+3q^{34}-2q^{32}+3q^{30}-3q^{28}+3q^{26}-2q^{24}+3q^{22}-q^{20}+3q^{18}+2q^{14}+2q^{10}+q^{6}+q^{2})A^{2}
\Big)$

\bigskip
\noindent
$\boxed{H^{T[4,4k+2]}_{[1]}=\frac{\{A\} }{ \{q\}^2 A^3 q^{12k+4}}
{\cal{H}}^{T[4,4k+2]}_{[1]}}$

\bigskip
\noindent
\begin{itemize}
\item{Alexander case}
\end{itemize}

\be
\begin{array}{l}
A^{T[4,2]}_{[1]}=\bf{q}^8-2q^6+2q^4-2q^2+1\\
\\
A^{4,6}_{[1]}=\bf{q}^{32}-2q^{30}+q^{28}+q^{24}-2q^{22}+2q^{20}-2q^{18}+2q^{16}-2q^{14}+2q^{12}-2q^{10}+q^8+q^4-2q^2+1\\
\\
A^{4,10}_{[1]}=\bf{q}^{56}-2q^{54}+q^{52}+q^{48}-2q^{46}+q^{44}+q^{40}-2q^{38}+2q^{36}-2q^{34}+2q^{32}-2q^{30}+2q^{28}-\\-2\bf{q}^{26}+2q^{24}-2q^{22}+2q^{20}-2q^{18}+q^{16}+q^{12}-2q^{10}+q^{8}+q^{4}-2q^{2}+1
\end{array}
\ee

\subsection{Case $(4,n)$, $n=4 k+3$ fundamental representation\label{4k+3}}
\be
\begin{array}{|c|}
\hline\\
P^{T[4,n]}_{[1]}=c^{[4]}_{[1]} M_{[4]}^{\ast} q^{-3n}+c^{[3,1]}_{[1]}
M_{[3,1]}^{\ast} q^{-\frac{3n}{2}} t^{\frac{n}{2}}
+c^{[2,2]}_{[1]} M_{[2,2]}^{\ast} q^{-n} t^{n} + c^{[2,1,1]}_{[1]}
M_{[2,1,1]}^{\ast} q^{-\frac{n}{2}} t^{\frac{3n}{2}} + c^{[1,1,1,1]}_{[1]}
M_{[1,1,1,1]}^{\ast} t^{3n}
\\
\\
\hline
\end{array}
\ee
with
\fr{
\nonumber
c^{[4]}_{[1]}=1, \ \ \ c^{[3,1]}_{[1]}=-{\frac {{q}^{3/2} \left(
{q}^{4}{t}^{2}+{t}^{2}+1+{q}^{2}{t}^{2}
 \right)  \left( -1+{t}^{2} \right) }{{t}^{3/2} \left( -1+{q}^{6}{t}^{
2} \right) }},\ \ \ c^{[2,2]}_{[1]}={\frac {{q}^{2} \left( -1+{t}^{2} \right) 
\left( 1+{t}^{2} \right)
 \left( {t}^{2}-{q}^{2} \right) }{t \left( -1+{q}^{4}{t}^{2} \right)
 \left( -1+{q}^{2}{t}^{2} \right) }}
\\
\nonumber
c^{[2,1,1]}_{[1]}={\frac {{q}^{5/2} \left( 1+{t}^{2} \right)  \left( -1+{t}^{2}
\right)
^{2} \left( {t}^{4}{q}^{2}+{t}^{4}+{t}^{2}+1 \right) }{{t}^{5/2}
 \left( {q}^{2}{t}^{2}+1 \right)  \left( -1+{q}^{2}{t}^{2} \right) ^{2
}}}, \ \ \ c^{[1,1,1,1]}_{[1]}=-{\frac {{q}^{3} \left( 1+{t}^{4} \right) 
\left( {t}^{2}+{t}^{4}+1
 \right)  \left( 1+{t}^{2} \right) ^{2} \left( -1+{t}^{2} \right) ^{3}
}{{t}^{3} \left( -1+{q}^{2}{t}^{2} \right)  \left( -1+{t}^{6}{q}^{2}
 \right)  \left( -1+{t}^{4}{q}^{2} \right) }}}
three first superpolynomials:

\bigskip
\noindent
$P^{T[4,3]}_{[1]}=\dfrac{\{A\} t^3 }{ \{t\} A^3 q^{9}} \Big( {q}^{6}{A}^{4}+
\left( {q}^{4}{t}^{2}+{q}^{6}{t}^{4}+{q}^{6}{t}^{2}+{
q}^{4}+{q}^{2} \right) (-A^{2})+{q}^{6}{t}^{6}+{q}^{4}{t}^{2}+{t}^{4}{q
}^{4}+1+{q}^{2}{t}^{2}\Big)=\\ = \bf{ -{\frac {{a}^{2}t+1}{ \left( -1+{q}^{2}
\right) {a}^{4}{t}^{11}{q}^{5}}}} \Big(  {q}^{6}{t}^{8}{a}^{4}+ \left(
{q}^{6}{t}^{5}+{q}^{10}{t}^{7}+{q}^{8}{t
}^{7}+{t}^{3}{q}^{2}+{t}^{5}{q}^{4} \right) {a}^{2}+{q}^{12}{t}^{6}+{q
}^{6}{t}^{4}+{t}^{4}{q}^{8}+1+{q}^{4}{t}^{2}
 \Big)$

\bigskip
\noindent
$P^{T[4,7]}_{[1]}=\dfrac{\{A\} t^3 }{ \{t\} A^3 q^{21}}
\Big(1+{t}^{4}{q}^{10}+( {t}^{2}{q}^{14}+{q}^{12}+{t}^{2}{q}^{16}+{q
}^{18}{t}^{6}+{t}^{4}{q}^{16} ) (-A^{2})^3+ ( {t}^{8}{q}^{18}+
{q}^{10}+3\,{t}^{4}{q}^{14}+{t}^{10}{q}^{18}+{t}^{12}{q}^{18}+2\,{q}^{
16}{t}^{8}+{q}^{6}+{t}^{2}{q}^{14}+{q}^{14}{t}^{8}+3\,{t}^{2}{q}^{12}+
{q}^{12}{t}^{6}+2\,{t}^{2}{q}^{10}+{t}^{4}{q}^{10}+2\,{t}^{4}{q}^{12}+
{q}^{8}+{t}^{10}{q}^{16}+{t}^{2}{q}^{8}+{t}^{4}{q}^{16}+2\,{t}^{6}{q}^
{14}+3\,{t}^{6}{q}^{16}) {A}^{4}+ ({q}^{6}+{q}^{2}+{q}^{
4}+4\,{t}^{4}{q}^{10}+{q}^{4}{t}^{2}+2\,{q}^{6}{t}^{2}+{q}^{6}{t}^{4}+
2\,{t}^{2}{q}^{10}+3\,{t}^{2}{q}^{8}+2\,{t}^{4}{q}^{8}+3\,{t}^{4}{q}^{
12}+{t}^{4}{q}^{14}+{t}^{6}{q}^{8}+4\,{q}^{12}{t}^{6}+3\,{t}^{6}{q}^{
14}+2\,{t}^{6}{q}^{10}+{t}^{12}{q}^{14}+2\,{t}^{10}{q}^{14}+3\,{t}^{10
}{q}^{16}+{t}^{12}{q}^{18}+{q}^{10}{t}^{8}+2\,{q}^{12}{t}^{8}+{t}^{10}
{q}^{12}+2\,{q}^{16}{t}^{8}+4\,{q}^{14}{t}^{8}+2\,{t}^{12}{q}^{16}+{t}
^{14}{q}^{16}+{t}^{14}{q}^{18}+{t}^{16}{q}^{18}) (-A^{2})+{q}^{2
}{t}^{2}+{t}^{4}{q}^{4}+{q}^{4}{t}^{2}+{q}^{6}{t}^{2}+{q}^{6}{t}^{6}+{
q}^{6}{t}^{4}+2\,{t}^{4}{q}^{8}+{t}^{6}{q}^{8}+{q}^{12}{t}^{6}+2\,{t}^
{6}{q}^{10}+{t}^{18}{q}^{18}+{t}^{12}{q}^{14}+2\,{t}^{10}{q}^{14}+{t}^
{8}{q}^{8}+{q}^{10}{t}^{8}+2\,{q}^{12}{t}^{8}+{t}^{12}{q}^{12}+{t}^{10
}{q}^{12}+{q}^{14}{t}^{8}+{t}^{12}{q}^{16}+{t}^{14}{q}^{16}+{t}^{16}{q
}^{16}+{t}^{14}{q}^{14}+{t}^{10}{q}^{10} \Big)=\\ =\bf{ -{\dfrac {{a}^{2}t+1}{
\left( -1+{q}^{2} \right) {a}^{4}{t}^{23}{q}^{17
}}}}\Big( 1+2\,{t}^{10}{q}^{16}+{t}^{12}{q}^{18}+ ( 2\,{q}^{16}{t}^{11}+4\,
{q}^{22}{t}^{15}+3\,{q}^{16}{t}^{13}+4\,{q}^{14}{t}^{11}+3\,{q}^{20}{t
}^{15}+4\,{q}^{18}{t}^{13}+2\,{q}^{12}{t}^{11}+{q}^{30}{t}^{17}+{q}^{
32}{t}^{19}+{q}^{34}{t}^{19}+2\,{q}^{24}{t}^{17}+{q}^{14}{t}^{9}+3\,{q
}^{26}{t}^{17}+{q}^{26}{t}^{15}+{t}^{3}{q}^{2}+{t}^{5}{q}^{4}+{q}^{6}{
t}^{5}+{q}^{6}{t}^{7}+{q}^{22}{t}^{13}+2\,{q}^{20}{t}^{13}+{q}^{18}{t}
^{11}+{q}^{18}{t}^{15}+2\,{q}^{28}{t}^{17}+2\,{q}^{24}{t}^{15}+{q}^{30
}{t}^{19}+2\,{q}^{8}{t}^{7}+{q}^{10}{t}^{7}+3\,{t}^{9}{q}^{10}+2\,{t}^
{9}{q}^{12} ) {a}^{2}+ ( {q}^{16}{t}^{17}+{q}^{20}{t}^{19}+
{q}^{18}{t}^{19}+{q}^{12}{t}^{15}+{q}^{24}{t}^{21} ) {a}^{6}+
( {q}^{28}{t}^{20}+{q}^{22}{t}^{16}+{q}^{6}{t}^{8}+{t}^{12}{q}^{
14}+2\,{q}^{20}{t}^{16}+2\,{q}^{24}{t}^{18}+{q}^{20}{t}^{18}+3\,{t}^{
14}{q}^{14}+{t}^{10}{q}^{8}+{q}^{30}{t}^{20}+2\,{t}^{14}{q}^{16}+{q}^{
26}{t}^{18}+{q}^{18}{t}^{14}+{t}^{10}{q}^{10}+{q}^{16}{t}^{16}+3\,{q}^
{18}{t}^{16}+3\,{q}^{22}{t}^{18}+2\,{t}^{12}{q}^{12}+{q}^{10}{t}^{12}+
{q}^{26}{t}^{20} ) {a}^{4}+{q}^{4}{t}^{2}+{q}^{6}{t}^{4}+{t}^{4}
{q}^{8}+{t}^{6}{q}^{8}+{q}^{12}{t}^{6}+{t}^{6}{q}^{10}+2\,{q}^{12}{t}^
{8}+{q}^{16}{t}^{8}+{q}^{14}{t}^{8}+{t}^{10}{q}^{18}+{t}^{10}{q}^{14}+
{q}^{22}{t}^{14}+{q}^{20}{t}^{10}+{q}^{24}{t}^{12}+{q}^{22}{t}^{12}+2
\,{q}^{20}{t}^{12}+2\,{q}^{24}{t}^{14}+{q}^{26}{t}^{14}+{q}^{28}{t}^{
14}+{q}^{28}{t}^{16}+{q}^{32}{t}^{16}+{q}^{30}{t}^{16}+{q}^{36}{t}^{18
}\Big)$

\bigskip
\noindent
$P^{T[4,11]}_{[1]}=\dfrac{\{A\} t^3 }{ \{t\} A^3 q^{33}} \Big(
1+{t}^{4}{q}^{10}+{t}^{20}{q}^{22}+{t}^{20}{q}^{20}+2\,{t}^{16}{q}^{20
}+2\,{q}^{20}{t}^{14}+3\,{q}^{20}{t}^{12}+2\,{q}^{20}{t}^{10}+{q}^{22}
{t}^{12}+{q}^{24}{t}^{14}+3\,{q}^{22}{t}^{14}+2\,{q}^{22}{t}^{18}+2\,{
q}^{24}{t}^{16}+2\,{q}^{22}{t}^{16}+2\,{q}^{24}{t}^{18}+{q}^{26}{t}^{
20}+{q}^{28}{t}^{24}+ ({q}^{12}+{q}^{22}{t}^{10}+{q}^{20}{t}^{8
}+{q}^{24}{t}^{10}+2\,{q}^{26}{t}^{10}+{q}^{28}{t}^{16}+{q}^{26}{t}^{
14}+{q}^{28}{t}^{14}+{t}^{2}{q}^{14}+{t}^{2}{q}^{16}+{q}^{18}{t}^{2}+{
q}^{18}{t}^{4}+{q}^{18}{t}^{6}+{t}^{6}{q}^{20}+{t}^{4}{q}^{16}+2\,{t}^
{4}{q}^{20}+2\,{t}^{6}{q}^{22}+{t}^{6}{q}^{24}+{q}^{22}{t}^{4}+{q}^{30
}{t}^{18}+{q}^{26}{t}^{12}+2\,{q}^{24}{t}^{8}+{q}^{24}{t}^{12}+{q}^{26
}{t}^{8}+{q}^{22}{t}^{8}+{q}^{28}{t}^{12} ) (-A^{2})^3+ ( {q}^
{6}+{t}^{4}{q}^{10}+{q}^{8}+{q}^{10}+2\,{q}^{28}{t}^{20}+{t}^{24}{q}^{
30}+{q}^{30}{t}^{20}+5\,{q}^{22}{t}^{10}+{q}^{20}{t}^{14}+5\,{q}^{20}{
t}^{8}+5\,{q}^{24}{t}^{10}+2\,{q}^{20}{t}^{12}+4\,{q}^{20}{t}^{10}+4\,
{q}^{22}{t}^{12}+4\,{q}^{24}{t}^{14}+2\,{q}^{22}{t}^{14}+{q}^{26}{t}^{
10}+4\,{q}^{26}{t}^{16}+2\,{q}^{24}{t}^{16}+{q}^{22}{t}^{16}+{q}^{24}{
t}^{18}+{q}^{26}{t}^{20}+2\,{q}^{28}{t}^{16}+4\,{q}^{26}{t}^{14}+2\,{t
}^{10}{q}^{18}+2\,{t}^{2}{q}^{10}+{t}^{2}{q}^{8}+{q}^{28}{t}^{14}+2\,{
t}^{2}{q}^{14}+3\,{t}^{2}{q}^{12}+2\,{t}^{4}{q}^{12}+4\,{t}^{4}{q}^{14
}+{t}^{2}{q}^{16}+4\,{q}^{18}{t}^{4}+{q}^{12}{t}^{6}+5\,{q}^{18}{t}^{6
}+5\,{t}^{6}{q}^{20}+4\,{t}^{6}{q}^{16}+2\,{t}^{6}{q}^{14}+4\,{t}^{4}{
q}^{16}+{t}^{4}{q}^{20}+2\,{t}^{6}{q}^{22}+4\,{t}^{8}{q}^{18}+{t}^{10}
{q}^{16}+{t}^{12}{q}^{18}+2\,{q}^{16}{t}^{8}+{q}^{14}{t}^{8}+4\,{q}^{
26}{t}^{12}+2\,{q}^{24}{t}^{8}+5\,{q}^{24}{t}^{12}+2\,{q}^{26}{t}^{18}
+6\,{q}^{22}{t}^{8}+{t}^{22}{q}^{28}+{q}^{30}{t}^{22}+3\,{t}^{18}{q}^{
28} ) {A}^{4}+ ( {q}^{6}+{q}^{2}+{q}^{4}+4\,{t}^{4}{q}^{10
}+2\,{q}^{28}{t}^{20}+{t}^{20}{q}^{22}+{t}^{24}{q}^{30}+5\,{q}^{22}{t}
^{10}+2\,{t}^{16}{q}^{20}+4\,{q}^{20}{t}^{14}+5\,{q}^{20}{t}^{8}+{q}^{
24}{t}^{10}+5\,{q}^{20}{t}^{12}+7\,{q}^{20}{t}^{10}+7\,{q}^{22}{t}^{12
}+6\,{q}^{24}{t}^{14}+5\,{q}^{22}{t}^{14}+4\,{q}^{26}{t}^{16}+2\,{q}^{
22}{t}^{18}+5\,{q}^{24}{t}^{16}+4\,{q}^{22}{t}^{16}+4\,{q}^{24}{t}^{18
}+4\,{q}^{26}{t}^{20}+2\,{q}^{26}{t}^{14}+2\,{q}^{28}{t}^{24}+{q}^{4}{
t}^{2}+2\,{q}^{6}{t}^{2}+{q}^{6}{t}^{4}+5\,{t}^{10}{q}^{18}+{t}^{26}{q
}^{28}+2\,{t}^{2}{q}^{10}+3\,{t}^{2}{q}^{8}+2\,{t}^{4}{q}^{8}+{t}^{2}{
q}^{12}+4\,{t}^{4}{q}^{12}+4\,{t}^{4}{q}^{14}+{t}^{6}{q}^{8}+4\,{q}^{
12}{t}^{6}+4\,{q}^{18}{t}^{6}+{t}^{6}{q}^{20}+6\,{t}^{6}{q}^{16}+5\,{t
}^{6}{q}^{14}+2\,{t}^{4}{q}^{16}+2\,{t}^{6}{q}^{10}+{t}^{12}{q}^{14}+7
\,{t}^{8}{q}^{18}+2\,{t}^{10}{q}^{14}+4\,{t}^{10}{q}^{16}+4\,{t}^{12}{
q}^{18}+{q}^{10}{t}^{8}+2\,{q}^{12}{t}^{8}+{t}^{28}{q}^{30}+{t}^{10}{q
}^{12}+5\,{q}^{16}{t}^{8}+4\,{q}^{14}{t}^{8}+2\,{t}^{12}{q}^{16}+{t}^{
14}{q}^{16}+2\,{t}^{14}{q}^{18}+{t}^{16}{q}^{18}+{t}^{24}{q}^{26}+4\,{
q}^{24}{t}^{12}+{t}^{18}{q}^{20}+4\,{q}^{26}{t}^{18}+{q}^{22}{t}^{8}+{
t}^{22}{q}^{24}+{t}^{26}{q}^{30}+3\,{t}^{22}{q}^{28}+2\,{t}^{20}{q}^{
24}+2\,{q}^{26}{t}^{22}+{t}^{18}{q}^{28}) (-A^{2})+{q}^{2}{t}^{2
}+{t}^{4}{q}^{4}+{q}^{4}{t}^{2}+{q}^{6}{t}^{2}+{q}^{6}{t}^{6}+{q}^{6}{
t}^{4}+3\,{t}^{10}{q}^{18}+{t}^{26}{q}^{28}+2\,{t}^{4}{q}^{8}+{t}^{4}{
q}^{12}+{t}^{6}{q}^{8}+2\,{q}^{12}{t}^{6}+{t}^{6}{q}^{16}+2\,{t}^{6}{q
}^{14}+2\,{t}^{6}{q}^{10}+{t}^{18}{q}^{18}+{t}^{12}{q}^{14}+{t}^{8}{q}
^{18}+2\,{t}^{10}{q}^{14}+2\,{t}^{10}{q}^{16}+2\,{t}^{12}{q}^{18}+{t}^
{8}{q}^{8}+{q}^{10}{t}^{8}+2\,{q}^{12}{t}^{8}+{t}^{12}{q}^{12}+{t}^{10
}{q}^{12}+3\,{q}^{16}{t}^{8}+2\,{q}^{14}{t}^{8}+{q}^{30}{t}^{30}+2\,{t
}^{12}{q}^{16}+{t}^{14}{q}^{16}+2\,{t}^{14}{q}^{18}+{t}^{16}{q}^{16}+{
t}^{16}{q}^{18}+{t}^{24}{q}^{26}+{t}^{24}{q}^{24}+{t}^{18}{q}^{20}+{q}
^{26}{t}^{18}+{t}^{22}{q}^{24}+{t}^{14}{q}^{14}+2\,{t}^{20}{q}^{24}+2
\,{q}^{26}{t}^{22}+{t}^{10}{q}^{10}+{t}^{28}{q}^{28}+{t}^{22}{q}^{22}+
{t}^{26}{q}^{26}
 \Big)=\\ = \bf{-{\dfrac {{a}^{2}t+1}{ \left( -1+{q}^{2} \right)
{a}^{4}{t}^{35}{q}^{29
}}}}
\Big( 1+2\,{t}^{18}{q}^{30}+2\,{t}^{16}{q}^{26}+2\,{t}^{10}{q}^{16}+2\,{t}^{
12}{q}^{18}+3\,{q}^{24}{t}^{16}+{q}^{46}{t}^{24}+{q}^{52}{t}^{28}+{q}^
{48}{t}^{24}+2\,{t}^{14}{q}^{20}+{t}^{12}{q}^{16}+2\,{q}^{30}{t}^{20}+
{q}^{4}{t}^{2}+{q}^{6}{t}^{4}+{q}^{26}{t}^{18}+{t}^{4}{q}^{8}+{q}^{22}
{t}^{16}+{t}^{6}{q}^{8}+{q}^{12}{t}^{6}+{t}^{6}{q}^{10}+2\,{q}^{44}{t}
^{24}+2\,{q}^{12}{t}^{8}+{q}^{34}{t}^{22}+3\,{q}^{32}{t}^{20}+2\,{q}^{
32}{t}^{18}+{q}^{38}{t}^{24}+2\,{q}^{34}{t}^{20}+3\,{q}^{36}{t}^{22}+{
q}^{34}{t}^{18}+2\,{q}^{40}{t}^{22}+{q}^{38}{t}^{20}+2\,{q}^{36}{t}^{
20}+2\,{q}^{38}{t}^{22}+2\,{q}^{42}{t}^{24}+2\,{q}^{40}{t}^{24}+{q}^{
16}{t}^{8}+{q}^{14}{t}^{8}+{t}^{10}{q}^{18}+{q}^{40}{t}^{20}+{q}^{54}{
t}^{28}+3\,{t}^{18}{q}^{28}+{q}^{60}{t}^{30}+{q}^{44}{t}^{22}+{q}^{46}
{t}^{26}+{q}^{56}{t}^{28}+{q}^{52}{t}^{26}+ ( {q}^{30}{t}^{25}+{q
}^{30}{t}^{27}+2\,{q}^{28}{t}^{25}+{q}^{16}{t}^{17}+{q}^{20}{t}^{21}+{
q}^{44}{t}^{31}+{q}^{22}{t}^{21}+{q}^{48}{t}^{33}+{q}^{40}{t}^{31}+{q}
^{34}{t}^{29}+2\,{q}^{36}{t}^{29}+{q}^{18}{t}^{19}+{q}^{12}{t}^{15}+{q
}^{20}{t}^{19}+{q}^{24}{t}^{21}+{q}^{26}{t}^{23}+{q}^{28}{t}^{23}+{q}^
{42}{t}^{31}+{q}^{32}{t}^{25}+{q}^{36}{t}^{27}+2\,{q}^{24}{t}^{23}+{q}
^{34}{t}^{27}+{q}^{40}{t}^{29}+{q}^{26}{t}^{25}+{q}^{38}{t}^{29}+2\,{q
}^{32}{t}^{27} ) {a}^{6}+ ( 6\,{t}^{24}{q}^{30}+2\,{q}^{32}
{t}^{22}+{q}^{50}{t}^{30}+{q}^{46}{t}^{28}+{q}^{42}{t}^{26}+2\,{q}^{32
}{t}^{26}+5\,{t}^{22}{q}^{26}+{q}^{24}{t}^{22}+2\,{t}^{24}{q}^{28}+4\,
{t}^{20}{q}^{22}+5\,{t}^{20}{q}^{24}+{q}^{54}{t}^{32}+{q}^{52}{t}^{32}
+{t}^{12}{q}^{14}+2\,{t}^{12}{q}^{12}+{q}^{30}{t}^{20}+4\,{q}^{26}{t}^
{20}+{q}^{6}{t}^{8}+{q}^{26}{t}^{18}+2\,{q}^{24}{t}^{18}+2\,{q}^{28}{t
}^{20}+{q}^{22}{t}^{16}+4\,{q}^{22}{t}^{18}+2\,{q}^{20}{t}^{16}+{q}^{
10}{t}^{12}+4\,{q}^{20}{t}^{18}+4\,{q}^{38}{t}^{26}+5\,{q}^{36}{t}^{26
}+4\,{q}^{34}{t}^{24}+4\,{q}^{42}{t}^{28}+2\,{q}^{40}{t}^{26}+2\,{q}^{
48}{t}^{30}+4\,{q}^{40}{t}^{28}+5\,{q}^{34}{t}^{26}+5\,{q}^{32}{t}^{24
}+{q}^{18}{t}^{14}+{q}^{34}{t}^{22}+{q}^{38}{t}^{24}+5\,{t}^{22}{q}^{
28}+4\,{q}^{38}{t}^{28}+3\,{q}^{46}{t}^{30}+2\,{q}^{36}{t}^{24}+{q}^{
50}{t}^{32}+{q}^{36}{t}^{28}+{t}^{10}{q}^{8}+{t}^{10}{q}^{10}+4\,{t}^{
22}{q}^{30}+3\,{t}^{14}{q}^{14}+2\,{q}^{16}{t}^{16}+{t}^{18}{q}^{18}+4
\,{q}^{18}{t}^{16}+2\,{t}^{14}{q}^{16}+2\,{q}^{44}{t}^{28}+2\,{q}^{44}
{t}^{30}+{q}^{42}{t}^{30} ) {a}^{4}+ ( 2\,{q}^{52}{t}^{29}+
{q}^{50}{t}^{27}+4\,{q}^{38}{t}^{23}+5\,{q}^{28}{t}^{21}+4\,{q}^{36}{t
}^{25}+{q}^{54}{t}^{31}+2\,{q}^{48}{t}^{27}+{q}^{46}{t}^{25}+{q}^{26}{
t}^{21}+5\,{q}^{32}{t}^{23}+{q}^{34}{t}^{25}+2\,{q}^{40}{t}^{23}+5\,{q
}^{28}{t}^{19}+6\,{q}^{22}{t}^{17}+2\,{q}^{16}{t}^{11}+4\,{q}^{22}{t}^
{15}+4\,{q}^{16}{t}^{13}+4\,{q}^{14}{t}^{11}+5\,{q}^{20}{t}^{15}+4\,{q
}^{46}{t}^{27}+2\,{q}^{36}{t}^{21}+2\,{q}^{48}{t}^{29}+7\,{q}^{30}{t}^
{21}+7\,{q}^{34}{t}^{23}+4\,{q}^{18}{t}^{13}+2\,{q}^{12}{t}^{11}+{q}^{
30}{t}^{17}+2\,{q}^{32}{t}^{19}+{q}^{34}{t}^{19}+5\,{q}^{24}{t}^{17}+{
q}^{14}{t}^{9}+4\,{q}^{26}{t}^{17}+{q}^{26}{t}^{15}+{t}^{3}{q}^{2}+{t}
^{5}{q}^{4}+{q}^{6}{t}^{5}+{q}^{6}{t}^{7}+{q}^{22}{t}^{13}+2\,{q}^{20}
{t}^{13}+{q}^{18}{t}^{11}+4\,{q}^{18}{t}^{15}+2\,{q}^{28}{t}^{17}+2\,{
q}^{24}{t}^{15}+4\,{q}^{30}{t}^{19}+2\,{q}^{8}{t}^{7}+{q}^{10}{t}^{7}+
5\,{q}^{36}{t}^{23}+3\,{t}^{9}{q}^{10}+2\,{t}^{9}{q}^{12}+4\,{q}^{44}{
t}^{27}+4\,{q}^{34}{t}^{21}+2\,{q}^{20}{t}^{17}+5\,{q}^{40}{t}^{25}+{q
}^{56}{t}^{31}+4\,{q}^{42}{t}^{25}+{q}^{46}{t}^{29}+4\,{q}^{42}{t}^{27
}+5\,{q}^{32}{t}^{21}+2\,{q}^{40}{t}^{27}+3\,{q}^{50}{t}^{29}+{q}^{38}
{t}^{21}+4\,{q}^{24}{t}^{19}+{q}^{58}{t}^{31}+{q}^{30}{t}^{23}+6\,{q}^
{38}{t}^{25}+{q}^{54}{t}^{29}+7\,{q}^{26}{t}^{19}+2\,{q}^{44}{t}^{25}+
{q}^{42}{t}^{23}+{t}^{13}{q}^{14} ) {a}^{2}+{q}^{44}{t}^{26}+{t}
^{10}{q}^{14}+2\,{q}^{22}{t}^{14}+{q}^{50}{t}^{26}+{q}^{42}{t}^{22}+2
\,{q}^{48}{t}^{26}+{q}^{20}{t}^{10}+{q}^{24}{t}^{12}+{q}^{22}{t}^{12}+
2\,{q}^{20}{t}^{12}+2\,{q}^{24}{t}^{14}+{q}^{26}{t}^{14}+{q}^{28}{t}^{
14}+2\,{q}^{28}{t}^{16}+{q}^{32}{t}^{16}+{q}^{30}{t}^{16}+{q}^{36}{t}^
{18}\Big)$

\bigskip
\noindent
$\boxed{P^{T[4,4k+3]}_{[1]}=\dfrac{\{A\} t^3}{\{t\} A^3
q^{12k+9}}{\cal{P}}^{T[4,4k+3]}_{[1]}}$

\begin{itemize}
\item{HOMFLY case}
\end{itemize}
\bigskip
\noindent
$H^{T[4,3]}_{[1]}=\dfrac{\{A\} }{ \{q\} A^3 q^{6}}\Big({q}^{6}{A}^{4}+ \left(
-{q}^{8}-{q}^{6}-{q}^{10}-{q}^{2}-{q}^{4}
 \right) {A}^{2}+{q}^{12}+{q}^{6}+{q}^{8}+{q}^{4}+1
 \Big)$

 \bigskip
\noindent
$H^{T[4,7]}_{[1]}=\dfrac{\{A\}}{ \{q\} A^3 q^{18}}\Big( (
-{q}^{20}-{q}^{18}-{q}^{16}-{q}^{24}-{q}^{12} ) {A}^{6}+
 ( {q}^{6}+2\,{q}^{26}+{q}^{8}+2\,{q}^{12}+2\,{q}^{10}+{q}^{30}+3
\,{q}^{20}+{q}^{28}+4\,{q}^{22}+4\,{q}^{14}+2\,{q}^{24}+4\,{q}^{18}+3
\,{q}^{16} ) {A}^{4}+( -4\,{q}^{26}-5\,{q}^{14}-4\,{q}^{12
}-5\,{q}^{22}-5\,{q}^{20}-2\,{q}^{28}-5\,{q}^{16}-4\,{q}^{10}-2\,{q}^{
8}-4\,{q}^{24}-2\,{q}^{30}-{q}^{32}-6\,{q}^{18}-{q}^{2}-{q}^{34}-{q}^{
4}-2\,{q}^{6} ) {A}^{2}+1+{q}^{26}+2\,{q}^{8}+{q}^{6}+{q}^{4}+2
\,{q}^{14}+3\,{q}^{16}+{q}^{10}+3\,{q}^{12}+3\,{q}^{24}+{q}^{32}+2\,{q
}^{18}+3\,{q}^{20}+{q}^{36}+2\,{q}^{28}+2\,{q}^{22}+{q}^{30}
\Big)$

\bigskip
\noindent
$H^{T[4,11]}_{[1]}=\frac{\{A\}}{ \{q\} A^3
q^{30}}\Big(1+{q}^{6}+{q}^{4}+2\,{q}^{8}+{q}^{10}+3\,{q}^{12}+2\,{q}^{14}+4\,{q}^{
16}+3\,{q}^{18}+5\,{q}^{20}+4\,{q}^{22}+6\,{q}^{24}+4\,{q}^{26}+6\,{q}
^{28}+{q}^{50}+{q}^{56}+{q}^{54}+2\,{q}^{52}+6\,{q}^{36}+2\,{q}^{46}+4
\,{q}^{44}+6\,{q}^{32}+3\,{q}^{48}+{q}^{60}+ ( -{q}^{42}-3\,{q}^{
36}-3\,{q}^{32}-{q}^{12}-2\,{q}^{20}-3\,{q}^{28}-{q}^{16}-3\,{q}^{24}-
2\,{q}^{40}-{q}^{18}-2\,{q}^{30}-{q}^{22}-{q}^{48}-2\,{q}^{26}-{q}^{38
}-{q}^{44}-2\,{q}^{34} ) {A}^{6}+( {q}^{6}+{q}^{8}+2\,{q}^
{10}+2\,{q}^{12}+4\,{q}^{14}+4\,{q}^{16}+6\,{q}^{18}+6\,{q}^{20}+9\,{q
}^{22}+8\,{q}^{24}+10\,{q}^{26}+9\,{q}^{28}+2\,{q}^{50}+{q}^{54}+{q}^{
52}+8\,{q}^{36}+4\,{q}^{46}+4\,{q}^{44}+9\,{q}^{32}+2\,{q}^{48}+6\,{q}
^{42}+10\,{q}^{34}+6\,{q}^{40}+9\,{q}^{38}+11\,{q}^{30}) {A}^{4
}+ ( -2\,{q}^{6}-{q}^{2}-{q}^{4}-2\,{q}^{8}-4\,{q}^{10}-4\,{q}^{
12}-6\,{q}^{14}-6\,{q}^{16}-9\,{q}^{18}-9\,{q}^{20}-11\,{q}^{22}-11\,{
q}^{24}-13\,{q}^{26}-12\,{q}^{28}-4\,{q}^{50}-{q}^{56}-2\,{q}^{54}-2\,
{q}^{52}-11\,{q}^{36}-{q}^{58}-6\,{q}^{46}-6\,{q}^{44}-12\,{q}^{32}-4
\,{q}^{48}-9\,{q}^{42}-13\,{q}^{34}-9\,{q}^{40}-11\,{q}^{38}-13\,{q}^{
30} ) {A}^{2}+3\,{q}^{42}+4\,{q}^{34}+5\,{q}^{40}+4\,{q}^{38}+5
\,{q}^{30}\Big)$

\bigskip
\noindent
$\boxed{H^{T[4,4k+3]}_{[1]}=\dfrac{\{A\}}{\{q\} A^3 q^{12k+6}}
{\cal{H}}^{T[4,4k+3]}_{[1]}=q^{-12k-9}s_{[4]}^{*}-q^{-4k-3}s_{[3,1]}^{*}+q^{4k+3}s_{[2,1,1]}^{*}-q^{12k+9}s_{[1,1,1,1]}^{*}}$

\bigskip

\noindent
and the results coincides with the well known HOMFLY polynomials, see
(\ref{Wrepfund}) and
(\ref{113}).

\bigskip
\noindent
\begin{itemize}
\item{Floer case}
\end{itemize}

\bigskip
\noindent
$F^{T[4,3]}_{[1]}={\textbf{q}}^{6}{\textbf{t}}^{8}+{\textbf{q}}^{6}{\textbf{t}}^{5}+{\textbf{q}}^{10}{\textbf{t}}^{7}
+{\textbf{q}}^{8}{\textbf{t}}^{7}+{\textbf{t}}^{5}{\textbf{q}
}^{4}+{\textbf{t}}^{3}{\textbf{q}}^{2}+{\textbf{q}}^{12}{\textbf{t}}^{6}+{\textbf{q}}^{6}{\textbf{t}}^{4}
+{\textbf{t}}^{4}{\textbf{q}}^{8}+1+{
\textbf{q}}^{4}{\textbf{t}}^{2}$

\bigskip
\noindent
$
F^{T[4,7]}_{[1]}=1+{\textbf{q}}^{26}{\textbf{t}}^{17}+{\textbf{q}}^{22}{\textbf{t}}^{15}+
{\textbf{q}}^{22}{\textbf{t}}^{14}+{\textbf{q}}^{22}{\textbf{t}}^{18}+
{\textbf{q}}^{28}{\textbf{t}}^{16}+{\textbf{q}}^{36}{\textbf{t}}^{18}+{\textbf{t}}^{6}{\textbf{q}}^{8}
+{\textbf{t}}^{10}{\textbf{q}}^{14}+{\textbf{t}}^
{3}{\textbf{q}}^{2}+{\textbf{t}}^{9}{\textbf{q}}^{10}+{\textbf{t}}^{11}{\textbf{q}}^{14}+{\textbf{q}}^{18}{\textbf{t}}^{13}
+{\textbf{q}}^{34}{
\textbf{t}}^{19}+{\textbf{q}}^{14}{\textbf{t}}^{14}
$

\bigskip
\noindent
$F^{T[4,11]}_{[1]}=1+2\,{\textbf{q}}^{30}{\textbf{t}}^{21}+2\,{\textbf{q}}^{34}{\textbf{t}}^{23}
+{\textbf{q}}^{42}{\textbf{t}}^{27}+{\textbf{q}}^{34}{\textbf{t}}
^{26}+2\,{\textbf{q}}^{30}{\textbf{t}}^{24}+{\textbf{q}}^{38}{\textbf{t}}^{28}+{\textbf{t}}^{3}{\textbf{q}}^{2}
+{\textbf{q}}^{52}{\textbf{t}}^
{28}+{\textbf{t}}^{6}{\textbf{q}}^{8}+{\textbf{q}}^{60}{\textbf{t}}^{30}+{\textbf{t}}^{9}{\textbf{q}}^{10}+
{\textbf{q}}^{38}{\textbf{t}}^{25}+
{\textbf{q}}^{22}{\textbf{t}}^{17}+{\textbf{q}}^{44}{\textbf{t}}^{26}+{\textbf{q}}^{16}{\textbf{t}}^{12}+{\textbf{q}}^{58}{\textbf{t}}^{31}
+{\textbf{q}
}^{18}{\textbf{t}}^{15}+{\textbf{q}}^{26}{t}^{22}+{\textbf{q}}^{22}{\textbf{t}}^{16}+{\textbf{q}}^{30}{\textbf{t}}^{20}+{\textbf{q}}^
{38}{\textbf{t}}^{24}+2\,{\textbf{q}}^{26}{\textbf{t}}^{19}+{\textbf{q}}^{22}{\textbf{t}}^{20}+{\textbf{q}}^{50}{\textbf{t}}^{29}$

\bigskip
\noindent
\begin{itemize}
\item{Alexander case}
\end{itemize}

\be
\begin{array}{l}
A^{T[4,3]}_{[1]}=1+{\textbf{q}}^{12}-{\textbf{q}}^{10}+{\textbf{q}}^{6}-{\textbf{q}}^{2}\\
\\
A^{4,7}_{[1]}=1+{\textbf{q}}^{36}-{\textbf{q}}^{34}+{\textbf{q}}^{28}-{\textbf{q}}^{26}+{\textbf{q}}^{22}
-{\textbf{q}}^{18}+{\textbf{q}}^{14}-{\textbf{q}}^{
10}+{\textbf{q}}^{8}-{\textbf{q}}^{2}\\
\\
A^{4,11}_{[1]}=1-{\textbf{q}}^{2}+{\textbf{q}}^{8}-{\textbf{q}}^{10}+{\textbf{q}}^{16}-{\textbf{q}}^{18}+{\textbf{q}}^{22}-
{\textbf{q}}^{26}-{\textbf{q}}^{50
}+{\textbf{q}}^{30}-{\textbf{q}}^{42}+{\textbf{q}}^{44}+{\textbf{q}}^{38}+{\textbf{q}}^{52}-{\textbf{q}}^{58}-{\textbf{q}}^{34}
+{\textbf{q}}^{
60}
\end{array}
\ee

\subsection{Case $(5,n)$, $n=5 k$ fundamental representation\label{5k}}
\be
\begin{array}{|c|}
\hline\\
P^{T[5,n]}_{[1]}=c^{[5]}_{[1]} M_{[5]}^{\ast} q^{-4n}+c^{[4,1]}_{[1]}
M_{[4,1]}^{\ast} q^{-\frac{12n}{5}} t^{\frac{2n}{5}}
+c^{[3,2]}_{[1]} M_{[3,2]}^{\ast} q^{-\frac{8n}{5}}
t^{\frac{4n}{5}}+c^{[3,1,1]}_{[1]} M_{[3,1,1]}^{\ast} q^{-\frac{6n}{5}}
t^{\frac{6n}{5}} +\\ \\+ c^{[2,2,1]}_{[1]} M_{[2,2,1]}^{\ast} q^{-\frac{4n}{5}}
t^{\frac{8n}{5}}+ c^{[2,1,1,1]}_{[1]} M_{[2,1,1,1]}^{\ast} q^{-\frac{2n}{5}}
t^{\frac{12n}{5}} + c^{[1,1,1,1,1]}_{[1]} M_{[1,1,1,1,1]}^{\ast} t^{4n}
\\
\\
\hline
\end{array}
\ee
\fr{ \\
c^{[5]}_{[1]}=1,\ \       c^{[4,1]}_{[1]}={\frac { (-1+q^2)
\left(4q^6t^2+q^6+3q^4t^2+2q^4+2t^2q^2+3q^2+t^2+4 \right)
}{\left(-1+{q}^{8}{t}^{2}  \right) }},\ \    \nonumber \\      
c^{[3,2]}_{[1]}={\frac { \left( -1+{q}^{2} \right)^2
(5q^6t^4+4q^6t^2+q^6+3q^4+11q^4t^2+6t^4q^4+3t^4q^2+11t^2q^2+6q^2+5+t^4+4t^2)}{
\left( -1+{q}^{4}{t}^{2} \right) \left( -1+{q}^{2}{t}^{2} \right)^2(1+t^2q^2)
}},  \  \nonumber \\    c^{[3,1,1]}_{[1]}={\frac {\left( -1+{q}^{2} \right)^2 
\left( 1+{t}^{2}
\right)\left(6t^4q^4+3q^4t^2+q^4+3t^4q^2+4t^2q^2+3q^2+t^4+3t^2+6\right)
}{\left(-1+ {q}^{2}{t}^{2} \right)  \left( -1+{q}^{6}{t}^{4} \right)}},\ \   
\nonumber \\   c^{[2,2,1]}_{[1]}={\frac {\left( -1+{q}^{2} \right)^3  \left(
1+{t}^{2}
\right)\left(5q^4t^6+6t^4q^4+3q^4t^2+q^4+4t^6q^2+11t^4q^2+11t^2q^2+4q^2+3t^4+5+6t^2+t^6\right)
}{ \left( {q}^{2}{t}^{2}+1 \right) \left(-1+{q}^{4}{t}^{2} \right) \left(
-1+{q}^{2}{t}^{2} \right) ^{2}}},\    \nonumber \\  \\
c^{[2,1,1,1]}_{[1]}={\frac {\left(-1+{q}^{2} \right)^3  \left( 1+{t}^{4}
\right)(1+t^2+t^4) \left( 4t^6q^2+3t^4q^2+2t^2q^2+q^2+t^6+2t^4+3t^2+4 \right)
}{\left( -1+{q}^{2}{t}^{2} \right)  \left( -1+{q}^{2}{t}^{4}
\right)(-1+q^4t^6)}},\ \ \nonumber \\      c^{[1,1,1,1,1]}_{[1]}={\frac
{(t^4+t^2+1)(t^8+t^6+t^4+t^2+1)(t^4+1)(t^2+1)^2(-1+q^2)^4}{\left(
-1+{q}^{2}{t}^{2} \right)   \left( -1+{t}^{4}{q}^{2} \right) \left( -1+t^6q^2
\right)(-1+t^8q^2) }}}

Similarly to the case of unknot, the obtained expressions for
$P^{T[5,n]}_{[1]}$ are not polynomials
with positive coefficients. Moreover, even ${\cal P}^{T[5,n]}_{[1]}$
(i.e. the expression normalized by unknot) is not a polynomial with positive
coefficients. However, it is a power series in ${\bf q,t}$ with positive
coefficients
as is seen from the examples:
$$P^{T[5,0]}_{[1]}=\dfrac{\{A\}}{\{t\}^5 A^4
}\Big(A^2-1\Big)^4=\bf{\dfrac{a^2t+1}{1-q^2}}\cdot\dfrac{({a}^{2}t+1)^4q^5}{a^5t^{5/2}}\Big(\summ{j=0}{\infty}\dfrac{1}{6}(1+j)(2+j)(3+j)q^{2j}\Big)$$
\bigskip \\
$P^{T[5,5]}_{[1]}=\dfrac{\{A\}}{\{t\}^5 A^4q^{20}} \Big(
q^{20}A^8+(-q^{18}t^4-t^6q^{20}-q^{18}t^2-q^{18}-11t^6q^{14}-q^{12}t^8-t^2q^{14}-6t^4q^{12}-q^{20}t^8-t^2q^{16}+4q^{12}t^6-t^4q^{20}+4q^{14}t^8-q^{14}-t^4q^{16}-q^{16}-q^{12}-t^2q^{20}+9t^4q^{14}+9t^6q^{16}-6q^{16}t^8+4q^{18}t^8-q^{18}t^6+4t^2q^{12})A^6+(q^{12}t^{14}+q^{20}t^8+2q^{20}t^{10}-2t^{10}q^{18}+3q^{18}t^8-6q^{16}t^{10}+q^6+2q^{10}+q^8-6t^4q^{10}+q^{12}+q^{14}+2t^{12}q^{14}+9t^{10}q^{14}-4q^{14}t^{14}+q^{10}t^{12}-4q^6t^2-4q^6t^6+6q^6t^4+q^6t^8-3q^{18}t^{12}-2t^2q^{10}-3t^2q^8+2t^4q^8+t^{14}q^{20}+3t^2q^{14}+3t^2q^{12}-11t^4q^{12}+4t^4q^{14}+2t^2q^{16}+q^{18}t^2+2q^{18}t^4+2t^6q^8+5q^{12}t^6+3q^{18}t^6+t^6q^{20}+4t^6q^{16}-20t^6q^{14}+3t^4q^{16}+9t^6q^{10}+5q^{14}t^8-3t^8q^8-q^{10}t^8+5q^{12}t^8+6q^{16}t^{14}-11q^{16}t^8-3t^{12}q^{12}-q^{12}t^{10}+t^{10}q^8+2q^{16}t^{12}+t^{12}q^{20}-3t^{10}q^{10}-4q^{18}t^{14})A^4+(3q^{12}t^{16}+q^{12}t^{14}-6q^{16}t^{18}-3t^{10}q^{18}-2q^{18}t^8-q^{20}t^{18}-q^{20}t^{16}-q^2-q^6-q^4-q^8-16t^{12}q^{14}+2q^{10}
t^{14}+4q^{14}t^{18}+18t^{10}q^{14}-4q^{14}t^{14}-2q^{16}t^{16}-t^{16}q^{10}+3q^{18}t^{16}-q^6t^{12}+4q^{18}t^{18}-2q^{14}t^{16}+4t^2q^2-2q^4t^6-6t^4q^2-2t^4q^4+3q^4t^2+2q^6t^2-4q^6t^6-t^8q^2+4t^6q^2+q^6t^4+q^6t^8+3q^4t^8+2q^6t^{10}-t^{10}q^4-3q^{18}t^{12}-3t^2q^{10}-3t^2q^8+15t^4q^8-t^{14}q^{20}-t^2q^{14}-2t^2q^{12}-4t^4q^{12}-3t^4q^{14}-16t^6q^8+9q^{12}t^6-q^{18}t^6-3t^6q^{16}-5t^6q^{14}-t^4q^{16}+18t^6q^{10}+9q^{14}t^8+4t^8q^8-23q^{10}t^8+13q^{12}t^8+q^{16}t^{14}-4q^{16}t^8-q^8t^{14}+4t^{12}q^{12}-23q^{12}t^{10}+2q^8t^{12}+15q^{16}t^{12}-t^{12}q^{20}+7t^{10}q^{10}+2q^{18}t^{14}-t^{18}q^{12})A^2+2q^{12}t^{16}-4t^2+6t^4-4t^6+t^8+t^4q^{10}-12t^{12}q^{14}+3q^{10}t^{14}+6q^{14}t^{18}+2t^{10}q^{14}+3q^{14}t^{14}-3q^{16}t^{16}-6q^{10}t^{12}+q^8t^{16}-3t^{16}q^{10}+q^{18}t^{16}-3q^6t^{12}+q^6t^{14}+q^{10}t^{18}+q^{18}t^{18}+2q^{14}t^{16}+t^2q^2+t^{10}q^2+2q^4t^6-4t^4q^2-3t^4q^4+q^4t^2+q^6t^2+3q^6t^6-4t^8q^2+6t^6q^2-3q^6t^4-2q^6t^8+2q^4t^8+3q^6t^{10}-3t^{10}q^4+1+q^{18}t^{12}-2q^{12}t^{14}-4q^{16}t^{18}+q^{16}t^{
10}+t^2q^8+2t^4q^8+t^4q^{12}-12t^6q^8+2q^{12}t^6+t^6q^{14}+2t^6q^{10}+2q^{14}t^8+14t^8q^8-12q^{10}t^8-2q^{12}t^8-3q^{16}t^{14}+6q^{16}t^{20}-4q^{14}t^{20}+q^{16}t^8-3q^8t^{14}+14t^{12}q^{12}+q^4t^{12}-12q^{12}t^{10}-4q^{18}t^{20}+q^{20}t^{20}-6t^{10}q^8+3q^8t^{12}+2q^{16}t^{12}+14t^{10}q^{10}+t^{20}q^{12}+q^{18}t^{14}-4t^{18}q^{12}
\Big)=\bf{\dfrac{a^2t+1}{1-q^2}}\cdot\Big({\dfrac
{1}{{a}^{5}{t}^{45/2}{q}^{15}}}+{\dfrac {1}{{a}^{3}{t}^{39/2}
{q}^{13}}}+{\dfrac {{a}^{2}{t}^{3}+1}{{t}^{41/2}{a}^{5} {q}^{11}}}
+{\dfrac {{t}^{4}{a}^{4}+{a}^{2}{t}^{3}+1+{a}^{2}t}{{t}^{37/2}{a}^{5}
{q}^{9}}}+{\dfrac
{1+{t}^{2}+{t}^{5}{a}^{2}+2\,{a}^{2}{t}^{3}+{t}^{6}{a}^{4}}{{t}^{37/2}{a}^{5}
{q}^{7}}}
+{\dfrac
{2\,{t}^{6}{a}^{4}+1+{t}^{2}+7\,{a}^{2}{t}^{3}+{a}^{2}t+{t}^{4}{a}^{4}}{{t}^{33/2}{a}^{5}
{q}^{5}}}+\dots\Big)$

\bigskip
\noindent
$\boxed{P^{T[5,5k]}_{[1]}=\dfrac{\{A\}}{\{t\}^5
A^4q^{20k}}{\cal{P}}^{T[5,5k]}_{[1]}}$

\begin{itemize}
\item{HOMFLY case}
\end{itemize}
$$H^{T[5,0]}_{[1]}=\dfrac{\{A\}}{\{q\}^5 A^4 }\Big(A^2-1\Big)^4$$ \bigskip \\
$H^{T[5,5]}_{[1]}=\dfrac{\{A\}}{\{q\}^5 A^4q^{20}} \Big(
q^{20}A^8+(3q^{26}+3q^{14}-8q^{24}+11q^{18}-8q^{16}-q^{12}-q^{28}-14q^{20}+11q^{22})A^6+(-2q^{16}-3q^{12}-3q^8+q^6+11q^{18}+5q^{10}+q^{14}-3q^{28}-2q^{24}-3q^{32}+11q^{22}+q^{26}+5q^{30}-14q^{20}+q^{34})A^4+(-18q^{22}-q^2-5q^{10}+11q^{12}-4q^{34}-18q^{18}-18q^{14}-4q^6+3q^4+3q^8-q^{38}+3q^{32}+19q^{16}-18q^{26}+11q^{28}+19q^{24}+16q^{20}+3q^{36}-5q^{30})A^2+7q^4+1-4q^2-16q^{14}-4q^{10}+5q^8-7q^6-19q^{22}+21q^{24}-16q^{26}+8q^{28}+7q^{36}-4q^{30}+5q^{32}-7q^{34}+8q^{12}+q^{40}+21q^{16}-19q^{18}+17q^{20}-4q^{38}
\Big)$

\bigskip
\noindent
$\boxed{H^{T[5,5k]}_{[1]}=\dfrac{\{A\}}{\{q\}^5 A^4 q^{20k}}
{\cal{H}}^{T[5,5k]}_{[1]}}$

\begin{itemize}
\item{Alexander case}
\end{itemize}
$$A^{T[5,0]}_{[1]}=0$$  \\
$A^{T[5,5]}_{[1]}=\bf{q}^{40}-5q^{38}+10q^{36}-10q^{34}+5q^{32}-4q^{30}+15q^{28}-30q^{26}+30q^{24}-15q^{22}+6q^{20}-15q^{18}+30q^{16}-30q^{14}+15q^{12}-4q^{10}+5q^8-10q^6+10q^4-5q^2+1$
 \\

\subsection{Case $(5,n)$, $n=5 k+1$ fundamental representation\label{5k+1}}
\be
\begin{array}{|c|}
\hline\\
P^{T[5,n]}_{[1]}=c^{[5]}_{[1]} M_{[5]}^{\ast} q^{-4n}+c^{[4,1]}_{[1]}
M_{[4,1]}^{\ast} q^{-\frac{12n}{5}} t^{\frac{2n}{5}}
+c^{[3,2]}_{[1]} M_{[3,2]}^{\ast} q^{-\frac{8n}{5}}
t^{\frac{4n}{5}}+c^{[3,1,1]}_{[1]} M_{[3,1,1]}^{\ast} q^{-\frac{6n}{5}}
t^{\frac{6n}{5}} +\\ \\+ c^{[2,2,1]}_{[1]} M_{[2,2,1]}^{\ast} q^{-\frac{4n}{5}}
t^{\frac{8n}{5}}+ c^{[2,1,1,1]}_{[1]} M_{[2,1,1,1]}^{\ast} q^{-\frac{2n}{5}}
t^{\frac{12n}{5}} + c^{[1,1,1,1,1]}_{[1]} M_{[1,1,1,1,1]}^{\ast} t^{4n}
\\
\\
\hline
\end{array}
\ee
\fr{ \\
c^{[5]}_{[1]}=1,\ \      
c^{[4,1]}_{[1]}=-{\Big(\frac{q^2}{t^2}\Big)^{\frac{1}{5}}\frac { (-1+t^2)
\left(1+q^2+q^4+q^6+t^2q^6 \right) }{\left(-1+{q}^{8}{t}^{2}  \right) }},\ \   
       c^{[3,2]}_{[1]}={\Big(\frac{q^2}{t}\Big)^{\frac{4}{5}}\frac { \left(
-1+{t}^{2} \right)\left( {t}^{2}-{q}^{2} \right) (q^4t^2+q^2t^2+q^4+q^2+1)}{
\left( -1+{q}^{4}{t}^{2} \right) \left( -1+{q}^{6}{t}^{2} \right) }},  \ 
\nonumber \\ \\ \nonumber      
c^{[3,1,1]}_{[1]}={\Big(\frac{q^2}{t^2}\Big)^{\frac{3}{5}}\frac {\left(
-1+{t}^{2} \right)  \left( -1+{t}^{4} \right)\left(q^4t^4+q^4t^2+q^4+q^2+1
\right) }{\left(-1+ {q}^{2}{t}^{2} \right)  \left( -1+{q}^{6}{t}^{4}
\right)}},\ \      
c^{[2,2,1]}_{[1]}=-{\Big(\frac{q^{14}}{t^8}\Big)^{\frac{1}{5}}\frac {\left(
-1+{t}^{2} \right)  \left( -1+{t}^{4}
\right)(t^2-q^2)\left(t^4q^2+q^2t^2+q^2+t^2+1\right) }{ \left( {q}^{2}{t}^{2}+1
\right) \left(-1+{q}^{4}{t}^{2} \right) \left( -1+{q}^{2}{t}^{2} \right)
^{2}}},\    \nonumber \\  \\
c^{[2,1,1,1]}_{[1]}=-{\Big(\frac{q^2}{t^2}\Big)^{\
\frac{6}{5}}\frac {\left(-1+{t}^{2} \right)  \left( -1+{t}^{4} \right)(-1+t^6)
\left( {t}^{6}{q}^{2}+{q}^{2}{t}^{4}+{q}^{2}t^2+q^2+1 \right) }{\left(
-1+{q}^{2}{t}^{2} \right)  \left( -1+{q}^{2}{t}^{4} \right)(-1+q^4t^6)}},\ \ \ 
    c^{[1,1,1,1,1]}_{[1]}={\frac{q^4}{t^4}\frac
{(-1+t^{10})(-1+t^8)(-1+t^6)(-1+t^4)}{\left( -1+{q}^{2}{t}^{2} \right)   \left(
-1+{t}^{4}{q}^{2} \right) \left( -1+{t}^{6}{q}^2 \right)(-1+t^8q^2) }}}
Several first superpolynomials are:
$$P^{T[5,1]}_{[1]}=\dfrac{\{A\} t^4}{\{t\} A^4 q^4}$$
\bigskip \\
$P^{T[5,6]}_{[1]}=\dfrac{\{A\} t^4}{\{t\} A^4 q^{24}} \Big(
q^{20}(-A^2)^4+(q^{18}+t^4q^{16}+q^{12}+t^6q^{20}+q^{18}t^4+q^{16}+q^{18}t^6+t^2q^{14}+q^{14}+q^{20}t^8+t^4q^{20}+t^2q^{20}+t^2q^{16}+q^{18}t^2)(-A^2)^3+(q^6+t^4q^{10}+q^8+2q^{10}+q^{12}+q^{14}+q^{20}t^{12}+t^{12}q^{18}+t^{10}q^{16}+2t^2q^{10}+t^2q^8+3t^2q^{14}+3t^2q^{12}+2t^4q^{12}+4t^4q^{14}+2t^2q^{16}+q^{18}t^2+2q^{18}t^4+q^{12}t^6+3q^{18}t^6+t^6q^{20}+4t^6q^{16}+2t^6q^{14}+3t^4q^{16}+2q^{20}t^{10}+2q^{18}t^{10}+q^{20}t^{14}+2q^{16}t^8+q^{14}t^8+q^{20}t^8+3q^{18}t^8)(-A^2)^2+(q^2+2q^{18}t^{14}+q^4+q^6+4t^4q^{10}+q^8+q^{20}t^{12}+3t^{12}q^{18}+t^{12}q^{14}+t^{16}q^{18}+t^{10}q^{12}+4t^{10}q^{16}+t^{14}q^{16}+q^4t^2+2q^6t^2+q^6t^4+3t^2q^{10}+3t^2q^8+2t^4q^8+t^2q^{14}+2t^2q^{12}+4t^4q^{12}+3t^4q^{14}+t^6q^8+4q^{12}t^6+q^{18}t^6+3t^6q^{16}+5t^6q^{14}+t^4q^{16}+2t^6q^{10}+q^{20}t^{18}+q^{20}t^{16}+3q^{18}t^{10}+q^{10}t^8+2q^{12}t^8+2q^{16}t^{12}+q^{20}t^{14}+4q^{16}t^8+4q^{14}t^8+2q^{18}t^8+2q^{14}t^{10})(-A^2)+t^{20}q^{20}+t^{12}q^{18}+t^{14}q^{18}
+t^{16}q^{18}+t^{18}q^{18}+t^8q^{16}+2t^{12}q^{16}+t^{14}q^{16}+t^{10}q^{16}+t^{16}q^{16}+t^{14}q^{14}+t^{12}q^{14}+2t^8q^{14}+2t^{10}q^{14}+t^6q^{14}+t^{10}q^{12}+2t^8q^{12}+t^{12}q^{12}+t^4q^{12}+2t^6q^{12}+t^4q^{10}+2t^6q^{10}+t^8q^{10}+t^{10}q^{10}+t^8q^8+2t^4q^8+t^6q^8+t^2q^8+t^4q^6+t^2q^6+t^6q^6+t^4q^4+t^2q^4+t^2q^2+1
\Big) = \\ = \bf{ -{\frac {{a}^{2}t+1}{ \left( -1+{q}^{2} \right)
{a}^{5}{t}^{53/2}{q}^{19}}}}\Big(
1+\qb^4 \tb^2+\qb^6 \tb^4+\qb^8 \tb^4+\qb^8 \tb^6+\qb^{10} \tb^6+\qb^{12}
\tb^6+\qb^{10} \tb^8+2 \qb^{12} \tb^8+\qb^{14} \tb^8+\qb^{16} \tb^8
+\qb^{14} \tb^{10}+2 \qb^{16} \tb^{10}+\qb^{18} \tb^{10}+ \qb^{20}
\tb^{10}+\qb^{16} \tb^{12}+2 \qb^{18} \tb^{12}+2 \qb^{20} \tb^{12} +\qb^{22}
\tb^{12}+\qb^{24} \tb^{12}+ \qb^{20} \tb^{14} +2 \qb^{22} \tb^{14}+2 \qb^{24}
\tb^{14}+\qb^{26} \tb^{14}+\qb^{28} \tb^{14}+\qb^{24} \tb^{16}+ \qb^{26}
\tb^{16}+2 \qb^{28} \tb^{16}+\qb^{30} \tb^{16}+\qb^{32} \tb^{16}+\qb^{30}
\tb^{18}+\qb^{32} \tb^{18}+\qb^{34} \tb^{18}+ \qb^{36} \tb^{18} +\qb^{40}
\tb^{20} +\ab^2  \big(\qb^2 \tb^3+\qb^4 \tb^5+\qb^6 \tb^5+\qb^6 \tb^7+ 2 \qb^8
\tb^7+\qb^{10} \tb^7+\qb^8 \tb^9+3 \qb^{10} \tb^9+2 \qb^{12} \tb^9+\qb^{14}
\tb^9 +3 \qb^{12} \tb^{11}+4 \qb^{14} \tb^{11}+2 \qb^{16} \tb^{11}+\qb^{18}
\tb^{11}+2 \qb^{14} \tb^{13}+4 \qb^{16} \tb^{13}+ 4 \qb^{18} \tb^{13}+2
\qb^{20} \tb^{13}+\qb^{22} \tb^{13}+\qb^{16} \tb^{15} +3 \qb^{18} \tb^{15}+5
\qb^{20} \tb^{15}+ 4 \qb^{22} \tb^{15}+2 \qb^{24} \tb^{15}+\qb^{26}
\tb^{15}+\qb^{20} \tb^{17}+3 \qb^{22} \tb^{17}+4 \qb^{24} \tb^{17}+ 4 \qb^{26}
\tb^{17}+2 \qb^{28} \tb^{17} +\qb^{30} \tb^{17}+\qb^{
24} \tb^{19}+2 \qb^{26} \tb^{19}+3 \qb^{28} \tb^{19}+3 \qb^{30} \tb^{19}+2
\qb^{32} \tb^{19}+\qb^{34} \tb^{19}+\qb^{32} \tb^{21}+\qb^{34}
\tb^{21}+\qb^{36} \tb^{21}+ \qb^{38} \tb^{21}\big)+\ab^4  \big(\qb^6
\tb^8+\qb^8 \tb^{10}+\qb^{10} \tb^{10}+2 \qb^{10} \tb^{12}+2 \qb^{12}
\tb^{12}+\qb^{14} \tb^{12}+\qb^{12} \tb^{14}+3 \qb^{14} \tb^{14}+2 \qb^{16}
\tb^{14}+\qb^{18} \tb^{14} +\qb^{14} \tb^{16}+ 3 \qb^{16} \tb^{16}+4 \qb^{18}
\tb^{16}+2 \qb^{20} \tb^{16}+\qb^{22} \tb^{16}+2 \qb^{18} \tb^{18}+3 \qb^{20}
\tb^{18}+4 \qb^{22} \tb^{18}+2 \qb^{24} \tb^{18}+\qb^{26} \tb^{18} +\qb^{20}
\tb^{20}+2 \qb^{22} \tb^{20}+3 \qb^{24} \tb^{20}+ 3 \qb^{26} \tb^{20}+2
\qb^{28} \tb^{20}+\qb^{30} \tb^{20}+\qb^{26} \tb^{22}+\qb^{28} \tb^{22}+2
\qb^{30} \tb^{22}+\qb^{32} \tb^{22}+ \qb^{34} \tb^{22}\big)+\ab^6 
\big(\qb^{12} \tb^{15}+\qb^{14} \tb^{17}+\qb^{16} \tb^{17}+ \qb^{16}
\tb^{19}+\qb^{18} \tb^{19}+\qb^{20} \tb^{19}+\qb^{18} \tb^{21}+\qb^{20}
\tb^{21}+\qb^{22} \tb^{21}+\qb^{24} \tb^{21}+\qb^{22} \tb^{23}+\qb^{24}
\tb^{23}+\
qb^{26}\tb^{23}+ \qb^{28} \tb^{23}\big)+\ab^8  \qb^{20} \tb^{24}\Big)\\
$

\bigskip
\noindent
$\boxed{P^{T[5,5k+1]}_{[1]}=\dfrac{\{A\} t^4}{\{t\} A^4
q^{20k+4}}{\cal{P}}^{T[5,5k+1]}_{[1]}}$

\begin{itemize}
\item{HOMFLY case}
\end{itemize}
$$H^{T[5,1]}_{[1]}=\dfrac{\{A\}}{\{q\} A^4}$$ \bigskip \\
$H^{T[5,6]}_{[1]}=\dfrac{\{A\}}{\{q\} A^4 q^{20}} \Big(q^{20}
(-A^{2})^4+(q^{28}+q^{26}+2 q^{24}+2 q^{22}+2 q^{20}+2 q^{18}+2
q^{16}+q^{14}+q^{12})(- A^{2})^3+(q^{34}+q^{32}+3 q^{30}+3 q^{28}+5 q^{26}+5
q^{24}+7 q^{22}+6 q^{20}+7 q^{18}+5 q^{16}+5 q^{14}+3 q^{12}+3
q^{10}+q^{8}+q^{6}) (-A^{2})^2+(q^{38}+q^{36}+2 q^{34}+3 q^{32}+4 q^{30}+5
q^{28}+7 q^{26}+7 q^{24}+8 q^{22}+8 q^{20}+8 q^{18}+7 q^{16}+7 q^{14}+5
q^{12}+4 q^{10}+3 q^{8}+2 q^{6}+q^{4}+q^{2}) (-A^{2})+q^{40}+q^{36}+q^{34}+2
q^{32}+2 q^{30}+3 q^{28}+2 q^{26}+4 q^{24}+3 q^{22}+4 q^{20}+3 q^{18}+4
q^{16}+2 q^{14}+3 q^{12}+2 q^{10}+2 q^{8}+q^{6}+q^{4}+1\Big)$

\bigskip
\noindent
$\boxed{H^{T[5,5k+1]}_{[1]}=\dfrac{\{A\}}{\{q\} A^4 q^{20k}}
{\cal{H}}^{T[5,5k+1]}_{[1]}=
q^{-20k-4}s_{[5]}^{*}-q^{-10k-2}s_{[4,1]}^{*}+s_{[3,1,1]}^{*}-q^{10k+2}s_{[2,1,1,1]}^{*}+q^{20k+4}s_{[1,1,1,1,1]}^{*}}$

\bigskip

\noindent
and the results coincides with well known HOMFLY polynomials, see
(\ref{Wrepfund}) and
(\ref{113}).

\begin{itemize}
\item{Floer case}
\end{itemize}
$$F^{T[5,1]}_{[1]}=1$$ \\
$F^{T[5,6]}_{[1]}=\bf{q}^{40}t^{20}+q^{38}t^{19}+q^{30}t^{18}+q^{26}t^{15}+q^{20}t^{14}+q^{14}t^{9}+q^{10}t^{8}+q^{2}t+1$
 \\
\begin{itemize}
\item{Alexander case}
\end{itemize}
$$A^{T[5,1]}_{[1]}=1$$  \\
$$A^{T[5,6]}_{[1]}=\bf{q}^{40}-q^{38}+q^{30}-q^{26}+q^{20}-q^{14}+q^{10}-q^2+1$$
 \\

\subsection{Case $(5,n)$, $n=5 k+2$ fundamental representation\label{5k+2}}
\be
\begin{array}{|c|}
\hline\\
P^{T[5,n]}_{[1]}=c^{[5]}_{[1]} M_{[5]}^{\ast} q^{-4n}+c^{[4,1]}_{[1]}
M_{[4,1]}^{\ast} q^{-\frac{12n}{5}} t^{\frac{2n}{5}}
+c^{[3,2]}_{[1]} M_{[3,2]}^{\ast} q^{-\frac{8n}{5}}
t^{\frac{4n}{5}}+c^{[3,1,1]}_{[1]} M_{[3,1,1]}^{\ast} q^{-\frac{6n}{5}}
t^{\frac{6n}{5}} +\\ \\+ c^{[2,2,1]}_{[1]} M_{[2,2,1]}^{\ast} q^{-\frac{4n}{5}}
t^{\frac{8n}{5}}+ c^{[2,1,1,1]}_{[1]} M_{[2,1,1,1]}^{\ast} q^{-\frac{2n}{5}}
t^{\frac{12n}{5}} + c^{[1,1,1,1,1]}_{[1]} M_{[1,1,1,1,1]}^{\ast} t^{4n}
\\
\\
\hline
\end{array}
\ee
\fr{ \\
c^{[5]}_{[1]}=1,\ \      
c^{[4,1]}_{[1]}=-{\Big(\frac{q^2}{t^2}\Big)^{\frac{2}{5}}\frac { (-1+t^2)
\left(1+q^2+q^4+t^2q^4+t^2q^6 \right) }{\left(-1+{q}^{8}{t}^{2}  \right) }},\ \
          c^{[3,2]}_{[1]}={\Big(\frac{q^3}{t^4}\Big)^{\frac{2}{5}}\frac {
\left( -1+{t}^{2} \right)\left( {t}^{2}-{q}^{2} \right)
(q^6t^4+q^4t^2+t^2q^2+q^2+1)}{ \left( -1+{q}^{4}{t}^{2} \right) \left(
-1+{q}^{6}{t}^{2} \right) }},  \  \nonumber \\ \\ \nonumber      
c^{[3,1,1]}_{[1]}={\Big(\frac{q^2}{t^2}\Big)^{\frac{6}{5}}\frac {\left(
-1+{t}^{2} \right)  \left( -1+{t}^{4} \right)\left(q^4t^4+q^2t^4+t^2+q^2t^2+1
\right) }{\left(-1+ {q}^{2}{t}^{2} \right)  \left( -1+{q}^{6}{t}^{4}
\right)}},\ \      
c^{[2,2,1]}_{[1]}=-{\Big(\frac{q}{t^2}\Big)^{\frac{8}{5}}\frac {\left(
-1+{t}^{2} \right)  \left( -1+{t}^{4}
\right)(t^2-q^2)\left(t^6q^4+q^4t^4+q^2t^4+q^2t^2+1\right) }{ \left(
{q}^{2}{t}^{2}+1 \right) \left(-1+{q}^{4}{t}^{2} \right) \left(
-1+{q}^{2}{t}^{2} \right) ^{2}}},\    \nonumber \\  \\
c^{[2,1,1,1]}_{[1]}=-{\Big(\frac{q^2}{
t^2}\Big)^{\frac{7}{5}}\frac {\left(-1+{t}^{2} \right)  \left( -1+{t}^{4}
\right)(-1+t^6) \left( {t}^{6}{q}^{2}+{q}^{2}{t}^{4}+{q}^{2}t^2+t^2+1 \right)
}{\left( -1+{q}^{2}{t}^{2} \right)  \left( -1+{q}^{2}{t}^{4}
\right)(-1+q^4t^6)}},\ \ \      c^{[1,1,1,1,1]}_{[1]}={\frac{q^4}{t^4}\frac
{(-1+t^{10})(-1+t^8)(-1+t^6)(-1+t^4)}{\left( -1+{q}^{2}{t}^{2} \right)   \left(
-1+{t}^{4}{q}^{2} \right) \left( -1+{t}^{6}{q}^2 \right)(-1+t^8q^2) }}}
Several first superpolynomials are:
$P^{T[5,2]}_{[1]}=\dfrac{\{A\} t^4}{\{t\} A^4
q^8}\Big((q^4t^2+q^2)(-A^2)+t^4q^4+q^2t^2+1\Big)= -\frac{{\bf{a}}^{2}\bf{t}+1}{
\left( -1+\bf{q}^{2} \right) \bf{a}^{5}\bf{t}^{21/2}\bf{q}^{3}}\Big(
\bf{a}^2t^5q^6+a^2t^3q^2+t^4q^8+q^4t^2+1 \Big)$
\bigskip \\
$P^{T[5,7]}_{[1]}=\dfrac{\{A\} t^4}{\{t\} A^4
q^{28}}\Big((t^4q^{24}+t^2q^{22}+q^{20})(-A^2)^4+(t^{12}q^{24}+t^{10}q^{24}+t^8q^{24}+t^6q^{24}+t^{10}q^{22}+2t^8q^{22}+2t^4q^{22}+2t^6q^{22}+t^8q^{20}+2t^6q^{20}+3t^4q^{20}+2t^2q^{20}+t^6q^{18}+2t^2q^{18}+q^{18}+2t^4q^{18}+t^4q^{16}+2t^2q^{16}+q^{16}+t^2q^{14}+q^{14}+q^{12})(-A^2)^3+(t^{18}q^{24}+t^{12}q^{24}+2t^{14}q^{24}+t^{10}q^{24}+t^{16}q^{24}+3t^8q^{22}+t^6q^{22}+2t^{14}q^{22}+t^{16}q^{22}+4t^{12}q^{22}+4t^{10}q^{22}+5t^6q^{20}+5t^{10}q^{20}+2t^{12}q^{20}+t^4q^{20}+t^{14}q^{20}+6t^8q^{20}+7t^6q^{18}+t^{12}q^{18}+5t^4q^{18}+5t^8q^{18}+t^2q^{18}+2t^{10}q^{18}+2t^8q^{16}+t^{10}q^{16}+5t^6q^{16}+6t^4q^{16}+3t^2q^{16}+2t^6q^{14}+5t^4q^{14}+t^8q^{14}+q^{14}+4t^2q^{14}+2t^4q^{12}+t^6q^{12}+q^{12}+4t^2q^{12}+2q^{10}+2t^2q^{10}+t^4q^{10}+t^2q^{8}+q^{8}+q^{6})(-A^2)^2+(t^{22}q^{24}+t^{20}q^{24}+t^{18}q^{24}+t^{16}q^{24}+2t^{12}q^{22}+t^{10}q^{22}+4t^{14}q^{22}+2t^{18}q^{22}+3t^{16}q^{22}+t^{20}q^{22}+6t^{12}q^{20}+t^{18}q^{20}+5t^{10}q^{20}+4t^{14}q^{20}+2t^{16}q^{
20}+2t^8q^{20}+7t^8q^{18}+2t^{14}q^{18}+t^{16}q^{18}+4t^{12}q^{18}+7t^{10}q^{18}+3t^6q^{18}+4t^{10}q^{16}+2t^4q^{16}+7t^8q^{16}+2t^{12}q^{16}+t^{14}q^{16}+7t^6q^{16}+t^{12}q^{14}+5t^4q^{14}+t^2q^{14}+7t^6q^{14}+2t^{10}q^{14}+4t^8q^{14}+6t^4q^{12}+2t^2q^{12}+4t^6q^{12}+t^{10}q^{12}+2t^8q^{12}+t^8q^{10}+2t^6q^{10}+4t^2q^{10}+4t^4q^{10}+q^{8}+3t^2q^{8}+2t^4q^{8}+t^6q^{8}+2t^2q^{6}+t^4q^{6}+q^{6}+q^{4}+t^2q^{4}+q^{2})(-A^2)+t^{24}q^{24}+t^{18}q^{22}+t^{16}q^{22}+t^{20}q^{22}+2t^{16}q^{20}+2t^{14}q^{20}+2t^{10}q^{18}+3t^{12}q^{18}+t^{18}q^{18}+2t^{14}q^{18}+t^{16}q^{18}+t^{16}q^{16}+3t^{10}q^{16}+t^{14}q^{16}+3t^8q^{16}+2t^{12}q^{16}+t^{14}q^{14}+2t^{10}q^{14}+2t^6q^{14}+t^{12}q^{14}+3t^8q^{14}+3t^6q^{12}+t^{10}q^{12}+2t^8q^{12}+t^{12}q^{12}+t^4q^{12}+t^{10}q^{10}+2t^4q^{10}+2t^6q^{10}+t^8q^{10}+2t^4q^8+t^8q^8+t^6q^8+t^2q^8+t^4q^6+t^2q^6+t^6q^6+t^4q^4+t^2q^4+t^2q^2+t^{22}q^{22}+t^{20}q^{20}+t^{12}q^{20}+t^{18}q^{20}+1\Big)=\\=-\frac{{\bf{a}}^{2}\bf{t}+1}{
\left( -1+\bf{q}^{2} \right) \bf{a}^{5}\bf{t}^{61/2}\bf{q}
^{23}}\Big(
\bf{1}+4\,{a}^{2}{t}^{11}{q}^{14}+4\,{a}^{2}{t}^{19}{q}^{30}+{q}^{20}{t}^{24}{a}^{8}+7\,{a}^{2}{t}^{19}{q}^{26}+{q}^{14}{t}^{12}{a}^{4}
\mbox{}+{a}^{6}{t}^{21}{q}^{24}+{q}^{28}{t}^{24}{a}^{4}+7\,{a}^{2}{t}^{17}{q}^{24}+{q}^{16}{t}^{15}{a}^{2}+{q}^{24}{t}^{22}{a}^{4}
\mbox{}+{q}^{22}{t}^{12}+6\,{a}^{2}{t}^{13}{q}^{16}+{t}^{4}{q}^{8}+5\,{a}^{4}{t}^{16}{q}^{18}+4\,{a}^{2}{t}^{15}{q}^{22}+2\,{a}^{6}{t}^{21}{q}^{22}
\mbox{}+{a}^{2}{t}^{13}{q}^{22}+{q}^{4}{t}^{2}+2\,{a}^{2}{t}^{13}{q}^{20}+{q}^{6}{t}^{4}+{q}^{28}{t}^{14}+{q}^{24}{t}^{12}+2\,{a}^{6}{t}^{25}{q}^{30}
\mbox{}+3\,{q}^{24}{t}^{19}{a}^{2}+2\,{a}^{2}{t}^{15}{q}^{24}+3\,{a}^{6}{t}^{23}{q}^{24}+{q}^{26}{t}^{18}{a}^{4}+{q}^{42}{t}^{26}{a}^{4}
\mbox{}+{a}^{2}{t}^{17}{q}^{30}+2\,{a}^{6}{t}^{23}{q}^{26}+{q}^{6}{t}^{8}{a}^{4}+{q}^{38}{t}^{24}{a}^{4}+2\,{q}^{28}{t}^{17}{a}^{2}+6\,{q}^{20}{t}^{18}{a}^{4}
\mbox{}+{a}^{6}{t}^{27}{q}^{32}+2\,{a}^{6}{t}^{25}{q}^{26}+2\,{a}^{2}{t}^{21}{q}^{36}+{a}^{6}{t}^{19}{q}^{20}+{a}^{8}{t}^{28}{q}^{28}+5\,{a}^{4}{t}^{20}{q}^{22}
\mbox{}+5\,{q}^{30}{t}^{22}{a}^{4}+{q}^{10}{t}^{10}{a}^{4}+5\,{a}^{2}{t}^{15}{q}^{18}+{a}^{6}{t}^{25}{q}^{32}+{a}^{2}{t}^{7}{q}^{10}+7\,{q}^{22}{t}^{17}{a}^{2}
\mbox{}+6\,{a}^{2}{t}^{21}{q}^{32}+{a}^{2}{t}^{25}{q}^{46}+{a}^{2}{t}^{21}{q}^{38}+{q}^{20}{t}^{20}{a}^{4}+7\,{q}^{20}{t}^{15}{a}^{2}
\mbox{}+{q}^{14}{t}^{16}{a}^{4}+4\,{q}^{34}{t}^{24}{a}^{4}+{a}^{4}{t}^{26}{q}^{40}+4\,{a}^{2}{t}^{17}{q}^{26}+2\,{a}^{2}{t}^{9}{q}^{12}+{q}^{34}{t}^{26}{a}^{4}
\mbox{}+{q}^{40}{t}^{25}{a}^{2}+{q}^{16}{t}^{8}+{q}^{26}{t}^{14}+{q}^{44}{t}^{22}+{q}^{48}{t}^{24}+{q}^{42}{t}^{22}+{q}^{10}{t}^{8}
\mbox{}+{t}^{6}{q}^{10}+5\,{q}^{26}{t}^{22}{a}^{4}+{q}^{12}{t}^{6}+{t}^{6}{q}^{8}+2\,{q}^{20}{t}^{14}+3\,{q}^{24}{t}^{16}+3\,{q}^{30}{t}^{18}
\mbox{}+{q}^{40}{t}^{20}+{q}^{34}{t}^{18}+{q}^{32}{t}^{16}+{q}^{30}{t}^{16}+{q}^{38}{t}^{22}+{q}^{38}{t}^{20}+{q}^{36}{t}^{18}
\mbox{}+{q}^{40}{t}^{22}+{t}^{12}{q}^{16}+{t}^{10}{q}^{20}+{t}^{10}{q}^{18}+{q}^{14}{t}^{8}+3\,{q}^{18}{t}^{12}+3\,{q}^{26}{t}^{16}
\mbox{}+2\,{q}^{24}{t}^{14}+2\,{q}^{16}{t}^{10}+2\,{q}^{34}{t}^{23}{a}^{2}+{a}^{6}{t}^{27}{q}^{34}+2\,{q}^{12}{t}^{8}+4\,{a}^{2}{t}^{21}{q}^{34}
\mbox{}+2\,{t}^{12}{q}^{20}+2\,{q}^{34}{t}^{20}+2\,{q}^{32}{t}^{18}+2\,{q}^{28}{t}^{18}+2\,{q}^{36}{t}^{20}+3\,{q}^{22}{t}^{14}+2\,{q}^{28}{t}^{16}
\mbox{}+{a}^{2}{t}^{19}{q}^{34}+{q}^{22}{t}^{16}{a}^{4}+2\,{t}^{10}{q}^{14}+{a}^{2}{t}^{23}{q}^{42}+2\,{a}^{2}{t}^{11}{q}^{16}+6\,{q}^{28}{t}^{22}{a}^{4}
\mbox{}+7\,{q}^{24}{t}^{20}{a}^{4}+{a}^{2}{t}^{25}{q}^{44}+{a}^{2}{t}^{9}{q}^{8}+2\,{q}^{38}{t}^{26}{a}^{4}+{q}^{32}{t}^{23}{a}^{2}+3\,{a}^{2}{t}^{9}{q}^{10}
\mbox{}+4\,{a}^{4}{t}^{24}{q}^{32}+2\,{q}^{24}{t}^{18}{a}^{4}+2\,{a}^{6}{t}^{21}{q}^{20}+{a}^{6}{t}^{17}{q}^{16}+2\,{a}^{2}{t}^{7}{q}^{8}+{a}^{2}{t}^{11}{q}^{18}
\mbox{}+{q}^{30}{t}^{27}{a}^{6}+4\,{q}^{14}{t}^{14}{a}^{4}+{a}^{6}{t}^{21}{q}^{18}+{a}^{2}{t}^{3}{q}^{2}+4\,{a}^{2}{t}^{13}{q}^{18}+2\,{a}^{6}{t}^{23}{q}^{22}
\mbox{}+{a}^{2}{t}^{5}{q}^{6}+{q}^{32}{t}^{20}+{q}^{18}{t}^{14}{a}^{4}+2\,{q}^{14}{t}^{13}{a}^{2}+{a}^{6}{t}^{27}{q}^{36}+7\,{a}^{2}{t}^{19}{q}^{28}
\mbox{}+2\,{a}^{6}{t}^{25}{q}^{28}+{a}^{2}{t}^{15}{q}^{26}+2\,{q}^{32}{t}^{22}{a}^{4}+2\,{q}^{36}{t}^{24}{a}^{4}+{q}^{34}{t}^{22}{a}^{4}
\mbox{}+{a}^{4}{t}^{26}{q}^{36}+5\,{q}^{26}{t}^{20}{a}^{4}+2\,{a}^{2}{t}^{19}{q}^{32}+2\,{a}^{2}{t}^{23}{q}^{40}+2\,{a}^{6}{t}^{19}{q}^{18}
\mbox{}+{q}^{30}{t}^{20}{a}^{4}+{a}^{2}{t}^{7}{q}^{6}+{a}^{4}{t}^{14}{q}^{12}+{a}^{6}{t}^{17}{q}^{14}+3\,{q}^{38}{t}^{23}{a}^{2}+{a}^{6}{t}^{15}{q}^{12}
\mbox{}+3\,{q}^{18}{t}^{18}{a}^{4}+2\,{a}^{4}{t}^{20}{q}^{28}+2\,{q}^{16}{t}^{14}{a}^{4}+4\,{a}^{4}{t}^{16}{q}^{16}+2\,{q}^{20}{t}^{17}{a}^{2}
\mbox{}+2\,{q}^{28}{t}^{21}{a}^{2}+{a}^{6}{t}^{19}{q}^{16}+{a}^{6}{t}^{23}{q}^{28}+{a}^{4}{t}^{10}{q}^{8}+4\,{a}^{2}{t}^{11}{q}^{12}+5\,{q}^{30}{t}^{21}{a}^{2}
\mbox{}+{a}^{2}{t}^{9}{q}^{14}+2\,{q}^{20}{t}^{16}{a}^{4}+{a}^{2}{t}^{5}{q}^{4}+3\,{a}^{4}{t}^{24}{q}^{30}+5\,{q}^{22}{t}^{18}{a}^{4}+2\,{q}^{12}{t}^{12}{a}^{4}
\mbox{}+2\,{q}^{10}{t}^{12}{a}^{4}+4\,{a}^{2}{t}^{23}{q}^{36}+{a}^{2}{t}^{25}{q}^{42}+{q}^{24}{t}^{26}{a}^{8}
\Big)$

\bigskip
\noindent
$\boxed{P^{T[5,5k+2]}_{[1]}=\dfrac{\{A\} t^4}{\{t\} A^4
q^{20k+8}}{\cal{P}}^{T[5,5k+2]}_{[1]}}$

\begin{itemize}
\item{HOMFLY case}
\end{itemize}
$$H^{T[5,2]}_{[1]}=-\dfrac{\{A\}}{\{q\} A^4
q^4}\Big(A^2q^8+A^2q^2-q^{12}-q^6-1\Big)$$ \bigskip \\
$H^{T[5,7]}_{[1]}=\dfrac{\{A\}}{\{q\} A^4 q^{24}}
\Big((q^{28}+q^{24}+q^{20})A^{8}+(-q^{36}-q^{34}-2q^{32}-3q^{30}-3q^{28}-4q^{26}-4q^{24}-4q^{22}-3q^{20}-3q^{18}-2q^{16}-q^{14}-q^{12})A^{6}+(q^{42}+q^{40}+3q^{38}+3q^{36}+6q^{34}+6q^{32}+9q^{30}+9q^{28}+11q^{26}+10q^{24}+11q^{22}+9q^{20}+9q^{18}+6q^{16}+6q^{14}+3q^{12}+3q^{10}+q^{8}+
q^{6})A^{4}+(-q^{46}-q^{44}-2q^{42}-3q^{40}-4q^{38}-6q^{36}-7q^{34}-9q^{32}-10q^{30}-11q^{28}-12q^{26}-12q^{24}-12q^{22}-11q^{20}-10q^{18}-9q^{16}-7q^{14}-6q^{12}-
4q^{10}-3q^{8}-
2q^{6}-q^{4}-q^{2})A^{2}+2q^{38}+q^{44}+q^{42}+2q^{40}+4q^{30}+3q^{36}+3q^{34}+4q^{32}+4q^{22}+q^{48}+4q^{18}+4q^{16}+3q^{14}+3q^{12}+2q^{10}+2q^{8}+
q^{6}+q^{4}+ 5q^{28}+4q^{26}+6q^{24}+5q^{20}+1\Big)$

\bigskip
\noindent
$\boxed{H^{T[5,5k+2]}_{[1]}=\dfrac{\{A\}}{\{q\} A^4 q^{20k+4}}
{\cal{H}}^{T[5,5k+2]}_{[1]}=
q^{-20k-8}s_{[5]}^{*}-q^{-10k-4}s_{[4,1]}^{*}+s_{[3,1,1]}^{*}-q^{10k+4}s_{[2,1,1,1]}^{*}+q^{20k+8}s_{[1,1,1,1,1]}^{*}}$

\bigskip

\noindent
and the results coincides with well known HOMFLY polynomials, see
(\ref{Wrepfund}) and
(\ref{113}).

\begin{itemize}
\item{Floer case}
\end{itemize}
$$F^{T[5,2]}_{[1]}=\bf{q}^8t^4+q^6t^3+q^4t^2+q^2t+1$$ \\
$F^{T[5,7]}_{[1]}=\bf{q}^{48}t^{24}+q^{46}t^{23}+q^{38}t^{22}+q^{36}t^{21}+q^{34}t^{20}+q^{32}t^{19}+q^{28}t^{18}+q^{26}t^{17}+q^{24}t^{16}+q^{22}t^{15}+q^{20}t^{14}+q^{16}t^{11}+q^{14}t^{10}+q^{12}t^9+q^{10}t^8+q^2t+1$
 \\
\begin{itemize}
\item{Alexander case}
\end{itemize}
$$A^{T[5,2]}_{[1]}=\bf{-q}^8+q^6-q^4+q^2-1$$  \\
$$A^{T[5,7]}_{[1]}=\bf{q}^{48}-q^{46}+q^{38}-q^{36}+q^{34}-q^{32}+q^{28}-q^{26}+q^{24}-q^{22}+q^{20}-q^{16}+q^{14}-q^{12}+q^{10}-q^2+1$$
 \\

\subsection{Case $(5,n)$, $n=5 k+3$ fundamental representation\label{5k+3}}
\be
\begin{array}{|c|}
\hline\\
P^{T[5,n]}_{[1]}=c^{[5]}_{[1]} M_{[5]}^{\ast} q^{-4n}+c^{[4,1]}_{[1]}
M_{[4,1]}^{\ast} q^{-\frac{12n}{5}} t^{\frac{2n}{5}}
+c^{[3,2]}_{[1]} M_{[3,2]}^{\ast} q^{-\frac{8n}{5}}
t^{\frac{4n}{5}}+c^{[3,1,1]}_{[1]} M_{[3,1,1]}^{\ast} q^{-\frac{6n}{5}}
t^{\frac{6n}{5}} +\\ \\+ c^{[2,2,1]}_{[1]} M_{[2,2,1]}^{\ast} q^{-\frac{4n}{5}}
t^{\frac{8n}{5}}+ c^{[2,1,1,1]}_{[1]} M_{[2,1,1,1]}^{\ast} q^{-\frac{2n}{5}}
t^{\frac{12n}{5}} + c^{[1,1,1,1,1]}_{[1]} M_{[1,1,1,1,1]}^{\ast} t^{4n}
\\
\\
\hline
\end{array}
\ee
\fr{ \\
c^{[5]}_{[1]}=1,\ \      
c^{[4,1]}_{[1]}=-{\Big(\frac{q^2}{t^2}\Big)^{\frac{3}{5}}\frac { (-1+t^2)
\left(1+q^2+q^2t^2+t^2q^4+t^2q^6 \right) }{\left(-1+{q}^{8}{t}^{2}  \right)
}},\ \           c^{[3,2]}_{[1]}={\Big(\frac{q}{t^3}\Big)^{\frac{4}{5}}\frac {
\left( -1+{t}^{2} \right)\left( {t}^{2}-{q}^{2} \right)
(q^6t^4+q^4t^4+t^2q^4+t^2q^2+1)}{ \left( -1+{q}^{4}{t}^{2} \right) \left(
-1+{q}^{6}{t}^{2} \right) }},  \  \nonumber \\ \\ \nonumber      
c^{[3,1,1]}_{[1]}={\Big(\frac{q}{t}\Big)^{\frac{8}{5}}\frac {\left( -1+{t}^{2}
\right)  \left( -1+{t}^{4} \right)\left(q^4t^4+q^4t^2+q^2t^2+q^2+1 \right)
}{\left(-1+ {q}^{2}{t}^{2} \right)  \left( -1+{q}^{6}{t}^{2} \right)}},\ \     
 c^{[2,2,1]}_{[1]}=-{\Big(\frac{q^6}{t^7}\Big)^{\frac{2}{5}}\frac {\left(
-1+{t}^{2} \right)  \left( -1+{t}^{4}
\right)(t^2-q^2)\left(t^6q^4+q^2t^4+q^2t^2+t^2+1\right) }{ \left(
{q}^{2}{t}^{2}+1 \right) \left(-1+{q}^{4}{t}^{2} \right) \left(
-1+{q}^{2}{t}^{2} \right) ^{2}}},\    \nonumber \\  \\
c^{[2,1,1,1]}_{[1]}=-{\Big(\frac{q^2}{t^2}
\Big)^{\frac{8}{5}}\frac {\left(-1+{t}^{2} \right)  \left( -1+{t}^{4}
\right)(-1+t^6) \left( {t}^{6}{q}^{2}+{q}^{2}{t}^{4}+t^4+t^2+1 \right) }{\left(
-1+{q}^{2}{t}^{2} \right)  \left( -1+{q}^{2}{t}^{4} \right)(-1+q^4t^6)}},\ \ \ 
    c^{[1,1,1,1,1]}_{[1]}={\frac{q^4}{t^4}\frac
{(-1+t^{10})(-1+t^8)(-1+t^6)(-1+t^4)}{\left( -1+{q}^{2}{t}^{2} \right)   \left(
-1+{t}^{4}{q}^{2} \right) \left( -1+{t}^{6}{q}^2 \right)(-1+t^8q^2) }}}
Several first superpolynomials are:
$P^{T[5,3]}_{[1]}=\dfrac{\{A\} t^4}{\{t\} A^4 q^{12}}
\Big((q^8t^2+q^6)(-A^2)^2+(q^8t^6+q^8t^4+q^6t^4+2q^6t^2+q^4t^2+q^4+q^2)(-A^2)+q^6t^6+q^4t^4+q^8t^8+q^2t^2+q^6t^4+q^4t^2+1\Big)
= \\ = -\frac{{\bf{a}}^{2}\bf{t}+1}{ \left( -1+\bf{q}^{2} \right)
\bf{a}^{5}\bf{t}^{29/2}\bf{q}^{7}}\Big(
\bf{q}^{10}t^{10}a^4+q^6t^8a^4+q^{14}t^9a^2+q^{12}t^9a^2+q^{10}t^7a^2+2q^8t^7a^2+q^6t^5a^2+q^4t^5a^2+q^2t^3a^2+q^{16}t^8+q^{10}t^6+q^{12}t^6+q^8t^4+q^6t^4+q^4t^2+1
\Big)$
\bigskip \\
$P^{T[5,8]}_{[1]}=\dfrac{\{A\} t^4}{\{t\} A^4 q^{32}}\Big(
(t^{8}q^{28}+t^{6}q^{26}+t^{4}q^{26}+t^{4}q^{24}+t^{2}q^{24}+t^{2}q^{22}+q^{20})(-A^{2})^4+(t^{16}q^{28}+t^{14}q^{28}+t^{12}q^{28}+t^{10}q^{28}+t^{14}q^{26}+2t^{12}q^{26}+3t^{10}q^{26}+3t^{8}q^{26}+t^{6}q^{26}+t^{12}q^{24}+2t^{10}q^{24}+4t^{8}q^{24}+2t^{4}q^{24}+4t^{6}q^{24}+t^{10}q^{22}+2t^{8}q^{22}+4t^{6}q^{22}+t^{2}q^{22}+4t^{4}q^{22}+4t^{4}q^{20}+2t^{6}q^{20}+t^{8}q^{20}+3t^{2}q^{20}+t^{6}q^{18}+3t^{2}q^{18}+q^{18}+2t^{4}q^{18}+2t^{2}q^{16}+t^{4}q^{16}+q^{16}+t^{2}q^{14}+q^{14}+q^{12})(-A^{2})^3+(t^{22}q^{28}+t^{20}q^{28}+t^{14}q^{28}+2t^{18}q^{28}+t^{16}q^{28}+t^{20}q^{26}+4t^{12}q^{26}+2t^{18}q^{26}+4t^{16}q^{26}+2t^{10}q^{26}+5t^{14}q^{26}+2t^{16}q^{24}+5t^{14}q^{24}+t^{18}q^{24}+4t^{8}q^{24}+7t^{12}q^{24}+t^{6}q^{24}+8t^{10}q^{24}+2t^{14}q^{22}+10t^{8}q^{22}+5t^{6}q^{22}+t^{16}q^{22}+5t^{12}q^{22}+8t^{10}q^{22}+t^{4}q^{22}+10t^{6}q^{20}+2t^{12}q^{20}+8t^{8}q^{20}+4t^{4}q^{20}+5t^{10}q^{20}+t^{14}q^{20}+t^{12}q^{18}+5t^{8}q^{18}+8t^{6}q^{18}
+2t^{2}q^{18}+8t^{4}q^{18}+2t^{10}q^{18}+t^{10}q^{16}+5t^{6}q^{16}+4t^{2}q^{16}+2t^{8}q^{16}+7t^{4}q^{16}+2t^{6}q^{14}+5t^{4}q^{14}+t^{8}q^{14}+5t^{2}q^{14}+q^{14}+2t^{4}q^{12}+4t^{2}q^{12}+q^{12}+t^{6}q^{12}+2t^{2}q^{10}+2q^{10}+t^{4}q^{10}+q^{8}+t^{2}q^{8}+q^{6})(-A^{2})^2+(t^{26}q^{28}+t^{24}q^{28}+t^{20}q^{28}+t^{22}q^{28}+2t^{22}q^{26}+t^{14}q^{26}+t^{24}q^{26}+4t^{18}q^{26}+3t^{20}q^{26}+3t^{16}q^{26}+2t^{20}q^{24}+t^{22}q^{24}+4t^{18}q^{24}+6t^{16}q^{24}+4t^{12}q^{24}+7t^{14}q^{24}+t^{10}q^{24}+2t^{18}q^{22}+4t^{16}q^{22}+7t^{14}q^{22}+t^{20}q^{22}+2t^{8}q^{22}+7t^{10}q^{22}+9t^{12}q^{22}+8t^{8}q^{20}+7t^{12}q^{20}+t^{18}q^{20}+10t^{10}q^{20}+2t^{6}q^{20}+4t^{14}q^{20}+2t^{16}q^{20}+2t^{14}q^{18}+t^{4}q^{18}+t^{16}q^{18}+4t^{12}q^{18}+7t^{6}q^{18}+10t^{8}q^{18}+7t^{10}q^{18}+t^{14}q^{16}+7t^{8}q^{16}+4t^{10}q^{16}+9t^{6}q^{16}+2t^{12}q^{16}+4t^{4}q^{16}+2t^{10}q^{14}+t^{2}q^{14}+t^{12}q^{14}+7t^{6}q^{14}+4t^{8}q^{14}+7t^{4}q^{14}+2t^{8}q^{12}+t^{10}q^{12}+6t^{4}q^{12}+4t^{6}q^{12}+3t^{2}q^{12}+4t^{4}
q^{10}+t^{8}q^{10}+4t^{2}q^{10}+2t^{6}q^{10}+t^{6}q^{8}+q^{8}+2t^{4}q^{8}+3t^{2}q^{8}+2t^{2}q^{6}+t^{4}q^{6}+q^{6}+t^{2}q^{4}+q^{4}+q^{2})(-A^{2})+t^{28}q^{28}+t^{20}q^{26}+t^{24}q^{26}+t^{22}q^{26}+t^{22}q^{24}+t^{26}q^{26}+t^{22}q^{22}+t^{20}q^{22}+2t^{18}q^{22}+3t^{16}q^{22}+3t^{14}q^{22}+t^{12}q^{22}+t^{20}q^{20}+2t^{16}q^{20}+t^{24}q^{24}+2t^{20}q^{24}+2t^{16}q^{24}+3t^{14}q^{20}+2t^{18}q^{24}+t^{16}q^{18}+2t^{14}q^{18}+2t^{8}q^{18}+4t^{10}q^{18}+3t^{12}q^{18}+t^{14}q^{16}+t^{16}q^{16}+4t^{8}q^{16}+2t^{12}q^{16}+3t^{10}q^{16}+t^{6}q^{16}+3t^{8}q^{14}+t^{12}q^{14}+t^{14}q^{14}+3t^{6}q^{14}+2t^{10}q^{14}+t^{12}q^{12}+3t^{6}q^{12}+2t^{8}q^{12}+t^{10}q^{12}+2t^{4}q^{12}+2t^{6}q^{10}+t^{8}q^{10}+t^{10}q^{10}+2t^{4}q^{10}+t^{6}q^{8}+t^{2}q^{8}+2t^{4}q^{8}+t^{8}q^{8}+t^{4}q^{6}+t^{6}q^{6}+t^{2}q^{6}+t^{2}q^{4}+t^{4}q^{4}+t^{2}q^{2}+t^{18}q^{20}+2t^{10}q^{20}+4t^{12}q^{20}+t^{18}q^{18}+1
\Big)=\\=-\frac{{\bf{a}}^{2}\bf{t}+1}{ \left( -1+\bf{q}^{2} \right)
\bf{a}^{5}\bf{t}^{69/2}\bf{q}^{27}}\Big( \bf{1}+{q}^{30}{
t}^{30}{a}^{8}+2\,{a}^{4}{t}^{30}{q}^{46}+{a}^{6}{t}^{17}{q}^{16}+2\,{q}^{28}{t}^{17}{a}^{2}+2\,{a}^{4}{t}^{26}{q}^{40}
\mbox{}+{q}^{34}{t}^{22}+4\,{q}^{14}{t}^{11}{a}^{2}+{q}^{34}{t}^{25}{a}^{2}+{a}^{6}{t}^{27}{q}^{36}+{q}^{14}{t}^{9}{a}^{2}+{q}^{14}{t}^{8}
\mbox{}+4\,{q}^{22}{t}^{15}{a}^{2}+{q}^{34}{t}^{22}{a}^{4}+{q}^{16}{t}^{15}{a}^{2}+{a}^{6}{t}^{19}{q}^{16}+{q}^{30}{t}^{17}{a}^{2}
\mbox{}+{q}^{22}{t}^{19}{a}^{2}+4\,{a}^{2}{t}^{25}{q}^{42}+10\,{a}^{2}{t}^{19}{q}^{26}+3\,{a}^{6}{t}^{29}{q}^{34}+{q}^{26}{t}^{18}{a}^{4}
\mbox{}+2\,{q}^{48}{t}^{27}{a}^{2}+{q}^{26}{t}^{24}{a}^{4}+2\,{a}^{2}{t}^{23}{q}^{40}+{q}^{4}{t}^{5}{a}^{2}+2\,{q}^{8}{t}^{7}{a}^{2}+10\,{a}^{4}{t}^{24}{q}^{30}
\mbox{}+5\,{a}^{4}{t}^{28}{q}^{40}+5\,{q}^{38}{t}^{26}{a}^{4}+{q}^{34}{t}^{18}+3\,{a}^{2}{t}^{9}{q}^{10}+2\,{q}^{24}{t}^{18}{a}^{4}+4\,{q}^{18}{t}^{18}{a}^{4}
\mbox{}+7\,{a}^{2}{t}^{15}{q}^{18}+4\,{q}^{20}{t}^{17}{a}^{2}+{q}^{22}{t}^{12}+{q}^{48}{t}^{24}+4\,{q}^{30}{t}^{27}{a}^{6}
\mbox{}+7\,{a}^{2}{t}^{19}{q}^{28}+{q}^{46}{t}^{26}+5\,{a}^{4}{t}^{16}{q}^{18}+2\,{a}^{6}{t}^{19}{q}^{18}+3\,{q}^{36}{t}^{22}+{q}^{6}{t}^{5}{a}^{2}
\mbox{}+2\,{q}^{24}{t}^{14}+{q}^{4}{t}^{2}+2\,{a}^{6}{t}^{23}{q}^{26}+7\,{q}^{38}{t}^{25}{a}^{2}+{a}^{8}{t}^{28}{q}^{28}+4\,{q}^{42}{t}^{28}{a}^{4}
\mbox{}+{q}^{54}{t}^{29}{a}^{2}+2\,{a}^{2}{t}^{21}{q}^{36}+3\,{a}^{6}{t}^{29}{q}^{36}+{q}^{40}{t}^{27}{a}^{2}+{q}^{32}{t}^{29}{a}^{6}
\mbox{}+5\,{q}^{22}{t}^{18}{a}^{4}+2\,{q}^{20}{t}^{13}{a}^{2}+2\,{a}^{6}{t}^{29}{q}^{38}+{a}^{6}{t}^{21}{q}^{24}+3\,{q}^{14}{t}^{13}{a}^{2}
\mbox{}+8\,{a}^{4}{t}^{24}{q}^{32}+8\,{a}^{4}{t}^{20}{q}^{22}+2\,{a}^{6}{t}^{21}{q}^{22}+{q}^{6}{t}^{8}{a}^{4}+4\,{a}^{6}{t}^{27}{q}^{32}+3\,{a}^{6}{t}^{23}{q}^{22}
\mbox{}+2\,{q}^{20}{t}^{20}{a}^{4}+4\,{a}^{2}{t}^{19}{q}^{30}+4\,{q}^{44}{t}^{27}{a}^{2}+{q}^{22}{t}^{13}{a}^{2}+{q}^{22}{t}^{16}
\mbox{}+{q}^{24}{t}^{26}{a}^{8}+4\,{q}^{24}{t}^{22}{a}^{4}+2\,{q}^{16}{t}^{14}{a}^{4}+{t}^{10}{q}^{20}+7\,{a}^{2}{t}^{23}{q}^{36}
\mbox{}+3\,{a}^{6}{t}^{21}{q}^{20}+7\,{a}^{4}{t}^{26}{q}^{36}+{a}^{6}{t}^{29}{q}^{40}+{a}^{2}{t}^{7}{q}^{6}+{q}^{42}{t}^{30}{a}^{4}+8\,{q}^{24}{t}^{20}{a}^{4}
\mbox{}+7\,{q}^{20}{a}^{2}{t}^{15}+{a}^{6}{t}^{31}{q}^{42}+{q}^{2}{t}^{3}{a}^{2}+10\,{q}^{30}{t}^{21}{a}^{2}+2\,{a}^{6}{t}^{25}{q}^{30}+2\,{q}^{44}{t}^{28}{a}^{4}
\mbox{}+7\,{a}^{2}{t}^{21}{q}^{32}+7\,{a}^{2}{t}^{17}{q}^{24}+{a}^{6}{t}^{31}{q}^{38}+{a}^{6}{t}^{15}{q}^{12}+{q}^{36}{t}^{18}+{a}^{6}{t}^{25}{q}^{32}
\mbox{}+2\,{q}^{36}{t}^{24}{a}^{4}+2\,{a}^{6}{t}^{27}{q}^{34}+{a}^{6}{t}^{17}{q}^{14}+4\,{q}^{14}{t}^{14}{a}^{4}+{q}^{56}{t}^{28}+{q}^{24}{t}^{12}
\mbox{}+{q}^{18}{t}^{10}+{q}^{32}{t}^{16}+{q}^{30}{t}^{16}+{q}^{16}{t}^{8}+{q}^{28}{t}^{14}+{q}^{46}{t}^{24}+{q}^{50}{t}^{26}+{q}^{18}{t}^{14}{a}^{4}
\mbox{}+4\,{a}^{6}{t}^{23}{q}^{24}+2\,{q}^{12}{t}^{9}{a}^{2}+{q}^{10}{t}^{8}+{q}^{48}{t}^{26}+{q}^{10}{t}^{6}+{q}^{12}{t}^{6}+{t}^{6}{q}^{8}+{q}^{8}{t}^{4}
\mbox{}+{q}^{6}{t}^{4}+7\,{q}^{20}{t}^{18}{a}^{4}+8\,{q}^{28}{t}^{21}{a}^{2}+2\,{q}^{40}{t}^{24}+2\,{q}^{30}{t}^{20}+4\,{q}^{32}{t}^{20}
\mbox{}+2\,{q}^{36}{t}^{20}+3\,{q}^{34}{t}^{20}+3\,{q}^{22}{t}^{14}+{q}^{26}{t}^{14}+{a}^{6}{t}^{19}{q}^{20}+5\,{q}^{26}{t}^{20}{a}^{4}
\mbox{}+{q}^{44}{t}^{22}+{q}^{10}{t}^{10}{a}^{4}+2\,{q}^{14}{t}^{10}+2\,{q}^{20}{t}^{12}+3\,{q}^{20}{t}^{14}+2\,{t}^{10}{q}^{16}
\mbox{}+2\,{q}^{16}{t}^{12}+2\,{q}^{12}{t}^{8}+2\,{q}^{44}{t}^{24}+2\,{q}^{42}{t}^{24}+4\,{q}^{28}{t}^{18}+2\,{q}^{28}{t}^{16}+4\,{q}^{24}{t}^{16}
\mbox{}+{a}^{4}{t}^{30}{q}^{50}+3\,{q}^{26}{t}^{16}+{q}^{14}{t}^{12}{a}^{4}+{a}^{2}{t}^{25}{q}^{46}+4\,{q}^{32}{t}^{26}{a}^{4}+2\,{a}^{6}{t}^{27}{q}^{28}
\mbox{}+{q}^{14}{t}^{16}{a}^{4}+{q}^{32}{t}^{30}{a}^{8}+{q}^{42}{t}^{26}{a}^{4}+2\,{q}^{40}{t}^{22}+9\,{q}^{34}{t}^{23}{a}^{2}
\mbox{}+{q}^{26}{t}^{15}{a}^{2}+2\,{q}^{12}{t}^{12}{a}^{4}+{q}^{38}{t}^{21}{a}^{2}+8\,{q}^{34}{t}^{26}{a}^{4}+5\,{q}^{28}{t}^{24}{a}^{4}
\mbox{}+{a}^{2}{t}^{9}{q}^{8}+3\,{q}^{46}{t}^{27}{a}^{2}+4\,{a}^{2}{t}^{17}{q}^{26}+{q}^{52}{t}^{29}{a}^{2}+2\,{a}^{2}{t}^{25}{q}^{44}+5\,{q}^{30}{t}^{22}{a}^{4}
\mbox{}+4\,{q}^{38}{t}^{28}{a}^{4}+4\,{q}^{36}{t}^{25}{a}^{2}+10\,{q}^{26}{t}^{22}{a}^{4}+{a}^{6}{t}^{21}{q}^{18}+7\,{q}^{24}{a}^{2}{t}^{19}
\mbox{}+{a}^{4}{t}^{10}{q}^{8}+{a}^{4}{t}^{14}{q}^{12}+2\,{q}^{24}{t}^{15}{a}^{2}+9\,{q}^{22}{t}^{17}{a}^{2}+{a}^{6}{t}^{31}{q}^{40}+2\,{q}^{32}{t}^{19}{a}^{2}
\mbox{}+{a}^{8}{t}^{32}{q}^{36}+6\,{a}^{2}{t}^{13}{q}^{16}+2\,{q}^{30}{t}^{23}{a}^{2}+2\,{q}^{10}{t}^{12}{a}^{4}+4\,{q}^{38}{t}^{23}{a}^{2}
\mbox{}+2\,{q}^{32}{t}^{22}{a}^{4}+{q}^{30}{t}^{26}{a}^{4}+{q}^{38}{t}^{20}+{q}^{48}{t}^{29}{a}^{2}+4\,{a}^{2}{t}^{21}{q}^{34}+{a}^{6}{t}^{23}{q}^{28}
\mbox{}+{q}^{52}{t}^{26}+8\,{q}^{28}{t}^{22}{a}^{4}+{q}^{30}{t}^{20}{a}^{4}+{a}^{6}{t}^{31}{q}^{44}+2\,{a}^{4}{t}^{20}{q}^{28}+4\,{a}^{6}{t}^{25}{q}^{26}
\mbox{}+6\,{q}^{40}{t}^{25}{a}^{2}+7\,{q}^{32}{t}^{23}{a}^{2}+2\,{q}^{16}{t}^{11}{a}^{2}+{q}^{38}{t}^{24}{a}^{4}+{a}^{6}{t}^{25}{q}^{24}
\mbox{}+{a}^{2}{t}^{19}{q}^{34}+{q}^{20}{t}^{24}{a}^{8}+5\,{q}^{34}{t}^{24}{a}^{4}+5\,{a}^{4}{t}^{16}{q}^{16}+3\,{q}^{18}{t}^{12}+3\,{q}^{42}{t}^{27}{a}^{2}
\mbox{}+4\,{a}^{6}{t}^{25}{q}^{28}+{q}^{42}{t}^{22}+{q}^{50}{t}^{29}{a}^{2}+{q}^{10}{t}^{7}{a}^{2}+4\,{q}^{18}{t}^{13}{a}^{2}+{q}^{22}{t}^{16}{a}^{4}
\mbox{}+{a}^{2}{t}^{27}{q}^{50}+2\,{q}^{26}{t}^{21}{a}^{2}+2\,{q}^{26}{t}^{18}+4\,{a}^{2}{t}^{11}{q}^{12}+2\,{q}^{32}{t}^{18}
\mbox{}+{a}^{4}{t}^{30}{q}^{48}+{q}^{26}{t}^{28}{a}^{8}+{q}^{44}{t}^{30}{a}^{4}+3\,{q}^{38}{t}^{22}+{q}^{40}{t}^{20}+{q}^{46}{t}^{28}{a}^{4}
\mbox{}+2\,{q}^{20}{t}^{16}{a}^{4}+{a}^{2}{t}^{23}{q}^{42}+2\,{q}^{36}{t}^{28}{a}^{4}+{q}^{18}{t}^{11}{a}^{2}+3\,{q}^{30}{t}^{18}
\Big)$

\bigskip
\noindent
$\boxed{P^{T[5,5k+3]}_{[1]}=\dfrac{\{A\} t^4}{\{t\} A^4
q^{20k+12}}{\cal{P}}^{T[5,5k+3]}_{[1]}}$

\begin{itemize}
\item{HOMFLY case}
\end{itemize}
$$H^{T[5,3]}_{[1]}=\dfrac{\{A\}}{\{q\} A^4
q^8}\Big((q^6+q^{10})A^4-(q^{10}+2q^8+q^{12}+q^6+q^4+q^{14}+q^2)A^2+q^6+q^8+q^{16}+q^{12}+q^{10}+q^4+1\Big)$$
\bigskip \\
$H^{T[5,8]}_{[1]}=\dfrac{\{A\}}{\{q\} A^4 q^{28}}
\Big((q^{36}+q^{32}+q^{30}+q^{28}+q^{26}+q^{24}+q^{20})A^{8}+(-q^{44}-q^{42}-2q^{40}-3q^{38}-4q^{36}-5q^{34}-6q^{32}-6q^{30}-7q^{28}-6q^{26}-6q^{24}-5q^{22}-4q^{20}-3q^{18}-2q^{16}-q^{14}-q^{12})A^{6}+(q^{50}+q^{48}+3q^{46}+3q^{44}+6q^{42}+7q^{40}+10q^{38}+11q^{36}+14q^{34}+14q^{32}+17q^{30}+15q^{28}+17q^{26}+14q^{24}+14q^{22}+11q^{20}+10q^{18}+7q^{16}+6q^{14}+3q^{12}+3q^{10}+q^{8}+q^{6}A^{4}+(-q^{54}-q^{52}-2q^{50}-3q^{48}-4q^{46}-6q^{44}-8q^{42}-9q^{40}-12q^{38}-13q^{36}-15q^{34}-16q^{32}-17q^{30}-17q^{28}-17q^{26}-16q^{24}-15q^{22}-13q^{20}-12q^{18}-9q^{16}-8q^{14}-6q^{12}-4q^{10}-3q^{8}-2q^{6}-q^{4}-q^{2})A^{2}+2q^{46}+3q^{44}+3q^{42}+5q^{40}+4q^{38}+6q^{36}+5q^{34}+7q^{32}+6q^{30}+q^{56}+q^{52}+q^{50}+2q^{48}+6q^{20}+4q^{18}+5q^{16}+3q^{14}+3q^{12}+2q^{10}+2q^{8}+q^{6}+q^{4}+7q^{28}+6q^{26}+7q^{24}+5q^{22}+1\Big)$

\bigskip
\noindent
$\boxed{H^{T[5,5k+3]}_{[1]}=\dfrac{\{A\}}{\{q\} A^4 q^{20k+8}}
{\cal{H}}^{T[5,5k+3]}_{[1]}=
q^{-20k-12}s_{[5]}^{*}-q^{-10k-6}s_{[4,1]}^{*}+s_{[3,1,1]}^{*}-q^{10k+6}s_{[2,1,1,1]}^{*}+q^{20k+12}s_{[1,1,1,1,1]}^{*}}$

\bigskip

\noindent
and the results coincides with well known HOMFLY polynomials, see
(\ref{Wrepfund}) and
(\ref{113}).

\begin{itemize}
\item{Floer case}
\end{itemize}
$$F^{T[5,3]}_{[1]}=\bf{q}^{16}t^8+q^{14}t^7+q^{10}t^6+q^8t^5+q^6t^4+q^2t+1$$ 
\\
$F^{T[5,8]}_{[1]}=\bf{q}^{56}{t}^{28}+{q}^{54}{t}^{27}+{q}^{46}{t}^{26}+{q}^{44}{t}^{25}+{q}^{40}{t}^{24}+{q}^{38}{t}^{23}+{q}^{36}{t}^{22}+{q}^{34}{t}^{21}+{q}^{30}{t}^{20}+{q}^{28}{t}^{19}+{q}^{26}{t}^{18}+{q}^{22}{t}^{15}+{q}^{20}{t}^{14}+{q}^{18}{t}^{13}
+{q}^{16}{t}^{12}+{q}^{12}{t}^{9}+{q}^{10}{t}^{8}+{q}^{2}t+1$ \\
\begin{itemize}
\item{Alexander case}
\end{itemize}
$$A^{T[5,3]}_{[1]}=\bf{q}^{16}-q^{14}+q^{10}-q^8+q^6-q^2+1$$  \\
$$A^{T[5,8]}_{[1]}=\bf{q}^{56}-q^{54}+q^{46}-q^{44}+q^{40}-q^{38}+q^{36}-q^{34}+q^{30}-q^{28}+q^{26}-q^{22}+q^{20}-q^{18}+q^{16}-q^{12}+q^{10}-q^2+1$$

\subsection{Case $(5,n)$, $n=5 k+4$ fundamental representation\label{5k+4}}
\be
\begin{array}{|c|}
\hline\\
P^{T[5,n]}_{[1]}=c^{[5]}_{[1]} M_{[5]}^{\ast} q^{-4n}+c^{[4,1]}_{[1]}
M_{[4,1]}^{\ast} q^{-\frac{12n}{5}} t^{\frac{2n}{5}}
+c^{[3,2]}_{[1]} M_{[3,2]}^{\ast} q^{-\frac{8n}{5}}
t^{\frac{4n}{5}}+c^{[3,1,1]}_{[1]} M_{[3,1,1]}^{\ast} q^{-\frac{6n}{5}}
t^{\frac{6n}{5}} +\\ \\+ c^{[2,2,1]}_{[1]} M_{[2,2,1]}^{\ast} q^{-\frac{4n}{5}}
t^{\frac{8n}{5}}+ c^{[2,1,1,1]}_{[1]} M_{[2,1,1,1]}^{\ast} q^{-\frac{2n}{5}}
t^{\frac{12n}{5}} + c^{[1,1,1,1,1]}_{[1]} M_{[1,1,1,1,1]}^{\ast} t^{4n}
\\
\\
\hline
\end{array}
\ee
\fr{ \\
c^{[5]}_{[1]}=1,\ \      
c^{[4,1]}_{[1]}=-{\Big(\frac{q^2}{t^2}\Big)^{\frac{4}{5}}\frac { (-1+t^2)
\left(1+t^2+q^2t^2+t^2q^4+t^2q^6 \right) }{\left(-1+{q}^{8}{t}^{2}  \right)
}},\ \           c^{[3,2]}_{[1]}={\Big(\frac{q^2}{t}\Big)^{\frac{6}{5}}\frac {
\left( -1+{t}^{2} \right)\left( {t}^{2}-{q}^{2} \right)
(q^4t^2+q^2t^2+t^2+q^2+1)}{ \left( -1+{q}^{4}{t}^{2} \right) \left(
-1+{q}^{6}{t}^{2} \right) }},  \  \nonumber \\ \\ \nonumber      
c^{[3,1,1]}_{[1]}={\Big(\frac{q}{t}\Big)^{\frac{14}{5}}\frac {\left( -1+{t}^{2}
\right)  \left( -1+{t}^{4} \right)\left(q^4t^4+q^2t^4+t^4+t^2+1 \right)
}{\left(-1+ {q}^{2}{t}^{2} \right)  \left( -1+{q}^{6}{t}^{4} \right)}},\ \     
 c^{[2,2,1]}_{[1]}=-{\Big(\frac{q^4}{t^3}\Big)^{\frac{4}{5}}\frac {\left(
-1+{t}^{2} \right)  \left( -1+{t}^{4}
\right)(t^2-q^2)\left(t^4q^2+q^2t^2+t^4+t^2+1\right) }{ \left( {q}^{2}{t}^{2}+1
\right) \left(-1+{q}^{4}{t}^{2} \right) \left( -1+{q}^{2}{t}^{2} \right)
^{2}}},\    \nonumber \\  \\
c^{[2,1,1,1]}_{[1]}=-{\Big(\frac{q^2}{t^2}\Big)^{\
\frac{9}{5}}\frac {\left(-1+{t}^{2} \right)  \left( -1+{t}^{4} \right)(-1+t^6)
\left( {t}^{6}{q}^{2}+{t}^{6}+t^4+t^2+1 \right) }{\left( -1+{q}^{2}{t}^{2}
\right)  \left( -1+{q}^{2}{t}^{4} \right)(-1+q^4t^6)}},\ \ \     
c^{[1,1,1,1,1]}_{[1]}={\frac{q^4}{t^4}\frac
{(-1+t^{10})(-1+t^8)(-1+t^6)(-1+t^4)}{\left( -1+{q}^{2}{t}^{2} \right)   \left(
-1+{t}^{4}{q}^{2} \right) \left( -1+{t}^{6}{q}^2 \right)(-1+t^8q^2) }}}
Several first superpolynomials are:

$P^{T[5,4]}_{[1]}=\dfrac{\{A\} t^4}{\{t\} A^4 q^{16}}
\Big((q^{12}(-A^2)^3+(t^2q^{12}+t^4q^{12}+t^6q^{12}+t^2q^{10}+t^4q^{10}+q^{10}+t^2q^8+q^8+q^6)(-A^2)^2+
(t^8q^{12}+t^{10}q^{12}+t^6q^{12}+t^8q^{10}+2t^6q^{10}+2t^4q^{10}+t^2q^{10}+2t^4q^8+t^6q^8+2t^2q^8+2t^2q^6+t^4q^6+q^6+t^2q^4+q^4+q^2)(-A^2)+t^{10}q^{10}+t^6q^{10}+t^8q^8+t^4q^8+t^6q^8+t^6q^6+t^4q^6+t^2q^6+t^2q^4+t^4q^4+t^2q^2+t^{12}q^{12}+t^8q^{10}+1\Big)=\\
=-\frac{{\bf{a}}^{2}\bf{t}+1}{ \left( -1+\bf{q}^{2} \right)
\bf{a}^{5}\bf{t}^{37/2}\bf{q}^{11}}\Big(
\bf{1}+{q}^{20}{t}^{13}{a}^{2}+{q}^{18}{t}^{11}{a}^{2}+{q}^{14}{t}^{9}{a}^{2}+{q}^{12}{t}^{15}{a}^{6}+{q}^{24}{t}^{12}
\mbox{}+{q}^{12}{t}^{11}{a}^{2}+{q}^{4}{t}^{2}+{q}^{6}{t}^{4}+{t}^{4}{q}^{8}+{t}^{6}{q}^{8}+{q}^{12}{t}^{6}+{t}^{6}{q}^{10}+{q}^{12}{t}^{8}+{q}^{16}{t}^{8}
\mbox{}+{q}^{14}{t}^{8}+{t}^{10}{q}^{16}+2\,{q}^{12}{t}^{9}{a}^{2}+{q}^{14}{t}^{12}{a}^{4}+{q}^{6}{t}^{5}{a}^{2}+{q}^{6}{t}^{8}{a}^{4}+{q}^{16}{t}^{14}{a}^{4}
\mbox{}+{q}^{20}{t}^{10}+{q}^{18}{t}^{10}+{q}^{12}{t}^{12}{a}^{4}+{q}^{6}{t}^{7}{a}^{2}+{q}^{18}{t}^{14}{a}^{4}+2\,{q}^{8}{t}^{7}{a}^{2}+{q}^{10}{t}^{7}{a}^{2}
\mbox{}+{q}^{10}{t}^{12}{a}^{4}+2\,{q}^{16}{t}^{11}{a}^{2}+{q}^{22}{t}^{13}{a}^{2}+{q}^{8}{t}^{10}{a}^{4}+{q}^{18}{t}^{13}{a}^{2}
\mbox{}+{q}^{10}{t}^{10}{a}^{4}+{q}^{14}{t}^{14}{a}^{4}+2\,{q}^{14}{t}^{11}{a}^{2}+2\,{q}^{10}{t}^{9}{a}^{2}+{q}^{4}{t}^{5}{a}^{2}+{q}^{2}{t}^{3}{a}^{2}
\Big)$
\bigskip \\
$P^{T[5,9]}_{[1]}=\dfrac{\{A\} t^4}{\{t\} A^4 q^{36}}\Big(
(t^{12}q^{32}+t^{10}q^{30}+t^{8}q^{30}+t^{6}q^{30}+t^{8}q^{28}+t^{6}q^{28}+t^{4}q^{28}+t^{6}q^{26}+t^{4}q^{26}+t^{2}q^{26}+t^{4}q^{24}+t^{2}q^{24}+t^{2}q^{22}+q^{20})(-A^{2})^4+(t^{20}q^{32}+t^{18}q^{32}+t^{16}q^{32}+t^{14}q^{32}+t^{18}q^{30}+2t^{16}q^{30}+3t^{14}q^{30}+t^{8}q^{30}+4t^{12}q^{30}+2t^{10}q^{30}+t^{16}q^{28}+2t^{14}q^{28}+4t^{8}q^{28}+4t^{12}q^{28}+2t^{6}q^{28}+5t^{10}q^{28}+4t^{10}q^{26}+2t^{12}q^{26}+6t^{8}q^{26}+t^{14}q^{26}+5t^{6}q^{26}+2t^{4}q^{26}+2t^{10}q^{24}+4t^{4}q^{24}+t^{12}q^{24}+4t^{8}q^{24}+6t^{6}q^{24}+t^{2}q^{24}+t^{10}q^{22}+4t^{6}q^{22}+5t^{4}q^{22}+2t^{8}q^{22}+2t^{2}q^{22}+t^{8}q^{20}+4t^{4}q^{20}+2t^{6}q^{20}+4t^{2}q^{20}+t^{6}q^{18}+3t^{2}q^{18}+2t^{4}q^{18}+q^{18}+t^{4}q^{16}+2t^{2}q^{16}+q^{16}+t^{2}q^{14}+q^{14}+q^{12})(-A^{2})^3+(t^{26}q^{32}+t^{24}q^{32}+2t^{22}q^{32}+t^{20}q^{32}+t^{18}q^{32}+t^{24}q^{30}+2t^{22}q^{30}+5t^{16}q^{30}+5t^{18}q^{30}+4t^{20}q^{30}+t^{12}q^{30}+3t^{14}q^{30}+4t^{10}q^{28}+t^{8}q^{
28}+7t^{12}q^{28}+2t^{20}q^{28}+9t^{14}q^{28}+t^{22}q^{28}+7t^{16}q^{28}+5t^{18}q^{28}+2t^{18}q^{26}+10t^{10}q^{26}+11t^{12}q^{26}+t^{20}q^{26}+6t^{8}q^{26}+8t^{14}q^{26}+t^{6}q^{26}+5t^{16}q^{26}+t^{4}q^{24}+2t^{16}q^{24}+8t^{12}q^{24}+5t^{14}q^{24}+6t^{6}q^{24}+12t^{10}q^{24}+t^{18}q^{24}+11t^{8}q^{24}+8t^{10}q^{22}+2t^{14}q^{22}+5t^{12}q^{22}+12t^{8}q^{22}+t^{16}q^{22}+4t^{4}q^{22}+10t^{6}q^{22}+5t^{10}q^{20}+t^{14}q^{20}+8t^{8}q^{20}+11t^{6}q^{20}+t^{2}q^{20}+2t^{12}q^{20}+7t^{4}q^{20}+5t^{8}q^{18}+9t^{4}q^{18}+2t^{10}q^{18}+t^{12}q^{18}+8t^{6}q^{18}+3t^{2}q^{18}+7t^{4}q^{16}+t^{10}q^{16}+2t^{8}q^{16}+5t^{6}q^{16}+5t^{2}q^{16}+2t^{6}q^{14}+q^{14}+5t^{4}q^{14}+t^{8}q^{14}+5t^{2}q^{14}+2t^{4}q^{12}+t^{6}q^{12}+4t^{2}q^{12}+q^{12}+2t^{2}q^{10}+t^{4}q^{10}+2q^{10}+t^{2}q^{8}+q^{8}+q^{6})(-A^{2})^2+(t^{30}q^{32}+t^{28}q^{32}+t^{24}q^{32}+t^{26}q^{32}+2t^{26}q^{30}+t^{28}q^{30}+2t^{18}q^{30}+3t^{24}q^{30}+4t^{22}q^{30}+3t^{20}q^{30}+t^{26}q^{28}+6t^{20}q^{28}+4t^{22}q^{28}+7t^{18}q^{28}+6t^{16}q^{28}+t^{12}q^{
28}+2t^{24}q^{28}+3t^{14}q^{28}+4t^{20}q^{26}+2t^{22}q^{26}+2t^{10}q^{26}+7t^{18}q^{26}+t^{24}q^{26}+9t^{14}q^{26}+6t^{12}q^{26}+9t^{16}q^{26}+7t^{16}q^{24}+t^{22}q^{24}+2t^{20}q^{24}+10t^{14}q^{24}+11t^{12}q^{24}+3t^{8}q^{24}+4t^{18}q^{24}+8t^{10}q^{24}+t^{20}q^{22}+4t^{16}q^{22}+7t^{14}q^{22}+2t^{6}q^{22}+12t^{10}q^{22}+10t^{12}q^{22}+2t^{18}q^{22}+8t^{8}q^{22}+t^{4}q^{20}+6t^{6}q^{20}+t^{18}q^{20}+4t^{14}q^{20}+10t^{10}q^{20}+2t^{16}q^{20}+7t^{12}q^{20}+11t^{8}q^{20}+9t^{6}q^{18}+7t^{10}q^{18}+4t^{12}q^{18}+3t^{4}q^{18}+2t^{14}q^{18}+10t^{8}q^{18}+t^{16}q^{18}+7t^{8}q^{16}+2t^{12}q^{16}+4t^{10}q^{16}+t^{14}q^{16}+6t^{4}q^{16}+9t^{6}q^{16}+4t^{8}q^{14}+7t^{6}q^{14}+2t^{10}q^{14}+t^{12}q^{14}+2t^{2}q^{14}+7t^{4}q^{14}+4t^{6}q^{12}+t^{10}q^{12}+2t^{8}q^{12}+6t^{4}q^{12}+3t^{2}q^{12}+2t^{6}q^{10}+4t^{4}q^{10}+t^{8}q^{10}+4t^{2}q^{10}+2t^{4}q^{8}+t^{6}q^{8}+3t^{2}q^{8}+q^{8}+t^{4}q^{6}+2t^{2}q^{6}+q^{6}+t^{2}q^{4}+q^{4}+q^{2})(-A^{2})+t^{28}q^{30}+2t^{24}q^{28}+t^{26}q^{28}+t^{26}q^{30}+t^{24}q^{30}+t^{18}q^{
28}+2t^{22}q^{28}+t^{28}q^{28}+2t^{20}q^{28}+3t^{20}q^{26}+t^{24}q^{26}+2t^{22}q^{26}+t^{14}q^{26}+3t^{18}q^{26}+2t^{20}q^{24}+2t^{12}q^{24}+t^{22}q^{24}+3t^{18}q^{24}+t^{24}q^{24}+3t^{14}q^{24}+t^{26}q^{26}+2t^{16}q^{26}+3t^{16}q^{22}+t^{22}q^{22}+t^{20}q^{22}+4t^{14}q^{22}+2t^{10}q^{22}+4t^{12}q^{22}+4t^{12}q^{20}+t^{20}q^{20}+2t^{16}q^{20}+t^{18}q^{20}+4t^{10}q^{20}+4t^{16}q^{24}+2t^{18}q^{22}+2t^{14}q^{18}+t^{32}q^{32}+t^{30}q^{30}+t^{16}q^{18}+t^{6}q^{18}+3t^{8}q^{18}+t^{18}q^{18}+t^{16}q^{16}+t^{14}q^{16}+2t^{12}q^{16}+4t^{8}q^{16}+3t^{10}q^{16}+2t^{6}q^{16}+3t^{8}q^{14}+3t^{6}q^{14}+t^{12}q^{14}+2t^{10}q^{14}+t^{4}q^{14}+t^{14}q^{14}+t^{12}q^{12}+2t^{8}q^{12}+3t^{6}q^{12}+2t^{4}q^{12}+t^{10}q^{12}+t^{8}q^{10}+2t^{4}q^{10}+2t^{6}q^{10}+t^{10}q^{10}+2t^{4}q^{8}+t^{8}q^{8}+t^{2}q^{8}+t^{6}q^{8}+t^{4}q^{6}+t^{2}q^{6}+t^{6}q^{6}+t^{4}q^{4}+t^{2}q^{4}+t^{2}q^{2}+3t^{14}q^{20}+2t^{8}q^{20}+4t^{10}q^{18}+3t^{12}q^{18}+1
\Big)=\\=-\frac{{\bf{a}}^{2}\bf{t}+1}{ \left( -1+\bf{q}^{2} \right)
\bf{a}^{5}\bf{t}^{77/2}
\bf{q}^{31}}\Big(
\bf{1}+4\,{a}^{6}{t}^{23}{q}^{24}+7\,{q}^{40}{t}^{30}{a}^{4}+2\,{q}^{16}{t}^{11}{a}^{2}+{q}^{32}{t}^{30}{a}^{8}+2\,{a}^{6}{t}^{27}{q}^{34}
\mbox{}+2\,{q}^{44}{t}^{25}{a}^{2}+{q}^{14}{t}^{9}{a}^{2}+4\,{q}^{46}{t}^{27}{a}^{2}+{q}^{18}{t}^{11}{a}^{2}+4\,{q}^{38}{t}^{23}{a}^{2}
\mbox{}+2\,{q}^{40}{t}^{33}{a}^{6}+2\,{q}^{32}{t}^{22}{a}^{4}+{q}^{6}{t}^{8}{a}^{4}+5\,{q}^{22}{t}^{18}{a}^{4}+{a}^{6}{t}^{17}{q}^{14}+7\,{q}^{32}{t}^{21}{a}^{2}
\mbox{}+{q}^{10}{t}^{10}{a}^{4}+{q}^{52}{t}^{34}{a}^{4}+4\,{q}^{22}{t}^{15}{a}^{2}+3\,{q}^{42}{t}^{29}{a}^{2}+2\,{q}^{44}{t}^{28}{a}^{4}
\mbox{}+2\,{q}^{12}{t}^{12}{a}^{4}+6\,{q}^{26}{t}^{21}{a}^{2}+5\,{q}^{48}{t}^{32}{a}^{4}+{q}^{36}{t}^{34}{a}^{8}+{q}^{12}{t}^{15}{a}^{6}
\mbox{}+{a}^{4}{t}^{30}{q}^{50}+2\,{q}^{16}{t}^{15}{a}^{2}+{a}^{6}{t}^{31}{q}^{44}+{a}^{6}{t}^{35}{q}^{50}+{q}^{42}{t}^{26}{a}^{4}+{a}^{6}{t}^{33}{q}^{48}
\mbox{}+7\,{q}^{24}{t}^{22}{a}^{4}+2\,{q}^{56}{t}^{31}{a}^{2}+2\,{q}^{32}{t}^{19}{a}^{2}+{a}^{6}{t}^{21}{q}^{24}+4\,{q}^{14}{t}^{11}{a}^{2}
\mbox{}+11\,{q}^{36}{t}^{25}{a}^{2}+{q}^{20}{t}^{24}{a}^{8}+2\,{q}^{26}{t}^{23}{a}^{6}+2\,{q}^{22}{t}^{21}{a}^{6}+{a}^{6}{t}^{23}{q}^{28}
\mbox{}+4\,{a}^{6}{t}^{31}{q}^{40}+2\,{q}^{28}{t}^{23}{a}^{2}+7\,{q}^{44}{t}^{27}{a}^{2}+4\,{q}^{14}{t}^{14}{a}^{4}+10\,{q}^{36}{t}^{28}{a}^{4}
\mbox{}+2\,{q}^{10}{t}^{12}{a}^{4}+5\,{q}^{42}{t}^{28}{a}^{4}+2\,{q}^{48}{t}^{31}{a}^{2}+{a}^{8}{t}^{32}{q}^{36}+4\,{q}^{26}{t}^{24}{a}^{4}
\mbox{}+2\,{q}^{36}{t}^{27}{a}^{2}+5\,{q}^{38}{t}^{26}{a}^{4}+3\,{a}^{6}{t}^{33}{q}^{44}+2\,{a}^{6}{t}^{29}{q}^{38}+6\,{q}^{30}{t}^{27}{a}^{6}
\mbox{}+7\,{q}^{46}{t}^{29}{a}^{2}+{q}^{34}{t}^{22}{a}^{4}+{q}^{38}{t}^{24}{a}^{4}+4\,{q}^{38}{t}^{30}{a}^{4}+2\,{q}^{30}{t}^{25}{a}^{6}
\mbox{}+7\,{q}^{44}{t}^{30}{a}^{4}+3\,{q}^{22}{t}^{19}{a}^{2}+2\,{a}^{6}{t}^{33}{q}^{46}+4\,{q}^{52}{t}^{31}{a}^{2}+2\,{q}^{40}{t}^{23}{a}^{2}
\mbox{}+{q}^{8}{t}^{10}{a}^{4}+8\,{q}^{24}{t}^{20}{a}^{4}+{q}^{18}{t}^{14}{a}^{4}+3\,{q}^{54}{t}^{31}{a}^{2}+8\,{a}^{4}{t}^{24}{q}^{32}
\mbox{}+{q}^{14}{t}^{12}{a}^{4}+2\,{q}^{8}{t}^{7}{a}^{2}+{q}^{32}{t}^{32}{a}^{8}+{q}^{34}{t}^{32}{a}^{8}+4\,{a}^{6}{t}^{31}{q}^{36}+8\,{q}^{28}{t}^{22}{a}^{4}
\mbox{}+{q}^{46}{t}^{25}{a}^{2}+2\,{q}^{34}{t}^{31}{a}^{6}+12\,{a}^{4}{t}^{24}{q}^{30}+10\,{q}^{26}{t}^{19}{a}^{2}+4\,{q}^{26}{t}^{17}{a}^{2}
\mbox{}+2\,{q}^{20}{t}^{16}{a}^{4}+{a}^{6}{t}^{35}{q}^{52}+2\,{q}^{48}{t}^{27}{a}^{2}+2\,{a}^{4}{t}^{30}{q}^{48}+{q}^{38}{t}^{34}{a}^{8}
\mbox{}+6\,{q}^{16}{t}^{13}{a}^{2}+{q}^{22}{t}^{16}{a}^{4}+10\,{q}^{30}{t}^{21}{a}^{2}+4\,{a}^{6}{t}^{25}{q}^{28}+{a}^{6}{t}^{19}{q}^{20}
\mbox{}+6\,{q}^{44}{t}^{29}{a}^{2}+10\,{q}^{34}{t}^{23}{a}^{2}+{q}^{38}{t}^{21}{a}^{2}+2\,{q}^{24}{t}^{18}{a}^{4}+12\,{q}^{34}{t}^{26}{a}^{4}
\mbox{}+2\,{q}^{28}{t}^{17}{a}^{2}+2\,{a}^{2}{t}^{21}{q}^{36}+4\,{q}^{42}{t}^{25}{a}^{2}+{q}^{34}{t}^{19}{a}^{2}+5\,{q}^{34}{t}^{24}{a}^{4}
\mbox{}+{q}^{36}{t}^{30}{a}^{4}+{q}^{24}{t}^{18}+{q}^{60}{t}^{30}+{q}^{46}{t}^{28}+{q}^{54}{t}^{28}+{q}^{40}{t}^{26}+{q}^{38}{t}^{20}
\mbox{}+{q}^{44}{t}^{22}+2\,{a}^{4}{t}^{26}{q}^{40}+5\,{a}^{6}{t}^{31}{q}^{38}+{q}^{42}{t}^{22}+{q}^{40}{t}^{20}+{q}^{36}{t}^{18}+{q}^{52}{t}^{26}
\mbox{}+4\,{q}^{50}{t}^{32}{a}^{4}+{q}^{32}{t}^{28}{a}^{4}+4\,{q}^{30}{t}^{19}{a}^{2}+{q}^{48}{t}^{24}+{q}^{34}{t}^{18}
\mbox{}+{q}^{26}{t}^{14}+{q}^{32}{t}^{16}+{q}^{30}{t}^{16}+{q}^{56}{t}^{28}+{q}^{58}{t}^{30}+2\,{q}^{20}{t}^{13}{a}^{2}+5\,{a}^{4}{t}^{30}{q}^{46}
\mbox{}+{q}^{14}{t}^{8}+{q}^{16}{t}^{8}+{q}^{28}{t}^{14}+{q}^{10}{t}^{8}+{t}^{6}{q}^{10}+{q}^{12}{t}^{6}+{t}^{6}{q}^{8}+{t}^{4}{q}^{8}+{q}^{6}{t}^{4}+{q}^{4}{t}^{2}
\mbox{}+{q}^{54}{t}^{30}+{q}^{46}{t}^{24}+9\,{q}^{42}{t}^{27}{a}^{2}+{q}^{22}{t}^{12}+{q}^{64}{t}^{32}+{q}^{40}{t}^{29}{a}^{6}
\mbox{}+5\,{q}^{46}{t}^{32}{a}^{4}+2\,{q}^{36}{t}^{24}{a}^{4}+{q}^{18}{t}^{10}+4\,{a}^{6}{t}^{27}{q}^{28}+{a}^{6}{t}^{35}{q}^{46}+8\,{a}^{4}{t}^{28}{q}^{40}
\mbox{}+3\,{q}^{44}{t}^{32}{a}^{4}+{q}^{20}{t}^{10}+{q}^{24}{t}^{12}+3\,{q}^{26}{t}^{18}+2\,{q}^{48}{t}^{28}+3\,{q}^{46}{t}^{26}
\mbox{}+2\,{q}^{48}{t}^{26}+{q}^{6}{t}^{7}{a}^{2}+3\,{q}^{44}{t}^{26}+2\,{q}^{22}{t}^{16}+3\,{t}^{12}{q}^{18}+11\,{q}^{32}{t}^{26}{a}^{4}
\mbox{}+9\,{q}^{24}{t}^{19}{a}^{2}+7\,{a}^{2}{t}^{15}{q}^{18}+2\,{q}^{32}{t}^{18}+9\,{q}^{42}{t}^{30}{a}^{4}+2\,{q}^{28}{t}^{20}{a}^{4}
\mbox{}+4\,{q}^{30}{t}^{20}+{q}^{46}{t}^{28}{a}^{4}+5\,{q}^{32}{t}^{29}{a}^{6}+4\,{q}^{32}{t}^{20}+10\,{q}^{28}{t}^{24}{a}^{4}
\mbox{}+{q}^{26}{t}^{18}{a}^{4}+{a}^{6}{t}^{27}{q}^{26}+{q}^{26}{t}^{15}{a}^{2}+{q}^{54}{t}^{29}{a}^{2}+3\,{a}^{6}{t}^{21}{q}^{20}
\mbox{}+4\,{a}^{6}{t}^{33}{q}^{42}+{q}^{6}{t}^{5}{a}^{2}+2\,{q}^{36}{t}^{20}+2\,{q}^{32}{t}^{22}+4\,{q}^{34}{t}^{22}+3\,{q}^{38}{t}^{22}+4\,{q}^{36}{t}^{22}
\mbox{}+7\,{q}^{20}{a}^{2}{t}^{15}+{q}^{42}{t}^{32}{a}^{4}+3\,{q}^{42}{t}^{24}+2\,{q}^{50}{t}^{28}+2\,{q}^{54}{t}^{34}{a}^{4}+4\,{q}^{18}{t}^{13}{a}^{2}
\mbox{}+2\,{q}^{42}{t}^{26}+2\,{q}^{40}{t}^{22}+3\,{q}^{26}{t}^{16}+{q}^{28}{t}^{30}{a}^{8}+2\,{q}^{28}{t}^{16}+4\,{q}^{28}{t}^{18}
\mbox{}+4\,{q}^{34}{t}^{21}{a}^{2}+2\,{q}^{52}{t}^{28}+4\,{q}^{40}{t}^{24}+2\,{q}^{20}{t}^{12}+3\,{q}^{32}{t}^{25}{a}^{2}+3\,{q}^{20}{t}^{14}
\mbox{}+2\,{q}^{44}{t}^{24}+3\,{q}^{38}{t}^{24}+2\,{q}^{24}{t}^{14}+{q}^{38}{t}^{33}{a}^{6}+3\,{q}^{34}{t}^{20}+{q}^{18}{t}^{14}
\mbox{}+2\,{q}^{12}{t}^{8}+6\,{q}^{30}{t}^{26}{a}^{4}+5\,{q}^{26}{t}^{20}{a}^{4}+2\,{q}^{28}{t}^{20}+2\,{t}^{10}{q}^{16}+3\,{q}^{22}{t}^{14}
\mbox{}+8\,{q}^{34}{t}^{25}{a}^{2}+11\,{q}^{26}{t}^{22}{a}^{4}+2\,{q}^{30}{t}^{29}{a}^{6}+7\,{q}^{40}{t}^{25}{a}^{2}+9\,{q}^{40}{t}^{27}{a}^{2}
\mbox{}+{a}^{6}{t}^{21}{q}^{18}+{q}^{44}{t}^{36}{a}^{8}+2\,{a}^{6}{t}^{19}{q}^{18}+6\,{q}^{34}{t}^{28}{a}^{4}+{q}^{26}{t}^{28}{a}^{8}
\mbox{}+7\,{a}^{2}{t}^{19}{q}^{28}+9\,{a}^{4}{t}^{20}{q}^{22}+7\,{q}^{20}{t}^{18}{a}^{4}+{q}^{50}{t}^{27}{a}^{2}+{q}^{36}{t}^{27}{a}^{6}
\mbox{}+10\,{q}^{38}{t}^{25}{a}^{2}+{q}^{58}{t}^{33}{a}^{2}+{q}^{24}{t}^{26}{a}^{8}+5\,{a}^{6}{t}^{25}{q}^{26}+{q}^{30}{t}^{20}{a}^{4}
\mbox{}+2\,{q}^{14}{t}^{10}+4\,{q}^{12}{t}^{11}{a}^{2}+2\,{a}^{6}{t}^{31}{q}^{42}+3\,{q}^{30}{t}^{18}+6\,{q}^{38}{t}^{27}{a}^{2}+9\,{q}^{22}{t}^{17}{a}^{2}
\mbox{}+3\,{q}^{20}{t}^{20}{a}^{4}+{q}^{60}{t}^{33}{a}^{2}+{q}^{14}{t}^{16}{a}^{4}+7\,{q}^{36}{t}^{23}{a}^{2}+{q}^{62}{t}^{33}{a}^{2}
\mbox{}+{q}^{50}{t}^{34}{a}^{4}+2\,{q}^{52}{t}^{32}{a}^{4}+{q}^{48}{t}^{35}{a}^{6}+2\,{q}^{24}{t}^{15}{a}^{2}+{q}^{22}{t}^{13}{a}^{2}
\mbox{}+{q}^{10}{t}^{7}{a}^{2}+8\,{q}^{30}{t}^{23}{a}^{2}+3\,{q}^{50}{t}^{31}{a}^{2}+2\,{q}^{52}{t}^{29}{a}^{2}+{q}^{2}{t}^{3}{a}^{2}+{q}^{56}{t}^{30}
\mbox{}+{q}^{4}{t}^{5}{a}^{2}+11\,{q}^{38}{t}^{28}{a}^{4}+6\,{q}^{48}{t}^{29}{a}^{2}+2\,{q}^{36}{t}^{24}+4\,{a}^{6}{t}^{29}{q}^{36}
\mbox{}+{q}^{58}{t}^{31}{a}^{2}+2\,{q}^{16}{t}^{12}+{q}^{28}{t}^{26}{a}^{4}+4\,{a}^{6}{t}^{27}{q}^{32}+12\,{q}^{32}{t}^{23}{a}^{2}
\mbox{}+5\,{q}^{30}{t}^{22}{a}^{4}+{a}^{4}{t}^{32}{q}^{54}+{q}^{56}{t}^{33}{a}^{2}+{a}^{6}{t}^{19}{q}^{16}+{q}^{56}{t}^{34}{a}^{4}
\mbox{}+8\,{a}^{4}{t}^{26}{q}^{36}+2\,{q}^{12}{t}^{9}{a}^{2}+7\,{q}^{24}{t}^{17}{a}^{2}+{q}^{30}{t}^{17}{a}^{2}+2\,{a}^{6}{t}^{25}{q}^{24}
\mbox{}+{q}^{32}{t}^{25}{a}^{6}+4\,{a}^{6}{t}^{23}{q}^{22}+6\,{q}^{20}{t}^{17}{a}^{2}+3\,{q}^{10}{t}^{9}{a}^{2}+{q}^{40}{t}^{34}{a}^{8}
\mbox{}+2\,{q}^{16}{t}^{14}{a}^{4}+5\,{q}^{18}{t}^{16}{a}^{4}+{q}^{40}{t}^{29}{a}^{2}+{a}^{8}{t}^{28}{q}^{28}+6\,{a}^{6}{t}^{29}{q}^{34}
\mbox{}+{q}^{42}{t}^{23}{a}^{2}+5\,{a}^{4}{t}^{16}{q}^{16}+{q}^{30}{t}^{30}{a}^{8}+{q}^{16}{t}^{17}{a}^{6}+{q}^{24}{t}^{21}{a}^{2}
\mbox{}+{t}^{26}{q}^{50}+5\,{q}^{18}{t}^{18}{a}^{4}+{q}^{22}{t}^{22}{a}^{4}+11\,{q}^{28}{t}^{21}{a}^{2}+{q}^{58}{t}^{34}{a}^{4}
\mbox{}+{a}^{4}{t}^{14}{q}^{12}+{a}^{2}{t}^{9}{q}^{8}+4\,{q}^{24}{t}^{16}+3\,{q}^{14}{t}^{13}{a}^{2}+4\,{q}^{50}{t}^{29}{a}^{2}
\Big)$

\bigskip
\noindent
$\boxed{P^{T[5,5k+4]}_{[1]}=\dfrac{\{A\} t^4}{\{t\} A^4
q^{20k+16}}{\cal{P}}^{T[5,5k+4]}_{[1]}}$

\begin{itemize}
\item{HOMFLY case}
\end{itemize}
$H^{T[5,4]}_{[1]}=\dfrac{\{A\}}{\{q\} A^4
q^{12}}\Big(-q^{12}A^{6}+(q^{18}+q^{16}+2q^{14}+q^{12}+2q^{10}+q^{8}+q^{6})A^{4}+(-q^{22}-q^{20}-2q^{18}-2q^{16}-3q^{14}-3q^{12}-3q^{10}-2q^{8}-2q^{6}-q^{4}-q^{2})A^{2}+q^{20}+q^{14}+2q^{12}+q^{18}+2q^{16}+2q^{8}+q^{6}+q^{4}+q^{10}+q^{24}+1\Big)$
\bigskip \\
$H^{T[5,9]}_{[1]}=\dfrac{\{A\}}{\{q\} A^4
q^{32}}\Big((q^{44}+q^{40}+q^{38}+2q^{36}+q^{34}+2q^{32}+q^{30}+2q^{28}+q^{26}+q^{24}+q^{20})A^{8}+(-q^{52}-q^{50}-2q^{48}-3q^{46}-4q^{44}-6q^{42}-7q^{40}-8q^{38}-9q^{36}-10q^{34}-10q^{32}-10q^{30}-9q^{28}-8q^{26}-7q^{24}-6q^{22}-4q^{20}-3q^{18}-2q^{16}-q^{14}-q^{12})A^{6}+(q^{58}+q^{56}+3q^{54}+3q^{52}+6q^{50}+7q^{48}+11q^{46}+12q^{44}+16q^{42}+17q^{40}+21q^{38}+21q^{36}+24q^{34}+22q^{32}+24q^{30}+21q^{28}+21q^{26}+17q^{24}+16q^{22}+12q^{20}+11q^{18}+7q^{16}+6q^{14}+3q^{12}+3q^{10}+q^{8}+q^{6})A^{4}+(-q^{62}-q^{60}-2q^{58}-3q^{56}-4q^{54}-6q^{52}-8q^{50}-10q^{48}-12q^{46}-15q^{44}-17q^{42}-19q^{40}-21q^{38}-22q^{36}-23q^{34}-24q^{32}-23q^{30}-22q^{28}-21q^{26}-19q^{24}-17q^{22}-15q^{20}-12q^{18}-10q^{16}-8q^{14}-6q^{12}-4q^{10}-3q^{8}-2q^{6}-q^{4}-q^{2})A^{2}+q^{64}+q^{60}+q^{58}+2q^{56}+5q^{46}+6q^{44}+6q^{42}+8q^{40}+7q^{38}+9q^{36}+8q^{34}+9q^{32}+8q^{30}+2q^{54}+3q^{52}+3q^{50}+5q^{48}+6q^{20}+5q^{18}+5q^{16}+3q^{14}+3q^{12}+2q^{10}+2q^{8}+q^{6}+q^{4}+9q^{28}
+7q^{26}+8q^{24}+6q^{22}+1\Big)$

\bigskip
\noindent
$\boxed{H^{T[5,5k+4]}_{[1]}=\dfrac{\{A\}}{\{q\} A^4 q^{20k+12}}
{\cal{H}}^{T[5,5k+4]}_{[1]}=
q^{-20k-16}s_{[5]}^{*}-q^{-10k-8}s_{[4,1]}^{*}+s_{[3,1,1]}^{*}-q^{10k+8}s_{[2,1,1,1]}^{*}+q^{20k+16}s_{[1,1,1,1,1]}^{*}}$

\bigskip

\noindent
and the results coincides with well known HOMFLY polynomials, see
(\ref{Wrepfund}) and
(\ref{113}).

\begin{itemize}
\item{Floer case}
\end{itemize}
$$F^{T[5,4]}_{[1]}=\bf{q}^{24}t^{12}+q^{22}t^{11}+q^{16}t^{10}+q^{12}t^9+q^8t^6+q^2t+1$$
 \\
$F^{T[5,9]}_{[1]}=\bf{q}^{64}{t}^{32}+{q}^{62}{t}^{31}+{q}^{54}{t}^{30}+{q}^{52}{t}^{29}+{q}^{46}{t}^{28}+{q}^{42}{t}^{25}+{q}^{36}{t}^{24}+{q}^{32}{t}^{21}+{q}^{28}{t}^{20}+{q}^{22}{t}^{17}+{q}^{18}{t}^{14}+{q}^{12}{t}^{9}+{q}^{10}{t}^{8}+{q}^{2}t+1$
 \\
\begin{itemize}
\item{Alexander case}
\end{itemize}
$$A^{T[5,4]}_{[1]}=\bf{q}^{24}-q^{22}+q^{16}-q^{12}+q^8-q^2+1$$  \\
$$A^{T[5,9]}_{[1]}=\bf{q}^{64}-q^{62}+q^{54}-q^{52}+q^{46}-q^{42}+q^{36}-q^{32}+q^{28}-q^{22}+q^{18}-q^{12}+q^{10}-q^2+1$$
 \\

\subsection{Case $(6,n)$, $n=6 k+1$ fundamental representation\label{6k+1}}
\be
\begin{array}{|c|}
\hline\\
P^{T[6,n]}_{[1]}=c^{[6]}_{[1]}M^{\ast}_{{6}}{q}^{-5\,n}+c^{[5,1]}_{[1]}M^{\ast}_{{5,1}}\,{q}^{-\frac{10}{3}\,n}{t}^{\frac{1}{3}\,n}+c^{[4,2]}_{[1]}M^{\ast}_{{4,2}}\,{q}^{-\frac{7}{3}\,n}{t}^{\frac{2}{3}\,n}+c^{[4,1,1]}_{[1]}M^{\ast}_{{4,1,1}}\,{q}^{-2n}{t}^{\,n}+c^{[3,3]}_{[1]}M^{\ast}_{{3,3}}\,{q}^{-2n}{t}^{\,n}+\\+c^{[3,2,1]}_{[1]}M^{\ast}_{{3,2,1}}\,{q}^{-\frac{4}{3}\,n}{t}^{\frac{4}{3}\,n}+c^{[3,1,1,1]}_{[1]}M^{\ast}_{{3,1,1,1}}\,{q}^{-\,n}{t}^{2n}+c^{[2,2,2]}_{[1]}M^{\ast}_{{2,2,2}}\,{q}^{-\,n}{t}^{2n}+c^{[2,2,1,1]}_{[1]}M^{\ast}_{{2,2,1,1}}\,{t}^{\frac{7}{3}\,n}{q}^{-\frac{2}{3}\,n}+\\+c^{[2,1,1,1,1]}_{[1]}M^{\ast}_{{2,1,1,1,1}}\,{t}^{\frac{10}{3}\,n}{q}^{-\frac{1}{3}\,n}+c^{[1,1,1,1,1,1]}_{[1]}M^{\ast}_{{1,1,1,1,1,1}}\,{t}^{5\,n}

\\
\\
\hline
\end{array}
\ee
\fr{ \\
c^{[6]}_{[1]}=1,\ \ c^{[5,1]}_{[1]} =-\frac{q ^{1/3}~\left(-1+t^{2}\right)~\left(1+q ^{2}+q ^{4}+q ^{6}+q ^{8}+q ^{8}t ^{2}\right)}{t^{1/3}~\left(-1+q ^{10}t ^{2}\right)} \\ 
c^{[4,2]}_{[1]} =-\frac{\left(-1+t ^{2}\right)~\left(1+q ^{2}+q ^{4}\right)~\left(q ^{4}~t ^{2}+q ^{4}+1\right)~\left(q ^{2}-t^{2}\right)~q ^{4/3}}{\left(-1+q ^{8}~t ^{2}\right)~\left(-1+q ^{6}~t^{2}\right)~t ^{2/3}}  \\
c^{[4,1,1]}_{[1]} =\frac{q ~\left(t ^{2}+1\right)~\left(-1+t^{2}\right)^{2}~\left(1+q ^{2}+q ^{4}+q ^{6}+q ^{6}~t ^{2}+q ^{6}~t
^{4}\right)}{t ~\left(-1+q ^{2}~t ^{2}\right)~\left(-1+q ^{8}~t^{4}\right)}\mathop{\rm  }\mathop{\rm  }    \\ 
c^{[3,3]}_{[1]} =-\frac{q ^{3}~\left(-1+t ^{2}\right)~\left(t ^{2}+1\right)~\left(q ^{2}-t^{2}\right)~\left(q ^{4}-t ^{2}\right)}{t ~\left(-1+q ^{6}~t^{2}\right)~\left(-1+q ^{4}~t ^{2}\right)~\left(-1+q ^{2}~t ^{2}\right)}  \\ 
c^{[3,2,1]}_{[1]} =\frac{q^{7/3}~\left(t ^{2}+1\right)~\left(-1+t ^{2}\right)^{2}~\left(1+q^{2}\right)~\left(q ^{2}-t ^{2}\right)~\left(q ^{4}~t ^{4}+q ^{4}~t ^{2}+q
^{4}+q ^{2}~t ^{2}+q ^{2}+1\right)}{t ^{4/3}~\left(-1+q ^{6}~t^{4}\right)~\left(-1+q ^{4}~t ^{2}\right)^{2}}\mathop{\rm  } \\ 
c^{[3,1,1,1]}_{[1]}=-\frac{q ^{2}~\left(-1+t^{2}\right)~\left(t ^{4}-1\right)~\left(t ^{6}-1\right)~\left(1+q ^{2}+q ^{4}+q^{4}~t ^{2}+q ^{4}~t ^{4}+q ^{4}~t ^{6}\right)}{t ^{2}~\left(-1+q ^{2}~t^{2}\right)~\left(t ^{4}~q ^{2}-1\right)~\left(-1+t ^{6}~q ^{6}\right)}  \\ 
c^{[2,2,2]}_{[1]} =\frac{q ^{4}~\left(t ^{2}+1\right)~\left(t ^{4}+t^{2}+1\right)~\left(-1+t ^{2}\right)^{2}~\left(q ^{2}-t ^{2}\right)~\left(q^{2}-t ^{4}\right)}{t^{2}~\left(t ^{4}~q ^{2}-1\right)~\left(q ^{2}~t ^{2}+1\right)~\left(-1+q ^{4}~t ^{2}\right)~\left(-1+q ^{2}~t^{2}\right)^{2}}\mathop{\rm  }   \\ 
c^{[2,2,1,1]}_{[1]}=-\frac{q ^{11/3}~\left(t ^{2}+1\right)~\left(t ^{4}+t ^{2}+1\right)~\left(-1+t^{2}\right)^{3}~\left(1+q ^{2}+t ^{2}+q ^{2}~t ^{2}+t ^{4}~q ^{2}+q ^{2}~t
^{6}\right)~\left(q ^{2}-t ^{2}\right)}{t ^{7/3}~\left(q ^{2}~t^{2}+1\right)~\left(q ^{4}~t ^{6}-1\right)~\left(-1+q ^{2}~t ^{2}\right)^{3}}  \\
c^{[2,1,1,1,1]}_{[1]} =\frac{q ^{10/3}~\left(t^{4}+1\right)~\left(t ^{4}+t ^{2}+1\right)~\left(t ^{2}+1\right)^{2}~\left(-1+t^{2}\right)^{4}~\left(1+q ^{2}+q ^{2}~t ^{2}+t ^{4}~q ^{2}+q ^{2}~t ^{6}+q^{2}~t ^{8}\right)}{t ^{10/3}~\left(t ^{8}~q ^{4}-1\right)~\left(-1+q ^{2}~t ^{2}\right)~\left(q ^{2}~t ^{6}-1\right)~\left(t ^{4}~q^{2}-1\right)}\mathop{\rm  } \\ 
c^{[1,1,1,1,1,1]}_{[1]} =-\frac{q ^{5}~\left(t ^{12}-1\right)~\left(-1+t ^{10}\right)~\left(-1+t ^{8}\right)~\left(t^{6}-1\right)~\left(t ^{4}-1\right)}{t ^{5}~\left(t ^{10}~q
^{2}-1\right)~\left(q ^{2}~t ^{8}-1\right)~\left(q ^{2}~t ^{6}-1\right)~\left(t^{4}~q ^{2}-1\right)~\left(-1+q ^{2}~t ^{2}\right)}}
Several first superpolynomials are:
$$P^{T[6,1]}_{[1]}=\dfrac{\{A\} t^5}{\{t\} A^5 q^5}$$
\bigskip \\
$P^{T[6,7]}_{[1]}=\dfrac{\{A\} t^5}{\{t\} A^5 q^{35}} \Big({q}^{30}(-A^{2})^5+
(
{q}^{28}{t}^{8}+{q}^{20}+{q}^{24}{t}^{4}+{q}^{24}+{t}^{6}{q}^{28}+{t}^{6}{q}^{26}+{q}^{30}{t}^{4}+{q}^{28}{t}^{4}+{q}^{30}{t}^{8}+{q}^{26}{t}^{2}
+{q}^{22}{t}^{2}+{q}^{26}{t}^{4}+{q}^{30}{t}^{10}+{q}^{28}{t}^{2}+{q}^{30}{t}^{2}+{q}^{22}+{t}^{2}{q}^{24}+{q}^{26}+{q}^{28}+{q}^{30}{t}^{6}
 ) (-A^{2})^4+ (
+{q}^{14}+{q}^{24}+2\,{t}^{8}{q}^{22}+{q}^{22}+{t}^{4}{q}^{16}+4\,{q}^{20}{t}^{2}+{q}^{30}{t}^{16}+{q}^{30}{t}^{6}+{t}^{8}{q}^{20}+2\,{q}^{16}
+2\,{q}^{20}+{t}^{14}{q}^{26}+{q}^{18}{t}^{6}+{t}^{12}{q}^{24}+2\,{q}^{18}+2\,{t}^{6}{q}^{20}+2\,{q}^{28}{t}^{14}+3\,{q}^{26}{t}^{4}
+6\,{t}^{6}{q}^{24}+2\,{q}^{24}{t}^{10}+4\,{t}^{10}{q}^{26}+{q}^{30}{t}^{8}+4\,{q}^{28}{t}^{8}+2\,{q}^{28}{t}^{4}+{q}^{28}{t}^{16}
+{q}^{14}{t}^{2}+3\,{q}^{28}{t}^{12}+2\,{q}^{30}{t}^{12}+2\,{q}^{30}{t}^{10}+3\,{t}^{6}{q}^{28}+5\,{t}^{4}{q}^{22}+4\,{q}^{24}{t}^{8}
+2\,{q}^{26}{t}^{2}+5\,{q}^{26}{t}^{8}+5\,{t}^{6}{q}^{26}+2\,{q}^{30}{t}^{14}+5\,{q}^{24}{t}^{4}+4\,{q}^{28}{t}^{10}+2\,{q}^{18}{t}^{4}
+3\,{t}^{2}{q}^{24}+3\,{t}^{2}{q}^{18}+2\,{q}^{16}{t}^{2}+{q}^{28}{t}^{2}+4\,{q}^{22}{t}^{6}+4\,{t}^{4}{q}^{20}+4\,{q}^{22}{t}^{2}+2\,{q}^{26}{t}^{12}
+{q}^{30}{t}^{18}+{q}^{12}+{t}^{10}{q}^{22} ) (-A^{2})^3+ (
2\,{q}^{26}{t}^{18}+{q}^{28}{t}^{22}+5\,{q}^{20}{t}^{10}+2\,{t}^{20}{q}^{28}+5\,{t}^{16}{q}^{26}+2\,{q}^{14}+{t}^{20}{q}^{26}
+5\,{t}^{4}{q}^{14}+11\,{t}^{8}{q}^{22}+2\,{t}^{10}{q}^{18}+{t}^{10}{q}^{16}+5\,{t}^{8}{q}^{18}+7\,{t}^{4}{q}^{16}+4\,{q}^{20}{t}^{2}+2\,{q}^{30}{t}^{20}
+2\,{q}^{30}{t}^{16}+{q}^{6}+{q}^{8}+8\,{t}^{8}{q}^{20}+{t}^{14}{q}^{20}+{q}^{18}{t}^{12}+{q}^{16}+2\,{q}^{22}{t}^{14}+{t}^{4}{q}^{10}+2\,{q}^{16}{t}^{8}
+{q}^{22}{t}^{16}+2\,{t}^{16}{q}^{24}+{q}^{14}{t}^{8}+2\,{t}^{12}{q}^{20}+7\,{t}^{14}{q}^{26}+{q}^{24}{t}^{18}+2\,{t}^{6}{q}^{14}
+5\,{t}^{6}{q}^{16}+{q}^{12}{t}^{6}+8\,{q}^{18}{t}^{6}+2\,{t}^{4}{q}^{12}+5\,{t}^{14}{q}^{24}+5\,{t}^{12}{q}^{22}+8\,{t}^{12}{q}^{24}
+{q}^{18}+11\,{t}^{6}{q}^{20}+6\,{q}^{28}{t}^{14}+{q}^{26}{t}^{4}+8\,{t}^{6}{q}^{24}+11\,{q}^{24}{t}^{10}+8\,{t}^{10}{q}^{26}+2\,{q}^{28}{t}^{8}
+5\,{q}^{28}{t}^{16}+5\,{q}^{14}{t}^{2}+5\,{q}^{28}{t}^{12}+4\,{q}^{28}{t}^{18}+{q}^{30}{t}^{12}+{t}^{6}{q}^{28}+6\,{t}^{4}{q}^{22}
+10\,{q}^{24}{t}^{8}+6\,{q}^{26}{t}^{8}+3\,{t}^{6}{q}^{26}+{q}^{30}{t}^{14}+3\,{q}^{24}{t}^{4}+4\,{q}^{28}{t}^{10}+9\,{q}^{18}{t}^{4}
+{t}^{2}{q}^{24}+5\,{t}^{2}{q}^{18}+{q}^{30}{t}^{22}+4\,{q}^{12}{t}^{2}+6\,{q}^{16}{t}^{2}+10\,{q}^{22}{t}^{6}+8\,{t}^{4}{q}^{20}+2\,{q}^{22}{t}^{2}
+9\,{q}^{26}{t}^{12}+{q}^{30}{t}^{24}+{q}^{8}{t}^{2}+2\,{q}^{10}{t}^{2}+2\,{q}^{10}+2\,{q}^{30}{t}^{18}+2\,{q}^{12}+8\,{t}^{10}{q}^{22}
 ) (-A^{2})^2+ (
+{q}^{6}{t}^{4}+6\,{q}^{26}{t}^{18}+{t}^{22}{q}^{24}+3\,{q}^{28}{t}^{22}+{t}^{10}{q}^{12}+10\,{q}^{20}{t}^{10}+2\,{t}^{24}{q}^{28}
+{q}^{22}{t}^{20}+{t}^{26}{q}^{28}+4\,{t}^{20}{q}^{28}+7\,{t}^{16}{q}^{26}+2\,{q}^{16}{t}^{12}+2\,{q}^{18}{t}^{14}+{q}^{4}{t}^{2}
+2\,{q}^{12}{t}^{8}+{t}^{24}{q}^{26}+4\,{t}^{20}{q}^{26}+7\,{t}^{4}{q}^{14}+{t}^{6}{q}^{8}+8\,{t}^{8}{q}^{22}+7\,{t}^{10}{q}^{18}+4\,{t}^{10}{q}^{16}
+10\,{t}^{8}{q}^{18}+6\,{t}^{4}{q}^{16}+{q}^{30}{t}^{20}+{q}^{4}+2\,{q}^{8}{t}^{4}+{q}^{10}{t}^{8}+{q}^{6}+{q}^{8}+10\,{t}^{8}{q}^{20}+{q}^{16}{t}^{14}
+{t}^{12}{q}^{14}+4\,{t}^{14}{q}^{20}+2\,{q}^{24}{t}^{20}+4\,{q}^{18}{t}^{12}+7\,{q}^{22}{t}^{14}+4\,{t}^{4}{q}^{10}+7\,{q}^{16}{t}^{8}
+4\,{q}^{22}{t}^{16}+7\,{t}^{16}{q}^{24}+2\,{t}^{22}{q}^{26}+2\,{q}^{6}{t}^{2}+4\,{q}^{14}{t}^{8}+7\,{t}^{12}{q}^{20}+6\,{t}^{14}{q}^{26}
+2\,{q}^{22}{t}^{18}+2\,{q}^{20}{t}^{16}+4\,{q}^{24}{t}^{18}+{q}^{2}+7\,{t}^{6}{q}^{14}+9\,{t}^{6}{q}^{16}+4\,{q}^{12}{t}^{6}+9\,{q}^{18}{t}^{6}
+6\,{t}^{4}{q}^{12}+{q}^{18}{t}^{16}+{q}^{20}{t}^{18}+9\,{t}^{14}{q}^{24}+10\,{t}^{12}{q}^{22}+9\,{t}^{12}{q}^{24}+7\,{t}^{6}{q}^{20}
+{q}^{30}{t}^{26}+2\,{q}^{28}{t}^{14}+{t}^{6}{q}^{24}+7\,{q}^{24}{t}^{10}+2\,{t}^{10}{q}^{26}+3\,{q}^{28}{t}^{16}+3\,{q}^{14}{t}^{2}
+{q}^{28}{t}^{12}+4\,{q}^{28}{t}^{18}+{t}^{4}{q}^{22}+4\,{q}^{24}{t}^{8}+{q}^{30}{t}^{28}+{q}^{26}{t}^{8}+5\,{q}^{18}{t}^{4}+{t}^{2}{q}^{18}
+{q}^{30}{t}^{22}+4\,{q}^{12}{t}^{2}+2\,{q}^{16}{t}^{2}+4\,{q}^{22}{t}^{6}+2\,{t}^{4}{q}^{20}+5\,{q}^{26}{t}^{12}+{q}^{30}{t}^{24}
+2\,{t}^{10}{q}^{14}+3\,{q}^{8}{t}^{2}+4\,{q}^{10}{t}^{2}+{q}^{10}+2\,{t}^{6}{q}^{10}+10\,{t}^{10}{q}^{22}
) (-A^{2})
+1+{q}^{6}{t}^{4}+2\,{q}^{26}{t}^{18}+{t}^{30}{q}^{30}+3\,{q}^{20}{t}^{10}+{t}^{28}{q}^{28}+{t}^{24}{q}^{28}+{q}^{22}{t}^{20}
+{t}^{26}{q}^{28}+{t}^{20}{q}^{28}+{t}^{16}{q}^{26}+{t}^{6}{q}^{6}+{t}^{22}{q}^{24}+{q}^{2}{t}^{2}+{q}^{28}{t}^{22}+{t}^{10}{q}^{12}
+2\,{t}^{20}{q}^{26}+{t}^{4}{q}^{14}+{t}^{6}{q}^{8}+{t}^{8}{q}^{22}+4\,{t}^{10}{q}^{18}+3\,{t}^{10}{q}^{16}+3\,{t}^{8}{q}^{18}+{q}^{4}{t}^{4}+{t}^{4}{q}^{16}
+2\,{q}^{18}{t}^{14}+{q}^{4}{t}^{2}+2\,{q}^{12}{t}^{8}+{t}^{24}{q}^{26}+{t}^{26}{q}^{26}+{t}^{8}{q}^{8}+2\,{q}^{8}{t}^{4}+{q}^{10}{t}^{8}+{t}^{12}{q}^{12}
+{q}^{10}{t}^{10}+{t}^{16}{q}^{16}+{q}^{18}{t}^{18}+2\,{t}^{8}{q}^{20}+{q}^{16}{t}^{14}+{t}^{12}{q}^{14}+{q}^{24}{t}^{24}+3\,{t}^{14}{q}^{20}
+{q}^{20}{t}^{20}+2\,{q}^{24}{t}^{20}+3\,{q}^{18}{t}^{12}+4\,{q}^{22}{t}^{14}+2\,{t}^{4}{q}^{10}+4\,{q}^{16}{t}^{8}+3\,{q}^{22}{t}^{16}
+3\,{t}^{16}{q}^{24}+2\,{t}^{22}{q}^{26}+2\,{q}^{16}{t}^{12}+{q}^{6}{t}^{2}+3\,{q}^{14}{t}^{8}+4\,{t}^{12}{q}^{20}+{t}^{14}{q}^{26}
+2\,{q}^{22}{t}^{18}+2\,{q}^{20}{t}^{16}+3\,{q}^{24}{t}^{18}+3\,{t}^{6}{q}^{14}+2\,{t}^{6}{q}^{16}+3\,{q}^{12}{t}^{6}+2\,{q}^{18}{t}^{6}
+2\,{t}^{4}{q}^{12}+{q}^{18}{t}^{16}+{q}^{14}{t}^{14}+{q}^{20}{t}^{18}+2\,{t}^{14}{q}^{24}+3\,{t}^{12}{q}^{22}+2\,{t}^{12}{q}^{24}
+2\,{t}^{10}{q}^{14}+{q}^{8}{t}^{2}+{q}^{10}{t}^{2}+2\,{t}^{6}{q}^{10}+{q}^{22}{t}^{22}+2\,{t}^{10}{q}^{22}
\Big) = \\ = \bf{ -{\frac {{a}^{2}t+1}{ \left( -1+{q}^{2} \right)
{a}^{6}{t}^{38}{q}^{35}}}} \Big( {q}^{30}{t}^{35}{a}^{10}+ (
{q}^{32}{t}^{34}+{q}^{34}{t}^{32}+{q}^{26}{t}^{28}+{q}^{32}{t}^{30}+{q}^{30}{t}^{32}+{q}^{36}{t}^{34}+{q}^{40}{t}^{34}
+{q}^{36}{t}^{32}+{q}^{34}{t}^{34}+{q}^{24}{t}^{26}+{t}^{28}{q}^{28}+{t}^{30}{q}^{30}+{q}^{38}{t}^{34}+{q}^{22}{t}^{26}
+{q}^{32}{t}^{32}+{q}^{20}{t}^{24}+{q}^{28}{t}^{32}+{q}^{24}{t}^{28}+{q}^{28}{t}^{30}+{q}^{26}{t}^{30}
) {a}^{8}+ (
2\,{q}^{20}{t}^{23}+3\,{q}^{34}{t}^{31}+2\,{q}^{30}{t}^{25}+{q}^{40}{t}^{29}+{q}^{22}{t}^{25}+2\,{q}^{18}{t}^{21}+{q}^{16}{t}^{17}
+3\,{q}^{30}{t}^{29}+5\,{q}^{32}{t}^{29}+2\,{q}^{26}{t}^{23}+{q}^{28}{t}^{23}+{q}^{12}{t}^{15}+5\,{q}^{28}{t}^{27}+{q}^{48}{t}^{33}
+4\,{q}^{24}{t}^{23}+{q}^{38}{t}^{33}+{q}^{14}{t}^{17}+4\,{q}^{28}{t}^{25}+{q}^{24}{t}^{27}+2\,{q}^{44}{t}^{33}+4\,{q}^{36}{t}^{31}
+2\,{q}^{28}{t}^{29}+4\,{q}^{22}{t}^{23}+2\,{q}^{40}{t}^{33}+2\,{q}^{34}{t}^{27}+4\,{q}^{32}{t}^{27}+2\,{q}^{16}{t}^{19}+{q}^{32}{t}^{25}
+6\,{q}^{30}{t}^{27}+5\,{q}^{26}{t}^{25}+4\,{q}^{36}{t}^{29}+4\,{q}^{38}{t}^{31}+3\,{q}^{40}{t}^{31}+4\,{q}^{24}{t}^{25}+3\,{q}^{26}{t}^{27}
+2\,{q}^{22}{t}^{21}+2\,{q}^{42}{t}^{33}+{q}^{30}{t}^{31}+5\,{q}^{34}{t}^{29}+3\,{q}^{20}{t}^{21}+{q}^{20}{t}^{19}+{q}^{36}{t}^{33}
+2\,{q}^{38}{t}^{29}+{q}^{46}{t}^{33}+{q}^{44}{t}^{31}+2\,{q}^{18}{t}^{19}+2\,{q}^{32}{t}^{31}+2\,{q}^{42}{t}^{31}+{q}^{24}{t}^{21}
+{t}^{27}{q}^{36} ) {a}^{6}
+ (
{q}^{26}{t}^{18}+8\,{t}^{22}{q}^{24}+8\,{q}^{28}{t}^{22}+{t}^{26}{q}^{42}+2\,{t}^{26}{q}^{40}+2\,{q}^{32}{t}^{22}+2\,{t}^{24}{q}^{36}
+10\,{t}^{24}{q}^{28}+9\,{q}^{22}{t}^{20}+3\,{t}^{26}{q}^{28}+2\,{t}^{20}{q}^{28}+{q}^{18}{t}^{14}+6\,{t}^{24}{q}^{26}+5\,{t}^{20}{q}^{26}
+{q}^{30}{t}^{20}+{t}^{26}{q}^{26}+5\,{q}^{42}{t}^{28}+2\,{t}^{12}{q}^{12}+{q}^{10}{t}^{10}+5\,{t}^{16}{q}^{16}+6\,{q}^{18}{t}^{18}
+2\,{q}^{16}{t}^{14}+{t}^{12}{q}^{14}+2\,{q}^{24}{t}^{24}+5\,{q}^{20}{t}^{20}+{t}^{24}{q}^{38}+8\,{q}^{24}{t}^{20}+{t}^{22}{q}^{34}
+{q}^{22}{t}^{16}+11\,{t}^{22}{q}^{26}+5\,{q}^{22}{t}^{18}+2\,{q}^{20}{t}^{16}+2\,{q}^{24}{t}^{18}+5\,{q}^{18}{t}^{16}+{q}^{18}{t}^{20}
+2\,{q}^{50}{t}^{32}+2\,{q}^{36}{t}^{30}+{q}^{44}{t}^{32}+5\,{q}^{40}{t}^{30}+8\,{q}^{36}{t}^{26}+{q}^{46}{t}^{28}+{q}^{8}{t}^{10}
+8\,{q}^{32}{t}^{24}+{t}^{30}{q}^{34}+4\,{q}^{14}{t}^{14}+7\,{q}^{20}{t}^{18}+6\,{q}^{42}{t}^{30}+2\,{q}^{44}{t}^{28}+2\,{q}^{14}{t}^{16}
+2\,{t}^{32}{q}^{48}+2\,{q}^{10}{t}^{12}+4\,{q}^{46}{t}^{30}+3\,{t}^{28}{q}^{32}+6\,{q}^{34}{t}^{28}+{q}^{16}{t}^{18}+{q}^{42}{t}^{32}
+11\,{q}^{34}{t}^{26}+8\,{q}^{36}{t}^{28}+2\,{q}^{46}{t}^{32}+{q}^{6}{t}^{8}+5\,{q}^{44}{t}^{30}+8\,{q}^{30}{t}^{26}+{q}^{30}{t}^{28}
+5\,{q}^{30}{t}^{22}+11\,{q}^{30}{t}^{24}+4\,{q}^{22}{t}^{22}+7\,{q}^{40}{t}^{28}+{q}^{54}{t}^{32}+4\,{q}^{38}{t}^{30}+5\,{q}^{34}{t}^{24}
+2\,{q}^{48}{t}^{30}+5\,{q}^{38}{t}^{26}+{q}^{50}{t}^{30}+10\,{q}^{32}{t}^{26}+2\,{q}^{12}{t}^{14}+{q}^{52}{t}^{32}+9\,{t}^{28}{q}^{38}
 ) {a}^{4}+ (
3\,{q}^{16}{t}^{15}+{q}^{30}{t}^{25}+{q}^{2}{t}^{3}+{q}^{40}{t}^{29}+{q}^{26}{t}^{23}+{q}^{50}{t}^{31}+4\,{q}^{22}{t}^{15}
+4\,{q}^{28}{t}^{23}+{q}^{56}{t}^{31}+{q}^{54}{t}^{29}+9\,{q}^{36}{t}^{25}+4\,{q}^{38}{t}^{23}+{q}^{14}{t}^{9}+5\,{q}^{38}{t}^{27}
+2\,{q}^{12}{t}^{9}+4\,{q}^{30}{t}^{19}+2\,{q}^{32}{t}^{19}+4\,{q}^{26}{t}^{17}+{q}^{30}{t}^{17}+9\,{q}^{22}{t}^{17}+{q}^{34}{t}^{19}
+10\,{q}^{34}{t}^{23}+{q}^{46}{t}^{25}+7\,{q}^{18}{t}^{15}+4\,{q}^{46}{t}^{27}+{q}^{52}{t}^{31}+{q}^{6}{t}^{5}+{q}^{34}{t}^{27}
+10\,{q}^{28}{t}^{21}+10\,{q}^{26}{t}^{19}+8\,{q}^{30}{t}^{23}+{q}^{8}{t}^{9}+2\,{q}^{48}{t}^{27}+4\,{q}^{48}{t}^{29}+4\,{q}^{32}{t}^{25}
+2\,{q}^{18}{t}^{17}+4\,{q}^{14}{t}^{13}+10\,{q}^{32}{t}^{23}+7\,{q}^{40}{t}^{25}+3\,{q}^{44}{t}^{29}+7\,{q}^{34}{t}^{25}
+4\,{q}^{14}{t}^{11}+6\,{q}^{16}{t}^{13}+4\,{q}^{46}{t}^{29}+5\,{q}^{22}{t}^{19}+3\,{q}^{10}{t}^{9}+4\,{q}^{12}{t}^{11}+7\,{q}^{42}{t}^{27}
+9\,{q}^{38}{t}^{25}+2\,{q}^{36}{t}^{21}+2\,{q}^{44}{t}^{25}+2\,{q}^{20}{t}^{13}+{q}^{10}{t}^{11}+4\,{q}^{34}{t}^{21}+2\,{q}^{40}{t}^{23}
+{q}^{6}{t}^{7}+2\,{q}^{28}{t}^{17}+{q}^{42}{t}^{23}+6\,{q}^{20}{t}^{17}+{q}^{26}{t}^{15}+2\,{q}^{42}{t}^{29}+4\,{q}^{42}{t}^{25}
+2\,{q}^{52}{t}^{29}+7\,{q}^{32}{t}^{21}+{q}^{50}{t}^{27}+{q}^{20}{t}^{19}+7\,{q}^{28}{t}^{19}+{q}^{22}{t}^{13}+{q}^{18}{t}^{11}
+{q}^{38}{t}^{21}+{q}^{4}{t}^{5}+{q}^{54}{t}^{31}+7\,{q}^{24}{t}^{17}+2\,{q}^{24}{t}^{15}+2\,{q}^{8}{t}^{7}+10\,{q}^{30}{t}^{21}+6\,{q}^{40}{t}^{27}
+6\,{q}^{44}{t}^{27}+7\,{q}^{36}{t}^{23}+4\,{q}^{18}{t}^{13}+7\,{q}^{20}{t}^{15}+2\,{q}^{24}{t}^{21}+9\,{t}^{19}{q}^{24}+7\,{t}^{21}{q}^{26}
+{q}^{10}{t}^{7}+2\,{q}^{16}{t}^{11}+2\,{t}^{27}{q}^{36}+{q}^{58}{t}^{31}+3\,{q}^{50}{t}^{29}
)
{a}^{2}+3\,{t}^{24}{q}^{42}+{q}^{52}{t}^{26}+{q}^{50}{t}^{26}+{t}^{18}{q}^{34}+{t}^{26}{q}^{42}
+{t}^{26}{q}^{40}+2\,{q}^{44}{t}^{26}+{q}^{48}{t}^{28}+1+{q}^{6}{t}^{4}+3\,{q}^{26}{t}^{18}+{t}^{16}{q}^{32}+4\,{q}^{32}{t}^{20}
+2\,{t}^{24}{q}^{36}+{q}^{40}{t}^{20}+{q}^{20}{t}^{10}+2\,{q}^{46}{t}^{26}+2\,{t}^{20}{q}^{28}+3\,{t}^{16}{q}^{26}+{q}^{60}{t}^{30}
+{q}^{56}{t}^{28}+{t}^{10}{q}^{12}+2\,{q}^{32}{t}^{22}+3\,{t}^{24}{q}^{40}+2\,{t}^{20}{q}^{36}+{q}^{48}{t}^{24}+{t}^{18}{q}^{36}
+{t}^{6}{q}^{8}+{t}^{10}{q}^{18}+2\,{t}^{10}{q}^{16}+3\,{t}^{20}{q}^{34}+{q}^{52}{t}^{28}+2\,{q}^{16}{t}^{12}+3\,{q}^{30}{t}^{20}
+{q}^{30}{t}^{16}+{q}^{50}{t}^{28}+2\,{t}^{22}{q}^{40}+{q}^{44}{t}^{22}+{q}^{8}{t}^{4}+{q}^{10}{t}^{8}+4\,{t}^{22}{q}^{36}+3\,{t}^{14}{q}^{20}
+3\,{t}^{22}{q}^{38}+{q}^{54}{t}^{28}+2\,{t}^{24}{q}^{38}+3\,{q}^{18}{t}^{12}+3\,{t}^{22}{q}^{34}+3\,{q}^{22}{t}^{14}+{q}^{16}{t}^{8}
+2\,{q}^{22}{t}^{16}+4\,{t}^{16}{q}^{24}+{q}^{14}{t}^{8}+2\,{t}^{12}{q}^{20}+{t}^{14}{q}^{26}+{q}^{20}{t}^{16}+2\,{q}^{24}{t}^{18}
+{q}^{12}{t}^{6}+2\,{q}^{48}{t}^{26}+{q}^{42}{t}^{22}+2\,{q}^{32}{t}^{18}+2\,{t}^{14}{q}^{24}+{q}^{18}{t}^{14}+{q}^{4}{t}^{2}+2\,{q}^{12}{t}^{8}
+{q}^{46}{t}^{24}+{t}^{12}{q}^{22}+{t}^{12}{q}^{24}+{q}^{28}{t}^{14}+2\,{q}^{28}{t}^{16}+4\,{q}^{28}{t}^{18}+{q}^{30}{t}^{22}
+2\,{t}^{10}{q}^{14}+{t}^{6}{q}^{10}+2\,{q}^{44}{t}^{24}+3\,{q}^{30}{t}^{18}+{t}^{20}{q}^{38}
\Big)\\
$

\bigskip
\noindent
$\boxed{P^{T[6,6k+1]}_{[1]}=\dfrac{\{A\} t^5}{\{t\} A^5
q^{30k+5}}{\cal{P}}^{T[6,6k+1]}_{[1]}}$

\begin{itemize}
\item{HOMFLY case}
\end{itemize}
$$H^{T[6,1]}_{[1]}=\dfrac{\{A\}}{\{q\} A^5}$$ \bigskip \\
$H^{T[6,7]}_{[1]}=\dfrac{\{A\}}{\{q\} A^5 q^{30}} \Big(
1-19\,{q}^{38}{A}^{2}-7\,{q}^{22}{A}^{6}+3\,{q}^{10}{A}^{4}+{q}^{54}{A}^{4}+2\,{q}^{50}-16\,{q}^{40}{A}^{2}+{q}^{60}-20\,{q}^{36}{A}^{2}
\mbox{}+13\,{q}^{42}{A}^{4}+6\,{q}^{38}+{q}^{54}-4\,{q}^{42}{A}^{6}+2\,{q}^{26}{A}^{8}+6\,{q}^{22}+5\,{q}^{44}+7\,{q}^{40}+19\,{q}^{38}{A}^{4}
\mbox{}-6\,{q}^{40}{A}^{6}-3\,{q}^{44}{A}^{6}-{q}^{30}{A}^{10}-3\,{q}^{8}{A}^{2}-{q}^{4}{A}^{2}+8\,{q}^{16}{A}^{4}-20\,{q}^{24}{A}^{2}-3\,{q}^{16}{A}^{6}
\mbox{}+2\,{q}^{52}-2\,{q}^{54}{A}^{2}-{q}^{2}{A}^{2}+{q}^{40}{A}^{8}-14\,{q}^{18}{A}^{2}-{q}^{14}{A}^{6}+23\,{q}^{28}{A}^{4}-6\,{q}^{12}{A}^{2}
\mbox{}-11\,{q}^{44}{A}^{2}+2\,{q}^{30}{A}^{8}+{q}^{4}+3\,{q}^{14}-3\,{q}^{52}{A}^{2}+9\,{q}^{28}+14\,{q}^{20}{A}^{4}+24\,{q}^{34}{A}^{4}+24\,{q}^{26}{A}^{4}
\mbox{}+{q}^{6}+5\,{q}^{42}-2\,{q}^{6}{A}^{2}+8\,{q}^{30}+14\,{q}^{40}{A}^{4}-14\,{q}^{42}{A}^{2}+{q}^{6}{A}^{4}+2\,{q}^{24}{A}^{8}-10\,{q}^{34}{A}^{6}
\mbox{}+20\,{q}^{24}{A}^{4}+9\,{q}^{24}+7\,{q}^{14}{A}^{4}-19\,{q}^{22}{A}^{2}-23\,{q}^{32}{A}^{2}+7\,{q}^{46}{A}^{4}+7\,{q}^{34}+7\,{q}^{20}
\mbox{}-{q}^{56}{A}^{2}-5\,{q}^{50}{A}^{2}-23\,{q}^{26}{A}^{2}+4\,{q}^{12}{A}^{4}+3\,{q}^{28}{A}^{8}-10\,{q}^{36}{A}^{6}+2\,{q}^{34}{A}^{8}
\mbox{}+9\,{q}^{36}+{q}^{56}-{q}^{58}{A}^{2}+{q}^{38}{A}^{8}+3\,{q}^{46}-12\,{q}^{30}{A}^{6}+4\,{q}^{12}-4\,{q}^{18}{A}^{6}-12\,{q}^{32}{A}^{6}
\mbox{}-{q}^{12}{A}^{6}+2\,{q}^{36}{A}^{8}-{q}^{46}{A}^{6}+23\,{q}^{32}{A}^{4}-{q}^{48}{A}^{6}+{q}^{20}{A}^{8}-16\,{q}^{20}{A}^{2}-10\,{q}^{24}{A}^{6}
\mbox{}-7\,{q}^{38}{A}^{6}+19\,{q}^{22}{A}^{4}-6\,{q}^{48}{A}^{2}-10\,{q}^{26}{A}^{6}+{q}^{22}{A}^{8}-23\,{q}^{28}{A}^{2}-24\,{q}^{30}{A}^{2}
\mbox{}+7\,{q}^{26}+4\,{q}^{48}{A}^{4}-23\,{q}^{34}{A}^{2}+3\,{q}^{50}{A}^{4}-11\,{q}^{16}{A}^{2}-9\,{q}^{14}{A}^{2}-12\,{q}^{28}{A}^{6}
\mbox{}+26\,{q}^{30}{A}^{4}+2\,{q}^{10}+20\,{q}^{36}{A}^{4}+13\,{q}^{18}{A}^{4}+{q}^{8}{A}^{4}-5\,{q}^{10}{A}^{2}-9\,{q}^{46}{A}^{2}+9\,{q}^{32}
\mbox{}+5\,{q}^{18}+{q}^{52}{A}^{4}+3\,{q}^{32}{A}^{8}+8\,{q}^{44}{A}^{4}-6\,{q}^{20}{A}^{6}+2\,{q}^{8}+4\,{q}^{48}+5\,{q}^{16}
\Big)$

\bigskip
\fr{
H^{T[6,6k+1]}_{[1]}=\dfrac{\{A\}}{\{q\} A^5 q^{30k}}
{\cal{H}}^{T[6,6k+1]}_{[1]}=
q^{-30k-5}s_{[6]}^{*}-q^{-18k-3}s_{[5,1]}^{*}+q^{-6k-1}s_{[4,1,1]}^{*}-q^{6k+1}s_{[3,1,1,1]}^{*}+q^{18k+3}s_{[2,1,1,1,1]}^{*}-\\-q^{30k+5}s_{[1,1,1,1,1,1]}^{*}}

and the results coincides with well known HOMFLY polynomials, see
(\ref{Wrepfund}) and
(\ref{113}).

\begin{itemize}
\item{Floer case}
\end{itemize}
$$F^{T[6,1]}_{[1]}=1$$ \\
$F^{T[6,7]}_{[1]}=\bf{q}^{60}{t}^{30}+{q}^{58}{t}^{29}+{q}^{48}{t}^{28}+{q}^{44}{t}^{25}+{q}^{36}{t}^{24}+{q}^{30}{t}^{19}+{q}^{24}{t}^{18}+{q}^{16}{t}^{11}+{q}^{12}{t}^{10}+{q}^{2}t+1$
 \\
\begin{itemize}
\item{Alexander case}
\end{itemize}
$$A^{T[6,1]}_{[1]}=1$$  \\
$$A^{T[6,7]}_{[1]}=\bf{q}^{60}-q^{58}+q^{48}-q^{44}+q^{36}-q^{30}+q^{24}-q^{16}+q^{12}-q^{2}+1$$
 \\

\subsection{Case $(7,n)$, $n=7 k+1$ fundamental representation\label{7k+1}}
\be
\begin{array}{|c|}
\hline\\
P^{T[7,n]}_{[1]}=c^{[7]}_{[1]}M^{\ast}_{{7}}{q}^{-6\,n}+c^{[6,1]}_{[1]}M^{\ast}_{{6,1}}\,{q}^{-{\frac
{30}{7}}\,n}
\mbox{}{t}^{2/7\,n}+c^{[5,2]}_{[1]}M^{\ast}_{{5,2}}\,{q}^{-{\frac {22}{7}}\,n}
\mbox{}{t}^{4/7\,n}+c^{[5,1,1]}_{[1]}M^{\ast}_{{5,1,1}}\,{q}^{-{\frac
{20}{7}}\,n}
\mbox{}{t}^{6/7\,n}+\\+c^{[4,3]}_{[1]}M^{\ast}_{{4,3}}\,{q}^{-{\frac
{18}{7}}\,n}
\mbox{}{t}^{6/7\,n}+c^{[4,2,1]}_{[1]}M^{\ast}_{{4,2,1}}\,{q}^{-2\,n}{t}^{{\frac
{8}{7}}\,n}
\mbox{}+c^{[4,1,1,1]}_{[1]}M^{\ast}_{{4,1,1,1}}\,{q}^{-{\frac {12}{7}}\,n}
\mbox{}{t}^{{\frac
{12}{7}}\,n}+c^{[3,3,1]}_{[1]}M^{\ast}_{{3,3,1}}\,{q}^{-{\frac {12}{7}}\,n}
\mbox{}{t}^{{\frac
{10}{7}}\,n}+\\+c^{[3,2,2]}_{[1]}M^{\ast}_{{3,2,2}}\,{t}^{{\frac {12}{7}}\,n}
\mbox{}{q}^{-{\frac {10}{7}}\,n}
\mbox{}+c^{[3,2,1,1]}_{[1]}M^{\ast}_{{3,2,1,1}}\,{t}^{2\,n}{q}^{-{\frac
{8}{7}}\,n}+c^{[3,1,1,1,1]}_{[1]}M^{\ast}_{{3,1,1,1,1}}\,{t}^{{\frac
{20}{7}}\,n}
\mbox{}{q}^{-\frac{6}{7}\,n}
\mbox{}+c^{[2,2,2,1]}_{[1]}M^{\ast}_{{2,2,2,1}}\,{t}^{{\frac {18}{7}}\,n}
\mbox{}{q}^{-\frac{6}{7}\,n}+\\+c^{[2,2,1,1,1]}_{[1]}M^{\ast}_{{2,2,1,1,1}}\,{t}^{{\frac
{22}{7}}\,n}
\mbox{}{q}^{-\frac{4}{7}\,n}+c^{[2,1,1,1,1,1]}_{[1]}M^{\ast}_{{2,1,1,1,1,1}}\,{t}^{{\frac
{30}{7}}\,n}
\mbox{}{q}^{-\frac{2}{7}\,n}+c^{[1,1,1,1,1,1,1]}_{[1]}M^{\ast}_{{1,1,1,1,1,1,1}}\,{t}^{6\,n}

\\
\\
\hline
\end{array}
\ee
\fr{ \\
c^{[7]}_{[1]}=1,\ \ c^{[6,1]}_{[1]}
=-\frac{q^{2/7}(-1+t^{2})(1+q^{2}+q^{4}+q^{6}+q^{8}+q^{10}+q^{10}t^{2})}{t^{2/7}(-1+q^{12}t^{2})},
\\
c^{[5,2]}_{[1]}
=-\frac{q^{8/7}(-1+t^{2})(q^{2}-t^{2})(q^{4}+1)(1+q^{2}+q^{4}+q^{6}+q^{8}+q^{6}t^{2}+q^{8}t^{2})}{t^{4/7}(q^{10}t^{2}-1)(q^{8}t^{2}-1)}
\\
c^{[5,1,1]}_{[1]}=\frac{q^{6/7}(t^{2}+1)(-1+t^{2})^{2}(1+q^{2}+q^{4}+q^{6}+q^{8}+q^{8}t^{2}+q^{8}t^{4})}{t^{6/7}(q^{2}t^{2}-1)(-1+q^{10}t^{4})},
 \\
c^{[4,3]}_{[1]}
=-\frac{(-1+t^{2})(q^{6}+q^{6}t^{2}+q^{4}+q^{4}t^{2}+q^{2}+q^{2}t^{2}+1)(-t^{2}+q^{4})(q^{2}-t^{2})q^{18/7}}{(q^{8}t^{2}-1)(q^{6}t
^{2}-1)(q^{4}t^{2}-1)t^{6/7}}\mathop{\rm  },  \\
c^{[4,2,1]}_{[1]}=\frac{q^{2}(t^{2}+1)(-1+t^{2})^{2}(q^{4}+q^{2}+1)(q^{6}t^{4}+q^{4}t^{2}+q^{6}t^{2}+q^{6}+q^{4}+q^{2}+1)
(q^{2}-t^{2})}{t^{8/7}(q^{4}t^{2}+1)(q^{6}t^{2}-1)(q^{4}t^{2}-1)^{2}}\mathop{\rm
 },  \\
c^{[4,1,1,1]}_{[1]}
=-\frac{q^{12/7}(-1+t^{2})(-1+t^{4})(t^{6}-1)(1+q^{2}+q^{4}+q^{6}+q^{6}t^{2}+q^{6}t^{4}+q^{6}t^{6})}{t^{12/7}(q^{2}t^{2}-1)
(q^{2}t
^{4}-1)(-1+t
^{6}q^{8})}\mathop{\rm  },  \\
c^{[3,3,1]}_{[1]}
=\frac{(t^{2}+1)(-1+t^{2})^{2}(q^{2}-t^{2})(-t^{2}+q^{4})(q^{4}t^{4
}+t^{2}+1+q^{2}t^{2}+q^{2}+q^{4}t^{2}+q^{4})q^{26/7}}{(q^{2}t^{2}-1)(q^{4}t^{2}-1)(q^{6}t^{2}-1)(-1+q^{6}t^{4})t^{10/7}}\mathop{\rm
 },\\ c^{[3,2,2]}_{[1]}
=\frac{q^{24/7}(t^{2}+1)(-1+t^{2})^{2}(q^{2}t^{4}+q^{4}t^{4}+q^{2
}t^{2}+q^{4}t^{2}+q^{4}+q^{2}+1)(q^{2}-t^{2})(q^{2}-t^{4})}{t^{12/7}(q^{2}t^{2}+1)(q^{4}t^{2}-1)(-1+q^{6}t^{4})(q^{2}t^{2}-1)^{2}}\mathop{\rm
 },  \\
c^{[3,2,1,1]}_{[1]}
=-\frac{(t^{2}+1)(1+t^{2}+t^{4})(-1+t^{2})^{3}(q^{2}+1)(q^{2}-t^{2
})(q^{4}t^{6}+q^{4}t^{4}+q^{4}t^{2}+q^{4}+q^{2}+1+q^{2}t^{2})q^{22/7}}{(q^{2}t^{2}+1)(q^{4}t^{2}-1)(q^{4}t^{4}+q^{2}t^{2}+1)(q^{2}t^{2}-1)^{3}t^{2}}\mathop{\rm
 },  \\
c^{[3,1,1,1,1]}_{[1]} =\frac{q^{20/7}(1+t^{4
})(1+t^{2}+t^{4})(t^{2}+1)^{2}(-1+t^{2})^{4}(1+q^{2}+q^{4}+q^{4}t^{2}+q^{4}t^{4}+q^{4}t^{6}+q^{4}t^{8})}{t^{20/7}(t^{8}q^{6}-1)(q^{2}t^{2}-1)(t^{6}q^{2}-1)(q^{2}t^{4}-1)}\mathop{\rm
 }\mathop{\rm  },  \\
c^{[2,2,2,1]}_{[1]} =-\frac{q^{34/7
}(t^{2}+1)(1+t^{2}+t^{4})(-1+t^{2})^{3}(q^{2}-t^{2})(t^{6}q^{2}+q^{2}t^{4}+q^{2}t^{2}+q^{2}+1+t^{2}+t^{4})(q^{2}-t^{4})}{t^{18/7}(q^{2}t^{2}+1)(q^{2}t^{4}-1)(q^{4}t^{2}-1)(-1+q^{4}t^{6})(q^{2}t^{2}-1)^{2}},
 \\
c^{[2,2,1,1,1]}_{[1]}
=\frac{q^{32/7}(1+t^{4})(1+t^{2}+t^{4})(t^{2}+1)^{2}(-1+t^{2})^{4}(q^{2}t^{8}+t^{6}q^{2}+q^{2}t^{4}+t^{2}+q^{2}t^{2}+q^{2}+1)(q^{2}-t^{2})}{t^{22/7}(q^{2}t^{4}+1)(-1+q^{4}t^{6})(q^{2}t^{4}-1)^{2}(q^{2}t^{2}-1)^{2}}\mathop{\rm
 },  \\
c^{[2,1,1,1,1,1]}_{[1]}
=-\frac{q^{30/7}(-1+t^{2})(-1+t^{4})(t^{6}-1)(-1+t^{8})(-1+t^{10})(1+q^{2}+q^{2}t^{2}+q^{2}t^{4}+t^{6}q^{2}+q^{2}t^{8}+q^{2}t^{10})}{t^{30/7}(t^{10}q^{4}-1)(q^{2}t^{8}-1)(q^{2}t^{2}-1)(t^{6}q^{2}-1)(q^{2}t^{4}-1)},
 \\
c^{[1,1,1,1,1,1,1]}_{[1]} =
\frac{q^{6}(t^{14}-1)(t^{12}-1)(-1+t^{10})(-1+t^{8})(t^{6}-1)(-1+t^{4})}{t^{6}(t^{12}q^{2}-1)(q^{2}t^{10}-1)(q^{2}t^{8}-1)(t^{6}q^{2}-1)(q^{2}t^{4}-1)(q^{2}t^{2}-1)}}
Several first superpolynomials are:
$$P^{T[7,1]}_{[1]}=\dfrac{\{A\} t^6}{\{t\} A^6 q^6}$$
\bigskip \\
$P^{T[7,8]}_{[1]}=\dfrac{\{A\} t^6}{\{t\} A^6 q^{48}} \Big( {q}^{42}(-A^2)^{6}+
(
+{q}^{42}{t}^{2}+{q}^{42}{t}^{10}+{q}^{34}{t}^{2}+{q}^{40}+{q}^{38}+{q}^{40}{t}^{8}+{q}^{34}+{q}^{38}{t}^{2}+{q}^{32}+{q}^{38}{t}^{4}
\mbox{}+{q}^{36}{t}^{4}+{q}^{42}{t}^{6}+{q}^{38}{t}^{6}+{q}^{40}{t}^{10}+{t}^{12}{q}^{42}+{q}^{42}{t}^{8}+{q}^{40}{t}^{2}+{q}^{38}{t}^{8}
\mbox{}+{t}^{4}{q}^{34}+{t}^{6}{q}^{40}+{t}^{2}{q}^{36}+{q}^{40}{t}^{4}+{q}^{36}+{q}^{32}{t}^{2}+{t}^{6}{q}^{36}+{q}^{42}{t}^{4}+{q}^{30}
) (-A^2)^{5}
\mbox{}+ (
4\,{t}^{14}{q}^{40}+4\,{t}^{14}{q}^{38}+{t}^{4}{q}^{24}+2\,{q}^{42}{t}^{10}+2\,{t}^{4}{q}^{26}+{q}^{42}{t}^{22}+2\,{q}^{26}+2\,{t}^{16}{q}^{42}
\mbox{}+4\,{q}^{34}{t}^{2}+{q}^{42}{t}^{20}+4\,{t}^{8}{q}^{32}+8\,{t}^{6}{q}^{34}+2\,{q}^{38}{t}^{16}+{t}^{6}{q}^{26}+{q}^{32}{t}^{12}+2\,{t}^{14}{q}^{36}
\mbox{}+4\,{q}^{40}{t}^{8}+5\,{q}^{30}{t}^{4}+6\,{t}^{6}{q}^{32}+5\,{q}^{40}{t}^{12}+{q}^{34}+{q}^{22}+7\,{q}^{38}{t}^{10}+{q}^{30}{t}^{10}
\mbox{}+{q}^{28}{t}^{8}+4\,{t}^{2}{q}^{28}+2\,{q}^{40}{t}^{18}+2\,{q}^{38}{t}^{2}+3\,{q}^{42}{t}^{14}+2\,{t}^{18}{q}^{42}+{q}^{34}{t}^{14}
\mbox{}+2\,{q}^{24}+{q}^{20}+2\,{q}^{32}+3\,{q}^{38}{t}^{4}+5\,{q}^{36}{t}^{4}+{q}^{42}{t}^{6}+5\,{q}^{38}{t}^{6}+5\,{q}^{40}{t}^{10}+{q}^{40}{t}^{20}
\mbox{}+5\,{q}^{38}{t}^{12}+2\,{t}^{12}{q}^{42}+{q}^{42}{t}^{8}+{q}^{40}{t}^{2}+{q}^{36}{t}^{16}+6\,{q}^{38}{t}^{8}+{q}^{38}{t}^{18}
\mbox{}+4\,{q}^{36}{t}^{12}+7\,{q}^{32}{t}^{4}+6\,{t}^{4}{q}^{34}+2\,{q}^{34}{t}^{12}+3\,{t}^{6}{q}^{40}+6\,{t}^{8}{q}^{34}+3\,{t}^{2}{q}^{36}+8\,{q}^{36}{t}^{8}
\mbox{}+2\,{q}^{40}{t}^{4}+{q}^{36}+5\,{q}^{32}{t}^{2}+4\,{t}^{4}{q}^{28}+7\,{t}^{6}{q}^{36}+4\,{t}^{6}{q}^{30}+2\,{t}^{6}{q}^{28}+4\,{t}^{10}{q}^{34}
\mbox{}+2\,{t}^{10}{q}^{32}+2\,{q}^{30}{t}^{8}+2\,{q}^{24}{t}^{2}+3\,{t}^{16}{q}^{40}+3\,{q}^{26}{t}^{2}+5\,{t}^{2}{q}^{30}+{t}^{2}{q}^{22}+6\,{q}^{36}{t}^{10}
\mbox{}+3\,{q}^{28}+2\,{q}^{30} ) (-A^2)^{4}+ (
+8\,{t}^{14}{q}^{40}+15\,{t}^{14}{q}^{38}+12\,{t}^{4}{q}^{24}+14\,{t}^{4}{q}^{26}+3\,{q}^{42}{t}^{22}+2\,{q}^{26}+2\,{t}^{16}{q}^{42}
\mbox{}+2\,{q}^{34}{t}^{2}+3\,{q}^{42}{t}^{20}+22\,{t}^{8}{q}^{32}+13\,{t}^{6}{q}^{34}+14\,{q}^{38}{t}^{16}+{q}^{40}{t}^{28}+14\,{t}^{6}{q}^{26}
\mbox{}+15\,{q}^{32}{t}^{12}+{q}^{42}{t}^{30}+18\,{t}^{14}{q}^{36}+2\,{q}^{40}{t}^{8}+3\,{q}^{42}{t}^{24}+13\,{q}^{30}{t}^{4}+18\,{t}^{6}{q}^{32}
\mbox{}+6\,{q}^{40}{t}^{12}+3\,{q}^{22}+2\,{q}^{16}+{q}^{14}+10\,{q}^{38}{t}^{10}+2\,{q}^{38}{t}^{24}+15\,{q}^{30}{t}^{10}+15\,{q}^{28}{t}^{8}
\mbox{}+{q}^{24}{t}^{12}+9\,{q}^{32}{t}^{14}+8\,{t}^{2}{q}^{28}+9\,{q}^{40}{t}^{18}+8\,{q}^{38}{t}^{20}+{q}^{42}{t}^{14}+8\,{q}^{22}{t}^{4}
\mbox{}+3\,{t}^{18}{q}^{42}+6\,{q}^{40}{t}^{22}+2\,{q}^{40}{t}^{26}+15\,{q}^{34}{t}^{14}+9\,{q}^{34}{t}^{16}+{q}^{34}{t}^{22}+{q}^{32}{t}^{20}
\mbox{}+{q}^{28}{t}^{16}+{q}^{22}{t}^{10}+5\,{q}^{38}{t}^{22}+2\,{q}^{26}{t}^{12}+9\,{q}^{28}{t}^{10}+5\,{q}^{30}{t}^{14}+5\,{q}^{32}{t}^{16}
\mbox{}+3\,{q}^{24}+5\,{q}^{28}{t}^{12}+{t}^{4}{q}^{16}+5\,{t}^{6}{q}^{22}+{q}^{20}{t}^{8}+5\,{q}^{26}{t}^{10}+3\,{q}^{18}+3\,{q}^{20}+{q}^{30}{t}^{18}
\mbox{}+{q}^{18}{t}^{6}+2\,{t}^{6}{q}^{20}+9\,{q}^{36}{t}^{18}+5\,{q}^{36}{t}^{20}+9\,{q}^{30}{t}^{12}+{q}^{38}{t}^{26}+{q}^{38}{t}^{4}
\mbox{}+3\,{q}^{36}{t}^{4}+3\,{q}^{38}{t}^{6}+4\,{q}^{40}{t}^{10}+{q}^{36}{t}^{24}+8\,{q}^{40}{t}^{20}+13\,{q}^{38}{t}^{12}+{t}^{12}{q}^{42}
\mbox{}+14\,{q}^{36}{t}^{16}+6\,{q}^{38}{t}^{8}+2\,{q}^{42}{t}^{26}+{q}^{14}{t}^{2}+{q}^{12}+12\,{q}^{38}{t}^{18}+20\,{q}^{36}{t}^{12}
\mbox{}+10\,{q}^{32}{t}^{4}+6\,{t}^{4}{q}^{34}+19\,{q}^{34}{t}^{12}+{t}^{6}{q}^{40}+19\,{t}^{8}{q}^{34}+{t}^{2}{q}^{36}+13\,{q}^{36}{t}^{8}+6\,{q}^{20}{t}^{2}
\mbox{}+4\,{q}^{18}{t}^{2}+2\,{q}^{18}{t}^{4}+2\,{q}^{32}{t}^{18}+5\,{t}^{4}{q}^{20}+9\,{t}^{6}{q}^{24}+4\,{q}^{32}{t}^{2}+5\,{q}^{24}{t}^{8}+15\,{t}^{4}{q}^{28}
\mbox{}+8\,{t}^{6}{q}^{36}+5\,{q}^{34}{t}^{18}+2\,{q}^{24}{t}^{10}+2\,{q}^{28}{t}^{14}+{q}^{26}{t}^{14}+20\,{t}^{6}{q}^{30}+18\,{t}^{6}{q}^{28}
\mbox{}+2\,{q}^{16}{t}^{2}+{q}^{42}{t}^{28}+22\,{t}^{10}{q}^{34}+20\,{t}^{10}{q}^{32}+19\,{q}^{30}{t}^{8}+2\,{q}^{34}{t}^{20}+2\,{q}^{22}{t}^{8}
\mbox{}+2\,{q}^{36}{t}^{22}+2\,{q}^{30}{t}^{16}+9\,{q}^{26}{t}^{8}+9\,{q}^{24}{t}^{2}+9\,{t}^{16}{q}^{40}+4\,{q}^{40}{t}^{24}+9\,{q}^{26}{t}^{2}
\mbox{}+6\,{t}^{2}{q}^{30}+8\,{t}^{2}{q}^{22}+18\,{q}^{36}{t}^{10}+{q}^{28}+{q}^{30}
) (-A^2)^{3}
\mbox{}+ (
2\,{t}^{14}{q}^{40}+4\,{q}^{40}{t}^{30}+9\,{t}^{14}{q}^{38}+2\,{q}^{10}{t}^{2}+{t}^{32}{q}^{38}+4\,{q}^{12}{t}^{2}+16\,{t}^{4}{q}^{24}
\mbox{}+2\,{q}^{42}{t}^{32}+14\,{t}^{4}{q}^{26}+{q}^{42}{t}^{22}+{q}^{42}{t}^{34}+{q}^{22}{t}^{16}+{q}^{42}{t}^{20}+19\,{t}^{8}{q}^{32}
\mbox{}+4\,{t}^{6}{q}^{34}+14\,{q}^{38}{t}^{16}+6\,{q}^{40}{t}^{28}+25\,{t}^{6}{q}^{26}+{q}^{8}{t}^{2}+33\,{q}^{32}{t}^{12}+{t}^{34}{q}^{40}
\mbox{}+{q}^{42}{t}^{36}+2\,{q}^{42}{t}^{30}+22\,{t}^{14}{q}^{36}+2\,{q}^{42}{t}^{24}+6\,{q}^{30}{t}^{4}+9\,{t}^{6}{q}^{32}+{q}^{40}{t}^{12}
\mbox{}+{q}^{10}{t}^{4}+{q}^{22}+2\,{q}^{16}+3\,{q}^{14}+{q}^{6}+{q}^{8}+2\,{q}^{10}+2\,{q}^{38}{t}^{10}+13\,{q}^{38}{t}^{24}+33\,{q}^{30}{t}^{10}
\mbox{}+2\,{t}^{32}{q}^{40}+30\,{q}^{28}{t}^{8}+9\,{q}^{24}{t}^{12}+31\,{q}^{32}{t}^{14}+2\,{t}^{2}{q}^{28}+6\,{q}^{40}{t}^{18}+18\,{q}^{38}{t}^{20}
\mbox{}+18\,{q}^{22}{t}^{4}+9\,{q}^{40}{t}^{22}+8\,{q}^{40}{t}^{26}+30\,{q}^{34}{t}^{14}+2\,{q}^{36}{t}^{28}+29\,{q}^{34}{t}^{16}
\mbox{}+5\,{t}^{14}{q}^{24}+9\,{q}^{34}{t}^{22}+9\,{q}^{32}{t}^{20}+9\,{q}^{28}{t}^{16}+9\,{q}^{22}{t}^{10}+15\,{q}^{38}{t}^{22}
\mbox{}+{t}^{10}{q}^{16}+16\,{q}^{26}{t}^{12}+31\,{q}^{28}{t}^{10}+5\,{q}^{28}{t}^{18}+23\,{q}^{30}{t}^{14}+23\,{q}^{32}{t}^{16}
\mbox{}+{q}^{14}{t}^{8}+{t}^{12}{q}^{18}+23\,{q}^{28}{t}^{12}+2\,{q}^{28}{t}^{20}+2\,{t}^{6}{q}^{14}+8\,{t}^{4}{q}^{16}+5\,{q}^{36}{t}^{26}
\mbox{}+20\,{t}^{6}{q}^{22}+2\,{q}^{16}{t}^{8}+2\,{q}^{30}{t}^{22}+9\,{q}^{20}{t}^{8}+23\,{q}^{26}{t}^{10}+5\,{q}^{30}{t}^{20}+2\,{q}^{18}{t}^{10}
\mbox{}+5\,{q}^{38}{t}^{28}+2\,{q}^{18}+{q}^{20}+9\,{q}^{30}{t}^{18}+2\,{t}^{4}{q}^{12}+5\,{t}^{4}{q}^{14}+{q}^{12}{t}^{6}+9\,{q}^{18}{t}^{6}+15\,{t}^{6}{q}^{20}
\mbox{}+5\,{t}^{6}{q}^{16}+25\,{q}^{36}{t}^{18}+5\,{q}^{26}{t}^{16}+5\,{q}^{18}{t}^{8}+20\,{q}^{36}{t}^{20}+2\,{q}^{32}{t}^{24}+2\,{q}^{34}{t}^{26}
\mbox{}+5\,{q}^{34}{t}^{24}+32\,{q}^{30}{t}^{12}+8\,{q}^{38}{t}^{26}+5\,{t}^{10}{q}^{20}+9\,{q}^{36}{t}^{24}+5\,{q}^{22}{t}^{12}
\mbox{}+{q}^{20}{t}^{14}+5\,{q}^{32}{t}^{22}+2\,{q}^{22}{t}^{14}+2\,{q}^{20}{t}^{12}+{t}^{24}{q}^{30}+2\,{q}^{26}{t}^{18}+2\,{q}^{24}{t}^{16}
\mbox{}+8\,{q}^{40}{t}^{20}+6\,{q}^{38}{t}^{12}+25\,{q}^{36}{t}^{16}+{q}^{38}{t}^{8}+2\,{q}^{42}{t}^{26}+6\,{q}^{14}{t}^{2}+2\,{q}^{12}
\mbox{}+16\,{q}^{38}{t}^{18}+16\,{q}^{36}{t}^{12}+2\,{q}^{32}{t}^{4}+{t}^{4}{q}^{34}+28\,{q}^{34}{t}^{12}+11\,{t}^{8}{q}^{34}+4\,{q}^{36}{t}^{8}
\mbox{}+9\,{q}^{20}{t}^{2}+9\,{q}^{18}{t}^{2}+13\,{q}^{18}{t}^{4}+16\,{q}^{32}{t}^{18}+15\,{t}^{4}{q}^{20}+25\,{t}^{6}{q}^{24}+22\,{q}^{24}{t}^{8}
\mbox{}+9\,{t}^{4}{q}^{28}+{t}^{6}{q}^{36}+22\,{q}^{34}{t}^{18}+16\,{q}^{24}{t}^{10}+{t}^{28}{q}^{34}+{q}^{24}{t}^{18}+16\,{q}^{28}{t}^{14}
\mbox{}+{t}^{30}{q}^{36}+{q}^{28}{t}^{22}+{q}^{26}{t}^{20}+9\,{q}^{26}{t}^{14}+16\,{t}^{6}{q}^{30}+22\,{t}^{6}{q}^{28}+8\,{q}^{16}{t}^{2}
\mbox{}+3\,{q}^{42}{t}^{28}+19\,{t}^{10}{q}^{34}+29\,{t}^{10}{q}^{32}+28\,{q}^{30}{t}^{8}+16\,{q}^{34}{t}^{20}+2\,{t}^{30}{q}^{38}
\mbox{}+{t}^{26}{q}^{32}+16\,{q}^{22}{t}^{8}+15\,{q}^{36}{t}^{22}+16\,{q}^{30}{t}^{16}+29\,{q}^{26}{t}^{8}+6\,{q}^{24}{t}^{2}+4\,{t}^{16}{q}^{40}
\mbox{}+9\,{q}^{40}{t}^{24}+4\,{q}^{26}{t}^{2}+{t}^{2}{q}^{30}+8\,{t}^{2}{q}^{22}+9\,{q}^{36}{t}^{10}
) (-A^2)^{2}+ (
+5\,{q}^{40}{t}^{30}+{q}^{2}+{q}^{40}{t}^{38}+{t}^{14}{q}^{38}+4\,{q}^{10}{t}^{2}+4\,{t}^{32}{q}^{38}+5\,{q}^{12}{t}^{2}+4\,{t}^{4}{q}^{24}
\mbox{}+{q}^{42}{t}^{32}+2\,{t}^{4}{q}^{26}+2\,{q}^{8}{t}^{4}+{t}^{6}{q}^{8}+{q}^{42}{t}^{34}+4\,{q}^{22}{t}^{16}+3\,{t}^{8}{q}^{32}+{q}^{42}{t}^{38}
\mbox{}+2\,{q}^{38}{t}^{16}+5\,{q}^{40}{t}^{28}+9\,{t}^{6}{q}^{26}+3\,{q}^{8}{t}^{2}+16\,{q}^{32}{t}^{12}+3\,{t}^{34}{q}^{40}+{q}^{42}{t}^{40}
\mbox{}+{q}^{42}{t}^{36}+{q}^{42}{t}^{30}+5\,{t}^{14}{q}^{36}+2\,{q}^{6}{t}^{2}+{q}^{4}{t}^{2}+{q}^{6}{t}^{4}+4\,{q}^{10}{t}^{4}+{q}^{6}+{q}^{8}+{q}^{10}
\mbox{}+10\,{q}^{38}{t}^{24}+{t}^{20}{q}^{22}+16\,{q}^{30}{t}^{10}+{t}^{30}{q}^{32}+{q}^{38}{t}^{36}+4\,{t}^{32}{q}^{40}+{q}^{14}{t}^{12}
\mbox{}+{q}^{16}{t}^{14}+14\,{q}^{28}{t}^{8}+{t}^{34}{q}^{36}+{t}^{10}{q}^{12}+17\,{q}^{24}{t}^{12}+22\,{q}^{32}{t}^{14}+7\,{q}^{38}{t}^{20}
\mbox{}+7\,{q}^{22}{t}^{4}+2\,{q}^{40}{t}^{22}+4\,{q}^{40}{t}^{26}+14\,{q}^{34}{t}^{14}+7\,{q}^{36}{t}^{28}+19\,{q}^{34}{t}^{16}
\mbox{}+12\,{t}^{14}{q}^{24}+16\,{q}^{34}{t}^{22}+17\,{q}^{32}{t}^{20}+17\,{q}^{28}{t}^{16}+17\,{q}^{22}{t}^{10}+9\,{q}^{38}{t}^{22}
\mbox{}+4\,{t}^{10}{q}^{16}+23\,{q}^{26}{t}^{12}+22\,{q}^{28}{t}^{10}+12\,{q}^{28}{t}^{18}+26\,{q}^{30}{t}^{14}+24\,{q}^{32}{t}^{16}
\mbox{}+4\,{q}^{14}{t}^{8}+4\,{t}^{12}{q}^{18}+26\,{q}^{28}{t}^{12}+7\,{q}^{28}{t}^{20}+7\,{t}^{6}{q}^{14}+10\,{t}^{4}{q}^{16}+2\,{t}^{6}{q}^{10}
\mbox{}+11\,{q}^{36}{t}^{26}+2\,{t}^{16}{q}^{20}+2\,{t}^{12}{q}^{16}+16\,{t}^{6}{q}^{22}+7\,{q}^{16}{t}^{8}+2\,{q}^{12}{t}^{8}+7\,{q}^{30}{t}^{22}
\mbox{}+16\,{q}^{20}{t}^{8}+24\,{q}^{26}{t}^{10}+12\,{q}^{30}{t}^{20}+7\,{q}^{18}{t}^{10}+9\,{q}^{38}{t}^{28}+2\,{q}^{38}{t}^{34}
\mbox{}+17\,{q}^{30}{t}^{18}+6\,{t}^{4}{q}^{12}+9\,{t}^{4}{q}^{14}+4\,{q}^{12}{t}^{6}+14\,{q}^{18}{t}^{6}+16\,{t}^{6}{q}^{20}+11\,{t}^{6}{q}^{16}
\mbox{}+13\,{q}^{36}{t}^{18}+{q}^{26}{t}^{24}+{q}^{34}{t}^{32}+2\,{q}^{34}{t}^{30}+12\,{q}^{26}{t}^{16}+12\,{q}^{18}{t}^{8}+2\,{t}^{14}{q}^{18}
\mbox{}+16\,{q}^{36}{t}^{20}+7\,{q}^{32}{t}^{24}+7\,{q}^{34}{t}^{26}+12\,{q}^{34}{t}^{24}+24\,{q}^{30}{t}^{12}+10\,{q}^{38}{t}^{26}
\mbox{}+12\,{t}^{10}{q}^{20}+14\,{q}^{36}{t}^{24}+2\,{q}^{14}{t}^{10}+12\,{q}^{22}{t}^{12}+4\,{q}^{20}{t}^{14}+12\,{q}^{32}{t}^{22}
\mbox{}+2\,{t}^{32}{q}^{36}+7\,{q}^{22}{t}^{14}+{q}^{4}+7\,{q}^{20}{t}^{12}+2\,{t}^{18}{q}^{22}+4\,{t}^{24}{q}^{30}+7\,{q}^{26}{t}^{18}
\mbox{}+{q}^{30}{t}^{28}+7\,{q}^{24}{t}^{16}+{q}^{40}{t}^{20}+9\,{q}^{36}{t}^{16}+5\,{q}^{14}{t}^{2}+{q}^{12}+4\,{q}^{38}{t}^{18}+2\,{q}^{36}{t}^{12}
\mbox{}+8\,{q}^{34}{t}^{12}+{t}^{8}{q}^{34}+2\,{q}^{20}{t}^{2}+3\,{q}^{18}{t}^{2}+10\,{q}^{18}{t}^{4}+22\,{q}^{32}{t}^{18}+9\,{t}^{4}{q}^{20}
\mbox{}+2\,{q}^{26}{t}^{22}+{q}^{28}{t}^{26}+13\,{t}^{6}{q}^{24}+{t}^{18}{q}^{20}+{q}^{18}{t}^{16}+{q}^{10}{t}^{8}+21\,{q}^{24}{t}^{8}
\mbox{}+{t}^{4}{q}^{28}+21\,{q}^{34}{t}^{18}+2\,{q}^{30}{t}^{26}+22\,{q}^{24}{t}^{10}+4\,{t}^{28}{q}^{34}+4\,{q}^{24}{t}^{18}+23\,{q}^{28}{t}^{14}
\mbox{}+{t}^{22}{q}^{24}+4\,{t}^{30}{q}^{36}+4\,{q}^{28}{t}^{22}+4\,{q}^{26}{t}^{20}+17\,{q}^{26}{t}^{14}+2\,{t}^{28}{q}^{32}+2\,{t}^{6}{q}^{30}
\mbox{}+5\,{t}^{6}{q}^{28}+4\,{q}^{16}{t}^{2}+3\,{t}^{10}{q}^{34}+8\,{t}^{10}{q}^{32}+8\,{q}^{30}{t}^{8}+20\,{q}^{34}{t}^{20}+6\,{t}^{30}{q}^{38}
\mbox{}+4\,{t}^{26}{q}^{32}+2\,{q}^{28}{t}^{24}+20\,{q}^{22}{t}^{8}+2\,{t}^{36}{q}^{40}+16\,{q}^{36}{t}^{22}+23\,{q}^{30}{t}^{16}
\mbox{}+19\,{q}^{26}{t}^{8}+2\,{q}^{24}{t}^{20}+3\,{q}^{40}{t}^{24}+{t}^{2}{q}^{22}
) (-A^2)+{q}^{36}{t}^{36}+{q}^{18}{t}^{18}+1+{q}^{12}{t}^{2}+{q}^{40}{t}^{30}
\mbox{}+2\,{q}^{8}{t}^{4}+{t}^{6}{q}^{8}+3\,{q}^{22}{t}^{16}+{q}^{16}{t}^{16}+{q}^{8}{t}^{2}+{q}^{32}{t}^{12}+{t}^{34}{q}^{40}+{q}^{6}{t}^{2}+{q}^{42}{t}^{42}
\mbox{}+{q}^{4}{t}^{2}+{q}^{6}{t}^{4}+{q}^{4}{t}^{4}+2\,{q}^{10}{t}^{4}+{q}^{6}{t}^{6}+{q}^{2}{t}^{2}+{t}^{22}{q}^{22}+{t}^{26}{q}^{26}+{q}^{32}{t}^{32}+{q}^{38}{t}^{24}
\mbox{}+{t}^{20}{q}^{22}+{q}^{30}{t}^{10}+{t}^{30}{q}^{32}+{q}^{12}{t}^{12}+{q}^{40}{t}^{40}+{q}^{38}{t}^{36}+{t}^{32}{q}^{40}
\mbox{}+{q}^{14}{t}^{12}+{q}^{16}{t}^{14}+{q}^{28}{t}^{8}+{t}^{34}{q}^{36}+{t}^{10}{q}^{12}+8\,{q}^{24}{t}^{12}+3\,{q}^{32}{t}^{14}
\mbox{}+{q}^{34}{t}^{14}+4\,{q}^{36}{t}^{28}+2\,{q}^{34}{t}^{16}+{q}^{28}{t}^{28}+6\,{t}^{14}{q}^{24}+6\,{q}^{34}{t}^{22}+7\,{q}^{32}{t}^{20}
\mbox{}+8\,{q}^{28}{t}^{16}+7\,{q}^{22}{t}^{10}+{q}^{38}{t}^{22}+{t}^{10}{q}^{10}+3\,{t}^{10}{q}^{16}+7\,{q}^{26}{t}^{12}+3\,{q}^{28}{t}^{10}
\mbox{}+6\,{q}^{28}{t}^{18}+6\,{q}^{30}{t}^{14}+5\,{q}^{32}{t}^{16}+3\,{q}^{14}{t}^{8}+3\,{t}^{12}{q}^{18}+6\,{q}^{28}{t}^{12}+5\,{q}^{28}{t}^{20}
\mbox{}+4\,{t}^{6}{q}^{14}+2\,{t}^{4}{q}^{16}+2\,{t}^{6}{q}^{10}+4\,{q}^{36}{t}^{26}+2\,{t}^{16}{q}^{20}+2\,{t}^{12}{q}^{16}+2\,{t}^{6}{q}^{22}+5\,{q}^{16}{t}^{8}
\mbox{}+2\,{q}^{12}{t}^{8}+5\,{q}^{30}{t}^{22}+6\,{q}^{20}{t}^{8}+5\,{q}^{26}{t}^{10}+6\,{q}^{30}{t}^{20}+5\,{q}^{18}{t}^{10}+2\,{q}^{38}{t}^{28}
\mbox{}+2\,{q}^{38}{t}^{34}+8\,{q}^{30}{t}^{18}+3\,{t}^{4}{q}^{12}+2\,{t}^{4}{q}^{14}+3\,{q}^{12}{t}^{6}+4\,{q}^{18}{t}^{6}+3\,{t}^{6}{q}^{20}
\mbox{}+4\,{t}^{6}{q}^{16}+{q}^{36}{t}^{18}+{q}^{26}{t}^{24}+{q}^{34}{t}^{32}+2\,{q}^{34}{t}^{30}+6\,{q}^{26}{t}^{16}+5\,{q}^{18}{t}^{8}
\mbox{}+2\,{t}^{14}{q}^{18}+2\,{q}^{36}{t}^{20}+5\,{q}^{32}{t}^{24}+5\,{q}^{34}{t}^{26}+5\,{q}^{34}{t}^{24}+3\,{q}^{30}{t}^{12}+2\,{q}^{38}{t}^{26}
\mbox{}+6\,{t}^{10}{q}^{20}+4\,{q}^{36}{t}^{24}+2\,{q}^{14}{t}^{10}+6\,{q}^{22}{t}^{12}+3\,{q}^{20}{t}^{14}+6\,{q}^{32}{t}^{22}+2\,{t}^{32}{q}^{36}
\mbox{}+5\,{q}^{22}{t}^{14}+{q}^{14}{t}^{14}+5\,{q}^{20}{t}^{12}+2\,{t}^{18}{q}^{22}+3\,{t}^{24}{q}^{30}+5\,{q}^{26}{t}^{18}+{t}^{30}{q}^{30}
\mbox{}+{q}^{30}{t}^{28}+{q}^{20}{t}^{20}+5\,{q}^{24}{t}^{16}+{q}^{18}{t}^{4}+6\,{q}^{32}{t}^{18}+{t}^{4}{q}^{20}+{q}^{34}{t}^{34}+2\,{q}^{26}{t}^{22}
\mbox{}+{q}^{28}{t}^{26}+{t}^{6}{q}^{24}+{q}^{40}{t}^{38}+{t}^{24}{q}^{24}+{q}^{10}{t}^{2}+{q}^{38}{t}^{38}+2\,{t}^{32}{q}^{38}+{t}^{18}{q}^{20}
\mbox{}+{q}^{18}{t}^{16}+{q}^{10}{t}^{8}+{t}^{8}{q}^{8}+4\,{q}^{24}{t}^{8}+4\,{q}^{34}{t}^{18}+2\,{q}^{30}{t}^{26}+6\,{q}^{24}{t}^{10}+3\,{t}^{28}{q}^{34}
\mbox{}+3\,{q}^{24}{t}^{18}+8\,{q}^{28}{t}^{14}+{t}^{22}{q}^{24}+3\,{t}^{30}{q}^{36}+3\,{q}^{28}{t}^{22}+3\,{q}^{26}{t}^{20}+8\,{q}^{26}{t}^{14}
\mbox{}+2\,{t}^{28}{q}^{32}+5\,{q}^{34}{t}^{20}+3\,{t}^{30}{q}^{38}+3\,{t}^{26}{q}^{32}+2\,{q}^{28}{t}^{24}+5\,{q}^{22}{t}^{8}+{t}^{36}{q}^{40}
\mbox{}+3\,{q}^{36}{t}^{22}+7\,{q}^{30}{t}^{16}+2\,{q}^{26}{t}^{8}+2\,{q}^{24}{t}^{20}
\Big) = \\ = \bf{ -{\frac {{a}^{2}t+1}{ ( -1+{q}^{2} )
{a}^{7}{t}^{103/2}{q}^{42}}}}\Big( {q}^{42}{t}^{48}{a}^{12}+ (
{q}^{34}{t}^{37}+{q}^{42}{t}^{41}+{q}^{50}{t}^{47}+{q}^{52}{t}^{47}+{q}^{54}{t}^{47}+{q}^{36}{t}^{39}+{q}^{38}{t}^{41}
\mbox{}+{q}^{46}{t}^{43}+{q}^{30}{t}^{35}+{q}^{32}{t}^{37}+{q}^{48}{t}^{45}+{q}^{44}{t}^{47}+{q}^{38}{t}^{39}+{q}^{50}{t}^{45}
\mbox{}+{q}^{36}{t}^{41}+{q}^{40}{t}^{41}+{q}^{44}{t}^{43}+{q}^{42}{t}^{43}+{q}^{46}{t}^{45}+{q}^{42}{t}^{45}+{t}^{45}{q}^{44}
\mbox{}+{q}^{46}{t}^{47}+{t}^{47}{q}^{48}+{q}^{38}{t}^{43}+{t}^{39}{q}^{34}+{q}^{40}{t}^{45}+{t}^{43}{q}^{40}
) {a}^{10}+ (
7\,{q}^{36}{t}^{36}+8\,{q}^{40}{t}^{38}+4\,{q}^{36}{t}^{38}+{q}^{64}{t}^{46}+{q}^{44}{t}^{36}+{q}^{48}{t}^{46}+{q}^{62}{t}^{46}
\mbox{}+3\,{q}^{28}{t}^{30}+2\,{q}^{24}{t}^{28}+2\,{q}^{58}{t}^{46}+{q}^{56}{t}^{42}+6\,{q}^{38}{t}^{38}+6\,{q}^{42}{t}^{38}+2\,{q}^{60}{t}^{46}
\mbox{}+{t}^{34}{q}^{40}+7\,{q}^{42}{t}^{40}+2\,{q}^{42}{t}^{36}+4\,{q}^{52}{t}^{42}+5\,{q}^{44}{t}^{42}+3\,{q}^{42}{t}^{42}+{q}^{42}{t}^{44}
\mbox{}+3\,{q}^{46}{t}^{44}+{q}^{60}{t}^{44}+{q}^{50}{t}^{46}+4\,{q}^{54}{t}^{44}+{q}^{52}{t}^{40}+2\,{q}^{54}{t}^{42}+4\,{q}^{32}{t}^{32}
\mbox{}+{t}^{30}{q}^{32}+5\,{q}^{40}{t}^{40}+6\,{q}^{38}{t}^{36}+4\,{t}^{34}{q}^{36}+{q}^{28}{t}^{28}+2\,{q}^{38}{t}^{34}+2\,{q}^{34}{t}^{32}
\mbox{}+5\,{q}^{34}{t}^{36}+{t}^{32}{q}^{36}+2\,{t}^{30}{q}^{30}+2\,{q}^{26}{t}^{30}+5\,{q}^{34}{t}^{34}+4\,{q}^{48}{t}^{44}+5\,{q}^{50}{t}^{44}
\mbox{}+{q}^{20}{t}^{24}+6\,{q}^{46}{t}^{42}+2\,{q}^{52}{t}^{46}+8\,{q}^{44}{t}^{40}+4\,{q}^{44}{t}^{38}+5\,{q}^{32}{t}^{34}+{q}^{22}{t}^{26}
\mbox{}+{q}^{36}{t}^{40}+7\,{q}^{48}{t}^{42}+2\,{q}^{58}{t}^{44}+2\,{q}^{32}{t}^{36}+{q}^{34}{t}^{38}+5\,{q}^{52}{t}^{44}+2\,{q}^{54}{t}^{46}
\mbox{}+2\,{q}^{44}{t}^{44}+2\,{q}^{30}{t}^{34}+3\,{q}^{28}{t}^{32}+4\,{t}^{36}{q}^{40}+6\,{q}^{46}{t}^{40}+4\,{q}^{30}{t}^{32}+2\,{q}^{40}{t}^{42}
\mbox{}+4\,{q}^{48}{t}^{40}+2\,{q}^{50}{t}^{40}+5\,{q}^{50}{t}^{42}+{q}^{48}{t}^{38}+2\,{q}^{46}{t}^{38}+{q}^{24}{t}^{26}+2\,{q}^{26}{t}^{28}
\mbox{}+3\,{q}^{38}{t}^{40}+3\,{q}^{56}{t}^{44}+3\,{q}^{56}{t}^{46} ) {a}^{8}
\mbox{}+ (
2\,{q}^{18}{t}^{19}+9\,{q}^{56}{t}^{43}+4\,{q}^{64}{t}^{43}+9\,{t}^{39}{q}^{54}+5\,{t}^{31}{q}^{40}+2\,{q}^{66}{t}^{43}
\mbox{}+9\,{q}^{42}{t}^{33}+5\,{t}^{29}{q}^{36}+{t}^{29}{q}^{40}+{t}^{39}{q}^{60}+3\,{q}^{64}{t}^{45}+2\,{t}^{41}{q}^{62}+{q}^{42}{t}^{41}
\mbox{}+5\,{q}^{24}{t}^{23}+20\,{q}^{42}{t}^{35}+5\,{t}^{37}{q}^{52}+14\,{q}^{54}{t}^{41}+19\,{q}^{38}{t}^{33}+19\,{q}^{42}{t}^{37}
\mbox{}+10\,{q}^{48}{t}^{41}+13\,{q}^{44}{t}^{39}+22\,{q}^{40}{t}^{35}+18\,{q}^{46}{t}^{39}+9\,{q}^{30}{t}^{27}+14\,{t}^{39}{q}^{52}
\mbox{}+2\,{t}^{39}{q}^{58}+{t}^{37}{q}^{56}+22\,{q}^{44}{t}^{37}+15\,{t}^{37}{q}^{48}+{q}^{46}{t}^{43}+{q}^{54}{t}^{45}+5\,{t}^{41}{q}^{60}
\mbox{}+9\,{q}^{26}{t}^{27}+{q}^{28}{t}^{23}+4\,{q}^{20}{t}^{21}+{t}^{33}{q}^{48}+15\,{q}^{52}{t}^{41}+6\,{q}^{38}{t}^{37}+10\,{q}^{36}{t}^{35}
\mbox{}+13\,{q}^{34}{t}^{33}+{q}^{14}{t}^{17}+15\,{q}^{32}{t}^{31}+5\,{t}^{35}{q}^{48}+{t}^{19}{q}^{20}+2\,{t}^{33}{q}^{46}+{q}^{28}{t}^{31}
\mbox{}+18\,{q}^{38}{t}^{35}+20\,{q}^{36}{t}^{33}+3\,{q}^{40}{t}^{39}+8\,{q}^{42}{t}^{39}+13\,{q}^{40}{t}^{37}+2\,{t}^{37}{q}^{54}
\mbox{}+9\,{q}^{58}{t}^{43}+3\,{q}^{44}{t}^{41}+2\,{q}^{58}{t}^{45}+2\,{t}^{35}{q}^{50}+{q}^{38}{t}^{39}+13\,{q}^{50}{t}^{41}+6\,{q}^{52}{t}^{43}
\mbox{}+{q}^{56}{t}^{45}+20\,{q}^{48}{t}^{39}+2\,{t}^{21}{q}^{22}+14\,{q}^{32}{t}^{29}+{q}^{72}{t}^{45}+8\,{q}^{24}{t}^{25}+18\,{q}^{50}{t}^{39}
\mbox{}+8\,{q}^{54}{t}^{43}+{t}^{41}{q}^{64}+9\,{t}^{31}{q}^{38}+2\,{t}^{23}{q}^{26}+{t}^{21}{q}^{24}+{t}^{31}{q}^{44}+15\,{q}^{44}{t}^{35}
\mbox{}+{q}^{30}{t}^{33}+2\,{q}^{68}{t}^{45}+3\,{q}^{66}{t}^{45}+4\,{q}^{50}{t}^{43}+{t}^{35}{q}^{52}+18\,{q}^{34}{t}^{31}+14\,{q}^{30}{t}^{29}
\mbox{}+{q}^{16}{t}^{17}+{q}^{12}{t}^{15}+6\,{q}^{62}{t}^{43}+3\,{q}^{62}{t}^{45}+2\,{q}^{36}{t}^{37}+4\,{q}^{34}{t}^{35}+6\,{q}^{32}{t}^{33}
\mbox{}+8\,{q}^{30}{t}^{31}+9\,{q}^{28}{t}^{29}+2\,{q}^{26}{t}^{29}+3\,{q}^{24}{t}^{27}+3\,{q}^{22}{t}^{25}+3\,{q}^{20}{t}^{23}+3\,{q}^{18}{t}^{21}
\mbox{}+8\,{q}^{60}{t}^{43}+2\,{q}^{16}{t}^{19}+5\,{q}^{28}{t}^{25}+15\,{q}^{36}{t}^{31}+9\,{q}^{34}{t}^{29}+5\,{q}^{32}{t}^{27}
\mbox{}+2\,{q}^{30}{t}^{25}+15\,{q}^{40}{t}^{33}+{q}^{70}{t}^{45}+{q}^{68}{t}^{43}+2\,{t}^{27}{q}^{34}+{t}^{25}{q}^{32}+12\,{q}^{56}{t}^{41}
\mbox{}+2\,{q}^{48}{t}^{43}+6\,{q}^{22}{t}^{23}+6\,{q}^{46}{t}^{41}+9\,{t}^{35}{q}^{46}+2\,{t}^{31}{q}^{42}+5\,{t}^{33}{q}^{44}+19\,{q}^{46}{t}^{37}
\mbox{}+2\,{t}^{29}{q}^{38}+{t}^{27}{q}^{36}+8\,{q}^{58}{t}^{41}+3\,{q}^{60}{t}^{45}+8\,{q}^{26}{t}^{25}+12\,{q}^{28}{t}^{27}+9\,{t}^{37}{q}^{50}
\mbox{}+5\,{t}^{39}{q}^{56} ) {a}^{6}+ (
25\,{q}^{54}{t}^{38}+13\,{q}^{62}{t}^{40}+22\,{q}^{52}{t}^{36}+6\,{q}^{68}{t}^{42}+3\,{q}^{70}{t}^{44}+23\,{q}^{40}{t}^{30}
\mbox{}+8\,{q}^{18}{t}^{18}+19\,{q}^{44}{t}^{36}+{q}^{76}{t}^{44}+9\,{q}^{64}{t}^{42}+2\,{q}^{66}{t}^{44}+2\,{q}^{68}{t}^{40}+2\,{q}^{72}{t}^{44}
\mbox{}+8\,{t}^{24}{q}^{24}+25\,{q}^{52}{t}^{38}+2\,{q}^{72}{t}^{42}+2\,{q}^{68}{t}^{44}+4\,{q}^{56}{t}^{42}+28\,{t}^{32}{q}^{38}
\mbox{}+9\,{q}^{52}{t}^{34}+16\,{q}^{54}{t}^{36}+5\,{q}^{62}{t}^{38}+{q}^{70}{t}^{40}+29\,{q}^{50}{t}^{36}+23\,{q}^{48}{t}^{34}
\mbox{}+32\,{q}^{42}{t}^{32}+23\,{q}^{44}{t}^{32}+9\,{q}^{60}{t}^{38}+29\,{q}^{42}{t}^{34}+{q}^{22}{t}^{16}+{q}^{42}{t}^{38}
\mbox{}+9\,{q}^{40}{t}^{28}+6\,{q}^{16}{t}^{16}+15\,{q}^{60}{t}^{40}+{q}^{66}{t}^{38}+19\,{t}^{34}{q}^{40}+{q}^{54}{t}^{32}+11\,{q}^{42}{t}^{36}
\mbox{}+9\,{q}^{48}{t}^{32}+16\,{q}^{42}{t}^{30}+{t}^{8}{q}^{6}+16\,{q}^{46}{t}^{32}+2\,{q}^{44}{t}^{28}+{q}^{52}{t}^{42}+9\,{q}^{62}{t}^{42}
\mbox{}+9\,{q}^{44}{t}^{30}+{q}^{62}{t}^{44}+{q}^{64}{t}^{44}+{q}^{50}{t}^{30}+{q}^{46}{t}^{28}+2\,{q}^{52}{t}^{32}+2\,{q}^{64}{t}^{38}
\mbox{}+2\,{q}^{60}{t}^{36}+31\,{q}^{46}{t}^{34}+5\,{q}^{50}{t}^{32}+2\,{q}^{56}{t}^{34}+{q}^{58}{t}^{34}+5\,{q}^{46}{t}^{30}+22\,{q}^{50}{t}^{38}
\mbox{}+9\,{q}^{52}{t}^{40}+2\,{q}^{54}{t}^{42}+9\,{q}^{56}{t}^{36}+5\,{q}^{54}{t}^{34}+{q}^{62}{t}^{36}+9\,{t}^{22}{q}^{22}+6\,{t}^{26}{q}^{26}
\mbox{}+{q}^{32}{t}^{32}+{q}^{38}{t}^{24}+13\,{t}^{20}{q}^{22}+9\,{t}^{30}{q}^{32}+2\,{q}^{12}{t}^{12}+{q}^{38}{t}^{36}+33\,{t}^{32}{q}^{40}
\mbox{}+{q}^{14}{t}^{12}+2\,{q}^{16}{t}^{14}+2\,{t}^{34}{q}^{36}+2\,{q}^{40}{t}^{26}+6\,{q}^{58}{t}^{42}+23\,{q}^{36}{t}^{28}+4\,{q}^{28}{t}^{28}
\mbox{}+{q}^{34}{t}^{22}+{t}^{10}{q}^{10}+2\,{q}^{28}{t}^{20}+9\,{q}^{36}{t}^{26}+2\,{t}^{16}{q}^{20}+5\,{q}^{30}{t}^{22}+{q}^{30}{t}^{20}
\mbox{}+16\,{q}^{38}{t}^{28}+9\,{q}^{38}{t}^{34}+18\,{q}^{26}{t}^{24}+6\,{q}^{34}{t}^{32}+22\,{q}^{34}{t}^{30}+{t}^{14}{q}^{18}
\mbox{}+9\,{q}^{32}{t}^{24}+16\,{q}^{34}{t}^{26}+5\,{q}^{34}{t}^{24}+5\,{q}^{38}{t}^{26}+2\,{q}^{36}{t}^{24}+2\,{q}^{32}{t}^{22}
\mbox{}+2\,{q}^{18}{t}^{20}+16\,{t}^{32}{q}^{36}+4\,{q}^{14}{t}^{14}+5\,{t}^{18}{q}^{22}+16\,{t}^{24}{q}^{30}+{q}^{26}{t}^{18}
\mbox{}+2\,{t}^{30}{q}^{30}+14\,{q}^{30}{t}^{28}+9\,{q}^{20}{t}^{20}+{q}^{42}{t}^{26}+14\,{q}^{54}{t}^{40}+5\,{q}^{58}{t}^{36}
\mbox{}+15\,{q}^{26}{t}^{22}+16\,{q}^{28}{t}^{26}+8\,{t}^{18}{q}^{20}+5\,{q}^{18}{t}^{16}+{q}^{20}{t}^{22}+4\,{q}^{44}{t}^{38}
\mbox{}+{q}^{22}{t}^{24}+2\,{q}^{10}{t}^{12}+33\,{q}^{44}{t}^{34}+2\,{q}^{12}{t}^{14}+25\,{q}^{30}{t}^{26}+29\,{t}^{28}{q}^{34}
\mbox{}+2\,{q}^{24}{t}^{18}+15\,{t}^{22}{q}^{24}+30\,{t}^{30}{q}^{36}+9\,{q}^{28}{t}^{22}+5\,{q}^{26}{t}^{20}+25\,{t}^{28}{q}^{32}
\mbox{}+5\,{q}^{42}{t}^{28}+31\,{t}^{30}{q}^{38}+2\,{q}^{48}{t}^{30}+22\,{t}^{26}{q}^{32}+2\,{q}^{16}{t}^{18}+3\,{q}^{14}{t}^{16}
\mbox{}+8\,{q}^{60}{t}^{42}+20\,{q}^{28}{t}^{24}+4\,{t}^{36}{q}^{40}+{q}^{46}{t}^{40}+{q}^{8}{t}^{10}+{q}^{78}{t}^{44}+16\,{q}^{50}{t}^{34}
\mbox{}+8\,{q}^{66}{t}^{42}+30\,{q}^{48}{t}^{36}+2\,{q}^{48}{t}^{40}+6\,{q}^{50}{t}^{40}+28\,{q}^{46}{t}^{36}+16\,{q}^{48}{t}^{38}
\mbox{}+2\,{q}^{74}{t}^{44}+16\,{q}^{56}{t}^{40}+5\,{q}^{66}{t}^{40}+20\,{q}^{56}{t}^{38}+18\,{q}^{58}{t}^{40}+9\,{q}^{46}{t}^{38}
\mbox{}+8\,{q}^{64}{t}^{40}+9\,{q}^{24}{t}^{20}+15\,{q}^{58}{t}^{38}+{q}^{74}{t}^{42}+4\,{q}^{70}{t}^{42}
) {a}^{4}+ (
{t}^{5}{q}^{4}+2\,{t}^{25}{q}^{44}+4\,{t}^{39}{q}^{70}+{t}^{39}{q}^{74}+3\,{t}^{41}{q}^{74}+2\,{t}^{39}{q}^{54}+10\,{t}^{39}{q}^{64}
\mbox{}+16\,{t}^{31}{q}^{40}+12\,{t}^{35}{q}^{58}+11\,{t}^{37}{q}^{62}+5\,{t}^{41}{q}^{70}+4\,{t}^{11}{q}^{14}+17\,{t}^{31}{q}^{48}
\mbox{}+{t}^{35}{q}^{66}+17\,{t}^{25}{q}^{36}+2\,{t}^{17}{q}^{28}+4\,{t}^{41}{q}^{72}+8\,{q}^{42}{t}^{33}+2\,{t}^{19}{q}^{32}+3\,{t}^{9}{q}^{10}
\mbox{}+{t}^{43}{q}^{80}+17\,{t}^{33}{q}^{52}+14\,{t}^{29}{q}^{36}+4\,{t}^{11}{q}^{12}+4\,{t}^{13}{q}^{18}+7\,{t}^{27}{q}^{44}
\mbox{}+4\,{t}^{37}{q}^{66}+26\,{t}^{29}{q}^{40}+23\,{t}^{27}{q}^{38}+16\,{t}^{35}{q}^{56}+9\,{t}^{39}{q}^{60}+5\,{t}^{41}{q}^{68}
\mbox{}+2\,{t}^{41}{q}^{62}+{q}^{24}{t}^{23}+{q}^{42}{t}^{35}+{t}^{17}{q}^{30}+9\,{t}^{37}{q}^{52}+7\,{t}^{35}{q}^{60}+{t}^{37}{q}^{70}
\mbox{}+17\,{t}^{29}{q}^{44}+9\,{t}^{39}{q}^{66}+{t}^{31}{q}^{58}+2\,{q}^{30}{t}^{27}+7\,{t}^{23}{q}^{36}+2\,{t}^{13}{q}^{20}+{t}^{39}{q}^{52}
\mbox{}+7\,{t}^{21}{q}^{32}+12\,{t}^{25}{q}^{38}+12\,{t}^{29}{q}^{46}+20\,{t}^{35}{q}^{54}+7\,{t}^{25}{q}^{40}+7\,{t}^{39}{q}^{58}
\mbox{}+16\,{t}^{37}{q}^{56}+4\,{t}^{29}{q}^{50}+2\,{t}^{35}{q}^{64}+2\,{t}^{37}{q}^{48}+16\,{t}^{37}{q}^{58}+{t}^{41}{q}^{60}
\mbox{}+23\,{t}^{31}{q}^{46}+2\,{t}^{29}{q}^{52}+{t}^{27}{q}^{50}+16\,{q}^{28}{t}^{23}+{t}^{11}{q}^{10}+{t}^{19}{q}^{34}+{t}^{29}{q}^{54}
\mbox{}+4\,{t}^{35}{q}^{62}+24\,{t}^{33}{q}^{48}+23\,{t}^{29}{q}^{42}+2\,{t}^{23}{q}^{40}+2\,{t}^{37}{q}^{68}+2\,{t}^{15}{q}^{24}
\mbox{}+14\,{t}^{37}{q}^{60}+{q}^{2}{t}^{3}+10\,{t}^{17}{q}^{20}+14\,{t}^{35}{q}^{48}+9\,{t}^{15}{q}^{18}+3\,{t}^{19}{q}^{20}+7\,{t}^{15}{q}^{20}
\mbox{}+6\,{t}^{13}{q}^{16}+11\,{t}^{17}{q}^{22}+{t}^{9}{q}^{14}+14\,{t}^{19}{q}^{24}+22\,{t}^{33}{q}^{46}+4\,{t}^{41}{q}^{66}+2\,{t}^{39}{q}^{72}
\mbox{}+16\,{t}^{21}{q}^{28}+12\,{t}^{21}{q}^{30}+12\,{t}^{19}{q}^{26}+4\,{t}^{15}{q}^{22}+2\,{t}^{31}{q}^{56}+13\,{t}^{37}{q}^{54}
\mbox{}+{t}^{11}{q}^{18}+19\,{t}^{35}{q}^{50}+20\,{t}^{23}{q}^{30}+17\,{t}^{23}{q}^{32}+2\,{t}^{21}{q}^{22}+22\,{t}^{25}{q}^{34}
\mbox{}+2\,{t}^{41}{q}^{76}+{q}^{32}{t}^{29}+{t}^{43}{q}^{78}+4\,{t}^{27}{q}^{46}+7\,{t}^{29}{q}^{48}+4\,{t}^{17}{q}^{26}+6\,{t}^{39}{q}^{68}
\mbox{}+{t}^{7}{q}^{6}+3\,{t}^{41}{q}^{64}+7\,{t}^{17}{q}^{24}+8\,{t}^{31}{q}^{38}+7\,{t}^{23}{q}^{26}+9\,{t}^{21}{q}^{24}+26\,{t}^{31}{q}^{44}
\mbox{}+{t}^{43}{q}^{76}+3\,{q}^{44}{t}^{35}+{t}^{43}{q}^{72}+{t}^{7}{q}^{10}+{t}^{13}{q}^{22}+21\,{t}^{35}{q}^{52}+22\,{t}^{33}{q}^{50}
\mbox{}+17\,{t}^{27}{q}^{40}+2\,{t}^{27}{q}^{48}+{t}^{9}{q}^{8}+4\,{t}^{25}{q}^{42}+16\,{t}^{21}{q}^{26}+10\,{t}^{19}{q}^{22}+4\,{t}^{33}{q}^{58}
\mbox{}+4\,{t}^{19}{q}^{30}+4\,{t}^{17}{q}^{18}+5\,{t}^{13}{q}^{14}+5\,{t}^{15}{q}^{16}+{t}^{41}{q}^{78}+12\,{t}^{27}{q}^{42}+12\,{t}^{33}{q}^{54}
\mbox{}+4\,{q}^{28}{t}^{25}+2\,{q}^{36}{t}^{31}+5\,{q}^{34}{t}^{29}+2\,{t}^{9}{q}^{12}+9\,{q}^{32}{t}^{27}+13\,{q}^{30}{t}^{25}+7\,{t}^{19}{q}^{28}
\mbox{}+7\,{t}^{31}{q}^{52}+3\,{q}^{40}{t}^{33}+2\,{t}^{33}{q}^{60}+4\,{t}^{31}{q}^{54}+19\,{t}^{27}{q}^{34}+21\,{t}^{25}{q}^{32}
\mbox{}+{t}^{23}{q}^{42}+{t}^{5}{q}^{6}+12\,{t}^{23}{q}^{34}+8\,{t}^{35}{q}^{46}+24\,{t}^{31}{q}^{42}+16\,{t}^{33}{q}^{44}+7\,{t}^{33}{q}^{56}
\mbox{}+22\,{t}^{29}{q}^{38}+2\,{t}^{7}{q}^{8}+{t}^{43}{q}^{82}+24\,{t}^{27}{q}^{36}+{t}^{25}{q}^{46}+{t}^{43}{q}^{74}+10\,{t}^{39}{q}^{62}
\mbox{}+7\,{t}^{37}{q}^{64}+4\,{t}^{23}{q}^{38}+2\,{t}^{11}{q}^{16}+{t}^{33}{q}^{62}+2\,{t}^{21}{q}^{36}+5\,{t}^{37}{q}^{50}+{t}^{21}{q}^{38}
\mbox{}+4\,{t}^{21}{q}^{34}+{t}^{13}{q}^{12}+{t}^{15}{q}^{26}+4\,{t}^{39}{q}^{56}+12\,{t}^{31}{q}^{50}
) {a}^{2}
\mbox{}+{q}^{40}{t}^{30}+{q}^{44}{t}^{22}+1+{q}^{56}{t}^{28}+{q}^{60}{t}^{30}+{q}^{72}{t}^{36}+{q}^{46}{t}^{24}+{q}^{58}{t}^{30}
\mbox{}+{q}^{62}{t}^{32}+{q}^{48}{t}^{24}+4\,{q}^{52}{t}^{34}+{q}^{52}{t}^{26}+{q}^{54}{t}^{36}+{q}^{62}{t}^{38}+{q}^{70}{t}^{40}
\mbox{}+{q}^{54}{t}^{28}+{q}^{50}{t}^{26}+{q}^{48}{t}^{34}+{q}^{42}{t}^{22}+{q}^{68}{t}^{34}+{q}^{76}{t}^{40}+{q}^{72}{t}^{40}
\mbox{}+{q}^{70}{t}^{36}+{q}^{76}{t}^{38}+{q}^{78}{t}^{40}+{q}^{80}{t}^{40}+{q}^{8}{t}^{4}+{q}^{74}{t}^{40}+{t}^{6}{q}^{8}+{q}^{64}{t}^{32}
\mbox{}+{q}^{44}{t}^{32}+{q}^{60}{t}^{38}+{q}^{84}{t}^{42}+4\,{q}^{22}{t}^{16}+6\,{q}^{40}{t}^{28}+{q}^{74}{t}^{38}+2\,{q}^{66}{t}^{38}
\mbox{}+6\,{q}^{54}{t}^{32}+5\,{q}^{48}{t}^{32}+3\,{q}^{42}{t}^{30}+3\,{q}^{46}{t}^{32}+8\,{q}^{44}{t}^{28}+6\,{q}^{44}{t}^{30}+3\,{q}^{42}{t}^{24}
\mbox{}+{q}^{4}{t}^{2}+{q}^{6}{t}^{4}+2\,{q}^{60}{t}^{32}+3\,{q}^{54}{t}^{30}+2\,{q}^{48}{t}^{26}+2\,{q}^{64}{t}^{34}+2\,{q}^{44}{t}^{24}
\mbox{}+4\,{q}^{64}{t}^{36}+6\,{q}^{50}{t}^{30}+5\,{q}^{56}{t}^{32}+5\,{q}^{44}{t}^{26}+6\,{q}^{46}{t}^{28}+2\,{q}^{70}{t}^{38}+3\,{q}^{58}{t}^{32}
\mbox{}+3\,{q}^{46}{t}^{26}+5\,{q}^{52}{t}^{30}+7\,{q}^{52}{t}^{32}+2\,{q}^{64}{t}^{38}+4\,{q}^{60}{t}^{36}+6\,{q}^{50}{t}^{32}+3\,{q}^{68}{t}^{38}
\mbox{}+6\,{q}^{56}{t}^{34}+2\,{q}^{72}{t}^{38}+2\,{q}^{68}{t}^{36}+5\,{q}^{58}{t}^{34}+5\,{q}^{48}{t}^{28}+5\,{q}^{60}{t}^{34}+3\,{q}^{62}{t}^{34}
\mbox{}+7\,{q}^{46}{t}^{30}+3\,{q}^{66}{t}^{36}+2\,{q}^{56}{t}^{36}+5\,{q}^{54}{t}^{34}+3\,{q}^{50}{t}^{28}+4\,{q}^{62}{t}^{36}+6\,{q}^{38}{t}^{24}
\mbox{}+{q}^{14}{t}^{12}+{t}^{10}{q}^{12}+{q}^{24}{t}^{12}+{q}^{38}{t}^{20}+2\,{q}^{40}{t}^{22}+8\,{q}^{40}{t}^{26}+{q}^{36}{t}^{28}
\mbox{}+2\,{t}^{14}{q}^{24}+6\,{q}^{34}{t}^{22}+5\,{q}^{32}{t}^{20}+2\,{q}^{28}{t}^{16}+3\,{q}^{38}{t}^{22}+2\,{t}^{10}{q}^{16}+5\,{q}^{28}{t}^{18}
\mbox{}+{q}^{32}{t}^{16}+{q}^{14}{t}^{8}+3\,{t}^{12}{q}^{18}+6\,{q}^{28}{t}^{20}+{t}^{6}{q}^{10}+5\,{q}^{36}{t}^{26}+2\,{t}^{16}{q}^{20}
\mbox{}+3\,{t}^{12}{q}^{16}+{q}^{16}{t}^{8}+2\,{q}^{12}{t}^{8}+5\,{q}^{30}{t}^{22}+6\,{q}^{30}{t}^{20}+{q}^{18}{t}^{10}+3\,{q}^{38}{t}^{28}
\mbox{}+3\,{q}^{30}{t}^{18}+{q}^{12}{t}^{6}+{q}^{36}{t}^{18}+3\,{q}^{26}{t}^{16}+2\,{t}^{14}{q}^{18}+2\,{q}^{36}{t}^{20}+4\,{q}^{32}{t}^{24}
\mbox{}+2\,{q}^{34}{t}^{26}+6\,{q}^{34}{t}^{24}+7\,{q}^{38}{t}^{26}+{t}^{10}{q}^{20}+8\,{q}^{36}{t}^{24}+2\,{q}^{14}{t}^{10}+{q}^{22}{t}^{12}
\mbox{}+4\,{q}^{20}{t}^{14}+7\,{q}^{32}{t}^{22}+3\,{q}^{22}{t}^{14}+2\,{q}^{20}{t}^{12}+{t}^{18}{q}^{22}+{t}^{24}{q}^{30}+5\,{q}^{26}{t}^{18}
\mbox{}+5\,{q}^{24}{t}^{16}+{q}^{40}{t}^{20}+6\,{q}^{42}{t}^{26}+2\,{q}^{56}{t}^{30}+2\,{q}^{32}{t}^{18}+2\,{q}^{52}{t}^{28}+3\,{q}^{58}{t}^{36}
\mbox{}+{q}^{10}{t}^{8}+{q}^{34}{t}^{18}+4\,{q}^{24}{t}^{18}+{q}^{28}{t}^{14}+2\,{q}^{28}{t}^{22}+3\,{q}^{26}{t}^{20}+{q}^{26}{t}^{14}
\mbox{}+8\,{q}^{42}{t}^{28}+3\,{q}^{34}{t}^{20}+8\,{q}^{48}{t}^{30}+{q}^{66}{t}^{34}+2\,{q}^{50}{t}^{34}+5\,{q}^{36}{t}^{22}
\mbox{}+{q}^{30}{t}^{16}+{q}^{24}{t}^{20}+5\,{q}^{40}{t}^{24} \Big)
$

\bigskip
\noindent
$\boxed{P^{T[7,7k+1]}_{[1]}=\dfrac{\{A\} t^6}{\{t\} A^6
q^{42k+6}}{\cal{P}}^{T[7,7k+1]}_{[1]}}$

\begin{itemize}
\item{HOMFLY case}
\end{itemize}
$$H^{T[7,1]}_{[1]}=\dfrac{\{A\}}{\{q\} A^6}$$ \bigskip \\
$H^{T[7,8]}_{[1]}=\dfrac{\{A\}}{\{q\} A^6 q^{42}} \Big( {q}^{42}{A}^{12}+ (
-{q}^{32}-3\,{q}^{38}-3\,{q}^{44}-3\,{q}^{46}-3\,{q}^{42}-2\,{q}^{50}-{q}^{52}-2\,{q}^{36}-3\,{q}^{40}-{q}^{30}-{q}^{54}-2\,{q}^{48}-2\,{q}^{34}
\mbox{} ) {A}^{10}
\mbox{}+ (
17\,{q}^{48}+7\,{q}^{56}+{q}^{64}+7\,{q}^{28}+8\,{q}^{54}+13\,{q}^{34}+17\,{q}^{46}+20\,{q}^{40}+3\,{q}^{24}+20\,{q}^{44}+19\,{q}^{42}
\mbox{}+{q}^{22}+17\,{q}^{36}+3\,{q}^{60}+17\,{q}^{38}+13\,{q}^{50}+12\,{q}^{32}+4\,{q}^{58}+4\,{q}^{26}+{q}^{20}+8\,{q}^{30}+{q}^{62}+12\,{q}^{52}
\mbox{} ) {A}^{8}
\mbox{}+ (
-3\,{q}^{68}-21\,{q}^{26}-59\,{q}^{42}-46\,{q}^{50}-59\,{q}^{40}-8\,{q}^{64}-55\,{q}^{38}-{q}^{70}-11\,{q}^{62}-17\,{q}^{60}-{q}^{72}
\mbox{}-5\,{q}^{66}-46\,{q}^{34}-53\,{q}^{48}-11\,{q}^{22}-3\,{q}^{16}-{q}^{14}-55\,{q}^{46}-17\,{q}^{24}-5\,{q}^{18}-8\,{q}^{20}-41\,{q}^{32}
\mbox{}-28\,{q}^{56}-{q}^{12}-53\,{q}^{36}-41\,{q}^{52}-59\,{q}^{44}-34\,{q}^{54}-21\,{q}^{58}-28\,{q}^{28}-34\,{q}^{30}
) {A}^{6}+ (
10\,{q}^{68}+45\,{q}^{26}+95\,{q}^{42}+79\,{q}^{50}+90\,{q}^{40}+20\,{q}^{64}+91\,{q}^{38}+8\,{q}^{70}+3\,{q}^{74}+29\,{q}^{62}
\mbox{}+34\,{q}^{60}+4\,{q}^{72}+16\,{q}^{66}+79\,{q}^{34}+82\,{q}^{48}+{q}^{78}+29\,{q}^{22}+10\,{q}^{16}+8\,{q}^{14}+91\,{q}^{46}+{q}^{6}+{q}^{8}
\mbox{}+3\,{q}^{10}+{q}^{76}+34\,{q}^{24}+16\,{q}^{18}+20\,{q}^{20}+68\,{q}^{32}+51\,{q}^{56}+4\,{q}^{12}+82\,{q}^{36}+68\,{q}^{52}+90\,{q}^{44}
\mbox{}+63\,{q}^{54}+45\,{q}^{58}+51\,{q}^{28}+63\,{q}^{30} ) {A}^{4}
\mbox{}+ (
-{q}^{2}-13\,{q}^{68}-40\,{q}^{26}-73\,{q}^{42}-63\,{q}^{50}-71\,{q}^{40}-22\,{q}^{64}-70\,{q}^{38}-10\,{q}^{70}-5\,{q}^{74}-28\,{q}^{62}
\mbox{}-33\,{q}^{60}-7\,{q}^{72}-18\,{q}^{66}-63\,{q}^{34}-66\,{q}^{48}-2\,{q}^{78}-28\,{q}^{22}-13\,{q}^{16}-10\,{q}^{14}-70\,{q}^{46}-2\,{q}^{6}
\mbox{}-3\,{q}^{8}-5\,{q}^{10}-3\,{q}^{76}-33\,{q}^{24}-18\,{q}^{18}-22\,{q}^{20}-57\,{q}^{32}-45\,{q}^{56}-{q}^{4}-7\,{q}^{12}-{q}^{80}-66\,{q}^{36}
\mbox{}-57\,{q}^{52}-71\,{q}^{44}-{q}^{82}-52\,{q}^{54}-40\,{q}^{58}-45\,{q}^{28}-52\,{q}^{30}
)
{A}^{2}+23\,{q}^{40}+9\,{q}^{64}+21\,{q}^{42}+18\,{q}^{50}+2\,{q}^{74}+9\,{q}^{62}+20\,{q}^{38}+4\,{q}^{70}+6\,{q}^{66}
\mbox{}+18\,{q}^{34}+12\,{q}^{26}+{q}^{78}+9\,{q}^{22}+6\,{q}^{16}+4\,{q}^{14}+20\,{q}^{46}+{q}^{6}+2\,{q}^{8}+2\,{q}^{10}+13\,{q}^{60}+4\,{q}^{72}+6\,{q}^{18}
\mbox{}+9\,{q}^{20}+22\,{q}^{48}+16\,{q}^{56}+{q}^{4}+4\,{q}^{12}+{q}^{80}+22\,{q}^{36}+{q}^{84}+19\,{q}^{52}+23\,{q}^{44}+2\,{q}^{76}+13\,{q}^{24}
\mbox{}+1+6\,{q}^{68}+19\,{q}^{32}+16\,{q}^{54}+12\,{q}^{58}+16\,{q}^{28}+16\,{q}^{30}
\Big)$

\bigskip
\fr{
H^{T[7,7k+1]}_{[1]}=\dfrac{\{A\}}{\{q\} A^6 q^{42k}}
{\cal{H}}^{T[7,7k+1]}_{[1]}=
q^{-42k-6}s_{[7]}^{*}-q^{-28k-4}s_{[6,1]}^{*}+q^{-14k-2}s_{[5,1,1]}^{*}-s_{[4,1,1,1]}^{*}+q^{14k+2}s_{[3,1,1,1,1]}^{*}-\\-q^{28k+4}s_{[2,1,1,1,1,1]}^{*}+q^{42k+6}s_{[1,1,1,1,1,1,1]}^{*}}

and the results coincides with well known HOMFLY polynomials, see
(\ref{Wrepfund}) and
(\ref{113}).

\begin{itemize}
\item{Floer case}
\end{itemize}
$$F^{T[7,1]}_{[1]}=1$$ \\
$F^{T[7,8]}_{[1]}=\bf{t}^{42}{q}^{84}+{a}^{2}{t}^{43}{q}^{82}+{t}^{40}{q}^{70}+{a}^{2}{t}^{39}{q}^{66}+{t}^{36}{q}^{56}+{a}^{2}{t}^{37}{q}^{50}+{t}^{30}{q}^{42}+{t}^{25}{a}^{2}{q}^{34}+{t}^{22}{q}^{28}+{a}^{2}{t}^{15}{q}^{18}+{t}^{12}{q}^{14}+{a}^{2}{t}^{3}{q}^{2}+1$
 \\
\begin{itemize}
\item{Alexander case}
\end{itemize}
$$A^{T[7,1]}_{[1]}=1$$  \\
$$A^{T[7,8]}_{[1]}=\bf{q}^{84}-q^{82}+q^{70}-q^{66}+q^{56}-q^{50}+q^{42}-q^{34}+q^{28}-q^{18}+q^{14}-q^{2}+1$$
 \\

\subsection{Case $(7,n)$, $n=7 k+2$ fundamental representation\label{7k+2}}
\be
\begin{array}{|c|}
\hline\\
P^{T[7,n]}_{[1]}=c^{[7]}_{[1]}M^{\ast}_{{7}}{q}^{-6\,n}+c^{[6,1]}_{[1]}M^{\ast}_{{6,1}}\,{q}^{-{\frac
{30}{7}}\,n}
\mbox{}{t}^{2/7\,n}+c^{[5,2]}_{[1]}M^{\ast}_{{5,2}}\,{q}^{-{\frac {22}{7}}\,n}
\mbox{}{t}^{4/7\,n}+c^{[5,1,1]}_{[1]}M^{\ast}_{{5,1,1}}\,{q}^{-{\frac
{20}{7}}\,n}
\mbox{}{t}^{6/7\,n}+\\+c^{[4,3]}_{[1]}M^{\ast}_{{4,3}}\,{q}^{-{\frac
{18}{7}}\,n}
\mbox{}{t}^{6/7\,n}+c^{[4,2,1]}_{[1]}M^{\ast}_{{4,2,1}}\,{q}^{-2\,n}{t}^{{\frac
{8}{7}}\,n}
\mbox{}+c^{[4,1,1,1]}_{[1]}M^{\ast}_{{4,1,1,1}}\,{q}^{-{\frac {12}{7}}\,n}
\mbox{}{t}^{{\frac
{12}{7}}\,n}+c^{[3,3,1]}_{[1]}M^{\ast}_{{3,3,1}}\,{q}^{-{\frac {12}{7}}\,n}
\mbox{}{t}^{{\frac
{10}{7}}\,n}+\\+c^{[3,2,2]}_{[1]}M^{\ast}_{{3,2,2}}\,{t}^{{\frac {12}{7}}\,n}
\mbox{}{q}^{-{\frac {10}{7}}\,n}
\mbox{}+c^{[3,2,1,1]}_{[1]}M^{\ast}_{{3,2,1,1}}\,{t}^{2\,n}{q}^{-{\frac
{8}{7}}\,n}+c^{[3,1,1,1,1]}_{[1]}M^{\ast}_{{3,1,1,1,1}}\,{t}^{{\frac
{20}{7}}\,n}
\mbox{}{q}^{-\frac{6}{7}\,n}
\mbox{}+c^{[2,2,2,1]}_{[1]}M^{\ast}_{{2,2,2,1}}\,{t}^{{\frac {18}{7}}\,n}
\mbox{}{q}^{-\frac{6}{7}\,n}+\\+c^{[2,2,1,1,1]}_{[1]}M^{\ast}_{{2,2,1,1,1}}\,{t}^{{\frac
{22}{7}}\,n}
\mbox{}{q}^{-\frac{4}{7}\,n}+c^{[2,1,1,1,1,1]}_{[1]}M^{\ast}_{{2,1,1,1,1,1}}\,{t}^{{\frac
{30}{7}}\,n}
\mbox{}{q}^{-\frac{2}{7}\,n}+c^{[1,1,1,1,1,1,1]}_{[1]}M^{\ast}_{{1,1,1,1,1,1,1}}\,{t}^{6\,n}

\\
\\
\hline
\end{array}
\ee
\fr{ \\
c^{[7]}_{[1]}=1,\ \ c^{[6,1]}_{[1]}
=-\frac{q^{4/7}(-1+t^{2})(1+q^{2}+q^{4}+q^{6}+q^{8}+q^{8}t^2+q^{10}t^{2})}{t^{4/7}(-1+q^{12}t^{2})},
\\
c^{[5,2]}_{[1]}
=-\frac{q^{16/7}(-1+t^{2})(q^{2}-t^{2})(t^4q^{12}+t^2q^{10}+2t^2q^8+q^8+2q^6t^2+q^6+2q^4+q^4t^2+2q^2+1)}{t^{8/7}(q^{10}t^{2}-1)(q^{8}t^{2}-1)}
\\
c^{[5,1,1]}_{[1]}=\frac{q^{12/7}(t^{2}+1)(-1+t^{2})^{2}(1+q^{2}+q^{4}+q^{4}t^2+q^{6}t^2+q^{6}t^{4}+q^{8}t^{4})}{t^{12/7}(q^{2}t^{2}-1)(-1+q^{10}t^{4})},
 \\
c^{[4,3]}_{[1]}
=-\frac{(-1+t^{2})(t^6q^{14}+t^4q^{12}-t^6q^{10}+t^2q^{10}-t^6q^8-t^4q^8+2t^2q^8+q^8-2q^6t^4+q^6+q^6t^2-t^4q^4-q^4t^2+q^4-q^2t^2-1)(q^{2}-t^{2})q^{8/7}}{(q^{8}t^{2}-1)(q^{6}t^{2}-1)(q^{4}t^{2}-1)t^{12/7}}\mathop{\rm
 },  \\
c^{[4,2,1]}_{[1]}=\frac{q^{2}(-1+t^{2})^{2}(q^{12}t^6+q^{12}t^8+3t^6q^{10}+t^4q^{10}+q^{10}t^8+3t^6q^8+4t^4q^8+t^2q^8+q^6t^6+5q^6t^4+4q^6t^2+q^6+2t^4q^4+
5q^4t^2+2q^4+3q^2t^2+3q^2+1)
(q^{2}-t^{2})}{t^{16/7}(q^{4}t^{2}+1)(q^{6}t^{2}-1)(q^{4}t^{2}-1)^{2}}\mathop{\rm
 },  \\
c^{[4,1,1,1]}_{[1]}
=-\frac{q^{24/7}(-1+t^{2})(-1+t^{4})(t^{6}-1)(1+t^2+q^2t^2+t^4q^2+t^4q^4+q^4t^6+q^6t^6)}{t^{24/7}(q^{2}t^{2}-1)
(q^{2}t^{4}-1)(-1+t^{6}q^{8})}\mathop{\rm  },  \\
c^{[3,3,1]}_{[1]}
=\frac{(t^{2}+1)(-1+t^{2})^{2}(q^{2}-t^{2})(q^{12}t^6+q^{12}t^8-q^{10}t^8+t^6q^{10}+t^4q^{10}-t^8q^8-t^6q^8+t^4q^8+t^2q^8-2q^6t^6-q^6t^4+2q^6t^2+q^6-2t^4q^4+q^4-q^2t^2-1)q^{10/7}}{(q^{2}t^{2}-1)(q^{4}t^{2}-1)(q^{6}t^{2}-1)(-1+q^{6}t^{4})t^{20/7}}\mathop{\rm
 },\\ c^{[3,2,2]}_{[1]}
=-\frac{q^{20/7}(t^{2}+1)(-1+t^{2})^{2}(q^8t^{12}-t^8q^8-t^6q^8+q^6t^{10}-2q^6t^6-q^6t^4+2q^4t^8+q^4t^6-t^4q^4-q^4t^2+2t^6q^2+t^4q^2-q^2t^2-q^2+t^4+t^2-1)(q^{2}-t^{2})}{t^{24/7}(q^{2}t^{2}+1)(q^{4}t^{2}-1)(-1+q^{6}t^{4})(q^{2}t^{2}-1)^{2}}\mathop{\rm
 },  \\
c^{[3,2,1,1]}_{[1]}
=-\frac{(t^{2}+1)(-1+t^{2})^{3}(q^{2}-t^{2})(q^8t^{12}+3t^{10}q^8+2t^8q^8+t^6q^8+3q^6t^{10}+5q^6t^8+4q^6t^6+q^6t^4+2q^4t^8+5q^4t^6+4t^4q^4+q^4t^2+t^6q^2+3t^4q^2+3q^2t^2+q^2+t^2+1)q^{16/7}}{(q^{2}t^{2}+1)(q^{4}t^{2}-1)(q^{4}t^{4}+q^{2}t^{2}+1)(q^{2}t^{2}-1)^{3}t^{4}}\mathop{\rm
 },  \\
c^{[3,1,1,1,1]}_{[1]} =\frac{q^{26/7}(1+t^{4
})(1+t^{2}+t^{4})(t^{2}+1)^{2}(-1+t^{2})^{4}(q^4t^8+q^4t^6+t^4q^2+t^4q^4+q^2t^2+t^2+1)}{t^{26/7}(t^{8}q^{6}-1)(q^{2}t^{2}-1)(t^{6}q^{2}-1)(q^{2}t^{4}-1)}\mathop{\rm
 }\mathop{\rm  },  \\
c^{[2,2,2,1]}_{[1]}
=-\frac{q^{26/7}(t^{2}+1)(1+t^{2}+t^{4})(-1+t^{2})^{3}(q^{2}-t^{2})(q^6t^{14}-q^6t^{10}-q^6t^8-q^6t^6+q^4t^{12}-2q^4t^6+t^{10}q^4-t^4q^4-q^4t^8+2t^8q^2+t^{10}q^2+t^6q^2-q^2t^2+t^6+t^4-1)}{t^{36/7}(q^{2}t^{2}+1)(q^{2}t^{4}-1)(q^{4}t^{2}-1)(-1+q^{4}t^{6})(q^{2}t^{2}-1)^{2}},
 \\
c^{[2,2,1,1,1]}_{[1]}
=\frac{q^{22/7}(1+t^{2}+t^{4})(t^{2}+1)^{2}(-1+t^{2})^{4}(q^4t^{12}+2t^{10}q^4+2q^4t^8+q^4t^6+t^4q^4+2t^6q^2+2t^4q^2+q^2t^2+t^8q^2+1)(q^{2}-t^{2})}{t^{30/7}(q^{2}t^{4}+1)(-1+q^{4}t^{6})(q^{2}t^{4}-1)^{2}(q^{2}t^{2}-1)^{2}}\mathop{\rm
 },  \\
c^{[2,1,1,1,1,1]}_{[1]}
=-\frac{q^{32/7}(-1+t^{2})(-1+t^{4})(t^{6}-1)(-1+t^{8})(-1+t^{10})(t^{10}q^2+t^8q^2+t^6q^2+t^4q^2+t^2+q^2t^2+1)}{t^{32/7}(t^{10}q^{4}-1)(q^{2}t^{8}-1)(q^{2}t^{2}-1)(t^{6}q^{2}-1)(q^{2}t^{4}-1)},
 \\
c^{[1,1,1,1,1,1,1]}_{[1]} =
\frac{q^{6}(t^{14}-1)(t^{12}-1)(-1+t^{10})(-1+t^{8})(t^{6}-1)(-1+t^{4})}{t^{6}(t^{12}q^{2}-1)(q^{2}t^{10}-1)(q^{2}t^{8}-1)(t^{6}q^{2}-1)(q^{2}t^{4}-1)(q^{2}t^{2}-1)}}

\subsection{Case $(7,n)$, $n=7 k+3$ fundamental representation\label{7k+3}}
\be
\begin{array}{|c|}
\hline\\
P^{T[7,n]}_{[1]}=c^{[7]}_{[1]}M^{\ast}_{{7}}{q}^{-6\,n}+c^{[6,1]}_{[1]}M^{\ast}_{{6,1}}\,{q}^{-{\frac
{30}{7}}\,n}
\mbox{}{t}^{2/7\,n}+c^{[5,2]}_{[1]}M^{\ast}_{{5,2}}\,{q}^{-{\frac {22}{7}}\,n}
\mbox{}{t}^{4/7\,n}+c^{[5,1,1]}_{[1]}M^{\ast}_{{5,1,1}}\,{q}^{-{\frac
{20}{7}}\,n}
\mbox{}{t}^{6/7\,n}+\\+c^{[4,3]}_{[1]}M^{\ast}_{{4,3}}\,{q}^{-{\frac
{18}{7}}\,n}
\mbox{}{t}^{6/7\,n}+c^{[4,2,1]}_{[1]}M^{\ast}_{{4,2,1}}\,{q}^{-2\,n}{t}^{{\frac
{8}{7}}\,n}
\mbox{}+c^{[4,1,1,1]}_{[1]}M^{\ast}_{{4,1,1,1}}\,{q}^{-{\frac {12}{7}}\,n}
\mbox{}{t}^{{\frac
{12}{7}}\,n}+c^{[3,3,1]}_{[1]}M^{\ast}_{{3,3,1}}\,{q}^{-{\frac {12}{7}}\,n}
\mbox{}{t}^{{\frac
{10}{7}}\,n}+\\+c^{[3,2,2]}_{[1]}M^{\ast}_{{3,2,2}}\,{t}^{{\frac {12}{7}}\,n}
\mbox{}{q}^{-{\frac {10}{7}}\,n}
\mbox{}+c^{[3,2,1,1]}_{[1]}M^{\ast}_{{3,2,1,1}}\,{t}^{2\,n}{q}^{-{\frac
{8}{7}}\,n}+c^{[3,1,1,1,1]}_{[1]}M^{\ast}_{{3,1,1,1,1}}\,{t}^{{\frac
{20}{7}}\,n}
\mbox{}{q}^{-\frac{6}{7}\,n}
\mbox{}+c^{[2,2,2,1]}_{[1]}M^{\ast}_{{2,2,2,1}}\,{t}^{{\frac {18}{7}}\,n}
\mbox{}{q}^{-\frac{6}{7}\,n}+\\+c^{[2,2,1,1,1]}_{[1]}M^{\ast}_{{2,2,1,1,1}}\,{t}^{{\frac
{22}{7}}\,n}
\mbox{}{q}^{-\frac{4}{7}\,n}+c^{[2,1,1,1,1,1]}_{[1]}M^{\ast}_{{2,1,1,1,1,1}}\,{t}^{{\frac
{30}{7}}\,n}
\mbox{}{q}^{-\frac{2}{7}\,n}+c^{[1,1,1,1,1,1,1]}_{[1]}M^{\ast}_{{1,1,1,1,1,1,1}}\,{t}^{6\,n}

\\
\\
\hline
\end{array}
\ee
\fr{ \\
c^{[7]}_{[1]}=1,\ \ c^{[6,1]}_{[1]}
=-\frac{q^{6/7}(-1+t^{2})(1+q^{2}+q^{4}+q^{6}+q^{6}t^2+q^{8}t^2+q^{10}t^{2})}{t^{6/7}(-1+q^{12}t^{2})},
\\
c^{[5,2]}_{[1]}
=-\frac{q^{10/7}(-1+t^{2})(q^{2}-t^{2})(t^4q^{14}+t^4q^{12}+t^4q^{10}+t^2q^{10}+2t^2q^8+2q^6t^2+q^6+q^4+q^4t^2+2q^2+1)}{t^{12/7}(q^{10}t^{2}-1)(q^{8}t^{2}-1)}
\\
c^{[5,1,1]}_{[1]}=\frac{q^{18/7}(t^{2}+1)(-1+t^{2})^{2}(t^4q^8+q^6t^4+t^4q^4+q^4t^2+q^2t^2+t^2+1)}{t^{18/7}(q^{2}t^{2}-1)(-1+q^{10}t^{4})},
 \\
c^{[4,3]}_{[1]}
=\frac{(-1+t^{2})(q^{12}t^8-q^{12}t^6-t^4q^{12}-2t^4q^{10}+t^6q^8-t^4q^8-t^2q^8+q^6t^6+q^6t^4-q^6t^2+2t^4q^4-q^4+q^2t^2+t^4q^2-q^2+t^2)(q^{2}-t^{2})q^{12/7}}{(q^{8}t^{2}-1)(q^{6}t^{2}-1)(q^{4}t^{2}-1)t^{18/7}}\mathop{\rm
 },  \\
c^{[4,2,1]}_{[1]}=\frac{q^{2}(-1+t^{2})^{2}(q^{12}t^8+3t^6q^{10}+t^4q^{10}+2q^{10}t^8+t^8q^8+5t^6q^8+3t^4q^8+4q^6t^6+6q^6t^4+q^6t^2+q^4t^6+4t^4q^4+3q^4t^2+t^4q^2+3q^2t^2+q^2+t^2+1)
(q^{2}-t^{2})}{t^{24/7}(q^{4}t^{2}+1)(q^{6}t^{2}-1)(q^{4}t^{2}-1)^{2}}\mathop{\rm
 },  \\
c^{[4,1,1,1]}_{[1]}
=-\frac{q^{22/7}(-1+t^{2})(-1+t^{4})(t^{6}-1)(q^6t^6+q^4t^6+q^4t^2+t^4q^4+q^2t^2+t^2+1)}{t^{22/7}(q^{2}t^{2}-1)
(q^{2}t^{4}-1)(-1+t^{6}q^{8})}\mathop{\rm  },  \\
c^{[3,3,1]}_{[1]}
=\frac{(t^{2}+1)(-1+t^{2})^{2}(q^{2}-t^{2})(q^{10}t^8-t^{10}q^8+2t^6q^8+t^4q^8-q^6t^8+2q^6t^4+q^6t^6-q^4t^8-2q^4t^6+t^4q^4+q^4t^2-t^6q^2-2t^4q^2-t^4-t^2+1)q^{22/7}}{(q^{2}t^{2}-1)(q^{4}t^{2}-1)(q^{6}t^{2}-1)(-1+q^{6}t^{4})t^{30/7}}\mathop{\rm
 },\\
c^{[3,2,2]}_{[1]}
=\frac{q^{16/7}(t^{2}+1)(-1+t^{2})^{2}(q^{10}t^{10}-t^{10}q^8+t^6q^8-q^6t^{10}-2q^6t^8+q^6t^6+2q^6t^4+q^6t^2-q^4t^8-2q^4t^6+2q^4t^2+t^4q^4-t^6q^2-t^4q^2+q^2-t^2)(q^{2}-t^{2})}{t^{22/7}(q^{2}t^{2}+1)(q^{4}t^{2}-1)(-1+q^{6}t^{4})(q^{2}t^{2}-1)^{2}}\mathop{\rm
 },  \\
c^{[3,2,1,1]}_{[1]}
=-\frac{(t^{2}+1)(-1+t^{2})^{3}(q^{2}-t^{2})(q^8t^{12}+t^{10}q^8+3q^6t^{10}+3q^6t^8+q^6t^6+t^{12}q^6+t^{10}q^4+4q^4t^8+6q^4t^6+3t^4q^4+q^4t^2+t^8q^2+4t^6q^2+5t^4q^2+3q^2t^2+t^4+2t^2+1)q^{24/7}}{(q^{2}t^{2}+1)(q^{4}t^{2}-1)(q^{4}t^{4}+q^{2}t^{2}+1)(q^{2}t^{2}-1)^{3}t^{4}}\mathop{\rm
 },  \\
c^{[3,1,1,1,1]}_{[1]} =\frac{q^{32/7}(1+t^{4
})(1+t^{2}+t^{4})(t^{2}+1)^{2}(-1+t^{2})^{4}(q^4t^8+t^6q^2+t^8q^2+t^4q^2+t^4+t^2+1)}{t^{32/7}(t^{8}q^{6}-1)(q^{2}t^{2}-1)(t^{6}q^{2}-1)(q^{2}t^{4}-1)}\mathop{\rm
 }\mathop{\rm  },  \\
c^{[2,2,2,1]}_{[1]}
=-\frac{q^{18/7}(t^{2}+1)(1+t^{2}+t^{4})(-1+t^{2})^{3}(q^{2}-t^{2})(t^{10}q^8+t^8q^8-t^{12}q^6-q^6t^{10}+q^6t^6+q^6t^4-t^{10}q^4-2q^4t^8-q^4t^6+t^4q^4+2q^4t^2+q^4-t^6q^2-t^4q^2+q^2-1)}{t^{26/7}(q^{2}t^{2}+1)(q^{2}t^{4}-1)(q^{4}t^{2}-1)(-1+q^{4}t^{6})(q^{2}t^{2}-1)^{2}},
 \\
c^{[2,2,1,1,1]}_{[1]}
=\frac{q^{26/7}(1+t^{2}+t^{4})(t^{2}+1)^{2}(-1+t^{2})^{4}(t^{14}q^4+2q^4t^{12}+t^{10}q^4+q^4t^8+2t^8q^2+2t^6q^2+t^4q^2+t^{10}q^2+t^4+t^2+1)(q^{2}-t^{2})}{t^{38/7}(q^{2}t^{4}+1)(-1+q^{4}t^{6})(q^{2}t^{4}-1)^{2}(q^{2}t^{2}-1)^{2}}\mathop{\rm
 },  \\
c^{[2,1,1,1,1,1]}_{[1]}
=-\frac{q^{34/7}(-1+t^{2})(-1+t^{4})(t^{6}-1)(-1+t^{8})(-1+t^{10})(t^{10}q^2+t^8q^2+t^6q^2+t^4q^2+t^4+t^2+1)}{t^{34/7}(t^{10}q^{4}-1)(q^{2}t^{8}-1)(q^{2}t^{2}-1)(t^{6}q^{2}-1)(q^{2}t^{4}-1)},
 \\
c^{[1,1,1,1,1,1,1]}_{[1]} =
\frac{q^{6}(t^{14}-1)(t^{12}-1)(-1+t^{10})(-1+t^{8})(t^{6}-1)(-1+t^{4})}{t^{6}(t^{12}q^{2}-1)(q^{2}t^{10}-1)(q^{2}t^{8}-1)(t^{6}q^{2}-1)(q^{2}t^{4}-1)(q^{2}t^{2}-1)}}

\subsection{Case $(7,n)$, $n=7 k+4$ fundamental representation\label{7k+4}}
\be
\begin{array}{|c|}
\hline\\
P^{T[7,n]}_{[1]}=c^{[7]}_{[1]}M^{\ast}_{{7}}{q}^{-6\,n}+c^{[6,1]}_{[1]}M^{\ast}_{{6,1}}\,{q}^{-{\frac
{30}{7}}\,n}
\mbox{}{t}^{2/7\,n}+c^{[5,2]}_{[1]}M^{\ast}_{{5,2}}\,{q}^{-{\frac {22}{7}}\,n}
\mbox{}{t}^{4/7\,n}+c^{[5,1,1]}_{[1]}M^{\ast}_{{5,1,1}}\,{q}^{-{\frac
{20}{7}}\,n}
\mbox{}{t}^{6/7\,n}+\\+c^{[4,3]}_{[1]}M^{\ast}_{{4,3}}\,{q}^{-{\frac
{18}{7}}\,n}
\mbox{}{t}^{6/7\,n}+c^{[4,2,1]}_{[1]}M^{\ast}_{{4,2,1}}\,{q}^{-2\,n}{t}^{{\frac
{8}{7}}\,n}
\mbox{}+c^{[4,1,1,1]}_{[1]}M^{\ast}_{{4,1,1,1}}\,{q}^{-{\frac {12}{7}}\,n}
\mbox{}{t}^{{\frac
{12}{7}}\,n}+c^{[3,3,1]}_{[1]}M^{\ast}_{{3,3,1}}\,{q}^{-{\frac {12}{7}}\,n}
\mbox{}{t}^{{\frac
{10}{7}}\,n}+\\+c^{[3,2,2]}_{[1]}M^{\ast}_{{3,2,2}}\,{t}^{{\frac {12}{7}}\,n}
\mbox{}{q}^{-{\frac {10}{7}}\,n}
\mbox{}+c^{[3,2,1,1]}_{[1]}M^{\ast}_{{3,2,1,1}}\,{t}^{2\,n}{q}^{-{\frac
{8}{7}}\,n}+c^{[3,1,1,1,1]}_{[1]}M^{\ast}_{{3,1,1,1,1}}\,{t}^{{\frac
{20}{7}}\,n}
\mbox{}{q}^{-\frac{6}{7}\,n}
\mbox{}+c^{[2,2,2,1]}_{[1]}M^{\ast}_{{2,2,2,1}}\,{t}^{{\frac {18}{7}}\,n}
\mbox{}{q}^{-\frac{6}{7}\,n}+\\+c^{[2,2,1,1,1]}_{[1]}M^{\ast}_{{2,2,1,1,1}}\,{t}^{{\frac
{22}{7}}\,n}
\mbox{}{q}^{-\frac{4}{7}\,n}+c^{[2,1,1,1,1,1]}_{[1]}M^{\ast}_{{2,1,1,1,1,1}}\,{t}^{{\frac
{30}{7}}\,n}
\mbox{}{q}^{-\frac{2}{7}\,n}+c^{[1,1,1,1,1,1,1]}_{[1]}M^{\ast}_{{1,1,1,1,1,1,1}}\,{t}^{6\,n}

\\
\\
\hline
\end{array}
\ee
\fr{ \\
c^{[7]}_{[1]}=1,\ \ c^{[6,1]}_{[1]}
=-\frac{q^{8/7}(-1+t^{2})(1+q^2+q^4+q^4t^2+q^6t^2+t^2q^8+t^2q^{10})}{t^{8/7}(-1+q^{12}t^{2})},
\\
c^{[5,2]}_{[1]}
=-\frac{q^{4/7}(-1+t^{2})(q^{2}-t^{2})(1+q^2+q^4+q^4t^2+2q^6t^2+2t^2q^8+t^4q^8+t^4q^{10}+t^2q^{10}+2t^4q^{12}+t^4q^{14})}{t^{16/7}(q^{10}t^{2}-1)(q^{8}t^{2}-1)}
\\
c^{[5,1,1]}_{[1]}=\frac{q^{10/7}(t^{2}+1)(-1+t^{2})^{2}(1+q^2+q^4+q^4t^2+q^6t^2+t^2q^8+t^4q^8)}{t^{10/7}(q^{2}t^{2}-1)(-1+q^{10}t^{4})},
 \\
c^{[4,3]}_{[1]}
=-\frac{(-1+t^{2})(1-t^2-t^4-2t^4q^2+q^4t^2-t^4q^4-q^4t^6+q^6t^2+q^6t^4-q^6t^6+2t^4q^8-t^8q^8+t^6q^{10}+t^4q^{10}-q^{10}t^8+q^{12}t^6)(q^{2}-t^{2})q^{16/7}}{(q^{8}t^{2}-1)(q^{6}t^{2}-1)(q^{4}t^{2}-1)t^{24/7}}\mathop{\rm
 },  \\
c^{[4,2,1]}_{[1]}=\frac{q^{2}(-1+t^{2})^{2}(1+3q^2t^2+t^4q^2+2q^2+q^4+5q^4t^2+3t^4q^4+4q^6t^2+6q^6t^4+q^6t^6+t^2q^8+4t^4q^8+3t^6q^8+t^4q^{10}+3t^6q^{10}+q^{10}t^8+q^{12}t^6+q^{12}t^8)
(q^{2}-t^{2})}{t^{18/7}(q^{4}t^{2}+1)(q^{6}t^{2}-1)(q^{4}t^{2}-1)^{2}}\mathop{\rm
 },  \\
c^{[4,1,1,1]}_{[1]}
=-\frac{q^{20/7}(-1+t^{2})(-1+t^{4})(t^{6}-1)(1+q^2+t^4q^2+q^2t^2+t^4q^4+q^6t^4+q^6t^6)}{t^{20/7}(q^{2}t^{2}-1)
(q^{2}t^{4}-1)(-1+t^{6}q^{8})}\mathop{\rm  },  \\
c^{[3,3,1]}_{[1]}
=-\frac{(t^{2}+1)(-1+t^{2})^{2}(q^{2}-t^{2})(t^2-q^2+2t^4q^2+t^6q^2-q^4t^2+2q^4t^6+t^4q^4-q^6t^2-2q^6t^4+q^6t^6+q^6t^8-t^4q^8-2t^6q^8-t^6q^{10}-q^{10}t^8+t^{10}q^{10})q^{20/7}}{(q^{2}t^{2}-1)(q^{4}t^{2}-1)(q^{6}t^{2}-1)(-1+q^{6}t^{4})t^{26/7}}\mathop{\rm
 },\\
c^{[3,2,2]}_{[1]}
=\frac{q^{12/7}(t^{2}+1)(-1+t^{2})^{2}(-1+q^2-t^4q^2+q^4+2q^4t^2-t^4q^4-2q^4t^6-q^4t^8+q^6t^2+2q^6t^4-2q^6t^8-q^6t^6+t^4q^8+t^6q^8-t^{10}q^8+q^{10}t^8)(q^{2}-t^{2})}{t^{20/7}(q^{2}t^{2}+1)(q^{4}t^{2}-1)(-1+q^{6}t^{4})(q^{2}t^{2}-1)^{2}}\mathop{\rm
 },  \\
c^{[3,2,1,1]}_{[1]}
=-\frac{(t^{2}+1)(-1+t^{2})^{3}(q^{2}-t^{2})(1+t^2+3q^2t^2+3t^4q^2+t^6q^2+q^2+q^4t^2+4t^4q^4+6q^4t^6+3q^4t^8+t^{10}q^4+q^6t^4+4q^6t^6+5q^6t^8+3q^6t^{10}+t^8q^8+2t^{10}q^8+q^8t^{12})q^{18/7}}{(q^{2}t^{2}+1)(q^{4}t^{2}-1)(q^{4}t^{4}+q^{2}t^{2}+1)(q^{2}t^{2}-1)^{3}t^{4}}\mathop{\rm
 },  \\
c^{[3,1,1,1,1]}_{[1]} =\frac{q^{24/7}(1+t^{4
})(1+t^{2}+t^{4})(t^{2}+1)^{2}(-1+t^{2})^{4}(1+q^2t^2+q^2+t^4q^2+t^4q^4+q^4t^6+q^4t^8)}{t^{24/7}(t^{8}q^{6}-1)(q^{2}t^{2}-1)(t^{6}q^{2}-1)(q^{2}t^{4}-1)}\mathop{\rm
 }\mathop{\rm  },  \\
c^{[2,2,2,1]}_{[1]}
=-\frac{q^{24/7}(t^{2}+1)(1+t^{2}+t^{4})(-1+t^{2})^{3}(q^{2}-t^{2})(-t^2-t^4+q^2+q^2t^2-t^6q^2-t^8q^2+q^4t^2+2t^4q^4+q^4t^6-q^4t^8-2t^{10}q^4-q^4t^{12}+q^6t^6+q^6t^8-q^6t^{12}+q^8t^{12})}{t^{30/7}(q^{2}t^{2}+1)(q^{2}t^{4}-1)(q^{4}t^{2}-1)(-1+q^{4}t^{6})(q^{2}t^{2}-1)^{2}},
 \\
c^{[2,2,1,1,1]}_{[1]}
=\frac{q^{30/7}(1+t^{2}+t^{4})(t^{2}+1)^{2}(-1+t^{2})^{4}(1+2t^2+t^4+t^6+2t^6q^2+2t^8q^2+t^{10}q^2+t^4q^2+t^{10}q^4+q^4t^{12}+q^4t^{14})(q^{2}-t^{2})}{t^{32/7}(q^{2}t^{4}+1)(-1+q^{4}t^{6})(q^{2}t^{4}-1)^{2}(q^{2}t^{2}-1)^{2}}\mathop{\rm
 },  \\
c^{[2,1,1,1,1,1]}_{[1]}
=-\frac{q^{36/7}(-1+t^{2})(-1+t^{4})(t^{6}-1)(-1+t^{8})(-1+t^{10})(1+t^2+t^4+t^6+t^6q^2+t^8q^2+t^{10}q^2)}{t^{36/7}(t^{10}q^{4}-1)(q^{2}t^{8}-1)(q^{2}t^{2}-1)(t^{6}q^{2}-1)(q^{2}t^{4}-1)},
 \\
c^{[1,1,1,1,1,1,1]}_{[1]} =
\frac{q^{6}(t^{14}-1)(t^{12}-1)(-1+t^{10})(-1+t^{8})(t^{6}-1)(-1+t^{4})}{t^{6}(t^{12}q^{2}-1)(q^{2}t^{10}-1)(q^{2}t^{8}-1)(t^{6}q^{2}-1)(q^{2}t^{4}-1)(q^{2}t^{2}-1)}}

\subsection{Case $(7,n)$, $n=7 k+5$ fundamental representation\label{7k+5}}
\be
\begin{array}{|c|}
\hline\\
P^{T[7,n]}_{[1]}=c^{[7]}_{[1]}M^{\ast}_{{7}}{q}^{-6\,n}+c^{[6,1]}_{[1]}M^{\ast}_{{6,1}}\,{q}^{-{\frac
{30}{7}}\,n}
\mbox{}{t}^{2/7\,n}+c^{[5,2]}_{[1]}M^{\ast}_{{5,2}}\,{q}^{-{\frac {22}{7}}\,n}
\mbox{}{t}^{4/7\,n}+c^{[5,1,1]}_{[1]}M^{\ast}_{{5,1,1}}\,{q}^{-{\frac
{20}{7}}\,n}
\mbox{}{t}^{6/7\,n}+\\+c^{[4,3]}_{[1]}M^{\ast}_{{4,3}}\,{q}^{-{\frac
{18}{7}}\,n}
\mbox{}{t}^{6/7\,n}+c^{[4,2,1]}_{[1]}M^{\ast}_{{4,2,1}}\,{q}^{-2\,n}{t}^{{\frac
{8}{7}}\,n}
\mbox{}+c^{[4,1,1,1]}_{[1]}M^{\ast}_{{4,1,1,1}}\,{q}^{-{\frac {12}{7}}\,n}
\mbox{}{t}^{{\frac
{12}{7}}\,n}+c^{[3,3,1]}_{[1]}M^{\ast}_{{3,3,1}}\,{q}^{-{\frac {12}{7}}\,n}
\mbox{}{t}^{{\frac
{10}{7}}\,n}+\\+c^{[3,2,2]}_{[1]}M^{\ast}_{{3,2,2}}\,{t}^{{\frac {12}{7}}\,n}
\mbox{}{q}^{-{\frac {10}{7}}\,n}
\mbox{}+c^{[3,2,1,1]}_{[1]}M^{\ast}_{{3,2,1,1}}\,{t}^{2\,n}{q}^{-{\frac
{8}{7}}\,n}+c^{[3,1,1,1,1]}_{[1]}M^{\ast}_{{3,1,1,1,1}}\,{t}^{{\frac
{20}{7}}\,n}
\mbox{}{q}^{-\frac{6}{7}\,n}
\mbox{}+c^{[2,2,2,1]}_{[1]}M^{\ast}_{{2,2,2,1}}\,{t}^{{\frac {18}{7}}\,n}
\mbox{}{q}^{-\frac{6}{7}\,n}+\\+c^{[2,2,1,1,1]}_{[1]}M^{\ast}_{{2,2,1,1,1}}\,{t}^{{\frac
{22}{7}}\,n}
\mbox{}{q}^{-\frac{4}{7}\,n}+c^{[2,1,1,1,1,1]}_{[1]}M^{\ast}_{{2,1,1,1,1,1}}\,{t}^{{\frac
{30}{7}}\,n}
\mbox{}{q}^{-\frac{2}{7}\,n}+c^{[1,1,1,1,1,1,1]}_{[1]}M^{\ast}_{{1,1,1,1,1,1,1}}\,{t}^{6\,n}

\\
\\
\hline
\end{array}
\ee
\fr{ \\
c^{[7]}_{[1]}=1,\ \ c^{[6,1]}_{[1]}
=-\frac{q^{10/7}(-1+t^{2})(t^2q^{10}+t^2q^8+q^6t^2+q^4t^2+q^2t^2+q^2+1)}{t^{10/7}(-1+q^{12}t^{2})},
\\
c^{[5,2]}_{[1]}
=-\frac{q^{12/7}(-1+t^{2})(q^{2}-t^{2})(1+q^2t^2+2q^4t^2+t^4q^4+2q^6t^2+q^6t^4+2t^4q^8+t^2q^8+2t^4q^{10}+t^4q^{12})}{t^{20/7}(q^{10}t^{2}-1)(q^{8}t^{2}-1)}
\\
c^{[5,1,1]}_{[1]}=\frac{q^{16/7}(t^{2}+1)(-1+t^{2})^{2}(t^4q^8+q^6t^4+t^4q^4+q^4t^2+q^2t^2+q^2+1)}{t^{16/7}(q^{2}t^{2}-1)(-1+q^{10}t^{4})},
 \\
c^{[4,3]}_{[1]}
=-\frac{(-1+t^{2})(-1-q^2t^2+q^4-t^4q^4+q^6+q^6t^2-2q^6t^4-q^6t^6+2t^2q^8-t^6q^8-t^4q^8+t^2q^{10}+t^4q^{10}-t^6q^{10}+t^4q^{12}+t^6q^{14})(q^{2}-t^{2})q^{6/7}}{(q^{8}t^{2}-1)(q^{6}t^{2}-1)(q^{4}t^{2}-1)t^{16/7}}\mathop{\rm
 },  \\
c^{[4,2,1]}_{[1]}=\frac{q^{2}(-1+t^{2})^{2}(t^2+1+3q^2t^2+t^4q^2+q^2+3q^4t^2+4t^4q^4+q^4t^6+q^6t^2+5q^6t^4+4q^6t^6+q^6t^8+2t^4q^8+5t^6q^8+2t^8q^8+3t^6q^{10}+3q^{10}t^8+q^{12}t^8)
(q^{2}-t^{2})}{t^{26/7}(q^{4}t^{2}+1)(q^{6}t^{2}-1)(q^{4}t^{2}-1)^{2}}\mathop{\rm
 },  \\
c^{[4,1,1,1]}_{[1]}
=-\frac{q^{18/7}(-1+t^{2})(-1+t^{4})(t^{6}-1)(q^6t^6+q^6t^4+t^4q^4+q^4t^2+q^2t^2+q^2+1)}{t^{18/7}(q^{2}t^{2}-1)
(q^{2}t^{4}-1)(-1+t^{6}q^{8})}\mathop{\rm  },  \\
c^{[3,3,1]}_{[1]}
=\frac{(t^{2}+1)(-1+t^{2})^{2}(q^{2}-t^{2})(-t^2-1+q^2-q^2t^2-t^4q^2+q^4+q^4t^2-t^4q^4-q^4t^6+2q^6t^2+q^6t^4-2q^6t^6-q^6t^8+2t^4q^8-t^8q^8+t^6q^{10}+q^{12}t^8)q^{18/7}}{(q^{2}t^{2}-1)(q^{4}t^{2}-1)(q^{6}t^{2}-1)(-1+q^{6}t^{4})t^{22/7}}\mathop{\rm
 },\\
c^{[3,2,2]}_{[1]}
=-\frac{q^{22/7}(t^{2}+1)(-1+t^{2})^{2}(-1+t^4+t^6-q^2t^2+2t^6q^2+t^8q^2-2t^4q^4-q^4t^6+q^4t^8+t^{10}q^4-2q^6t^6-q^6t^8+q^6t^{10}+t^{12}q^6-t^8q^8-t^{10}q^8+q^8t^{12})(q^{2}-t^{2})}{t^{32/7}(q^{2}t^{2}+1)(q^{4}t^{2}-1)(-1+q^{6}t^{4})(q^{2}t^{2}-1)^{2}}\mathop{\rm
 },  \\
c^{[3,2,1,1]}_{[1]}
=-\frac{(t^{2}+1)(-1+t^{2})^{3}(q^{2}-t^{2})(1+3t^2+2t^4+t^6+3q^2t^2+5t^4q^2+4t^6q^2+t^8q^2+2t^4q^4+5q^4t^6+4q^4t^8+t^{10}q^4+q^6t^6+3q^6t^8+3q^6t^{10}+t^{12}q^6+t^{10}q^8+q^8t^{12})q^{26/7}}{(q^{2}t^{2}+1)(q^{4}t^{2}-1)(q^{4}t^{4}+q^{2}t^{2}+1)(q^{2}t^{2}-1)^{3}t^{4}}\mathop{\rm
 },  \\
c^{[3,1,1,1,1]}_{[1]} =\frac{q^{30/7}(1+t^{4
})(1+t^{2}+t^{4})(t^{2}+1)^{2}(-1+t^{2})^{4}(1+t^2+t^4q^2+t^4+t^6q^2+q^4t^6+q^4t^8)}{t^{30/7}(t^{8}q^{6}-1)(q^{2}t^{2}-1)(t^{6}q^{2}-1)(q^{2}t^{4}-1)}\mathop{\rm
 }\mathop{\rm  },  \\
c^{[2,2,2,1]}_{[1]}
=\frac{q^{30/7}(t^{2}+1)(1+t^{2}+t^{4})(-1+t^{2})^{3}(q^{2}-t^{2})(-1+t^4+t^6+t^8-q^2t^2+2t^8q^2-t^4q^2+t^{10}q^2+t^6q^2-2q^4t^6-t^4q^4-q^4t^8+q^4t^{12}-q^6t^8-q^6t^{10}+q^6t^{14})}{t^{34/7}(q^{2}t^{2}+1)(q^{2}t^{4}-1)(q^{4}t^{2}-1)(-1+q^{4}t^{6})(q^{2}t^{2}-1)^{2}},
 \\
c^{[2,2,1,1,1]}_{[1]}
=\frac{q^{34/7}(1+t^{2}+t^{4})(t^{2}+1)^{2}(-1+t^{2})^{4}(1+2t^2+2t^4+t^6+t^8+2t^6q^2+2t^8q^2+t^{10}q^2+t^4q^2+q^4t^{12})(q^{2}-t^{2})}{t^{26/7}(q^{2}t^{4}+1)(-1+q^{4}t^{6})(q^{2}t^{4}-1)^{2}(q^{2}t^{2}-1)^{2}}\mathop{\rm
 },  \\
c^{[2,1,1,1,1,1]}_{[1]}
=-\frac{q^{38/7}(-1+t^{2})(-1+t^{4})(t^{6}-1)(-1+t^{8})(-1+t^{10})(1+t^2+t^4+t^6+t^8q^2+t^8+t^{10}q^2)}{t^{38/7}(t^{10}q^{4}-1)(q^{2}t^{8}-1)(q^{2}t^{2}-1)(t^{6}q^{2}-1)(q^{2}t^{4}-1)},
 \\
c^{[1,1,1,1,1,1,1]}_{[1]} =
\frac{q^{6}(t^{14}-1)(t^{12}-1)(-1+t^{10})(-1+t^{8})(t^{6}-1)(-1+t^{4})}{t^{6}(t^{12}q^{2}-1)(q^{2}t^{10}-1)(q^{2}t^{8}-1)(t^{6}q^{2}-1)(q^{2}t^{4}-1)(q^{2}t^{2}-1)}}

\subsection{Case $(7,n)$, $n=7 k+6$ fundamental representation\label{7k+6}}
\be
\begin{array}{|c|}
\hline\\
P^{T[7,n]}_{[1]}=c^{[7]}_{[1]}M^{\ast}_{{7}}{q}^{-6\,n}+c^{[6,1]}_{[1]}M^{\ast}_{{6,1}}\,{q}^{-{\frac
{30}{7}}\,n}
\mbox{}{t}^{2/7\,n}+c^{[5,2]}_{[1]}M^{\ast}_{{5,2}}\,{q}^{-{\frac {22}{7}}\,n}
\mbox{}{t}^{4/7\,n}+c^{[5,1,1]}_{[1]}M^{\ast}_{{5,1,1}}\,{q}^{-{\frac
{20}{7}}\,n}
\mbox{}{t}^{6/7\,n}+\\+c^{[4,3]}_{[1]}M^{\ast}_{{4,3}}\,{q}^{-{\frac
{18}{7}}\,n}
\mbox{}{t}^{6/7\,n}+c^{[4,2,1]}_{[1]}M^{\ast}_{{4,2,1}}\,{q}^{-2\,n}{t}^{{\frac
{8}{7}}\,n}
\mbox{}+c^{[4,1,1,1]}_{[1]}M^{\ast}_{{4,1,1,1}}\,{q}^{-{\frac {12}{7}}\,n}
\mbox{}{t}^{{\frac
{12}{7}}\,n}+c^{[3,3,1]}_{[1]}M^{\ast}_{{3,3,1}}\,{q}^{-{\frac {12}{7}}\,n}
\mbox{}{t}^{{\frac
{10}{7}}\,n}+\\+c^{[3,2,2]}_{[1]}M^{\ast}_{{3,2,2}}\,{t}^{{\frac {12}{7}}\,n}
\mbox{}{q}^{-{\frac {10}{7}}\,n}
\mbox{}+c^{[3,2,1,1]}_{[1]}M^{\ast}_{{3,2,1,1}}\,{t}^{2\,n}{q}^{-{\frac
{8}{7}}\,n}+c^{[3,1,1,1,1]}_{[1]}M^{\ast}_{{3,1,1,1,1}}\,{t}^{{\frac
{20}{7}}\,n}
\mbox{}{q}^{-\frac{6}{7}\,n}
\mbox{}+c^{[2,2,2,1]}_{[1]}M^{\ast}_{{2,2,2,1}}\,{t}^{{\frac {18}{7}}\,n}
\mbox{}{q}^{-\frac{6}{7}\,n}+\\+c^{[2,2,1,1,1]}_{[1]}M^{\ast}_{{2,2,1,1,1}}\,{t}^{{\frac
{22}{7}}\,n}
\mbox{}{q}^{-\frac{4}{7}\,n}+c^{[2,1,1,1,1,1]}_{[1]}M^{\ast}_{{2,1,1,1,1,1}}\,{t}^{{\frac
{30}{7}}\,n}
\mbox{}{q}^{-\frac{2}{7}\,n}+c^{[1,1,1,1,1,1,1]}_{[1]}M^{\ast}_{{1,1,1,1,1,1,1}}\,{t}^{6\,n}

\\
\\
\hline
\end{array}
\ee
\fr{ \\
c^{[7]}_{[1]}=1,\ \ c^{[6,1]}_{[1]}
=-\frac{q^{12/7}(-1+t^{2})(q^{10}t^2+q^8t^2+q^6t^2+q^4t^2+q^2t^2+t^2+1)}{t^{12/7}(-1+q^{12}t^{2})},
\\
c^{[5,2]}_{[1]}
=-\frac{q^{20/7}(-1+t^{2})(q^{2}-t^{2})(q^{4}+1)(q^8t^2+q^6t^2+q^4t^2+q^2t^2+t^2+q^2+1)}{t^{10/7}(q^{10}t^{2}-1)(q^{8}t^{2}-1)}
\\
c^{[5,1,1]}_{[1]}=\frac{q^{22/7}(t^{2}+1)(-1+t^{2})^{2}(q^8t^4+q^6t^4+q^4t^4+q^2t^4+t^4+t^2+1)}{t^{22/7}(q^{2}t^{2}-1)(-1+q^{10}t^{4})},
 \\
c^{[4,3]}_{[1]}
=-\frac{(-1+t^{2})(q^6t^2+q^4t^2+q^2t^2+t^2+q^4+q^2+1)(-t^{2}+q^{4})(q^{2}-t^{2})q^{24/7}}{(q^{8}t^{2}-1)(q^{6}t
^{2}-1)(q^{4}t^{2}-1)t^{8/7}}\mathop{\rm  },  \\
c^{[4,2,1]}_{[1]}=\frac{q^{4}(t^{2}+1)(-1+t^{2})^{2}(q^{4}+q^{2}+1)(q^2t^4+q^4t^4+q^6t^4+t^4+q^2t^2+t^2+1)
(q^{2}-t^{2})}{t^{20/7}(q^{4}t^{2}+1)(q^{6}t^{2}-1)(q^{4}t^{2}-1)^{2}}\mathop{\rm
 },  \\
c^{[4,1,1,1]}_{[1]}
=-\frac{q^{30/7}(-1+t^{2})(-1+t^{4})(t^{6}-1)(q^6t^6+q^4t^6+q^2t^6+t^6+t^4+t^2+1)}{t^{30/7}(q^{2}t^{2}-1)
(q^{2}t
^{4}-1)(-1+t^{6}q^{8})}\mathop{\rm  },  \\
c^{[3,3,1]}_{[1]}
=\frac{(t^{2}+1)(-1+t^{2})^{2}(q^{2}-t^{2})(-t^{2}+q^{4})(q^4t^4+q^2t^4+t^4+q^4t^2+q^2t^2+t^2+1)q^{30/7}}{(q^{2}t^{2}-1)(q^{4}t^{2}-1)(q^{6}t^{2}-1)(-1+q^{6}t^{4})t^{18/7}}\mathop{\rm
 },\\
c^{[3,2,2]}_{[1]}
=\frac{q^{32/7}(t^{2}+1)(-1+t^{2})^{2}(t^4+q^2t^4+q^4t^4+t^2+q^2t^2+q^2+1)(q^{2}-t^{2})(q^{2}-t^{4})}{t^{16/7}(q^{2}t^{2}+1)(q^{4}t^{2}-1)(-1+q^{6}t^{4})(q^{2}t^{2}-1)^{2}}\mathop{\rm
 },  \\
c^{[3,2,1,1]}_{[1]}
=-\frac{(t^{2}+1)(1+t^{2}+t^{4})(-1+t^{2})^{3}(q^{2}+1)(q^{2}-t^{2})
(q^4t^6+q^2t^6+t^6+t^4+q^2t^4+t^2+1)q^{34/7}}{(q^{2}t^{2}+1)(q^{4}t^{2}-1)(q^{4}t^{4}+q^{2}t^{2}+1)(q^{2}t^{2}-1)^{3}t^{4}}\mathop{\rm
 },  \\
c^{[3,1,1,1,1]}_{[1]} =\frac{q^{36/7}(1+t^{4
})(1+t^{2}+t^{4})(t^{2}+1)^{2}(-1+t^{2})^{4}(q^4t^8+q^2t^8+t^8+t^6+t^4+t^2+1)}{t^{36/7}(t^{8}q^{6}-1)(q^{2}t^{2}-1)(t^{6}q^{2}-1)(q^{2}t^{4}-1)}\mathop{\rm
 }\mathop{\rm  },  \\
c^{[2,2,2,1]}_{[1]} =-\frac{q^{36/7
}(t^{2}+1)(1+t^{2}+t^{4})(-1+t^{2})^{3}(q^{2}-t^{2})(q^2t^6+t^6+q^2t^4+t^4+t^2+q^2t^2+1)(q^{2}-t^{4})}{t^{24/7}(q^{2}t^{2}+1)(q^{2}t^{4}-1)(q^{4}t^{2}-1)(-1+q^{4}t^{6})(q^{2}t^{2}-1)^{2}},
 \\
c^{[2,2,1,1,1]}_{[1]}
=\frac{q^{38/7}(1+t^{4})(1+t^{2}+t^{4})(t^{2}+1)^{2}(-1+t^{2})^{4}(t^8+q^2t^8+t^6+q^2t^6+t^4+t^2+1)(q^{2}-t^{2})}{t^{34/7}(q^{2}t^{4}+1)(-1+q^{4}t^{6})(q^{2}t^{4}-1)^{2}(q^{2}t^{2}-1)^{2}}\mathop{\rm
 },  \\
c^{[2,1,1,1,1,1]}_{[1]}
=-\frac{q^{40/7}(-1+t^{2})(-1+t^{4})(t^{6}-1)(-1+t^{8})(-1+t^{10})(q^2t^{10}+t^{10}+t^8+t^6+t^4+t^2+1)}{t^{40/7}(t^{10}q^{4}-1)(q^{2}t^{8}-1)(q^{2}t^{2}-1)(t^{6}q^{2}-1)(q^{2}t^{4}-1)},
 \\
c^{[1,1,1,1,1,1,1]}_{[1]} =
\frac{q^{6}(t^{14}-1)(t^{12}-1)(-1+t^{10})(-1+t^{8})(t^{6}-1)(-1+t^{4})}{t^{6}(t^{12}q^{2}-1)(q^{2}t^{10}-1)(q^{2}t^{8}-1)(t^{6}q^{2}-1)(q^{2}t^{4}-1)(q^{2}t^{2}-1)}}

\subsection{Case $(m,km)$, fundamental representation\label{mk}}
For the $n$-component links of the form $(m,km)$ the general answer is:
\be
\begin{array}{|c|}
\hline\\
P^{[m,k m]}_{[1]}=\sum\limits_{|Q|=m} c^{Q}_{[1]} M_{Q}^{\ast} q^{-2
\nu(Q^{\prime}) k} t^{2 \nu(Q) k}
\\
\\
\hline
\end{array}
\ee
the coefficients $c^{Q}_{[1]}$ is determined as unique solutions of the system
of linear equations:
\be
M_{[1]}^m=p_{1}^m=\sum_{|Q|=m} c^{Q}_{[1]} M_{Q}(p_{k})
\ee
explicitly they are given by:
\be
c^{Q}_{[1]}= M_{Q}(\delta_{k,1})\,|Q|!\dfrac{(1-q^2)^{|Q|}}{(1-t^2)^{|Q|}}
\prod\limits_{(i,j)\in
Q}
 \frac{1-t^2 t^{2(Q_{j}^{\prime}-i)} q^{2(Q_{i}-j) }}{1-q^2
t^{2(Q_{j}^{\prime}-i)}
q^{2(Q_{i}-j) }}
\ee
where $M_{\lambda}(\delta_{k,1})$ is the value MacDonald polynomial at the
point $p_{k}=\delta_{k,1}$:

\be
\begin{array}{|c|c|}
\hline &\\
M_{1}(\delta_{k,1})&1 \\ &\\
\hline
\hline &\\
M_{2}(\delta_{k,1})&\frac{1}{2}\frac{(t-1)(q+1)}{-1+tq} \\ &\\
\hline &\\
M_{11}(\delta_{k,1})&\frac{1}{2} \\ &\\
\hline
\hline &\\
M_{3}(\delta_{k,1})&\frac{1}{6}\frac{(t-1)^2(q+1)(q^2+q+1)}{(-1+tq)(-1+tq^2)}
\\ &\\
\hline &\\
M_{21}(\delta_{k,1})&\frac{1}{6}\frac{(t-1)(2tq+t+q+2)}{-1+t^2q} \\ &\\
\hline &\\
M_{111}(\delta_{k,1})&\frac{1}{6} \\ &\\
\hline
\hline &\\
M_{4}(\delta_{k,1})&\frac{1}{24}\frac{(t-1)^3(q+1)(q^2+q+1)(q^3+q^2+q+1)}{(-1+tq)(-1+tq^2)(-1+tq^3)}
\\ &\\
\hline &\\
M_{31}(\delta_{k,1})&\frac{1}{24}\frac{(t-1)^2(q+1)(3q^2t+q^2+2tq+2q+t+3)}{(-1+tq)(-1+t^2q^2)}
\\ &\\
\hline &\\
M_{22}(\delta_{k,1})&\frac{1}{24}\frac{(t-1)^2(q+1)(2tq+q+t+2)}{(-1+tq)(-1+t^2q)}
\\ &\\
\hline &\\
M_{211}(\delta_{k,1})&\frac{1}{24}\frac{(t-1)(t^2+3t^2q+2t+2tq+3+q)}{-1+t^3q}
\\ &\\
\hline &\\
M_{1111}(\delta_{k,1})&\frac{1}{24} \\ &\\
\hline
\end{array}
\ee
\be
\begin{array}{|c|c|}
\hline &\\
M_{5}(\delta_{k,1})&\frac{1}{120}\frac{(t-1)^4(q+1)(q^2+q+1)(q^3+q^2+q+1)(q^4+q^3+q^2+q+1)}{(-1+tq)(-1+tq^2)(-1+tq^3)((-1+tq^4))}
\\ &\\
\hline &\\
M_{41}(\delta_{k,1})&\frac{1}{120}\frac{(t-1)^3(q+1)(q^2+q+1)(4q^3t+q^3+3q^2t+2q^2+2tq+3q+t+4)}{(-1+tq)(-1+tq^2)(-1+t^2q^3)}
\\ &\\
\hline &\\
M_{32}(\delta_{k,1})&\frac{1}{120}\frac{(t-1)^2(q+1)(5q^3t^2+4q^3t+q^3+6q^2t^2+11q^2t+3q^2+3t^2q+11tq+6q+t^2+4t+5)}{(-1+tq)(-1+t^2q)(-1+t^2q^2)}
\\ &\\
\hline &\\
M_{311}(\delta_{k,1})&\frac{1}{120}\frac{(t-1)^2(q+1)(6q^2t^2+3q^2t+q^2+3t^2q+4tq+3q+t^2+3t+6)}{(-1+tq)(-1+t^3q^2)}
\\ &\\
\hline &\\
M_{221}(\delta_{k,1})&\frac{1}{120}\frac{(t-1)^2(5t^3q^2+t^3+4qt^3+6q^2t^2+11t^2q+3t^2+11tq+3q^2t+6t+4q+5+q^2)}{(-1+t^2q)(-1+t^3q)}
\\ &\\
\hline &\\
M_{2111}(\delta_{k,1})&\frac{1}{120}\frac{(t-1)(4qt^3+t^3+3t^2q+2t^2+2tq+3t+q+4)}{-1+t^4q}
\\ &\\
\hline &\\
M_{11111}(\delta_{k,1})&\frac{1}{120} \\  &\\
\hline
\end{array}
\ee

\bigskip

\begin{itemize}
\item{HOMFLY case}
\end{itemize}
Taking the formulae above at the point $t=q$ we find a simple expression for
HOMFLY polynomial of
the links:
\be
H^{T[m,k m]}_{[1]}=|m|!  \sum\limits_{|Q|=m} s_{\lambda}(\delta_{k,1}) 
s_{\lambda}^{\ast}
q^{ \varkappa(Q) k}
\ee
where
\be
\varkappa(Q)=2\Big( \nu(Q) -  \nu(Q^{\prime})\Big) =2 \sum\limits_{(i,j)\in Q}
(i-j)
\ee
\begin{itemize}
\item{Alexander case}
\end{itemize}
\be
A^{T[m,k m]}_{[1]}=|m|!  \sum\limits_{|\lambda|=m} s_{Q}(\delta_{k,1}) 
s_{Q}^{\ast} (A^2=-1)
q^{ \varkappa(Q) k}
\ee

\subsection{Case $(m,km+1)$, fundamental representation\label{km+1}}
\be
\begin{array}{|c|}
\hline\\
P^{T[m,k m+1]}_{[1]}=\sum\limits_{|Q|=m} c^{Q}_{[1]} M_{Q}^{\ast} q^{-2
\nu(Q^{\prime})k} t^{2 \nu(Q)k}
\\
\\
\hline
\end{array}
\ee
the coefficients $c^{Q}_{[1]}=\bar{c}^{Q}_{[1]} \gamma^{Q}_{[1]}$ are defined
from system of
linear equations:
\be
p_{m}=\sum_{|\lambda|=m} \bar{c}^{Q}_{[1]} M_{Q}(p_{k}),
\ee
and
\be
\gamma^{Q}_{[1]}=\dfrac{q^{\alpha_{Q}} }{[m]_{q}}
\sum\limits_{(i,j)\in Q} t^{2 (i-1)} q^{2Q_{1}-2Q_{i}+2 j-2}
\ee
with the proper choice of integers $\alpha_{Q}$
\begin{itemize}
\item{HOMFLY case (see (\ref{Wrepfund}))}
\end{itemize}

\be
\boxed{H^{T[m,k m+1]}_{[1]}=\sum\limits_{|Q|=m} (-1)^j s_{h_{j}}^{\ast}
q^{2(2j+1-m)(km+1)}}
\ee

\begin{itemize}
\item{Alexander case}
\end{itemize}
\be
\boxed{A^{T[m,k m+1]}_{[1]}=\dfrac{1}{[m]_{q}}\sum\limits_{|Q|=m} (-1)^j
q^{2(2j+1-m)(km+1)}}
\ee

\end{small}

\subsection{Case $(m,km-1)$, fundamental representation\label{km-1}}
\begin{small}
\be
\begin{array}{|c|}
\hline\\
P^{T[m,k m-1]}_{[1]}=\sum\limits_{|Q|=m} c^{Q}_{[1]} M_{Q}^{\ast} q^{2
\nu(Q^{\prime})k} t^{-2 \nu(Q)k}
\\
\\
\hline
\end{array}
\ee
the coefficients $c^{Q}_{[1]}=\bar{c}^{Q}_{[1]} \tilde\gamma^{Q}_{[1]}$
are defined from system of
linear equations:
\be
p_{m}=\sum_{|\lambda|=m} \bar{c}^{Q}_{[1]} M_{Q}(p_{k}),
\ee
and
\be
\tilde\gamma^{Q}_{[1]}=\dfrac{q^{-\alpha_{Q}} }{[m]_{q}}
\sum\limits_{(i,j)\in Q} t^{2 (1-i)} q^{2Q_{i}-2Q_{1}+2-2j}
\ee
with the proper choice of integers $\alpha_{Q}$
\begin{itemize}
\item{HOMFLY case (see (\ref{Wrepfund}))}
\end{itemize}

\be
\boxed{H^{T[m,k m-1]}_{[1]}=\sum\limits_{|Q|=m} (-1)^j s_{h_{j}}^{\ast}
q^{2(2j+1-m)(km-1)}}
\ee

\begin{itemize}
\item{Alexander case}
\end{itemize}
\be
\boxed{A^{T[m,k m-1]}_{[1]}=\dfrac{1}{[m]_{q}}\sum\limits_{|Q|=m} (-1)^j
q^{2(2j+1-m)(km-1)}}
\ee

\end{small}

\subsection{Cases $(m,mk+2), \ (m,mk+3), \ (m,mk+4)$, fundamental
representation\label{km+2}
\cite{MMSS}}
The key observation is that the extended superpolynomial at $k=0$ has a nice
decomposition in terms of the Hall-Littlewood polynomials $L_Q\{p\} =
\left.M_Q\{p\}\right|_{q=0}$:
\be\label{dec}
{\cal P}_{m,r}\{p\} = \sum_{Q\vdash m} c^Q_{(m,r)}M_Q\{p\}
= \sum_{\stackrel{Q\vdash m}{l(Q)\leq r}} h^Q_{(m,r)} L_Q\{p\}
\ee
It is convenient to separate a simple overall factor
$$
h^Q = q^{\nu(Q)} (1-t)^{l(Q)-1} \hat h^Q
$$
which makes $\hat h^Q$ normalized to unity: $\hat h^Q = 1 + O(q,t)$.
Then the coefficients $\hat h^Q$ for $r=1,\ 2, \ 3$ are given by the following
simple
formulas:
\be
\boxed{
\hat h^Q_{(m,1)} = \left\{ \begin{array}{ll} 1 & l(Q) = 1 \\
0 & {\rm otherwise} \end{array} \right.
}
\ee
\be\label{h2}
\boxed{
\hat h^Q_{(m,2)} = \left\{ \begin{array}{ll}
1 & l(Q) = 1 \\
1 & l(Q) = 2 \\
0 & {\rm otherwise} \end{array} \right.
}
\ee
\be
\boxed{
\hat h^Q_{(m,3)} = \left\{ \begin{array}{ll}
1 & l(Q) = 1 \\
1 + t + (q-t)[\alpha]_q & l(Q) = 2,3 \\
0 & {\rm otherwise} \end{array} \right.
}
\ee
with $\alpha = \min(Q_1-Q_2,Q_2-Q_3)$. These examples clearly show the
important role of the Hall-Littlewood basis. However to get series for
arbitrary $k$, one needs to return to MacDonald basis and coefficients $c^Q$.
It can be extracted from (\ref{dec}) by the linear transformations from
$h^Q_{(m,r)}$ to $c^Q_{(m,r)}$.

Similar formulas for $\hat h^Q_{(m,4)}$ are more tedious, and can be found in
the
explicit form in \cite{MMSS}.

\section{Torus knots: higher representations}

\subsection{Case $(2,n)$, series $n=2 k+1$, symmetric representation [2]
\label{2k+12}}
\begin{small}
\be
\begin{array}{|c|}
\hline\\
P^{T[2,n]}=c^{[4]}_{[2]} M_{[4]}^{\ast} q^{-6n} + \Big(\frac{q}{t}\Big)
c^{[3,1]}_{[2]} M_{[3,1]}^{\ast} q^{-3n}t^{n} + \Big(\frac{q}{t}\Big)^2
c^{[2,2]}_{[2]} M_{[2,2]}^{\ast} q^{-2n}t^{2n}
\\
\\
\hline
\end{array}
\ee
with the coefficients:
\be
\boxed{
c^{[4]}_{[2]}=1,\ \ \ c^{[3,1]}_{[2]}=-\frac{(1-t^2)(1+q^2)(1+t^2q^2)}{1-q^6
t^2} \ \ \ c^{[2,2]}_{[2]}=\frac{(1-t^4)(1-t^4q^2)}{(1-q^4 t^2)(1-q^2t^2)}
}
\ee
such that one obtains
\be
\begin{array}{l}
P^{T[2,1]}_{[2]}=\dfrac{\{A\}\{qA\} t^2}{\{t\}\{tq\} A^2 q^6} = \bf{\frac{ (
\bf{a}^{2}t+1 )  ( {q}^{2}{t}^{3}{a}^{2}+1 ) }{{q}^{5} \{q\} \{q^{2}t\}\mbox{} 
{t}^{9}{a}^{4}}} \\
P^{T[2,3]}_{[2]}=\dfrac{\{A\}\{qA\} t^2}{\{t\}\{tq\} A^2 q^{18}}
\Big(q^{10}A^4+(q^{10}t^2+q^8t^2+q^6+q^4)(-A^2)+q^8t^4+q^6t^2+q^4t^2+1\Big)
=\\= \bf{\frac { \left( {a}^{2}t+1 \right)  \left( {q}^{2}{t}^{3}{a}^{2}+1
\right)}{{q}^{17}
\mbox{} \{q\}  \{{q}^{2}t\} {a}^{4}{t}^{21}}\left( {q}^{10}{t}^{12}{a}^{4}+
\left( {t}^{5}{q}^{4}+{t}^{11}{q}^{12}+{t}^{9}{q}^{10}+{t}^{7}{q}^{6} \right)
{a}^{2}
\mbox{}+{t}^{6}{q}^{8}+1+{q}^{6}{t}^{4}+{q}^{12}{t}^{8} \right)} \\
P^{T[2,5]}_{[2]}=\dfrac{\{A\}\{qA\} t^2}{\{t\}\{tq\} A^2 q^{30}}
\Big((q^{18}t^4+q^{14}t^2+q^{16}t^2+q^{10})(-A^2)^2+(q^{16}t^6+q^{18}t^6+q^{16}t^4+q^{12}t^4+2q^{14}t^4+q^{12}t^2+\\+2q^{10}t^2+q^8t^2+q^6+q^4)(-A^2)+q^{16}t^8+q^{14}t^6+q^{12}t^6+q^{10}t^4+q^8t^4+q^{12}t^4+q^4t^2+q^6t^2+1\Big)
=\\= \bf{\frac { \left( {a}^{2}t+1 \right)  \left( {q}^{2}{t}^{3}{a}^{2}+1
\right)}{{q}^{29}
\mbox{} \{q\}  \{{q}^{2}t\} {a}^{4}{t}^{33}}\Big(\left(
{q}^{16}{t}^{16}+{q}^{10}{t}^{12}+{t}^{20}{q}^{22}+{q}^{18}{t}^{18} \right)
{a}^{4}+ (
{t}^{5}{q}^{4}+2\,{t}^{11}{q}^{12}+{t}^{9}{q}^{10}+{t}^{7}{q}^{6}+{q}^{24}{t}^{19}+{q}^{22}{t}^{17}+}\\
\bf{+{q}^{20}{t}^{17}+2\,{q}^{18}{t}^{15}\mbox{}+{t}^{13}{q}^{16}+{q}^{14}{t}^{13}
) {a}^{2}+{t}^{12}{q}^{16}+{q}^{24}{t}^{16}+{q}^{20}{t}^{14}
\mbox{}+{t}^{12}{q}^{18}+{q}^{6}{t}^{4}+{q}^{14}{t}^{10}+{q}^{12}{t}^{8}+{t}^{6}{q}^{8}+1\Big)}
\\
P^{T[2,7]}_{[2]}=\dfrac{\{A\}\{qA\} t^2}{\{t\}\{tq\} A^2 q^{42}}
\Big((2q^{14}t^4+q^{18}t^4+q^6+q^4+q^{12}t^2+2q^{10}t^2+q^8t^2+q^{12}t^4+q^{26}t^{10}+q^{24}t^{10}+2q^{22}t^8+\\+q^{20}t^8+q^{24}t^8+q^{16}t^6+2q^{20}t^6+q^{22}t^6+2q^{18}t^6+2q^{16}t^4)(-A^2)+q^{24}t^{12}+q^{20}t^{10}+q^{22}t^{10}+q^{26}A^4t^8+q^{18}t^8+\\+q^{16}t^8+q^{20}t^8+q^{22}A^4t^6+q^{16}t^6+q^{24}A^4t^6+q^{18}t^6+q^{14}t^6+q^{12}t^6+q^{18}A^4t^4+q^{22}A^4t^4+q^{20}A^4t^4+q^8t^4+\\+q^{10}t^4+q^{12}t^4+q^{16}A^4t^2+q^{14}A^4t^2+q^6t^2+q^4t^2+q^{10}A^4+1\Big)
=\\= \bf{\frac { \left( {a}^{2}t+1 \right)  \left( {q}^{2}{t}^{3}{a}^{2}+1
\right)}{{q}^{41} \mbox{} \{q\}  \{{q}^{2}t\} {a}^{4}{t}^{45}}
\Big(({q}^{16}{t}^{16}+{t}^{22}{q}^{24}+{t}^{20}{q}^{22}+{q}^{28}{t}^{24}+{q}^{26}{t}^{24}+{t}^{28}{q}^{34}+{q}^{18}{t}^{18}
\mbox{}+{q}^{30}{t}^{26}+{q}^{10}{t}^{12} ) {a}^{4}+}\\ \bf{+
({q}^{14}{t}^{13}+2\,{q}^{26}{t}^{21}+{q}^{32}{t}^{25}+{t}^{7}{q}^{6}+
{q}^{28}{t}^{23}+
2\,{t}^{11}{q}^{12}+2\,{q}^{24}{t}^{19}\mbox{}+{t}^{9}{q}^{10}+2\,{q}^{18}{t}^{15}+{q}^{36}{t}^{27}
+{t}^{5}{q}^{4}+ 2\,{q}^{20}{t}^{17}+}\\
\bf{+2\,{q}^{30}{t}^{23}+{q}^{22}{t}^{19}+{q}^{28}{t}^{21}\mbox{}+{t}^{13}{q}^{16}+{q}^{22}{t}^{17}+
{q}^{34}{t}^{25} )
{a}^{2}\mbox{}+1+{q}^{22}{t}^{16}+{q}^{12}{t}^{8}+{q}^{30}{t}^{20}+{q}^{20}{t}^{14}+{q}^{32}{t}^{22}+}\\
\bf{+
{q}^{36}{t}^{24}+{q}^{6}{t}^{4}+{q}^{28}{t}^{20}\mbox{}+{t}^{6}{q}^{8}+{q}^{26}{t}^{18}+{t}^{12}{q}^{18}+{q}^{14}{t}^{10}+{t}^{12}{q}^{16}
+{q}^{24}{t}^{16}+{q}^{24}{t}^{18}\Big) }\\
....\\
\boxed{P^{T[2,2k+1]}_{[2]}=\dfrac{\{A\}\{qA\} t^2}{\{t\}\{tq\} A^2 q^{12k+6}}
{\cal{P}}^{T[2,2k+1]}_{[2]}}
\end{array}
\ee
Our answer for  $P^{T[2,3]}_{[2]}$ is in full agreement with the known results
\cite{ASh} and generalizes it to $n>3$.
\begin{itemize}
\item{HOMFLY case}
\end{itemize}
At the point $t=q$ one has:
\be
H^{T[2,n]}_{[2]}=s_{[4]}^{\ast} q^{-6n}- s_{[3,1]}^{\ast}
q^{-2n}+s_{[2,2]}^{\ast}
\ee
Several first answers are:
\be
\begin{array}{l}
H^{T[2,1]}_{[2]}=\dfrac{\{A\}\{qA\}}{\{q\}\{q^2\} A^2 q^4} \\
H^{T[2,3]}_{[2]}=\dfrac{\{A\}\{qA\}}{\{q\}\{q^2\} A^2 q^{16}}
\Big(q^{12}-A^2q^{12}+A^4q^{10}-A^2q^{10}+q^8-A^2q^6+q^6-A^2q^4+1\Big)\\
H^{T[2,5]}_{[2]}=\dfrac{\{A\}\{qA\}}{\{q\}\{q^2\} A^2 q^{28}}
\Big(q^{24}-A^2q^{24}+A^4q^{22}-A^2q^{22}-A^2q^{20}+q^{20}+A^4q^{18}-2A^2q^{18}
+
q^{18}+A^4q^{16}-\\-A^2q^{16}+q^{16}-A^2q^{14}+q^{14}-2A^2q^{12}+q^{12}+A^4q^{10}-A^2q^{10}+q^8-A^2q^6+q^6-A^2q^4+1
\Big)\\
H^{T[2,7]}_{[2]}=\dfrac{\{A\}\{qA\}}{\{q\}\{q^2\} A^2 q^{40}}
\Big(q^{36}-A^2q^{36}-A^2q^{34}+A^4q^{34}-A^2q^{32}+q^{32}-2A^2q^{30}+
A^4q^{30}+ q^{30}+q^{28}+A^4q^{28}-\\
-2A^2q^{28}+A^4q^{26}+q^{26}-2A^2q^{26}+2q^{24}-2A^2q^{24}+A^4q^{24}+A^4q^{22}-2A^2q^{22}+q^{22}+q^{20}
-2A^2q^{20}-2A^2q^{18}+\\+q^{18}+A^4q^{18}+q^{16}+A^4q^{16}-A^2q^{16}-A^2q^{14}+q^{14}-2A^2q^{12}+q^{12}+A^4q^{10}-A^2q^{10}+q^8-A^2q^6+q^6-A^2q^4+1\Big)\\
....\\
\boxed{H^{T[2,2k+1]}_{[2]}=\dfrac{\{A\}\{qA\}}{\{q\}\{q^2\} A^2 q^{12k+4}}
{\cal{H}}^{T[2,2k+1]}_{[2]}}
\end{array}
\ee
and the results coincide with the well known HOMFLY polynomials.
\begin{itemize}
\item{Floer case}
\end{itemize}
 \be
\begin{array}{l}
F^{T[2,1]}_{[2]}=1\\
F^{T[2,3]}_{[2]}={\bf{q}}^{8}{\textbf{t}}^{6}+{\textbf{t}}^{3}{\textbf{q}}^{4}+1\\
F^{T[2,5]}_{[2]}=\bf{t}^{12}q^{16}+t^9q^{12}+t^6q^8+t^3q^4+1\\
F^{T[2,7]}_{[2]}=\bf{t}^{18}q^{24}+t^{15}q^{20}+t^{12}q^{16}+t^9q^{12}+t^6q^8+t^3q^4+1
\\
F^{T[2,9]}_{[2]}=\bf{t}^{24}q^{32}+t^{21}q^{28}+t^{18}q^{24}+t^{15}q^{20}+t^{12}q^{16}+t^9q^{12}+t^6q^8+t^3q^4+1
\\
....\\
\end{array}
\ee
\begin{itemize}
\item{Alexander case}
\end{itemize}
  \be
\begin{array}{l}
A^{T[2,1]}_{[2]}=1 \\
A^{T[2,3]}_{[2]}=\bf{q}^8-q^4+1 \\
A^{T[2,5]}_{[2]}=\bf{q}^{16}-q^{12}+q^8-q^4+1 \\
A^{T[2,7]}_{[2]}=\bf{q}^{24}-q^{20}+q^{16}-q^{12}+q^8-q^4+1 \\
A^{T[2,9]}_{[2]}=\bf{q}^{32}-q^{28}+q^{24}-q^{20}+q^{16}-q^{12}+q^8-q^4+1 \\
....\\
\end{array}
\ee

\subsection{Case $(2,n)$, series $n=2 k+1$, antisymmetric representation
[1,1]\label{2k+111} }
\be
\begin{array}{|c|}
\hline\\
P^{T[2,n]}_{[1,1]}=c^{[2,2]}_{[1,1]} M_{[2,2]}^{\ast} q^{-2n}t^{2n} +
\Big(\frac{q}{t}\Big) c^{[2,1,1]}_{[1,1]} M_{[2,1,1]}^{\ast} q^{-n}t^{3n} +
\Big(\frac{q}{t}\Big)^2 c^{[1,1,1,1]}_{[1,1]} M_{[1,1,1,1]}^{\ast} t^{6n}
\\
\\
\hline
\end{array}
\ee
with the coefficients:
\be
\boxed{
c^{[2,2]}_{[1,1]}=1,\ \ \ c^{[2,1,1]}_{[1,1]}=-\frac{1-t^4}{1-q^2 t^2} \ \ \
c^{[1,1,1,1]}_{[1,1]}=\frac{(1-t^8)(1-t^6)}{(1-q^2 t^4)(1-q^2t^6)}
}
\ee
such that one obtains
\be
\begin{array}{l}
P^{T[2,1]}_{[1,1]}=\dfrac{\{A\}\{\frac{A}{t}\} t^6}{\{t\}\{t^2\} A^2 q^2}
=\bf{\frac { \left( {a}^{2}t+1 \right)  \left( {a}^{2}t+{q}^{2} \right)
{q}^{2}}{\{q\}\{q^2\}{t}^{4}{a}^{4}}} \\
P^{T[2,3]}_{[1,1]}=\dfrac{\{A\}\{\frac{A}{t}\}t^8}{\{t\}\{t^2\} A^2
q^6}\Big((q^4t^6+q^4t^4+q^2t^2+q^2)(-A^2)+q^4t^{10}+q^2t^6+q^2t^4+t^2+ A^4q^4
\Big) = \\ = \bf{\frac { \left( {a}^{2}t+1 \right)  \left( {a}^{2}t+{q}^{2}
\right) {q}^{2}}{\{q\}\{q^2\}{t}^{8}{a}^{4}}\Big( {t}^{6}{q}^{2}{a}^{4}+ \left(
{t}^{3}+{t}^{5}{q}^{8}+{q}^{6}{t}^{5}+{q}^{2}{t}^{3} \right)
{a}^{2}+{t}^{4}{q}^{12}+{q}^{6}{t}^{2}+q^4t^2+1 \Big)} \\
P^{T[2,5]}_{[1,1]}=\dfrac{\{A\}\{\frac{A}{t}\} t^{12}}{\{t\}\{t^2\} A^2 q^{10}}
\Big((q^8t^{14}+q^8t^{12}+q^6t^{10}+2q^6t^8+q^6t^6+q^4t^6+2q^4t^4+q^2t^2+q^4t^2+q^2)(-A^2)+q^8t^{18}+\\+q^6t^{14}+q^6t^{12}+q^4t^{10}+q^4t^8+q^8A^4t^8+q^4t^6+q^2t^6+q^2t^4+q^6A^4t^4+q^6A^4t^2+t^2+A^4q^4\Big)
= \\ = \bf{\frac { \left( {a}^{2}t+1 \right)  \left( {a}^{2}t+{q}^{2} \right)
{q}^{2}}{\{q\}\{q^2\}{t}^{12}{a}^{4}}\Big(  (
{t}^{8}{q}^{6}+{q}^{14}{t}^{10}+{t}^{6}{q}^{2}+{t}^{8}{q}^{8} ) {a}^{4}+ (
{t}^{3}+2\,{q}^{6}{t}^{5}+{t}^{5}{q}^{8}+{q}^{2}{t}^{3}+{t}^{9}{q}^{18}+{q}^{20}{t}^{9}+}\\
\bf{+{t}^{7}{q}^{10}+{q}^{14}{t}^{7}+2\,{q}^{12}{t}^{7}+{t}^{5}{q}^{4}
\mbox{} ) {a}^{2}
\mbox{}+{t}^{4}{q}^{12}+{q}^{24}{t}^{8}+{t}^{6}{q}^{16}+{q}^{18}{t}^{6}+{q}^{6}{t}^{2}+{q}^{8}{t}^{4}+{q}^{10}{t}^{4}+{q}^{4}{t}^{2}+1
\Big)} \\
P^{T[2,7]}_{[1,1]}=\dfrac{\{A\}\{\frac{A}{t}\} t^{16}}{\{t\}\{t^2\} A^2
q^{14}}
\Big((q^4t^2+q^2+q^2t^2+2q^4t^4+q^6t^4+2q^6t^6+q^4t^6+2q^6t^8+q^8t^8+2q^8t^{10}+q^6t^{10}+2q^8t^{12}+\\+q^{10}t^{14}+q^8t^{14}+2q^{10}t^{16}+q^{10}t^{18}+q^{12}t^{22}+q^{12}t^{20})(-A^2)+q^{12}t^{26}+q^{10}t^{22}+
q^{10}t^{20}+q^8t^{18}+q^{12}A^4t^{16}+\\+q^8t^{16}+q^6t^{14}+q^8t^{14}+q^{10}A^4t^{12}+q^6t^{12}+q^{10}A^4t^{10}+q^6t^{10}+q^4t^{10}+q^4t^8+q^6t^8+q^8A^4t^8+q^4t^6+q^8A^4t^6+q^2t^6+\\+q^6A^4t^4+q^2t^4+q^8A^4t^4+q^6A^4t^2+t^2+A^4q^4\Big)=
\\ = \bf{\frac { \left( {a}^{2}t+1 \right)  \left( {a}^{2}t+{q}^{2} \right)
{q}^{2}}{\{q\}\{q^2\}{t}^{16}{a}^{4}}\Big( (
{t}^{12}{q}^{18}+{q}^{14}{t}^{10}+{t}^{10}{q}^{10}+{t}^{8}{q}^{8}+{t}^{6}{q}^{2}+{q}^{20}{t}^{12}+{t}^{8}{q}^{6}+{t}^{10}{q}^{12}
+{q}^{26}{t}^{14} ) {a}^{4}+ ( 2\,{q}^{24}{t}^{11}+}\\
\bf{+{t}^{11}{q}^{22}+{t}^{5}{q}^{8}+{t}^{5}{q}^{4}+2\,{q}^{12}{t}^{7}+
{t}^{13}{q}^{32}+{t}^{9}{q}^{14}+{t}^{7}{q}^{8}+{t}^{3}
+{q}^{20}{t}^{9}+2\,{t}^{9}{q}^{18}+{q}^{14}{t}^{7}+2\,{t}^{7}{q}^{10}+2\,{q}^{16}{t}^{9}+}\\
\bf{+2\,{q}^{6}{t}^{5}+{q}^{2}{t}^{3}+{t}^{11}{q}^{26}+{t}^{13}{q}^{30}
 ) {a}^{2}
+1+{t}^{6}{q}^{14}+{q}^{22}{t}^{8}+{q}^{6}{t}^{2}+{q}^{4}{t}^{2}+{q}^{18}{t}^{6}+{q}^{10}{t}^{4}+{q}^{30}{t}^{10}
+{q}^{8}{t}^{4}+{q}^{24}{t}^{8}+}\\
\bf{+{q}^{36}{t}^{12}+{q}^{20}{t}^{8}+{t}^{4}{q}^{12}+{q}^{12}{t}^{6}+{t}^{6}{q}^{16}+{q}^{28}{t}^{10}
\Big) } \\
....\\
\boxed{P^{T[2,2k+1]}_{[1,1]}=\dfrac{\{A\}\{\frac{A}{t}\} t^{4k+4}}{\{t\}\{t^2\}
A^2 q^{4k+2}} {\cal{P}}^{T[2,2k+1]}_{[1,1]}}
\end{array}
\ee

\begin{itemize}
\item{HOMFLY case}
\end{itemize}
At the point $t=q$ we have:
\be
H^{T[2,n]}_{[1,1]}= s_{[2,2]}^{\ast} -s_{[2,1,1]}^{\ast} q^{2n} +
s_{[1,1,1,1]}^{\ast} q^{6n}
\ee
Several first answers are:
\be
\begin{array}{l}
H^{T[2,1]}_{[1,1]}=\dfrac{\{A\}\{\frac{A}{q}\} q^4}{\{q\}\{q^2\} A^2} \\
H^{T[2,3]}_{[1,1]}=\dfrac{\{A\}\{\frac{A}{q}\} q^4}{\{q\}\{q^2\} A^2}
\Big(q^{12}-A^2q^8-A^2q^6+q^6+q^4+A^4q^2-A^2q^2-A^2+1\Big)\\
H^{T[2,5]}_{[1,1]}=\dfrac{\{A\}\{\frac{A}{q}\} q^{4}}{\{q\}\{q^2\} A^2}
\Big(q^{24}-A^2q^{20}+q^{18}-A^2q^{18}+q^{16}-A^2q^{14}+A^4q^{14}+q^{12}-2A^2q^{12}+q^{10}-A^2q^{10}-A^2q^8+\\+A^4q^8+q^8+q^6-2A^2q^6+A^4q^6-A^2q^4+q^4+A^4q^2-A^2q^2-A^2+1\Big)\\
H^{T[2,7]}_{[1,1]}=\dfrac{\{A\}\{\frac{A}{q}\} q^{4}}{\{q\}\{q^2\} A^2}
\Big(q^{36}-A^2q^{32}-A^2q^{30}+q^{30}+q^{28}+A^4q^{26}-A^2q^{26}-2A^2q^{24}
+q^{24}+q^{22}-A^2q^{22}+q^{20}+\\+A^4q^{20}-A^2q^{20}+A^4q^{18}-2A^2q^{18}+q^{18}-2A^2q^{16}+q^{16}+A^4q^{14}+q^{14}-2A^2q^{14}+A^4q^{12}+2q^{12}-2A^2q^{12}+\\+q^{10}+A^4q^{10}-2A^2q^{10}-2A^2q^8+q^8+A^4q^8+A^4q^6+q^6-2A^2q^6-A^2q^4+q^4-A^2q^2+A^4q^2-A^2+1\Big)\\
....\\
\boxed{H^{T[2,2k+1]}_{[1,1]}=\dfrac{\{A\}\{\frac{A}{q}\} q^4}{\{q\}\{q^2\} A^2}
{\cal{H}}^{T[2,2k+1]}_{[1,1]}}
\end{array}
\ee
and the results coincide with the well known HOMFLY polynomials.
\begin{itemize}
\item{Floer case}
\end{itemize}
 \be
\begin{array}{l}
F^{T[2,1]}_{[1,1]}=1\\
F^{T[2,3]}_{[1,1]}=\bf{t}^4q^{12}+t^3q^8+t^2q^4\\
F^{T[2,5]}_{[1,1]}=\bf{t}^8q^{24}+t^7q^{20}+t^6q^{16}+t^5q^{12}+t^4q^8\\
F^{T[2,7]}_{[1,1]}=\bf{t}^{12}q^{36}+t^{11}q^{32}+t^{10}q^{28}+t^9q^{24}+t^8q^{20}+t^7q^{16}+t^6q^{12}
\\
F^{T[2,9]}_{[1,1]}=\bf{t}^{16}q^{48}+t^{15}q^{44}+t^{14}q^{40}+t^{13}q^{36}+t^{12}q^{32}+t^{11}q^{28}+t^{10}q^{24}+t^9q^{20}+t^8q^{16}
\\
....\\
\end{array}
\ee
\begin{itemize}
\item{Alexander case}
\end{itemize}
  \be
\begin{array}{l}
A^{T[2,1]}_{[1,1]}=1 \\
A^{T[2,3]}_{[1,1]}=\bf{q}^{12}-q^8+q^4 \\
A^{T[2,5]}_{[1,1]}=\bf{q}^{24}-q^{20}+q^{16}-q^{12}+q^8 \\
A^{T[2,7]}_{[1,1]}=\bf{q}^{36}-q^{32}+q^{28}-q^{24}+q^{20}-q^{16}+q^{12} \\
A^{T[2,9]}_{[1,1]}=\bf{q}^{48}-q^{44}+q^{40}-q^{36}+q^{32}-q^{28}+q^{24}-q^{20}+q^{16}
\\
....\\
\end{array}
\ee
\end{small}

\subsection{Case $(2,2k+1)$, all symmetric and antisymmetric representation
\cite{FGS}}

Let us introduce the standard notation for the $q$-Pochhammer symbol
\be
(x;q)_n \; = \; \prod_{j=0}^{n-1} (1 - x q^j) = (1-x) (1-xq) (1-xq^2) \ldots
(1-x q^{n-1}) \,,
\ee
Then the explicit expressions for $P_{R}^{T[2,2k+1]}$ with $R=S^r$ and
$R=\Lambda^r$:
\begin{eqnarray}
P_{S^r}^{T[2,2k+1]}
&=&\sum_{\ell=0}^r
\frac{(t^2;q^2)_{\ell}(q^2;q^2)_r(A^2;q^2)_{r+\ell}(A^2t^{-2};q^2)_{r-\ell}}
{({q}^2;q^2)_{\ell}(A^2;q^2)_r(q^2t^2;q^2)_{r+\ell}(q^2;q^2)_{r-\ell}}
\nonumber \\
&\times& \frac{(1-{q}^{4\ell}t^2)}{(1-{t}^2)}
(-1)^{r^2-\ell^2}q^{r(r+1)-\ell(\ell+1)}t^{2(r-\ell)}
\left[ (-1)^r A^{2r}q^{2r^2-2\ell^2-r}t^{r-2\ell} \right]^{k} \nonumber \\
&=&\sum_{\ell=0}^r
\frac{(\bf{q}^2;q^2t^2)_{\ell}(q^2t^2;q^2t^2)_r(-a^2t;q^2t^2)_{r+\ell}(-a^2q^{-2}t;q^2t^2)_{r-\ell}}
{(\bf{q}^2t^2;q^2t^2)_{\ell}(-a^2t;q^2t^2)_r(q^4t^2;q^2t^2)_{r+\ell}(q^2t^2;q^2t^2)_{r-\ell}}
\nonumber \\
&\times&
\frac{(1-\bf{q}^{4\ell+2}t^{4\ell})}{(1-\bf{q}^2)}\bf{a}^{-r}q^{3r-2\ell}t^{r-\ell}
\left[ (-1)^{r-\ell}a^rq^{r^2-\ell(\ell+1)}t^{r^2-\ell^2} \right]^{2k+1}
\label{Symmetric_DGR_refined}
\end{eqnarray}
\begin{eqnarray}
P_{\Lambda^r}^{T[2,2k+1]}
&=& \sum_{\ell=0}^r
\frac{({q}^2;t^2)_{\ell}(A^{-2};t^2)_{r+\ell}(A^{-2}q^{-2};t^2)_{r-\ell}(t^2;t^2)_r}{({t}^2;t^2)_{\ell}(q^2t^2;t^2)_{r+\ell}(t^2;t^2)_{r-\ell}(A^{-2};t^2)_r}
\nonumber \\
&\times& \frac{(1-{t}^{4\ell}q^2)}{(1-{q}^2)} (-1)^{\ell-r}
A^{2r}t^{\ell(\ell-1)-r(r-1)}q^{2r}\left[(-1)^{r}A^{2r}q^{2\ell-r}
t^{-2r^2+2\ell^2+r}\right]^{k}\nonumber \\
&=& \sum_{\ell=0}^r
\frac{(\bf{q}^2t^2;q^2)_{\ell}(-a^{-2}t^{-1};q^2)_{r+\ell}(-a^{-2}q^{-2}t^{-3};q^2)_{r-\ell}(q^2;q^2)_r}{(\bf{q}^2;q^2)_{\ell}(q^4t^2;q^2)_{r+\ell}(q^2;q^2)_{r-\ell}(-a^{-2}t^{-1};q^2)_r}
\nonumber \\
&\times& \frac{(1-\bf{q}^{4\ell+2}t^2)}{(1-\bf{q}^2t^2)}\bf{a}^r
q^{3r-2\ell}t^{3r-\ell}\left[(-1)^{\ell}a^{r}q^{-r^2+\ell(\ell+1)}t^{\ell}\right]^{2k+1}
\label{Anti-Symmetric_DGR_refined}
\end{eqnarray}

\begin{itemize}
\item{HOMFLY case}
\end{itemize}
\begin{eqnarray}
P_{S^r}^{T[2,2k+1]}
&=&\sum_{\ell=0}^r
\frac{(q^2;q^2)_r(A^2;q^2)_{r+\ell}(A^2q^{-2};q^2)_{r-\ell}}
{(A^2;q^2)_r(q^4;q^2)_{r+\ell}(q^2;q^2)_{r-\ell}}
\nonumber \\
&\times& \frac{(1-{q}^{4\ell+2})}{(1-{q}^2)}
(-1)^{r^2-\ell^2}q^{r(r+1)-\ell(\ell+1)}
\left[ (-1)^r A^{2r}q^{2r^2-2\ell(\ell+1)} \right]^{k}
\end{eqnarray}
\begin{eqnarray}
P_{\Lambda^r}^{T[2,2k+1]}
&=& \sum_{\ell=0}^r
\frac{(A^{-2};q^2)_{r+\ell}(A^{-2}q^{-2};q^2)_{r-\ell}(q^2;q^2)_r}{(q^4;q^2)_{r+\ell}(q^2;q^2)_{r-\ell}(A^{-2};q^2)_r}
\nonumber \\
&\times& \frac{(1-{q}^{4\ell+2})}{(1-{q}^2)} (-1)^{\ell-r}
A^{2r}q^{\ell(\ell-1)-r(r-3)}\left[(-1)^{r}A^{2r}
q^{-2r^2+2\ell(\ell+1)}\right]^{k}
\end{eqnarray}

\begin{itemize}
\item{Alexander case}
\end{itemize}
\begin{eqnarray}
P_{S^r}^{T[2,2k+1]}
=\dfrac{1+q^{4rp+2r}}{(1+q^{2r})q^{2rk}}
\end{eqnarray}
\begin{eqnarray}
P_{\Lambda^r}^{T[2,2k+1]} = \dfrac{1+q^{4rp+2r}}{(1+q^{2r})q^{2rk}}
\end{eqnarray}

\subsection{Case $(2,n)$ series $n=2 k$, representations
[1]$\otimes$[1,1]\label{2k1+11}}
\begin{small}

\be
\begin{array}{|c|}
\hline\\
P^{T[2,n]}_{[1],[1,1]}=c^{[2,1]}_{[1],[1,1]} M_{[2,1]}^{\ast} q^{-n}t^{n}+
c^{[1,1,1]}_{[1],[1,1]} M_{[1,1,1]}^{\ast} t^{3n}
\\
\\
\hline
\end{array}
\ee
with the coefficients:
\be
\boxed{
c^{[2,1]}_{[1],[1,1]}=1,\ \ \
c^{[1,1,1]}_{[1],[1,1]}=\frac{(1-q^2)(1+t^2+t^4)}{1-q^2 t^4}
}
\ee
\be
\begin{array}{l}
P^{T[2,0]}_{[1],[1,1]}=\dfrac{\{A\}\{\frac{A}{t}\}}{\{t\}^2\{t^2\}
A}\Big(A^2-1\Big) =
\bf{-\dfrac{(a^2t+1)(a^2t+q^2)}{a^3qt^{3/2}\{q^2\}\{q\}^2}}\Big( a^2t+1 \Big) =
 \bf{\dfrac{(a^2t+1)(a^2t+q^2)}{a^3t^{3/2}\{q^2\}\{q\}}}\Big( a^2t+1
\Big)\summ{j=0}{\infty}q^{2j} \\ \\

P^{T[2,2]}_{[1],[1,1]}=\dfrac{\{A\}\{\frac{A}{t}\} t^2}{\{t\}^2\{t^2\} A
q^2}\Big(t^4-q^2t^4+A^2q^2-1\Big) =
\bf{-\dfrac{(a^2t+1)^2(a^2t+q^2)}{a^3qt^{7/2}\{q^2\}\{q\}^2}}\Big(
t^3q^2a^2+q^6t^2-q^4+1 \Big) = \\ =
\bf{\dfrac{(a^2t+1)^2(a^2t+q^2)}{a^3t^{7/2}\{q^2\}\{q\}}}\Big(\dfrac{t^3q^2a^2+q^6t^2}{1-q^2}+q^2+1
\Big) =
\bf{\dfrac{(a^2t+1)^2(a^2t+q^2)}{a^3t^{7/2}\{q^2\}\{q\}}}\Big(q^2t^2(a^2t+q^4)\summ{j=0}{\infty}q^{2j}+q^2+1
\Big) \\ \\

P^{T[2,4]}_{[1],[1,1]}=\dfrac{\{A\}\{\frac{A}{t}\} t^4}{\{t\}^2\{t^2\} A
q^4}\Big(q^2t^8-q^4t^8+q^4A^2t^4-A^2t^4q^2+t^4-q^2t^4+A^2q^2-1\Big) =\\=
\bf{-\dfrac{(a^2t+1)^2(a^2t+q^2)}{a^3qt^{11/2}\{q^2\}\{q\}^2}}\Big(
(q^8t^5-q^6t^3+q^2t^3)a^2+q^{12}t^4-q^{10}t^2+q^6t^2-q^4+1 \Big) = \\ =
\bf{\dfrac{(a^2t+1)^2(a^2t+q^2)}{a^3t^{11/2}\{q^2\}\{q\}}}\Big(
q^8t^4(a^2t+q^{4})\summ{j=0}{\infty}q^{2j}+(a^2q^2t^3+q^6t^2+1)(q^2+1) \Big) \\
\\
P^{T[2,6]}_{[1],[1,1]}=\dfrac{\{A\}\{\frac{A}{t}\} t^6}{\{t\}^2\{t^2\} A
q^6}\Big(t^{12}q^4-t^{12}q^6-q^4t^8A^2+t^8q^6A^2+q^2t^8-q^4t^8+ q^4A^2t^4-
A^2t^4q^2+ t^4-q^2t^4+A^2q^2-1\Big) = \\ =
\bf{-\dfrac{(a^2t+1)^2(a^2t+q^2)}{a^3qt^{15/2}\{q^2\}\{q\}^2}}\Big(
(q^{14}t^7+q^8t^5-t^5q^{12}+q^2t^3-q^6t^3)a^2+q^{18}t^6-q^{16}t^4-q^{10}t^2+q^{12}t^4-q^4+q^6t^2+1
\Big) =\\= \bf{\dfrac{(a^2t+1)^2(a^2t+q^2)}{a^3t^{15/2}\{q^2\}\{q\}}}\Big(
q^{14}t^6(a^2t+q^{4})\summ{j=0}{\infty}q^{2j}+(q^8t^5a^2+q^2t^3a^2+q^{12}t^4+q^6t^2+1)(1+q^2)
\Big) \\
...\\
\boxed{P^{T[2,2k]}_{[1],[1,1]}=\dfrac{\{A\}\{\frac{A}{t}\}
t^{2k}}{\{t\}^2\{t^2\} A q^{2k}}{\cal{P}}^{T[2,2k]}_{[1],[1,1]}}
\end{array}
\ee
Our answer for the Hopf link $P^{T[2,2]}_{[1],[1,1]}$ is in full agreement with
the
known results, for example  \cite{IGV} and generalizes it to $n>2$.
\begin{itemize}
\item{HOMFLY case}
\end{itemize}
At the point $t=q$ one has:
\be
H^{T[2,n]}=s_{[2,1]}^{\ast} + s_{[1,1,1]}^{\ast} q^{2n}
\ee
Several first answers are:
\be
\begin{array}{l}
H^{T[2,0]}_{[1],[1,1]}=\dfrac{\{A\}\{\frac{A}{q}\}}{\{q\}^2\{q^2\}
A}\Big(A^2-1\Big) \\ \\
H^{T[2,2]}_{[1],[1,1]}=\dfrac{\{A\}\{\frac{A}{q}\}}{\{q\}^2\{q^2\}
A}\Big(q^4-q^6+A^2q^2-1\Big) \\ \\
H^{T[2,4]}_{[1],[1,1]}=\dfrac{\{A\}\{\frac{A}{q}\}}{\{q\}^2\{q^2\}
A}\Big(q^{10}-q^{12}+q^8A^2-A^2q^6+q^4-q^6+A^2q^2-1\Big) \\ \\
H^{T[2,6]}_{[1],[1,1]}=\dfrac{\{A\}\{\frac{A}{q}\}}{\{q\}^2\{q^2\}
A}\Big(q^{16}-q^{18}+q^{14}A^2-q^{12}A^2+q^{10}-q^{12}-A^2q^6+q^8A^2+q^4-q^6
+A^2q^2- 1\Big) \\
...\\
\boxed{H^{T[2,2k]}_{[1],[1,1]}=\dfrac{\{A\}\{\frac{A}{q}\}}{\{q\}^2\{q^2\}
A}{\cal{H}}^{T[2,2k]}_{[1],[1,1]}}
\end{array}
\ee
\begin{itemize}
\item{Alexander case}
\end{itemize}
\be
\begin{array}{l}
A^{T[2,0]}_{[1],[1,1]}=0  \\
A^{T[2,2]}_{[1],[1,1]}=\bf{q}^6-q^4-q^2+1  \\
A^{T[2,4]}_{[1],[1,1]}=\bf{q}^{12}-q^{10}-q^8+2q^6-q^4-q^2+1  \\
A^{T[2,6]}_{[1],[1,1]}=\bf{q}^{18}-q^{16}-q^{14}+2q^{12}-q^{10}-q^8+2q^6-q^4-q^2+1
 \\
....\\
\end{array}
\ee
\end{small}

\subsection{Case $(2,n)$ series $n=2 k$, representations
[1]$\otimes$[2]\label{2k1+2}}
\begin{small}

\be
\begin{array}{|c|}
\hline\\
P^{T[2,n]}_{[1],[2]}=c^{[3]}_{[1],[2]} M_{[3]}^{\ast} q^{-3n}+
c^{[2,1]}_{[1],[2]} M_{[2,1]}^{\ast} q^{-n}t^{n}
\\
\\
\hline
\end{array}
\ee
with the coefficients:
\be
\boxed{
c^{[3]}_{[1],[2]}=1,\ \ \ c^{[2,1]}_{[1],[2]}=\frac{(1-q^4)(1-t^4q^2)}{(1-q^4
t^2)(1-q^2t^2)}
}
\ee
\be
\begin{array}{l}
P^{T[2,0]}_{[1],[2]}=\dfrac{\{A\}^2\{Aq\}}{\{t\}^2\{tq\} } =
\bf{-\dfrac{(a^2t+1)^2(a^2t^3q^2+1)}{a^3qt^{3/2}\{q^2t\}\{q\}^2}} = 
\bf{\dfrac{(a^2t+1)(a^2t^3q^2+1)}{a^3qt^{3/2}\{q^2t\}\{q\}^2}}\Big( a^2t+1
\Big)\summ{j=0}{\infty}q^{2j} \\ \\

P^{T[2,2]}_{[1],[2]}=\dfrac{\{A\}\{Aq\}}{\{t\}^2\{tq\}q^{12}A
}\Big(q^8A^2t^2-t^4q^8-A^2q^4t^2+A^2q^4+t^4q^4-q^4t^2+t^2-1\Big) = \\ 
\bf{=-\dfrac{(a^2t+1)(a^2t^3q^2+1)}{a^3q^{13}t^{27/2}\{q^2t\}\{q\}^2}}\Big(
q^{10}t^9a^2+q^{12}t^8-a^2t^5q^6+a^2t^5q^4-t^4q^8+q^6t^4-q^2+1 \Big)  = \\
\bf{=\dfrac{(a^2t+1)(a^2t^3q^2+1)}{a^3q^{12}t^{27/2}\{q^2t\}\{q\}}}\Big(
(q^{10}t^8(a^2t+q^{2})\summ{j=0}{\infty}q^{2j}+(a^2t^5q^4+t^4q^6+1) \Big) \\
\\

P^{T[2,4]}_{[1],[2]}=\dfrac{\{A\}\{Aq\}}{\{t\}^2\{tq\}q^{24}A
}\Big({q}^{16}{t}^{6}{A}^{2}-{q}^{16}{t}^{8}-{A}^{2}{q}^{12}{t}^{6}
+{q}^{12}{t}^{4}{A}^{2}+
{q}^{12}{t}^{8}-{q}^{12}{t}^{6}-{t}^{4}{q}^{8}{A}^{2}+{q}^{8}{A}^{2}{t}^{2}
+{t}^{6}{q}^{8}- {t}^{4}{q}^{8}- \\
-{A}^{2}{q}^{4}{t}^{2}+{A}^{2}{q}^{4}+{t}^{4}{q}^{4}-{q}^{4}{t}^{2}+{t}^{2}-1\Big)
=\\ \bf{=-\dfrac{(a^2t+1)(a^2t^3q^2+1)}{a^3q^{25}t^{51/2}\{q^2t\}\{q\}^2}}\Big(
{q}^{22}{t}^{17}{a}^{2}+{q}^{24}{t}^{16}-{a}^{2}{t}^{13}{q}^{18}+
{q}^{16}{t}^{13}{a}^{2}-
{q}^{20}{t}^{12}+{q}^{18}{t}^{12}-{a}^{2}{t}^{9}{q}^{12}+{q}^{10}{t}^{9}{a}^{2}-\\
\bf{-} {q}^{14}{t}^{8}+ {q}^{12}{t}^{8}
-{a}^{2}{t}^{5}{q}^{6}+{a}^{2}{t}^{5}{q}^{4}-{t}^{4}{q}^{8}+{q}^{6}{t}^{4}-{q}^{2}+1
\Big)
= \\ \bf{=\dfrac{(a^2t+1)(a^2t^3q^2+1)}{a^3q^{24}t^{27/2}\{q^2t\}\{q\}}}
\Big(q^{22}t^{16}(a^2t+q^{2})\summ{j=0}{\infty}q^{2j}+(q^{18}t^{12}+q^{16}t^{13}a^2+q^{12}t^8+q^{10}t^9a^2+q^6t^4+a^2t^5q^4+1)
\Big) \\ \\

P^{T[2,6]}_{[1],[2]}=\dfrac{\{A\}\{Aq\}}{\{t\}^2\{tq\}q^{36}A }\Big(
{q}^{24}{t}^{10}{A}^{2}-{q}^{24}{t}^{12}-{A}^{2}{t}^{10}{q}^{20}+{q}^{20}{t}^{8}{A}^{2}+{q}^{20}{t}^{12}-{t}^{10}{q}^{20}
-{A}^{2}{t}^{8}{q}^{16}+{q}^{16}{t}^{6}{A}^{2}+{t}^{10}{q}^{16}-\\-{q}^{16}{t}^{8}-{A}^{2}{q}^{12}{t}^{6}+{q}^{12}{t}^{4}{A}^{2}+{q}^{12}{t}^{8}-{q}^{12}{t}^{6}-{t}^{4}{q}^{8}{A}^{2}+{q}^{8}{A}^{2}{t}^{2}+{t}^{6}{q}^{8}-{t}^{4}{q}^{8}-{A}^{2}{q}^{4}{t}^{2}+{A}^{2}{q}^{4}+{t}^{4}{q}^{4}-{q}^{4}{t}^{2}+{t}^{2}-1
\Big) =\\
\bf{=-\dfrac{(a^2t+1)(a^2t^3q^2+1)}{a^3q^{36}t^{75/2}\{q^2t\}\{q\}^2}}\Big({q}^{34}{t}^{25}{a}^{2}+{q}^{36}{t}^{24}-{a}^{2}{t}^{21}{q}^{30}+{q}^{28}{t}^{21}{a}^{2}-{q}^{32}{t}^{20}+{q}^{30}{t}^{20}-{a}^{2}{t}^{17}{q}^{24}+{q}^{22}{t}^{17}{a}^{2}-\\
\bf{-}{q}^{26}{t}^{16}+
{q}^{24}{t}^{16}-{a}^{2}{t}^{13}{q}^{18}+{q}^{16}{t}^{13}{a}^{2}-{q}^{20}{t}^{12}+{q}^{18}{t}^{12}-{a}^{2}{t}^{9}{q}^{12}+{q}^{10}{t}^{9}{a}^{2}-{q}^{14}{t}^{8}+{q}^{12}{t}^{8}-{a}^{2}{t}^{5}{q}^{6}+\\
\bf{+}{a}^{2}{t}^{5}{q}^{4}
-{t}^{4}{q}^{8}+{q}^{6}{t}^{4}-{q}^{2}+1 \Big)
= \dfrac{(a^2t+1)(a^2t^3q^2+1)}{a^3q^{36}t^{75/2}\{q^2t\}\{q\}}
\Big(q^{34}t^{24}(a^2t+q^{2})\summ{j=0}{\infty}q^{2j}+ \\
\bf{+}(q^{30}t^{20}+q^{28}t^{21}a^2+q^{24}t^{16}+q^{22}t^{17}a^2+q^{18}t^{12}+q^{16}t^{13}a^2+q^{12}t^8+q^{10}t^9a^2+q^6t^4+a^2t^5q^4+1)
\Big) \\
...\\
\boxed{P^{T[2,2k]}_{[1],[2]}=\dfrac{\{A\}\{Aq\}}{\{t\}^2\{tq\}
q^{12k}A}{\cal{P}}^{T[2,2k]}_{[1],[2]}=\dfrac{\{A\}\{Aq\}}{\{t\}^2\{tq\}
q^{12k}}{\cal{P}}^{T[2,2k]}_{[1],[2]}}
\end{array}
\ee

\begin{itemize}
\item{HOMFLY case}
\end{itemize}
At the point $t=q$ one has:
\be
H^{T[2,n]}=s_{[3]}^{\ast} + s_{[2,1]}^{\ast} q^{3n}
\ee
Several first answers are:
\be
\begin{array}{l}
H^{T[2,0]}_{[1],[2]}=\dfrac{\{A\}^2\{Aq\}}{\{q\}^2\{q^2\} } \\ \\
H^{T[2,2]}_{[1],[2]}=\dfrac{\{A\}\{Aq\}}{\{q\}^2\{q^2\}q^{12}
A}\Big(q^{10}A^2-q^{12}-A^2q^6+A^2q^4+q^8-q^6+q^2-1\Big) \\ \\
H^{T[2,4]}_{[1],[2]}=\dfrac{\{A\}\{Aq\}}{\{q\}^2\{q^2\}q^{24} A}\Big((
{q}^{22}-{q}^{18}+{q}^{16}-{q}^{12}+{q}^{10}-{q}^{6}+{q}^{4} )
{A}^{2}-{q}^{24}+{q}^{20}-{q}^{18}+{q}^{14}-{q}^{12}+{q}^{8}-{q}^{6}+{q}^{2}-1\Big)
\\ \\
H^{T[2,6]}_{[1],[2]}=\dfrac{\{A\}\{Aq\}}{\{q\}^2\{q^2\}q^{36} A} \Big(
({q}^{34}-{q}^{30}+{q}^{28}-{q}^{24}+{q}^{22}-{q}^{18}+{q}^{16}-{q}^{12}+{q}^{10}-{q}^{6}+{q}^{4})
{A}^{2}-{q}^{36}+{q}^{32}-{q}^{30}+{q}^{26}-\\-{q}^{24}+{q}^{20}-{q}^{18}+{q}^{14}-{q}^{12}+{q}^{8}-{q}^{6}+{q}^{2}-1
\Big) \\
...\\
\boxed{H^{T[2,2k]}_{[1],[2]}=\dfrac{\{A\}\{Aq\}}{\{q\}^2\{q^2\}q^{12k}
A}{\cal{H}}^{T[2,2k]}_{[1],[2]}}
\end{array}
\ee
\begin{itemize}
\item{Alexander case}
\end{itemize}
\be
\begin{array}{l}
A^{T[2,0]}_{[1],[2]}=0  \\
A^{T[2,2]}_{[1],[2]}=\bf{q}^{10}-q^{12}-2q^6+q^4+q^8+q^2-1  \\
A^{T[2,4]}_{[1],[2]}=\bf{q}^{22}-q^{24}-2q^{18}+q^{16}+q^{20}-2q^{12}+q^{10}+q^{14}-2q^6+q^4+q^8+q^2-1
 \\
A^{T[2,6]}_{[1],[2]}=\bf{-{q}^{36}}+{q}^{34}+{q}^{32}-2\,{q}^{30}+{q}^{28}+{q}^{26}-2\,{q}^{24}+{q}^{22}+{q}^{20}-2\,{q}^{18}+{q}^{16}+{q}^{14}-2\,{q}^{12}+{q}^{10}+{q}^{8}-\\
\bf{-}2\,{q}^{6}+{q}^{4}+{q}^{2}-1  \\
....\\
\end{array}
\ee
\end{small}

\subsection{Case $(2,n)$ series $n=2 k$, representations
[1,1]$\otimes$[1,1]\label{2k11+11}}
\be
\begin{array}{|c|}
\hline\\
P^{T[2,n]}_{[1,1],[1,1]}=c^{[2,2]}_{[1,1],[1,1]} M_{[2,2]}^{\ast}
q^{-2n}t^{2n}+c^{[2,1,1]}_{[1,1],[1,1]} M_{[2,1,1]}^{\ast} q^{-n}t^{3n}+
c^{[1,1,1,1]}_{[1,1],[1,1]} M_{[1,1,1,1]}^{\ast} t^{6n}
\\
\\
\hline
\end{array}
\ee
with the coefficients:
\be
\boxed{
c^{[2,1,1]}_{[1,1],[1,1]}=\frac{(1-q^2)(1+t^2)}{1-q^2 t^2},\ \ \
c^{[1,1,1,1]}_{[1,1],[1,1]}=\frac{(1-q^2)(1-t^2q^2)(1+t^4)(1+t^2+t^4)}{(1-q^2
t^4)(1-q^2t^6)}
}
\ee
\be
\begin{array}{l}
P^{T[2,0]}_{[1,1],[1,1]}=\dfrac{\{A\}^2\{\frac{A}{t}\}^2}{\{t\}^2\{t^2\}^2 A} =
\bf{\dfrac{(a^2t+1)(a^2t+q^2)}{a^4q^2t^2\{q^2\}^2\{q\}^2}}\Big(
(a^2t+1)(a^2t+q^2) \Big)=
\bf{\dfrac{(a^2t+1)^2(a^2t+q^2)^2}{a^4qt^2\{q^2\}^2\{q\}}}
\Big(\summ{j=0}{\infty}q^{2j} \Big)^2 \\ \\

P^{T[2,2]}_{[1,1],[1,1]}=\dfrac{\{A\}\{\frac{A}{t}\} t^3}{\{t\}^2\{t^2\}^2 A^2
q^4}\Big(q^4t^{10}-q^2t^{10}+t^8-q^2t^8-q^4A^2t^6+q^2A^2t^6-t^6 +q^2t^6-
q^4A^2t^4 +q^2A^2t^4 -t^4 +q^2t^4-\\ -q^2A^2t^2+ t^2 +q^4A^4 -A^2q^2\Big)=\\=
\bf{\dfrac{(a^2t+1)(a^2t+q^2)}{a^4t^6\{q^2\}^2\{q\}^2}}
\Big(t^6q^2a^4+(-q^6t^3-q^4t^3+t^5q^8+t^3q^2+t^5q^6+t^3)a^2+t^4q^{12}-q^{10}t^2+q^6-q^8t^2-\\
\bf{-}q^4+q^6t^2-q^2+q^4t^2+1 \Big)
=\\=\bf{\dfrac{(a^2t+1)(a^2t+q^2)q^2}{a^4t^6\{q^2\}^2\}}} \Big(
1+{t}^{3}{a}^{2}+ ( 1+3\,{t}^{3}{a}^{2}+{t}^{6}{a}^{4} ) {q}^{2}+ (
4\,{t}^{3}{a}^{2}+{t}^{2}+2\,{t}^{6}{a}^{4} ) {q}^{4}+ \\ \bf{+} (
3\,{t}^{2}+4\,{t}^{3}{a}^{2}+{t}^{5}{a}^{2}+3\,{t}^{6}{a}^{4} ) {q}^{6}+ (
4\,{t}^{3}{a}^{2}+4\,{t}^{2}+3\,{t}^{5}{a}^{2}+4\,{t}^{6}{a}^{4} ) {q}^{8}+ (
4\,{t}^{2}+5\,{t}^{6}{a}^{4}+4\,{t}^{3}{a}^{2}+5\,{t}^{5}{a}^{2} ) {q}^{10} +
... \Big) \\ \\

P^{T[2,4]}=\dfrac{\{A\}\{\frac{A}{t}\} t^7}{\{t\}^2\{t^2\}^2 A^2
q^8}\Big(q^8t^{18}-q^6t^{18}-q^6t^{16}+q^4t^{16}+A^2q^6t^{14}-q^4t^{14}
-A^2q^8t^{14}+q^6t^{14}+2A^2q^6t^{12}-A^2q^4t^{12}+\\+q^6t^{12}-A^2q^8t^{12}-2q^4t^{12}+q^2t^{12}-A^4q^6t^8+A^4q^8t^8-2A^2q^6t^8-2q^2t^8+3A^2q^4t^8+t^8-A^2q^2t^8+q^4t^8+q^4t^6-\\-A^4q^6t^6+A^4q^4t^6+A^2q^4t^6-A^2q^6t^6-t^6+2A^2q^2t^4-2A^2q^4t^4+q^2t^4-A^4q^4t^4+A^4q^6t^4-t^4+A^4q^6t^2-\\-A^4q^4t^2-A^2q^4t^2+t^2+A^4q^4-A^2q^2\Big)
=\\= \bf{\dfrac{(a^2t+1)(a^2t+q^2)}{a^4t^{10}\{q^2\}^2\{q\}^2}}
\Big((t^6q^8+t^6q^2-t^8q^{12}-t^6q^6-q^{10}t^8+q^{14}t^{10}
\bf{+}q^8t^8+q^6t^8-t^6q^4)a^4+(-t^7q^{18}+t^3+\\ \bf{+}t^5q^4-2t^7q^{16}+
q^{10}t^7-2q^4t^3+2t^5q^6+q^{20}t^9-t^5q^8+2q^{12}t^7
\bf{+}q^{18}t^9-3t^5q^{10}+q^8t^3+t^5q^{14})a^2-\\ \bf{-}q^{22}t^6- q^{16}t^4
\bf{+} q^4t^2- q^2-q^{20}t^6+ q^{18}t^6+t^4q^{18}-2t^4q^{14}+t^4q^8
\bf{+}q^6-2q^8t^2+q^{24}t^8-q^4+q^{10}t^4+\\ \bf{+}q^{12}t^2+q^{16}t^6+1 \Big)
=\\ \bf{=} \bf{\dfrac{(a^2t+1)(a^2t+q^2)q^2}{a^4t^{10}\{q^2\}^2}} \Big(
1+t^3a^2+(1+2t^3a^2+t^6a^4)q^2+(t^3a^2+t^6a^4+t^2+t^5a^2)q^4+(2t^2+ 4t^5a^2+\\
\bf{+}t^8a^4)q^6
\bf{+}(6t^5a^2+3t^8a^4+t^2+t^4)q^8+(3t^4+5t^5a^2+t^7a^2+4t^8a^4)q^{10}+...
\Big)
\end{array}
\ee
\be
\begin{array}{l}
P^{T[2,6]}_{[1,1],[1,1]}=\dfrac{\{A\}\{\frac{A}{t}\} t^{11}}{\{t\}^2\{t^2\}^2
A^2
q^{12}}\Big(q^{12}t^{26}-q^{10}t^{26}+q^8t^{24}-q^{10}t^{24}-A^2q^{12}t^{22}
-q^8t^{22}+A^2q^{10}t^{22}+q^{10}t^{22}-A^2q^8t^{20}-\\-2q^8t^{20}+2A^2q^{10}t^{20}+q^{10}t^{20}-A^2q^{12}t^{20}+q^6t^{20}-2A^2q^{10}t^{16}+A^4q^{12}t^{16}-A^2q^6t^{16}+q^4t^{16}-A^4q^{10}t^{16}+\\+3A^2q^8t^{16}+q^8t^{16}-2q^6t^{16}-A^4q^{10}t^{14}-A^2q^{10}t^{14}-q^6t^{14}+A^4q^8t^{14}+q^8t^{14}-A^2q^6t^{14}+2A^2q^8t^{14}-q^4t^{12}+\\+q^2t^{12}-A^2q^8t^{12}-A^4q^8t^{12}+A^4q^{10}t^{12}-A^2q^4t^{12}+2A^2q^6t^{12}+A^4q^{10}t^{10}-2A^4q^8t^{10}-q^4t^{10}+q^6t^{10}+A^4q^6t^{10}-\\-2A^2q^8t^{10}-A^2q^4t^{10}+3A^2q^6t^{10}+q^6t^8-A^2q^8t^8+2A^2q^4t^8+t^8-A^2q^2t^8-2q^2t^8+A^4q^4t^6+2A^2q^4t^6+A^4q^8t^6-\\-t^6-2A^4q^6t^6+q^4t^6-2A^2q^6t^6+2A^2q^2t^4-t^4-A^4q^4t^4+q^2t^4-A^2q^6t^4-A^2q^4t^4+A^4q^8t^4-A^2q^4t^2+A^4q^6t^2-\\-A^4q^4t^2+t^2+A^4q^4-A^2q^2\Big)
=\\=
\bf{\dfrac{(a^2t+1)(a^2t+q^2)}{a^4t^{14}\{q^2\}^2\{q\}^2}}\Big(({q}^{26}{t}^{14}-{q}^{24}{t}^{12}-{q}
^{22}{t}^{12}+{q}^{20}{t}^{12}+{t}^{12}{q}^{18}+ {t}^{10}{q}^{20}-\\ \bf{-}
{q}^{18}{t}^{10}-2\,{t}^{10}{q}^{16}+{t}^{10}{q}^{12}+{t}^{10}{q}^{10}+{q}^{14}{t}^{8}
-2\,{q}^{10}{t}^{8}+
{t}^{8}{q}^{6}+{t}^{6}{q}^{8}-{q}^{6}{t}^{6}-{q}^{4}{t}^{6}+{t}^{6}{q}^{2} )
{a}^{4}+({t}^{5}{q}^{4}-\\ \bf{-} 2\,{q}^{20}{t}^{9}-
3\,{q}^{14}{t}^{7}+{t}^{13}{q}^{32}-2\,{t}^{5}{q}^{8}-  {q}^{30}{t}^{11} + 
{q}^{26}{t}^{9}+
{q}^{20}{t}^{7}+{q}^{18}{t}^{7}+{q}^{14}{t}^{5}+{t}^{3}+{q}^{12}{t}^{5}+{q}^{8}{t}^{3}-\\
\bf{-}2\,{q}^{28}{t}^{11}- 3\,{q}^{22}{t}^{9}-
2\,{q}^{16}{t}^{7}-2\,{q}^{10}{t}^{5}+2\,{q}^{24}{t}^{11}+{t}^{7}{q}^{8}+{t}^{9}{q}^{14}+{t}^{13}{q}^{30}+
2\,{q}^{16}{t}^{9}+{t}^{11}{q}^{22}-2\,{t}^{3}{q}^{4}+\\
\bf{+}{q}^{6}{t}^{5}+{t}^{9}{q}^{18}
+2\,{t}^{7}{q}^{10} )
{a}^{2}+1+{q}^{12}{t}^{2}-2\,{q}^{8}{t}^{2}-{q}^{2}+{q}^{6}+{q}^{8}{t}^{4}-{q}^{28}{t}^{8}+{q}^{4}{t}^{2}+{t}^{6}{q}^{14}
+{q}^{20}{t}^{8}-{t}^{4}{q}^{12}-\\
\bf{-}{t}^{4}{q}^{14}+{q}^{12}{t}^{6}-{q}^{18}{t}^{6}-2\,{t}^{6}{q}^{20}-{q}^{4}+{q}^{36}{t}^{12}+{q}^{18}{t}^{4}
+{q}^{30}{t}^{10}+{t}^{6}{q}^{24}+{q}^{28}{t}^{10}-{t}^{10}{q}^{34}-{t}^{10}{q}^{32}+\\
\bf{+}{q}^{30}{t}^{8}+{q}^{22}{t}^{8}-2\,{q}^{26}{t}^{8} \Big)=\\
\bf{=}\dfrac{(a^2t+1)(a^2t+q^2)q^2}{a^4t^{14}\{q^2\}^2}\Big(
1+t^3a^2+(1+2t^3a^2+t^6a^4)q^2+(t^3a^2+t^6a^4+t^2+t^5a^2)q^4+(3t^5a^2+\\
\bf{+}2t^2+t^8a^4)q^6
\bf{+}(3t^5a^2+2t^8a^4+t^2+t^7a^2+t^4)q^8+(4t^7a^2+t^5a^2+t^8a^4+t^{10}a^4+2t^4)q^{10}
+ ...\Big)\\
...\\
\boxed{P^{T[2,2k]}_{[1,1],[1,1]}=\dfrac{\{A\}\{\frac{A}{t}\}
t^{4k-1}}{\{t\}^2\{t^2\}^2 A^2 q^{4k}}{\cal{P}}^{T[2,2k]}_{[1,1],[1,1]}}
\end{array}
\ee
Our answer for Hopf link $K_{2,2}$ is in full agreement with the known
results,
for example  \cite{IGV} and generalizes it to $n>2$.
\begin{itemize}
\item{HOMFLY case}
\end{itemize}
At the point $t=q$ one has:
\be
H^{T[2,n]}_{[1,1],[1,1]}= s_{[2,2]}^{\ast}+s_{[2,1,1]}^{\ast} q^{2n}+
s_{[1,1,1,1]}^{\ast} q^{6n}
\ee
Several first answers are:
\be
\begin{array}{l}
H^{T[2,0]}_{[1,1],[1,1]}=\dfrac{\{A\}^2\{\frac{A}{q}\}^2}{\{q\}^2\{q^2\}^2 A}
\\ \\
H^{T[2,2]}_{[1,1],[1,1]}=\dfrac{\{A\}\{\frac{A}{q}\}q}{\{q\}^2\{q^2\}^2
A^2}\Big(q^{12}-q^{10}-A^2q^8-q^8+2q^6+A^2q^4+A^4q^2-q^2-A^2q^2-A^2+1\Big) \\
\\
H^{T[2,4]}_{[1,1],[1,1]}=\dfrac{\{A\}\{\frac{A}{q}\} q}{\{q\}^2\{q^2\}^2
A^2}\Big(q^{24}-q^{22}-q^{20}-A^2q^{20}+2q^{18}+2A^2q^{16}-A^2q^{14}+A^4q^{14}-
2q^{14}-2A^2q^{12}+q^{12}-\\-A^4q^{12}+2A^2q^{10}-A^4q^{10}+q^{10}+2A^4q^8-q^8-2A^2q^6+q^6-A^4q^4+A^2q^4+A^4q^2-q^2-A^2+1\Big)
\\ \\
H^{T[2,6]}_{[1,1],[1,1]}=\dfrac{\{A\}\{\frac{A}{q}\} q}{\{q\}^2\{q^2\}^2 A^2
}\Big(q^{36}-q^{34}-A^2q^{32}-q^{32}+2q^{30}+2A^2q^{28}-2q^{26}-A^2q^{26}
+A^4q^{26}-2A^2q^{24}-A^4q^{24}+q^{24}+\\+q^{22}+2A^2q^{22}-A^4q^{22}-q^{20}+2A^4q^{20}+A^2q^{20}-2A^2q^{18}-2A^4q^{16}+A^4q^{14}+A^2q^{14}-A^2q^{12}+q^{12}+A^4q^{12}-\\-A^4q^{10}+A^4q^8-q^8-A^2q^6+q^6+A^2q^4-A^4q^4+A^4q^2-q^2+1-A^2\Big)
\\
...\\
\boxed{H^{T[2,2k]}_{[1,1],[1,1]}=\dfrac{\{A\}\{\frac{A}{q}\}
q}{\{q\}^2\{q^2\}^2 A^2 }{\cal{H}}^{T[2,2k]}_{[1,1],[1,1]}}
\end{array}
\ee
\begin{itemize}
\item{Alexander case}
\end{itemize}
\be
\begin{array}{l}
A^{T[2,0]}_{[1,1],[1,1]}=1  \\
A^{T[2,2]}_{[1,1],[1,1]}=\bf{q}^{10}-q^8-2q^6+2q^4+q^2-1  \\
A^{T[2,4]}_{[1,1],[1,1]}=\bf{q}^{18}-q^{16}-2q^{14}+2q^{12}+2q^{10}-2q^8-2q^6+2q^4+q^2-1
 \\
A^{T[2,6]}_{[1,1],[1,1]}=\bf{q}^{26}-q^{24}-2q^{22}+2q^{20}+2q^{18}-2q^{16}-2q^{14}+2q^{12}+2q^{10}-2q^8-2q^6+2q^4+q^2-1
 \\
....\\
\end{array}
\ee

\subsection{Case $(2,n)$ series $n=2 k$, representations
[2]$\otimes$[1,1]\label{2k2+11}}
\be
\begin{array}{|c|}
\hline\\
P^{T[2,n]}_{[2],[1,1]}=c^{[3,1]}_{[2],[1,1]} M_{[3,1]}^{\ast}
q^{-3n}t^{n}+c^{[2,1,1]}_{[2],[1,1]} M_{[2,1,1]}^{\ast} q^{-n}t^{3n}
\\
\\
\hline
\end{array}
\ee
with the coefficients:
\be
\boxed{
c^{[3,1]}_{[2],[1,1]}=1,\ \ \
c^{[2,1,1]}_{[2],[1,1]}=\frac{(1-q^4)(1-t^6q^2)}{(1-q^2 t^2)(1-q^4t^4)}
}
\ee
\be
\begin{array}{l}
P^{T[2,0]}_{[2],[1,1]}=\dfrac{\{A\}\{\frac{A}{t}\}\{Aq\}}{\{t\}^2\{t^2\}\{tq\}
A}\Big(A^2-1\Big) =
\bf{-\dfrac{(a^2t+1)(a^2t+q^2)(a^2t^3q^2+1)}{a^4qt^3\{q^2\}\{q\}\{q^2t\}}}\Big(
a^2t+1 \Big)\summ{j=0}{\infty}q^{2j} \\ \\

P^{T[2,2]}_{[2],[1,1]}=\dfrac{\{A\}\{\frac{A}{t}\}\{Aq\}t^2}{\{t\}^2\{t^2\}\{tq\}
Aq^{6}} \Big(q^4A^2-t^4q^4+t^4-1\Big)=\\=
\bf{-\dfrac{(a^2t+1)(a^2t+q^2)(a^2t^3q^2+1)}{a^4q^5t^{8}\{q^2\}\{q\}\{q^2t\}}}\Big(
(q^2+1)(a^2t^5q^4+1) + q^{8}t^4(a^2t+1)\summ{j=0}{\infty}q^{2j}\Big)
\\ \\

P^{T[2,4]}_{[2],[1,1]}=\dfrac{\{A\}\{\frac{A}{t}\}\{Aq\}t^4}{\{t\}^2\{t^2\}\{tq\}
Aq^{12}} \Big(t^4q^8A^2-t^8q^8-q^4t^4A^2+q^4A^2+q^4t^8-t^4q^4+t^4-1\Big)=\\=
\bf{-\dfrac{(a^2t+1)(a^2t+q^2)(a^2t^3q^2+1)}{a^4q^9t^{14}\{q^2\}\{q\}\{q^2t\}}}\Big(
(q^2+1)(t^4q^8+1)(a^2t^5q^4+1) + q^{16}t^8(a^2t+1)\summ{j=0}{\infty}q^{2j}\Big)
\end{array}
\ee
\be
\begin{array}{l}
P^{T[2,6]}_{[2],[1,1]}=\dfrac{\{A\}\{\frac{A}{t}\}\{Aq\}t^{6}}{\{t\}^2\{t^2\}\{tq\}
Aq^{18}} \Big(  (q^{12}t^8+t^4q^8-t^8q^8-q^4t^4+q^4)A^2-q^{12}t^{12}+q^8t^{12}-t^8q^8-t^4q^4+q^4t^8+t^4-1
\Big) =\\=
\bf{-\dfrac{(a^2t+1)(a^2t+q^2)(a^2t^3q^2+1)}{a^4q^{13}t^{20}\{q^2\}\{q\}\{q^2t\}}}\Big(
(q^2+1)(t^5q^4a^2+1)(q^{16}t^8+t^4q^8+1) +q^{24}t^{12}(a^2t+1)\summ{j=0}{\infty}q^{2j}\Big) \\
...\\
\boxed{P^{T[2,2k]}_{[2],[1,1]}=\dfrac{\{A\}\{\frac{A}{t}\}\{Aq\}t^{2k}}{\{t\}^2\{t^2\}\{tq\}
Aq^{6k}}{\cal{P}}^{T[2,2k]}_{[2],[1,1]}}
\end{array}
\ee

\begin{itemize}
\item{HOMFLY case}
\end{itemize}
At the point $t=q$ one has:
\be
H^{T[2,n]}_{[2],[1,1]}= s_{[3,1]}^{\ast}+s_{[2,1,1]}^{\ast} q^{4n}
\ee
Several first answers are:
\be
\begin{array}{l}
H^{T[2,0]}_{[2],[1,1]}=\dfrac{\{A\}\{\frac{A}{q}\}\{Aq\}}{\{q\}^2\{q^2\}^2
A}\Big(A^2-1\Big) \\
\\
H^{T[2,2]}_{[2],[1,1]}=\dfrac{\{A\}\{\frac{A}{q}\}\{Aq\}}{\{q\}^2\{q^2\}^2
Aq^{4}} \Big(q^4A^2-q^8+q^4-1\Big) \\
\\
H^{T[2,4]}_{[2],[1,1]}=\dfrac{\{A\}\{\frac{A}{q}\}\{Aq\}}{\{q\}^2\{q^2\}^2
Aq^{8}} \Big(
q^{12}A^2-q^{16}-q^8A^2+q^4A^2+q^{12}-q^8+q^4-1
\Big) \\
\\
H^{T[2,6]}_{[2],[1,1]}=\dfrac{\{A\}\{\frac{A}{q}\}\{Aq\}}{\{q\}^2\{q^2\}^2
Aq^{12}} \Big( (q^{20}-q^{16}+q^{12}-q^8+q^4)A^2-q^{24}+q^{20}-q^{16}+q^{12}-q^8+q^4-1 \Big) \\
...\\
\boxed{H^{T[2,2k]}_{[2],[1,1]}=\dfrac{\{A\}\{\frac{A}{q}\}\{Aq\}}{\{q\}^2\{q^2\}^2
Aq^{4k}}{\cal{H}}^{T[2,2k]}_{[2],[1,1]}}
\end{array}
\ee
\begin{itemize}
\item{Alexander case}
\end{itemize}
\be
\begin{array}{l}
A^{T[2,0]}_{[2],[1,1]}=0  \\ \\
A^{T[2,2]}_{[2],[1,1]}=\bf{-}q^8+2q^4-1  \\ \\
A^{T[2,4]}_{[2],[1,1]}=\bf{-}q^{16}+2q^{12}-2q^8+2q^4-1
 \\ \\
A^{T[2,6]}_{[2],[1,1]}=\bf{-}q^{24}+2q^{20}-2q^{16}+2q^{12}-2q^8+2q^4-1
 \\
....\\
\end{array}
\ee

\subsection{Case $(2,n)$ series $n=2 k$, representations
[2]$\otimes$[2]\label{2k2+2}}
\be
\begin{array}{|c|}
\hline\\
P^{T[2,n]}_{[2],[2]}=c^{[4]}_{[2],[2]} M_{[4]}^{\ast}
q^{-6n}+c^{[3,1]}_{[2],[2]} M_{[3,1]}^{\ast} q^{-3n}t^{n}+c^{[2,2]}_{[2],[2]}
M_{[2,2]}^{\ast} q^{-2n}t^{2n}
\\
\\
\hline
\end{array}
\ee
with the coefficients:
\be
\boxed{
c^{[4]}_{[2],[2]}=1,\ \ \
c^{[3,1]}_{[2],[2]}=\frac{(1+q^2)(1-q^4)(1+t^2q^2)(1-t^2)}{(1-q^2
t^2)(1-q^6t^2)}, \ \ \
c^{[2,2]}_{[2],[2]}=\frac{(1-q^2)(1-q^4)(1-t^4q^2)(1+t^2)}{(1-q^2
t^2)^2(1-q^4t^2)}
}
\ee
\be
\begin{array}{l}
P^{T[2,0]}_{[2],[2]}=\dfrac{\{A\}\{Aq\}}{\{t\}^2\{tq\}^2
A^2q}\Big((A^2-1)(A^2q^2-1)\Big) =
\bf{\dfrac{(a^2t+1)^2(1+q^2t^3a^2)^2q}{a^4t^3\{q\}\{q^2t\}}}\summ{j=0}{\infty}q^{2j}\cdot
\summ{j=0}{\infty}q^{2j}t^j \\ \\

P^{T[2,2]}_{[2],[2]}=\dfrac{\{A\}\{Aq\}}{\{t\}^2\{tq\}^2 A^2q^{13}} \Big( q^{10}A^4+(-q^6+q^6t^2-q^4+q^4t^2-t^2q^{10}-t^2q^8)A^2+\\+t^4q^2-t^2+t^4q^8-q^2t^2+q^6t^2-q^6t^4+q^4t^2-t^4q^4+1 \Big)
=
\\= \bf{\dfrac{(a^2t+1)^2(1+q^2t^3a^2)^2}{a^4t^{15}q^{11}\{q\}\{q^2t\}}}
\Big(1+q^2t+q^4t^5a^2+(a^2t^6+t^4+t^7a^2)q^6+(t^6+t^7a^2+t^5+a^2t^8)q^8+\\\bf{+}(t^6+t^7+a^2t^8+t^{12}a^4+2t^9a^2)q^{10}+(t^7+2t^8+2a^2t^{10}+t^{13}a^4+t^{12}a^4+t^{11}a^2+2t^9a^2)q^{12}+\\\bf{+}(2t^9+2t^8+2a^2t^{10}+t^{13}a^4+t^{12}a^4+3t^{11}a^2+t^9a^2+t^{14}a^4+t^{12}a^2)q^{14}+...\Big) \\ \\

P^{T[2,4]}_{[2],[2]}=\dfrac{\{A\}\{Aq\}}{\{t\}^2\{tq\}^2 A^2q^{25}} \Big(
({q}^{18}{t}^{4}-{q}^{16}{t}^{4}-{q}^{14}{t}^{4}+{q}^{12}{t}^{4}+
{q}^{16}{t}^{2}+ {q}^{14}{t}^{2}- {q}^{12}{t}^{2}-{q}^{10}{t}^{2}+{q}^{10})
{A}^{4}+ \\+(-{q}^{18}{t}^{6}+2\,{q}^{14}{t}^{6}- {q}^{10}{t}^{6}-
{q}^{16}{t}^{4}+{q}^{12}{t}^{4}+3\,{q}^{10}{t}^{4}-2\,{q}^{14}{t}^{4}-{q}^{6}{t}^{4}-{q}^{12}{t}^{2}-
2\,{q}^{10}{t}^{2}+ {q}^{4}{t}^{2}+2\,{q}^{6}{t}^{2}-\\-{q}^{4}-{q}^{6}
){A}^{2}+{q}^{16}{t}^{8}-{q}^{14}{t}^{8}+{q}^{10}{t}^{8}- {q}^{12}{t}^{8}-
2\,{q}^{10}{t}^{6}+
{q}^{14}{t}^{6}+{q}^{6}{t}^{6}+{q}^{12}{t}^{4}+{q}^{10}{t}^{4}-2\,{q}^{6}{t}^{4}-
{q}^{4}{t}^{4}+\\ +{q}^{2}{t}^{4} +{q}^{6}{t}^{2}
+{q}^{4}{t}^{2}-{q}^{2}{t}^{2}-{t}^{2}+1\Big)=
\\ =\bf{\dfrac{(a^2t+1)^2(1+q^2t^3a^2)^2}{a^4t^{27}q^{23}\{q\}^2\{q^2t\}^2}}
\Big(({q}^{22}{t}^{20}+{q}^{18}{t}^{18}- {q}^{20}{t}^{18}-
{q}^{18}{t}^{16}+{q}^{16}{t}^{16}+{q}^{16}{t}^{14}-{q}^{14}{t}^{14}+{q}^{10}{t}^{12}-\\
\bf{-}{q}^{12}{t}^{12} ) {a}^{4}+ ({q}^{24}{t}^{19}+
{q}^{20}{t}^{17}-2\,{q}^{20}{t}^{15}+2\,{q}^{18}{t}^{15}-{q}^{16}{t}^{13}+{q}^{14}{t}^{13}+2\,{q}^{12}{t}^{11}-3\,{q}^{14}{t}^{11}+\\
\bf{+}{q}^{16}{t}^{11}-2\,{q}^{8}{t}^{7}+{q}^{6}{t}^{7}+{q}^{10}{t}^{7}-{q}^{6}{t}^{5}+{q}^{4}{t}^{5}
){a}^{2}+ {q}^{24}{t}^{16}+ {q}^{20}{t}^{14} -{q}^{22}{t}^{14}
-{q}^{20}{t}^{12}+{q}^{16}{t}^{12}+\\
\bf{+}{q}^{18}{t}^{10}-2\,{q}^{16}{t}^{10}+{q}^{14}{t}^{10}+{q}^{12}{t}^{6}+
{q}^{8}{t}^{6}-2\,{q}^{10}{t}^{6}+{q}^{6}{t}^{4}-{q}^{8}{t}^{4}+{q}^{6}{t}^{2}-{q}^{4}{t}^{2}-{q}^{2}+1\Big)
=
\\= \bf{\dfrac{(a^2t+1)^2(1+q^2t^3a^2)^2}{a^4t^{27}q^{23}\{q\}\{q^2t\}}}
\Big(1+{q}^{2}t+{t}^{5}{a}^{2}{q}^{4}+ ( {t}^{7}{a}^{2}+{t}^{4}+{t}^{6}{a}^{2}
) {q}^{6}+ ( {t}^{8}{a}^{2}+{t}^{5}+{t}^{6} ) {q}^{8}
+\\ \bf{+} ( {t}^{12}{a}^{4}+{t}^{7}+{t}^{9}{a}^{2} ) {q}^{10}+ (
{t}^{8}+{t}^{10}{a}^{2}+{t}^{13}{a}^{4}+2\,{t}^{11}{a}^{2} ) {q}^{12}
+ ( {t}^{10}+{t}^{9}+2\,{t}^{12}{a}^{2}+{t}^{13}{a}^{2} ) {q}^{14}+\\ \bf{+} (
{t}^{11}+{t}^{12}+{t}^{14}{a}^{2}+2\,{t}^{13}{a}^{2}+{t}^{16}{a}^{4} )
{q}^{16}+ (
{t}^{18}{a}^{4}+3\,{t}^{15}{a}^{2}+2\,{t}^{12}+2\,{t}^{14}{a}^{2}+{t}^{17}{a}^{4}+{t}^{13}
) {q}^{18}+...\Big) \\ \\
\end{array}
\ee
\\
...\\
$\boxed{P^{T[2,2k]}_{[2],[2]}=\dfrac{\{A\}\{Aq\}}{\{t\}^2\{tq\}^2
A^2q^{12k+1}}{\cal{P}}^{T[2,2k]}_{[2],[2]}}$

\begin{itemize}
\item{HOMFLY case}
\end{itemize}
At the point $t=q$ one has:
\be
H^{T[2,n]}_{[2],[2]}= s_{[4]}^{\ast}q^{-6n}+s_{[3,1]}^{\ast}
q^{-2n}+s_{[2,2]}^{\ast}
\ee
Several first answers are:
\be
\begin{array}{l}
H^{T[2,0]}_{[2],[2]}=\dfrac{\{A\}\{Aq\}}{\{q\}^2\{q^2\}^2
A^2q}\Big((A^2-1)(A^2q^2-1)\Big) \\
\\
H^{T[2,2]}_{[2],[2]}=\dfrac{\{A\}\{Aq\}}{\{q\}^2\{q^2\}^2 A^2q^{13}} \Big( q^{10}A^4+(q^8-q^{10}-q^{12}-q^4)A^2+q^{12}-q^{10}+2q^6-q^4-q^2+1 \Big)
\\
\\
H^{T[2,4]}_{[2],[2]}=\dfrac{\{A\}\{Aq\}}{\{q\}^2\{q^2\}^2 A^2q^{25}} \Big((
-{q}^{24}+{q}^{20}-2\,{q}^{18}+2\,{q}^{14}-2\,{q}^{12}-{q}^{10}+2\,{q}^{8}-{q}^{4}
) {A}^{2}+ ( {q}^{22}-{q}^{20}+\\+2\,{q}^{16}-{q}^{14}-{q}^{12}+{q}^{10} )
{A}^{4}+{q}^{24}-{q}^{22}+{q}^{18}-{q}^{16}+{q}^{14}+{q}^{12}-2\,{q}^{10}+2\,{q}^{6}-{q}^{4}-{q}^{2}+1\Big) \\
\\
...\\
\boxed{H^{T[2,2k]}_{[2],[2]}=\dfrac{\{A\}\{Aq\}}{\{q\}^2\{q^2\}^2
A^2q^{12k+1}}{\cal{H}}^{T[2,2k]}_{[2],[2]}}
\end{array}
\ee
\begin{itemize}
\item{Alexander case }
\end{itemize}
\be
\begin{array}{l}
A^{T[2,0]}_{[2],[2]}=0  \\ \\
A^{T[2,2]}_{[2],[2]}=\bf{-}q^{10}+q^8+2q^6-2q^4-q^2+1 \\ \\
A^{T[2,4]}_{[2],[2]}=\bf{-}q^{18}+q^{16}+2q^{14}-2q^{12}-2q^{10}+2q^8+2q^6-2q^4-q^2+1  \\ \\
 \\
....\\
\end{array}
\ee

\section{Non-torus knots}

\begin{small}

\subsection{$5_2$ and descendants: fundamental representation  \label{52}}
\be
\begin{array}{|c|}
\hline
\\
P^{5_2,n}_{[1]}=c_{[1]}^{[3]} M_{3}^{\ast} q^{-2n} + c_{[1]}^{[2,1]}
M_{2,1}^{\ast} q^{-\frac{2n}{3}} t^{ \frac{2n}{3}}+c_{[1]}^{[1,1,1]}
M_{1,1,1}^{\ast} t^{ 2n }
\\
\\
\hline
\end{array}
\ee
with the coefficients:
\fr{
c_{[1]}^{[3]}=1, \ \ \ c_{[1]}^{[2,1]}=-{\frac {q \left( -1+t \right)  \left(
t+1 \right)  \left( {q}^{6}{t}^
{6}-{t}^{4}{q}^{2}+{q}^{5}{t}^{3}+{q}^{3}{t}^{3}-{q}^{4}{t}^{2}+{q}^{3
}t+1 \right) }{t \left( -1+{q}^{2}t \right)  \left( {q}^{2}t+1
 \right) }}
\\
c_{[1]}^{[1,1,1]}={\frac {{q}^{6} \left( 1+{t}^{2} \right)  \left( t+{t}^{2}+1
\right)
 \left( {t}^{2}-t+1 \right)  \left( -1+t \right) ^{2} \left( t+1
 \right) ^{2}{t}^{2}}{ \left( -1+tq \right)  \left( -1+{t}^{2}q
 \right)  \left( {t}^{2}q+1 \right)  \left( tq+1 \right) }}}
First two superpolynomials in the series are:
$
P^{5_2,0}_{[1]}= P^{5_2}_{[1]}=\dfrac{\{A\}t}{ \{t\} A^2 }\Big(
-{q}^{3}{A}^{4}+ \left( {q}^{3}{t}^{2}-{q}^{2}t+{q} \right) {A}^{2
}+{q}^{2}{t}^{3}-{q}{t}^{2}+t
 \Big) =\\ =\bf{{\dfrac {1+{a}^{2}t}{ \left( -1+{q}^{2} \right)
{a}^{3}{t}^{9/2}{q}^{3}
}}}
 \Big({t}^{8}{q}^{6}{a}^{4}+ ( {q}^{8}{t}^{7}+{t}^{5}{q}^{4}+{q}^{6}{t
}^{6} ) {a}^{2}+{t}^{5}{q}^{8}+{q}^{6}{t}^{4}+{t}^{3}{q}^{4}
 \Big)
$

\bigskip
\noindent
$P^{5_2,3}_{[1]}=P^{10_{139}}=\dfrac{\{A\}t^2}{ \{t\} A^2q^6 }\Big(
(t^2q^8+q^6-q^7t)A^4+(-q^2-2q^6t^2-q^4-q^6t^4+q^5t-t^4q^8-q^4t^2-t^6q^8+q^7t^3)A^2+q^4t^2-q^5t^3+t^4q^4+q^6t^4+t^8q^8+q^2t^2+q^6t^6+1
\Big)=\\ =\bf{{\dfrac {1+{a}^{2}t}{ \left( -1+{q}^{2} \right)
{a}^{3}{t}^{15/2}{q}^{3}
}}} \Big(
(t^{10}q^{10}+q^6t^8+t^9q^8)a^4+(t^3q^2+2q^8t^7+t^5q^4+q^{10}t^7+q^6t^6+t^9q^{12}+q^6t^5+t^9q^{14}+q^{10}t^8)a^2+q^6t^4+t^5q^8+t^4q^8+t^6q^{10}+q^{16}t^8+q^4t^2+q^{12}t^6+1
\Big)$

\bigskip
\noindent
$P^{5_2,6}_{[1]}=\dfrac{\{A\}t^2}{ \{t\} A^2 q^{12}} \Big((
-{t}^{4}{q}^{11}+{q}^{14}{t}^{9}+{t}^{5}{q}^{12}+{t}^{5}{q}^{10}+{q}^{6}t+{t}^{3}{q}^{8}+{t}^{7}{q}^{12}+{t}^{3}{q}^{10}
) {A
}^{4}+ ( -2\,{q}^{6}{t}^{3}-{t}^{3}{q}^{4}-{q}^{4}t-{t}^{3}{q}^{8
}-2\,{t}^{5}{q}^{10}-{q}^{14}{t}^{11}+{t}^{6}{q}^{11}-2\,{t}^{9}{q}^{
12}-{t}^{9}{q}^{10}-2\,{t}^{5}{q}^{8}-2\,{q}^{10}{t}^{7}-{t}^{13}{q}^{
14}-{t}^{7}{q}^{12}-{q}^{6}{t}^{5}-{q}^{2}t+{t}^{4}{q}^{9}-{q}^{12}{t}
^{11}-{q}^{8}{t}^{7} ) {A}^{2}+t+{q}^{6}{t}^{5}+{q}^{6}{t}^{7}+{
t}^{5}{q}^{8}-{t}^{6}{q}^{9}+{q}^{8}{t}^{7}+{q}^{10}{t}^{7}+{t}^{9}{q}
^{10}+{t}^{9}{q}^{8}+{q}^{10}{t}^{11}+{t}^{13}{q}^{12}+{t}^{3}{q}^{2}+
{t}^{3}{q}^{4}+{t}^{5}{q}^{4}+{q}^{14}{t}^{15}+{q}^{12}{t}^{11}
 \Big) = \\ = \bf{{\dfrac {1+{a}^{2}t}{ \left( -1+{q}^{2} \right)
{a}^{3}{t}^{27/2}{q}^{3}
}}} \Big(( {q}^{17}{t}^{14}+{q}^{7}{t}^{8}+{t}^{16}{q}^{23}+{q}^{15}{t}^{
12}+{q}^{19}{t}^{14}+{t}^{13}{q}^{15}+{q}^{11}{t}^{10}+{q}^{13}{t}^{12
} ) {a}^{4}+ ( {q}^{23}{t}^{13}+{q}^{11}{t}^{9}+{t}^{15}{q}
^{27}+2\,{q}^{21}{t}^{13}+{q}^{11}{t}^{7}+{q}^{15}{t}^{9}+{t}^{12}{q}^
{17}+2\,{t}^{11}{q}^{17}+2\,{q}^{13}{t}^{9}+{t}^{15}{q}^{25}+2\,{t}^{
11}{q}^{15}+{t}^{11}{q}^{19}+{q}^{5}{t}^{5}+2\,{q}^{9}{t}^{7}+{q}^{19}
{t}^{13}+{t}^{10}{q}^{13}+{q}^{3}{t}^{3}+{t}^{5}{q}^{7}) {a}^{2
}+q+{q}^{17}{t}^{8}+{t}^{6}{q}^{11}+{t}^{6}{q}^{13}+{q}^{15}{t}^{9}+{t
}^{10}{q}^{17}+{q}^{25}{t}^{12}+{t}^{12}{q}^{23}+{t}^{14}{q}^{29}+{q}^
{5}{t}^{2}+{t}^{4}{q}^{9}+{q}^{13}{t}^{8}+{q}^{7}{t}^{4}+{t}^{10}{q}^{
21}+{q}^{19}{t}^{10}+{q}^{15}{t}^{8}
\Big) $

\bigskip
\noindent
$\boxed{P^{5_2,n}_{[1]}=\dfrac{\{A\}t^2}{\{t\} A^2
q^{2n}}{\cal{P}}^{5_2,n}_{[1]}}$

\begin{itemize}
\item{HOMFLY case}
\end{itemize}
$H_{[1]}^{5_2,0}=-{q}^{6}{A}^{4}+ \left( {q}^{8}-{q}^{6}+{q}^{4} \right)
{A}^{2}+{q}^{8
}-{q}^{6}+{q}^{4}$

\bigskip
\noindent
$H_{[1]}^{5_2,3}=(q^6-q^8+q^{10})A^4+(-q^{14}-q^{12}-2q^8-q^4-q^2)A^2+q^{16}+q^{12}+q^{10}+q^6+q^4+1$

\bigskip
\noindent
$H_{[1]}^{5_2,6}= ( {q}^{11}+{q}^{17}+{q}^{23}+{q}^{19}+{q}^{13}+{q}^{7}) {
A}^{4}+ ( -2\,{q}^{11}-3\,{q}^{15}-{q}^{17}-2\,{q}^{19}-{q}^{13}-
{q}^{27}-2\,{q}^{9}-{q}^{5}-{q}^{7}-{q}^{25}-{q}^{3}-2\,{q}^{21}-{q}^{
23} ) {A}^{2}+q+2\,{q}^{17}+{q}^{21}+{q}^{23}+{q}^{5}+{q}^{7}+{q
}^{9}+{q}^{11}+2\,{q}^{13}+{q}^{29}+{q}^{19}+{q}^{25}$

\begin{itemize}
\item{Floer case}
\end{itemize}
$
F^{5_2,0}_{[1]}=\bf{2}q^4t^2+3q^2t+2
$

\bigskip
\noindent
$
F^{5_2,3}_{[1]}=\bf{t}^8q^{16}+t^7q^{14}+2t^6q^{10}+3t^5q^8+2t^4q^6+tq^2+1
$

\bigskip
\noindent
$
F^{5_2,6}_{[1]}=\bf{t}^{14}q^{28}+t^{13}q^{26}+t^{12}q^{22}+t^{11}q^{20}+2t^{10}q^{16}+3t^9q^{14}+2t^8q^{12}+t^5q^8+t^4q^6+tq^2+1
$
\begin{itemize}
\item{Alexander case}
\end{itemize}
\be
\begin{array}{l}
A_{[1]}^{5_2,0}=2\,\bf{{q}}^{4}-3\,{q}^{2}+2
\\
\\
A_{[1]}^{5_2,3}=\bf{q}^{16}-q^{14}+2q^{10}-3q^8+2q^6-q^2+1
\\
\\
A_{[1]}^{5_2,6}=\bf{q}^{28}-q^{26}+q^{22}-q^{20}+2q^{16}-3q^{14}+2q^{12}-q^8+q^6-q^2+1
\end{array}
\ee
\end{small}

\subsection{Figure eight knot and descendants: fundamental
representation\label{41}}
\be
\boxed{
P^{4_1,n}_{[1]}=c^{[3]}_{[1]} M_{[3]}^{\ast} q^{-2n}+c^{[2,1]}_{[1]}
M_{[2,1]}^{\ast} q^{\frac{-2n}{3}}t^{\frac{2n}{3}}+ c^{[1,1,1]}_{[1]}
M_{[1,1,1]}^{\ast} t^{2n}
}
\ee
with the coefficients:
\fr{
c^{[3]}_{[1]}=tq\ \ \
c^{[2,1]}_{[1]}=\frac{(t-1)(1+t)(t^6q^6-t^5q^5-t^4q^2-t^3q^3-q^4t^2-tq+1)}{(-1+q^2t)(1+q^2t)t^2}
\ \\ c^{[1,1,1]}_{[1]} =
\frac{q^3(1+t^2)(t^2+t+1)(t^2-t+1)(t-1)^2(1+t)^2}{t(-1+t^2q)(1+t^2q)(-1+tq)(1+tq)}
}
\be
\begin{array}{l}
P^{4_1,0}_{[1]}=\dfrac{\{A\}}{\{t\}A^2}\Big(q^2A^4-q^2t^2A^2+tqA^2-A^2+
t^2\Big) =\\
=-\dfrac{(\bf{a}^2t+1)}{(-1+\bf{q}^2)a^3t^{\frac{3}{2}}}\Big(\bf{q}^2t^4a^4+ta^2+q^2t^2a^2+q^4t^3a^2+q^2\Big)
   \\ \\
P^{4_1,3}_{[1]}=\dfrac{\{A\}t^2}{\{t\}A^2q^5}\Big(q^5A^4-t^2q^5A^2-t^3q^4A^2-tq^2A^2-q^3A^2+t^2q^3+t^3q^2+t^5q^4+t\Big)\\=\dfrac{(\bf{a}^2t+1)}{(-1+\bf{q}^2)q^5a^3t^{\frac{9}{2}}}\Big(\bf{t}^7q^4a^4+q^6t^6a^2-q^6t^5a^2+q^2t^4a^2-q^2t^3a^2-t^4q^8+t^3q^4-q^4t^2-1\Big)
 \\ \\
... \\
\boxed{P^{4_1,n}_{[1]}=\dfrac{\{A\} t^{\frac{2}{3}n}}{\{t\} A^2
q^{\frac{5}{3}n}}{\cal{P}}^{4_1,n}_{[1]}}

\end{array}
\ee
\begin{itemize}
\item{HOMFLY case}
\end{itemize}
\be
\begin{array}{l}
H^{4_1,0}_{[1]}=\dfrac{\{A\}}{\{q\}A^2}\Big(q^2A^4-q^4A^2+q^2A^2-A^2+ q^2\Big) 
\\ \\
H^{4_1,3}_{[1]}=\dfrac{\{A\}}{\{q\}A^2q^2}\Big(q^4A^4-q^6A^2-q^6A^2-q^2A^2-q^2A^2+q^8+2q^4+1\Big)
 \\
... \\
\boxed{H^{4_1,n}_{[1]}=\dfrac{\{A\}}{\{q\} A^2 q^{\frac{2}{3}n}}
{\cal{H}}^{4_1,n}_{[1]}}
\end{array}
\ee
\begin{itemize}
\item{Floer case}
\end{itemize}
\be
\begin{array}{l}
F^{4_1,0}_{[1]}=\bf{t}^2q^4+3tq^2+1  \\
F^{4_1,3}_{[1]}= \bf{t}^4q^8+t^3q^6-t^4q^6-2t^3q^4+t^2q^4-t^2q^2+tq^2+1\\
....\\
\end{array}
\ee

\begin{itemize}
\item{Alexander case}
\end{itemize}
\be
\begin{array}{l}
A^{4_1,0}_{[1]}= \bf{q}^4-3q^2+1   \\
A^{4_1,3}_{[1]}= \bf{q}^8-2q^6+3q^4-2q^2+1 \\
....\\
\end{array}
\ee

\subsubsection{Symmetric and antisymmetric representations for figure eight
knot
\cite{IMMMfe}}

\be
\boxed{
\frac{^*{\cal P}^{4_1}_{[1^p\,]}(A|\,q,t)}{ ^*\!M_{[1^p\,]}(A|\,q,t)}\ \ =\
\sum_{k=0}^p\ \ \sum_{1\leq i_1\leq\ldots \leq i_k\leq p}\!\!\!\!\!
\bar{\mathfrak Z}_{i_1}(A)\bar{\mathfrak Z}_{i_2}(At^{-1})\bar{\mathfrak
Z}_{i_3}(At^{-2})\ldots
\bar{\mathfrak Z}_{i_k}(At^{-k+1})
}
\label{mainPa}
\ee
with
\be
\!\!\!\!\!\!\!\!
\bar{\mathfrak Z}_i(At^{-s}) =
\frac{\left(1-A^2t^{-4(p-i)-2-2s}\right)\left(q^{-2}-A^2t^{-2s}\right)}
{(A^2/tq)\cdot t^{-2(p-i+s)}} = \frac{\left(1+{\bf a}^2{\bf t}{\bf
q}^{-4(p-i)-2-2s}\right)
\left(1+{\bf a}^2{\bf t}^3{\bf q}^{2(1-s)}\right)}
{ {\bf a}^2\cdot {\bf t}^2{\bf q}^{-2(p-i+s)}}
\ee

and

\be
\boxed{
\frac{^*{\cal P}^{4_1}_{\rp }(A|\,q,t)}{ ^*\!M_{\rp }(A|\,q,t)}\ \ =\
 \sum_{k=0}^p\ \ \sum_{1\leq i_1\leq\ldots \leq i_k\leq p}\!\!\!\!\!
{\mathfrak Z}_{i_1}(A){\mathfrak Z}_{i_2}(Aq){\mathfrak Z}_{i_3}(Aq^2)\ldots
{\mathfrak Z}_{i_k}(Aq^{k-1})
}
\label{mainP}
\ee
with
\be
{\mathfrak Z}_i(A)=\{Aq^{2(p-i)+1}\}\{At^{-1}\}
\ee

\be
\begin{array}{l}
\dfrac{{\cal P}^{4_1}_{[1]}}
{ \!M^{*}_{[1]}}\
= \ 1+\{Aq\}\{At^{-1}\} =
1+{\bf t^2a^2}+{\bf q^{-2}t^{-1}}+{\bf
q^2t}+{\bf t^{-2}a^{-2}} \\ \\
\dfrac{{\cal P}^{4_1}_{[2]}}
{ \!M^{*}_{[2]}}\ =
1+\{Aq\}\{At^{-1}\}+\{Aq^3\}\{At^{-1}\}+\{Aq^3\}\{At^{-1}\}\{Aq^2\}\{Aqt^{-1}\}=
\\
={\bf a}^4{\bf q}^4{\bf t}^8+{\bf a}^2({\bf q}^{-2}{\bf t}+{\bf t}^2+{\bf
t}^3+{\bf q}^2{\bf t}^4+{\bf q}^4{\bf t}^5+
{\bf q}^6{\bf t}^7)+{\bf q}^6{\bf t}^4+{\bf q}^4{\bf t}^3
+{\bf q}^2{\bf t}^2+{\bf q}^2{\bf t}+3+{\bf q}^{-2}{\bf t}^{-1}+{\bf
q}^{-2}{\bf t}^{-2}+\\+
{\bf q}^{-4}{\bf t}^{-3}+{\bf q}^{-6}{\bf t}^{-4}+
{\bf a}^{-2}({\bf q}^2{\bf t}^{-1}+{\bf t}^{-2}+{\bf t}^{-3}+{\bf q}^{-2}{\bf
t}^{-4}+
{\bf q}^{-4}{\bf t}^{-5}+{\bf q}^{-6}{\bf t}^{-7})+{\bf a}^{-4}{\bf q}^{-4}{\bf
t}^{-8}
 \\ \\
\dfrac{{\cal P}^{4_1}_{[11]}}
{ \!M^{*}_{[11]}}\ = 
1+\{Aq\}\{At^{-1}\}+\{Aq\}\{At^{-3}\}+\{Aq\}\{At^{-3}\}\{At^{-2}\}\{Aqt^{-1}\}=
 \\
={\bf a}^4{\bf q}^{-4}{\bf t}^4+{\bf a}^2({\bf q}^{2}{\bf t}^3+{\bf t}^2+{\bf
t}^3+{\bf q}^{-2}{\bf t}^2+
{\bf q}^{-4}{\bf t}+
{\bf q}^{-6}{\bf t})+{\bf q}^{-6}{\bf t}^{-2}+{\bf q}^{-4}{\bf t}^{-1}
+{\bf q}^{-2}+{\bf q}^{-2}{\bf t}^{-1}+3+{\bf q}^{2}{\bf t}+\\+{\bf q}^{2}+
{\bf q}^{4}{\bf t}+{\bf q}^{6}{\bf t}^{2}+
{\bf a}^{-2}({\bf q}^{-2}{\bf t}^{-3}+{\bf t}^{-2}+{\bf t}^{-3}+{\bf q}^{2}{\bf
t}^{-2}+
{\bf q}^{4}{\bf t}^{-1}+{\bf q}^{6}{\bf t}^{-1})+{\bf a}^{-4}{\bf q}^{4}{\bf
t}^{-4}\\ \\
\dfrac{{\cal P}^{4_1}_{[3]}}
{ \!M^{*}_{[3]}}\ = {\bf a}^6{\bf q}^{12}{\bf t}^{18}+{\bf a}^4({\bf t}^7{\bf
q}^2+{\bf t}^8{\bf q}^4+
{\bf t}^9{\bf q}^4+{\bf t}^{10}{\bf q}^6+{\bf t}^{11}{\bf q}^6+{\bf t}^{12}{\bf
q}^8+
{\bf t}^{13}{\bf q}^{10}+{\bf t}^{15}{\bf q}^{12}+{\bf t}^{17}{\bf q}^{14})+\\
+{\bf a}^2({\bf t}^{-2}{\bf q}^{-6}+{\bf t}^{-1}{\bf q}^{-4}+{\bf q}^{-4}+2{\bf
t}{\bf q}^{-2}+
{\bf t}^2{\bf q}^{-2}+2{\bf t}^2+2{\bf t}^3+3{\bf t}^4{\bf q}^2+{\bf t}^5{\bf
q}^2+
{\bf t}^5{\bf q}^4+4{\bf t}^6{\bf q}^4+
2{\bf t}^7{\bf q}^6+2{\bf t}^8{\bf q}^6+
2{\bf t}^9{\bf q}^8+\\+
{\bf t}^{10}{\bf q}^8+
{\bf t}^{10}{\bf q}^{10}+{\bf t}^{11}{\bf q}^{10}+{\bf t}^{12}{\bf q}^{12}+{\bf
t}^{14}{\bf q}^{14})+
{\bf t}^{-9}{\bf q}^{-12}+{\bf t}^{-8}{\bf q}^{-10}+{\bf t}^{-7}{\bf
q}^{-8}+{\bf t}^{-6}{\bf q}^{-8}+
3{\bf t}^{-5}{\bf q}^{-6}+{\bf t}^{-4}{\bf q}^{-6}+{\bf t}^{-4}{\bf q}^{-4}+\\
+4{\bf t}^{-3}{\bf q}^{-4}+
3{\bf t}^{-2}{\bf q}^{-2}+3{\bf t}^{-1}{\bf q}^{-2}+{\bf t}^{-1}+5+{\bf
t}+3{\bf t}{\bf q}^2+
3{\bf t}^2{\bf q}^2+4{\bf t}^3{\bf q}^4+{\bf t}^4{\bf q}^4+{\bf t}^4{\bf
q}^6+3{\bf t}^5{\bf q}^6+
{\bf t}^6{\bf q}^8+
{\bf t}^7{\bf q}^8+{\bf t}^8{\bf q}^{10}+\\
+{\bf t}^9{\bf q}^{12}
+{\bf a}^{-2}({\bf t}^{-14}{\bf q}^{-14}+{\bf t}^{-12}{\bf q}^{-12}+{\bf
t}^{-11}{\bf q}^{-10}+
{\bf t}^{-10}{\bf q}^{-10}+{\bf t}^{-10}{\bf q}^{-8}+2{\bf t}^{-9}{\bf
q}^{-8}+2{\bf t}^{-8}{\bf q}^{-6}+
2{\bf t}^{-7}{\bf q}^{-6}+
4{\bf t}^{-6}{\bf q}^{-4}+\\
+{\bf t}^{-5}{\bf q}^{-4}+{\bf t}^{-5}{\bf q}^{-2}+3{\bf t}^{-4}{\bf q}^{-2}+
2{\bf t}^{-3}+2{\bf t}^{-2}+{\bf t}^{-2}{\bf q}^2+2{\bf t}^{-1}{\bf q}^2+{\bf
q}^4+{\bf t}{\bf q}^4+
{\bf t}^2{\bf q}^6)+\\+
{\bf a}^{-4}({\bf t}^{-17}{\bf q}^{-14}+{\bf t}^{-15}{\bf q}^{-12}+{\bf
t}^{-13}{\bf q}^{-10}+
{\bf t}^{-12}{\bf q}^{-8}+{\bf t}^{-11}{\bf q}^{-6}+{\bf t}^{-10}{\bf
q}^{-6}+{\bf t}^{-9}{\bf q}^{-4}+
{\bf t}^{-8}{\bf q}^{-4})+{\bf t}^{-7}{\bf q}^{-2})+{\bf a}^{-6}{\bf
t}^{-18}{\bf q}^{-12}
\end{array}
\ee

\begin{itemize}
\item{HOMFLY case}
\end{itemize}
\be
\begin{array}{l}
\dfrac{{\cal H}^{4_1}_{[0]}}{\SS^*_{[0]}(A|\, q)} = 1, \\
\dfrac{{\cal H}^{4_1}_{[1]}}{\SS^*_{[1]}(A|\, q)} =
A^2-q^2+1-q^{-2}+A^{-2} = 1 +
\{Aq\}\{Aq^{-1}\}, \\
\dfrac{{\cal H}^{4_1}_{[2]}}{\SS^*_{[2]}(A|\, q)} =
q^4A^4-(q^6+q^4-q^2+q^{-2})\,A^2+(q^6-q^4+3-q^{-4}+q^{-6})\,-\\
-\,(q^2-q^{-2}+q^{-4}+q^{-6})\,A^{-2}+q^{-4}A^{-4} = 1 +
[2]\{Aq^2\}\{Aq^{-1}\}+\{Aq^3\}\{Aq^2\}\{A\}\{Aq^{-1}\}, \\
\dfrac{{\cal H}^{4_1}_{[3]}}{\SS^*_{[3]}(A|\, q)} = 1 +
[3]\{Aq^3\}\{Aq^{-1}\}+[3]\{Aq^4\}\{Aq^3\}\{A\}\{Aq^{-1}\}
+\{Aq^5\}\{Aq^4\}\{Aq^3\}\{Aq\}\{A\}\{Aq^{-1}\}, \\
\dfrac{{\cal H}^{4_1}_{[11]}(A|\,q)}{\SS_{[11]}(A|\, q)} =  1 +
[2]\{Aq^{-2}\}\{Aq\}+\{Aq^{-3}\}\{Aq^{-2}\}\{A\}\{Aq\}, \\
\dfrac{{\cal H}^{4_1}_{[111]}(A|\,q)}{\SS_{[111]}(A|\, q)} = 1 +
[3]\{Aq^{-3}\}\{Aq^{1}\}+[3]\{Aq^{-4}\}\{Aq^{-3}\}\{A\}\{Aq\}
+\{Aq^{-5}\}\{Aq^{-4}\}\{Aq^{-3}\}\{Aq^{-1}\}\{A\}\{Aq\}, \\
\ldots
\end{array}
\ee


\begin{thebibliography}{12}

\bibitem{knots} E.Witten, Comm.Math.Phys. {\bf 121} (1989) 351-399

\bibitem{knotstoday} E.Witten,  arXiv:1001.2933; arXiv:1101.3216;\\
T.Dimofte, S.Gukov and L.Hollands, arXiv:1006.0977

\bibitem{topver} M.Aganagic, M.Mari\~no and C.Vafa, Commun. Math. Phys. {\bf
247} (2004) 467,
arXiv:hep-th/0206164;\\
M. Aganagic, A. Klemm, M. Mari.no and C. Vafa, Commun. Math. Phys.
{\bf 254} (2005) 425, arXiv:hep-th/0305132;\\
A.Iqbal, C.Kozcaz and C.Vafa, JHEP {\bf 0910} (2009) 069,
hep-th/0701156;\\
H.Awata and H.Kanno, Int.J.Mod.Phys. {\bf A24} (2009) 2253-2306,
arXiv:0805.0191

\bibitem{McD} I.G.Macdonald, {\sl Symmetric functions and Hall polynomials},
Second Edition, Oxford
University Press, 1995

\bibitem{Ruij} S.N.M.Ruijsenaars and H.Schneider,
Ann.Phys. (NY), {\bf 170} (1986) 370;\\
S.N.M.Ruijsenaars, Comm.Math.Phys., {\bf 110} (1987) 191-213;
Comm.Math.Phys., {\bf 115} (1988) 127-165

\bibitem{KhR} M.Khovanov and L.Rozhansky, Fund.
Math. {\bf 199} (2008) 1, math.QA/0401268; Geom. Topol. {\bf 12}
(2008) 1387, math.QA/0505056

\bibitem{GSV} S.Gukov, A.Schwarz and C.Vafa, Lett.Math.Phys. {\bf 74} (2005)
53-74, arXiv:hep-th/0412243

\bibitem{DGR}
N.M.Dunfield, S.Gukov, and J.Rasmussen,
math/0505662

\bibitem{IGV} S.Gukov, A.Iqbal, C.Kozcaz and C.Vafa, arXiv:0705.1368

\bibitem{ACJK}
J.W.Alexander, 
Trans.Amer.Math.Soc. {\bf 30} (2) (1928) 275?306;\\
J.H.Conway, 
Algebraic Properties, In: John Leech (ed.), {\sl Computational
Problems in Abstract Algebra}, Proc. Conf.
Oxford, 1967, Pergamon Press, Oxford-New York, 329-358, 1970;\\
V.F.R.Jones, 
Invent.Math. {\bf 72} (1983) 1
Bull.AMS {\bf 12} (1985) 103
Ann.Math. {\bf 126} (1987) 335;\\
L.Kauffman,
Topology {\bf 26} (1987) 395

\bibitem{HOMFLY} P.Freyd, D.Yetter, J.Hoste, W.B.R.Lickorish, K.Millet,
A.Ocneanu,
Bull. AMS. {\bf 12} (1985) 239\\
J.H.Przytycki and K.P.Traczyk, 
Kobe J. Math. {\bf 4} (1987) 115-139

\bibitem{knotinv} See also S. Chmutov, S. Duzhin, J. Mostovoy,
arXiv:1103.5628;\\
math/0406190 and references therein

\bibitem{AGT} L.Alday, D.Gaiotto and Y.Tachikawa,
Lett.Math.Phys. {\bf 91} (2010) 167-197, arXiv:0906.3219;\\
N.Wyllard,
JHEP {\bf 0911} (2009) 002, arXiv:0907.2189;\\
A.Mironov and A.Morozov, Phys.Lett. {\bf B680} (2009) 188-194,
arXiv:0908.2190; Nucl.Phys. {\bf B825} (2009) 1-37, arXiv:0908.2569

\bibitem{TY} Yu.Terashima and M.Yamazaki, arXiv:1103.5748

\bibitem{3dAGT} D.Galakhov, A.Mironov, A.Morozov and A.Smirnov,
arXiv:1104.2589

\bibitem{AGTmamo} R.Dijkgraaf and C.Vafa, arXiv:0909.2453;\\
H.Itoyama, K.Maruyoshi and T.Oota,
Prog.Theor.Phys. {\bf 123} (2010) 957-987, arXiv:0911.4244;\\
T.Eguchi and K.Maruyoshi,
arXiv:0911.4797;
arXiv:1006.0828;\\
R.Schiappa and N.Wyllard,
arXiv:0911.5337;\\
A.Mironov, A.Morozov, Sh.Shakirov,
JHEP {\bf 02} (2010) 030, arXiv:0911.5721;
Int.J.Mod.Phys. {\bf A25} (2010) 3173-3207, arXiv:1001.0563

\bibitem{ASh} M.Aganagic and Sh.Shakirov,
arXiv: 1105.5117

\bibitem{chi}
M. Rosso and V. F. R. Jones, J. Knot Theory Ramifications, {\bf 2} (1993)
97-112;\\
X.-S.Lin and H.Zheng,
Trans. Amer. Math. Soc. {\bf 362} (2010) 1-18
math/0601267;\\
S.Stevan, Annales Henri Poincare, {\bf 11} (2010) 1201-1224, arXiv:
1003.2861;\\
A.Brini, B.Eynard and M.Mari\~no, arXiv:1105.2012;\\
see also \cite{Wreps}

\bibitem{MSm} A.Morozov and A.Smirnov,
the
Nucl.Phys. {\bf B835} (2010) 284-313, arXiv:1001.2003\\
A.Smirnov,
hep-th/0910.5011, to appear in the Proceedings of International
School of Subnuclar Phys. in Erice, Italy, 2009

\bibitem{EGth} E.Gorsky,  Journal of Singularities, {\bf 3} (2011), 48-82,
arXiv:0807.0491

\bibitem{MMN} A.Mironov, A.Morozov and S.Natanzon,  Theor.Math.Phys. {\bf 166}
(2011) 1-22,
arXiv:0904.4227; Journal of Geometry and Physics, {\bf 62} (2012) 148-155,
arXiv:1012.0433

\bibitem{Wreps} A.Morozov and Sh.Shakirov,
JHEP {\bf 0904} (2009) 064, arXiv:0902.2627; Mod.Phys.Lett. {\bf
A24} (2009) 2659-2666,
arXiv:0906.2573;\\
G.Borot, B.Eynard, M.Mulase and B.Safnuk, arXiv:0906.1206p;\\
A.Alexandrov,
arXiv:1005.5715, arXiv:1009.4887

\bibitem{AK} H.Awata and H.Kanno,  arXiv:0903.5383; arXiv:0910.0083

\bibitem{EG} E.Gorsky,
{\it $q,t$-Catalan numbers and knot homology}, arXiv: 1003.0916

\bibitem{GSu} S.Gukov and P.Su\l kowski, arXiv:1108.0002

\bibitem{Cor} N.Carqueville and D.Murfet, arXiv:1108.1081

\bibitem{Ch} I.Cherednik, arXiv:1111.6195

\bibitem{Sh} Sh.Shakirov, arXiv:1111.7035\footnote{
A remarkably simple suggestion for the $\gamma$-factors in this paper
does not seem literally applicable to the torus knots,
it does not even reproduce the HOMFLY polynomials.
It is so attractive, however, and it is non-trivial that
such a simple ansatz provides polynomials with positive integer
coefficients. Therefore, there should be a prominent place for it
in the future theory of superpolynomials.
}

\bibitem{ORSG} A.Oblomkov, J.Rasmussen and V.Shende with an Appendix by
E.Gorsky
{\it The Hilbert scheme of a plane curve singularity and the HOMFLY
homology of its link}, arXiv: 1201.2115

\bibitem{MMSS} A.Mironov, A.Morozov, Sh.Shakirov and A.Sleptsov,
arXiv:1201.3339

\bibitem{MMS} A.Mironov, A.Morozov and Sh.Shakirov, arXiv:1203.0667

\bibitem{FGS} H.Fuji, S.Gukov and P.Su\l kowski (with an appendix by Hidetoshi
Awata), arXiv:1203.2182

\bibitem{IMMMfe} H.Itoyama, A.Mironov, A.Morozov and An.Morozov,
arXiv:1203.5978

\bibitem{FGS2} H.Fuji, S.Gukov and P.Su\l kowski, arXiv:1205.1515

\bibitem{RT} E.Guadagnini, M.Martellini and M.Mintchev, In Clausthal 1989,
Proceedings, Quantum groups, 307-317;
Phys.Lett. B235 (1990) 275;\\
N.Yu.Reshetikhin and V.G.Turaev, 
Comm. Math. Phys. {\bf 127} (1990) 1-26

\bibitem{KnotInt} A.Mironov, A.Morozov and And.Morozov, arXiv:1112.5754;\\
A.Mironov, A.Morozov, A.Sleptsov and A.Smirnov, to appear

\end{thebibliography}
\end{document}